\documentclass[12pt]{iopart} 

\usepackage{iopams}  

\expandafter\let\csname equation*\endcsname\relax
\expandafter\let\csname endequation*\endcsname\relax

\usepackage{amsmath}

\usepackage{mathrsfs}
\usepackage{bm}
\usepackage{graphicx}
\usepackage{dcolumn}
\usepackage{multirow}
\usepackage{braket}
\usepackage{natbib}
\usepackage{hyperref}
\usepackage{xcolor}
\usepackage{etoolbox}

\graphicspath{{.}{./figs/}}

\makeatletter
\def\cat@comma@active{\catcode`\,12}%
\newrobustcmd{\fixappendix}{%
  \patchcmd{\l@section}{1.5em}{7em}{}{}%
  \patchcmd{\l@subsection}{2.3em}{7em}{}{}%
}
\makeatother

\begin{document}
\title[Effective field theories for collective excitations]{Effective field theories for collective excitations of atomic nuclei}

\author{E.~A.~Coello~P{\'e}rez$^{1}$ and T.~Papenbrock$^{2,3}$}
\address{$^{1}$National Center for Computational Sciences, Oak Ridge National Laboratory, Oak Ridge, Tennessee 37831, USA}
\address{$^{2}$Department of Physics and Astronomy, University of Tennessee, Knoxville, Tennessee 37996, USA}
\address{$^{3}$Physics Division, Oak Ridge National Laboratory, Oak Ridge, Tennessee 37831, USA}

\ead{coellopereea@ornl.gov, tpapenbr@utk.edu}

\begin{abstract}
Collective modes emerge as the relevant degrees of freedom that govern low-energy excitations of atomic nuclei. These modes -- rotations, pairing rotations, and vibrations --  are separated in energy from non-collective excitations, making it possible to describe them in the framework of effective field theory. Rotations and pairing rotations are the remnants of Nambu-Goldstone modes from the emergent breaking of rotational symmetry and phase symmetries in finite deformed and finite superfluid nuclei, respectively. The symmetry breaking severely constrains the structure of low-energy Lagrangians and thereby clarifies what is essential and simplifies the description. The approach via effective field theories exposes the essence of nuclear collective excitations and is defined with a breakdown scale in mind. This permits one to make systematic improvements and to estimate and quantify uncertainties. Effective field theories of collective excitations have been used to compute spectra, transition rates, and other matrix elements of interest. In particular, predictions of the nuclear matrix element for neutrinoless double beta decay then come with quantified uncertainties. This review summarizes these results and also compares the approach via effective field theories to well-known models and {\it ab initio} computations.    
 
\end{abstract}

\newpage
\setcounter{tocdepth}{2} 
\tableofcontents

\newpage

\section{Introduction}
\label{sec:intro}
\begin{figure}[htb]
\centering
\includegraphics[width=0.75\textwidth]{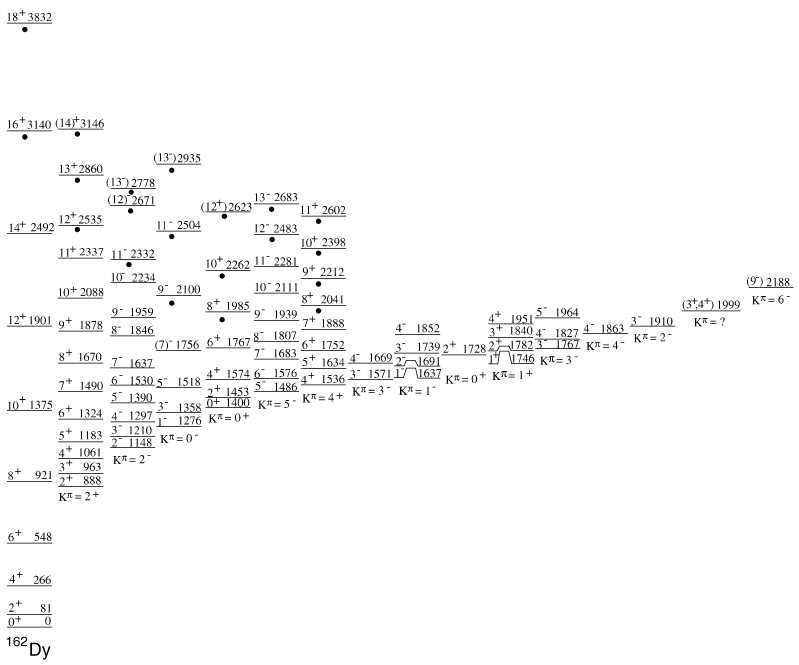}
\caption{Spectrum of $^{162}$Dy, with levels ordered into rotational bands. Energies are in keV and spin/parity $I^\pi$ as indicated. The intrinsic excitations are identified by their spin projection $K$ onto the nuclear symmetry axis and the parity $\pi$. Figure taken from~\citep{aprahamian2006} with permission from the authors. Reprinted from Nuclear Physics A, volume 764, Aprahamian {\it et al.}, Complete Spectroscopy of $^{162}$Dy, pages 42-78, Copyright (2006), with permission from Elsevier.}
\label{fig:dy162-exp}
\end{figure}

Deformation and superfluidity are key properties of nuclei. The corresponding low-energy excitations are collective modes, namely rotations and pairing rotations. Figure~\ref{fig:dy162-exp} shows the level scheme of the nucleus $^{162}$Dy as an example for rotations~\citep{aprahamian2006}. Levels are grouped into rotational bands with energies, spins, and parities as indicted. Also shown are the spin projections $K$ of the intrinsic excitations with respect to the nuclear symmetry axis and their parity $\pi$. One sees a large number of different intrinsic excitations, each of which is the head for a rotational band from the corresponding rotation of the whole nucleus. A separation of scales is clearly visible: The energy spacings between the lowest levels in a rotational band are much smaller than the energy of the band heads (i.e the intrinsic excitations) with respect to the ground state.    

A large number of models are available to describe the physics of such systems. Examples are collective models that are based on the surface rotations and vibrations of a liquid drop~\citep{bohr1952, bohr1953, bohr1975}, algebraic models of interacting $s$ and $d$ bosons~\citep{arima1975,iachello1987}, pairing models~\citep{kerman1961,richardson1963,dukelsky2004, brink-broglia2005}, and mean-field models based on pairing-plus-quadrupole interactions~\citep{kumar1968,frauendorf2001}. While these models explain some observations they leave out other low-energy phenomena. Many deformed rare earth nuclei and actinides, for instance, exhibit low-lying rotational bands with negative parity. In the celebrated models by \citet{bohr1975,iachello1987} only positive parity states appear at low energies and negative-parity states are expected to be much higher in energies. (In contrast, symmetry-breaking mean-field calculations with octupole deformation capture some negative-parity states~\citep{nazarewicz1984}.) 
This points to a general challenge: It is not clear where models break down, how to systematically improve them, or how to assign uncertainties to calculated results. 

Recently, collective phenomena also emerged from \textit{ab initio} no-core shell model computations~\citep{caprio2015, dytrych2013,dytrych2020} where nucleons interact via  realistic nucleon-nucleon potentials. While these computations of nuclei are at the highest resolution scale possible today, the interpretation of such {\it ab initio} results was based on collective models. In other words, it would be hard to see collective phenomena emerge from the no-core shell-model calculations without knowing what to look for in spectra and transition matrix elements. 

This can be contrasted to \textit{ab initio} computations that start from symmetry-breaking reference states~\citep{frosini2022, hagen2022,sun2024}. Then, the collective excitations arise naturally from symmetry projections~\citep{sheikh2021}. This relates nuclear deformation and superfluidity to the emergent breaking~\citep{yannouleas2007} of the rotational and phase symmetry, respectively. Fortunately, physicists know since long how to  construct effective Lagrangians for such systems:~\citet{weinberg1968} pioneered this approach for the breaking of chiral symmetry, and~\citet{coleman1969}, and~\citet{callan1969} generalized it to other cases, see reference~\citep{brauner2010} for a recent review. This ``coset approach'' via nonlinear realizations of the symmetry identifies the relevant degrees of freedom (that is, the Nambu-Goldstone modes) and severely constrains their interactions. It is the basis of chiral perturbation theory~\citep{gasser1984} and chiral effective field theory~\citep{weinberg1990, vankolck1994, epelbaum2009, machleidt2011, hammer2020}. It explains the low-lying excitations in magnets~\citep{leutwyler1994, roman1999, hofmann1999, baer2004, kampfer2005} and 
the universal fluctuation properties of complex quantum systems~\citep{altland2021}. 

The same model-independent approach allows one to describe rotations and pairing rotations in atomic nuclei, and to view venerable collective nuclear models as the leading-order Hamiltonians of corresponding effective theories~\citep{papenbrock2011, papenbrock2014, coelloperez2015, papenbrock2020, chen2017, chen2018, chen2020, alnamlah2021, alnamlah2022, papenbrock2022}. There are many commonalities between the effective theories for deformed nuclei and those for magnets~\citep{roman1999, hofmann1999, baer2004, kampfer2005}. In contrast to those infinite systems, however, atomic nuclei are finite. This introduces modifications to the standard field theoretical approach~\citep{papenbrock2014} and leads to quantum mechanics (rather than quantum field theory). The emergent breaking of spherical symmetry in deformed nuclei leads -- at leading order -- to the physics of the axially symmetric and triaxially deformed rotors. Odd-mass nuclei, described by coupling a nucleon to an even-even core, introduce Abelian and non-Abelian gauge potentials and this connects them to topological systems such as quantum hall fluids~\citep{estienne2011} and to the physics of geometric phases~\citep{berry1984,wilczek1989}. 

In superfluid nuclei, the emergent breaking of phase symmetries leads to the physics of coupled superfluids and describes pairing rotational bands. These govern differences in binding energies for neighboring nuclei that are quadratic in the differences of nucleon numbers~\citep{broglia1968,bohr1969,brink-broglia2005}.  

Vibrations in spherical nuclei can also be approached via effective theories. No symmetries are broken in this case, and one has to identify the relevant degrees of freedom from data.  Such an approach to nuclear vibrations~\citep{coelloperez2015b, coelloperez2016} is conceptually somewhat similar to pionless~\citep{bedaque2002} or halo~\citep{hammer2017} effective field theory. Effective theories allow one to quantify theoretical uncertainties~\citep{schindler2009, furnstahl2014c}. This made it possible to employ effective theories of nuclear vibrations to make quantified predictions of electroweak processes and neutrinoless double beta decay~\citep{coelloperez2018, brase2022}.

In this article, we review the developments and applications of effective theories for collective phenomena (rotations, pairing rotations, and vibrations) in atomic nuclei. We contrast them to collective models and highlight the commonalities and differences to effective field theories in other fields of physics. This  focused review is not meant to survey or summarize the vast literature about collective nuclear models. Instead, we made an attempt to cite at least some of the relevant original literature and otherwise refer the reader to reviews and textbooks. 

This review is organized as follows. Sections~\ref{sec:found} to \ref{sec:pair} are dedicated to collective phenomena associated with emergent symmetry breaking. We start with the microscopic foundations of the employed effective theories in section~\ref{sec:found} and discuss emergent symmetry breaking in section~\ref{sec:ssb}. The effective field theory of axially symmetric deformed nuclei is reviewed in sections~\ref{sec:even} to \ref{sec:electro}. We  start with even-even nuclei in section~\ref{sec:even}, discuss the coupling to intrinsic degrees of freedom in section~\ref{sec:coup} and present details for odd-mass nuclei in section~\ref{sec:odd}. Finally we review electromagnetic transitions in deformed even-even nuclei in section~\ref{sec:electro}. Section~\ref{sec:triax} reviews works on triaxially deformed nuclei and section~\ref{sec:pair} is dedicated to pairing rotational bands.    
In sections~\ref{sec:vib} and \ref{sec:weak} we review effective theories for nuclear vibrations and their use in computing matrix elements for weak decays and neutrinoless double beta decay. We finally compare and contrast  the effective theories with other models and {\it ab initio} computations in section~\ref{sec:compare}. The review ends with a summary and outlook in section~\ref{sec:summary}.

\section{Microscopic foundation}
\label{sec:found}
\subsection{Deformed nuclei}
\label{sub:deformed}

Let us start from a microscopic Hamiltonian $H_{\rm mic}$ whose degrees of freedom are the positions, spins, and isospin projections of nucleons. For simplicity, we think of an even-even nucleus and further assume that neither proton nor neutron numbers are magic numbers. Thus, we deal with an open-shell nucleus. It is profitable to break down the computation of its ground state $|\psi\rangle$ into several steps, each of which gives increasingly better approximations. We denote the states, energies, and angular-momentum expectation values at the $n^{\rm th}$ step as $\ket{\psi_{(n)}}$, $E_{(n)}$, and $J_{(n)}^2$.

The first step usually consists of a Hartree-Fock calculation~\footnote{We discuss the more general case of Hartree-Fock-Bogoliubov calculation in section~\ref{sub:superfluid}}. Let us assume that this calculation is based on single-particle states with good angular momentum projection $j_z$, and that we seek a state with total angular momentum projection $J_z=0$. In practice this is achieved by starting from a trial product states where pairs of time-reversed single-particle states are occupied. The Hartree-Fock computation then yields an axially symmetric state $\ket{\psi_{(1)}}$. The deformation results from a competition between short and long-range correlations~\citep{lipkin1960}. The state has zero angular momentum projection with respect to the symmetry axis (which we choose as the laboratory $z$ axis) but not good angular momentum. We have $J_{(1)}^2 \equiv \braket{\psi_{(1)}|J^2|\psi_{(1)}} > 0$. While the energy $E_{(1)} \equiv \braket{\psi_{(1)}|H_{\rm mic}|\psi_{(1)}}$ generally is a poor approximation of the true ground-state energy, the Hartree-Fock state is a great starting point for more refined calculations. 

The second step could, for instance, consist of a coupled-cluster computation where particle-hole excitations of the Hartree-Fock reference are included. Typically, such calculations are limited to up to two-particle--two-hole or three-particle-three-hole excitations~\citep{hagen2014}. If one were to include four-particle--four-hole excitations, the treatment of short-range physics would be complete: spin-isospin degrees of freedom permit up to four nucleons to be very close. This then yields the refined approximation $|\psi_{(2)}\rangle$ of the ground state. The energy $E_{(2)}$ of this state is a much lower than the Hartree-Fock energy $E_{(1)}$ and usually already close to the exact ground-state energy. However, the state  $|\psi_{(2)}\rangle$ still does not have good angular momentum. Typically one finds $J_{(2)}^2 < J_{(1)}^2$ but one still has $J_{(2)}^2 > 0$. This is not surprising: It would require $A$-particle--$A$-hole excitations to get good angular momentum for a nucleus with mass-number $A$. These missing correlations are long range. So one realistically deals with a situation where rotational symmetry is broken down to axial symmetry at any step $n$ that can be achieved in practical calculations. Results that illustrate these arguments are shown in table~\ref{tab:micro}.

\begin{table}
\caption{\label{tab:micro}Energies $E_{(n)}$ and angular-momentum expectation values $J_{(n)}^2$ from Hartree-Fock ($n=1$) and coupled-cluster theory including up to three-particle--three hole excitations ($n=2$) compared to the estimated energies $E_0$ from symmetry restoration, for the ground-states of nuclei as indicated, and compared to data $E_{\rm exp}$. Computed results taken from \citep{hagen2022}.} 
\begin{indented}
\lineup
\item[]\begin{tabular}{@{}*{7}{l}}
\br                              
          & \m$E_{(1)}$& \m$E_{(2)}$ & \m$E_0$ & \m$E_{\rm exp}$    &$J_{(1)}^2$&$J_{(2)}^2$\\
\mr
$^8$Be    & $-16.74$ & \0$ -50.24$ &\0$ -53.57$ &\0$ -56.50$  &  $11.17$  &    \0$5.82$ \\
$^{20}$Ne & $-59.62$ & $-161.95$ &$-164.21$ &$-160.64$  &  $21.26$  &   $12.09$  \\
$^{34}$Mg & $-90.21$ & $-264.34$ &$-265.84$ &$-256.71$  &  $22.62$  &   $15.03$   \\
\br
\end{tabular}
\end{indented}
\end{table}

The energy  after symmetry restoration is denoted by $E_0$. (That the energies deviate somewhat from data is not important here; it mainly points to an inaccuracy of the employed interaction.)
We see that the symmetry breaking is not very costly in energy as $E_{(2)}$ is close to $E_0$. 
Symmetry restoration is about capturing low-energy (or long wavelength) physics. The small energy gain comes from lowering the kinetic energy and not from improving the contributions from the short-ranged potential. 

One can construct the corresponding low-energy or collective Hamiltonian. To do this, we follow the well known approach described, for example, by~\citet{ringschuck}. 
We first identify the relevant Hilbert space.
The symmetry axis of the deformed nucleus is along the $z$ axis. Because of this symmetry, it is not the full group of rotations with elements $\exp{(-i\phi J_z)}\exp{(-i\theta J_y)}\exp{(-i\gamma J_z)}$ and Euler angles $(\phi,\theta,\gamma)$ that we need  to consider when restoring the symmetry, but rather those group elements that are in the coset SO(3)/SO(2); these are the rotations
\begin{equation}
\label{rotationop}
    R(\phi,\theta)\equiv e^{-i\phi J_z}e^{-i\theta J_y} \ ,
\end{equation}
where the angles ($\phi,\theta$) parameterize the sphere. When acting onto the state $|\psi_{(n)}\rangle$ these rotations yield states $\ket{\psi_{(n)},\Omega} \equiv R(\phi,\theta)\ket{\psi_{(n)}}$. Here we combined $\Omega \equiv (\phi,\theta)$ to keep a compact notation for these states. We have $|\psi_{(n)}\rangle= |\psi_{(n)},0\rangle$. The states $\ket{\psi_{(n)},\Omega}$ all have identical energy expectation values because the Hamiltonian $H_{\rm mic}$ is invariant under rotations, i.e. $R^{-1} H_{\rm mic}R =H_{\rm mic}$. 
Thus, one needs to diagonalize the microscopic Hamiltonian in this basis~\citep{peierls1957}. The basis set is not orthogonal, and one computes the Hamiltonian and norm kernels
\begin{equation}
\begin{aligned}
\label{kernels}
    {\cal H}(\Omega',\Omega)&\equiv \braket{\psi_{(n)},\Omega'|H_{\rm mic}|\psi_{(n)},\Omega} \ , \\
    {\cal N}(\Omega',\Omega)&\equiv \braket{\psi_{(n)},\Omega'|\psi_{(n)},\Omega} \ .
    \end{aligned}
\end{equation}
Diagonalization of the norm kernel yields an orthonormal basis. One can re-express the Hamiltonian kernel in this basis set. In practice, one computes the Hamiltonian
\begin{equation}
\label{H_omega}
    H(\Omega',\Omega)\equiv {\cal N}^{-\frac{1}{2}}{\cal H}{\cal N}^{-\frac{1}{2}} \ ,
\end{equation}
The key point is that a diagonalization of the matrix $H(\Omega',\Omega)$ yields a rotational band, i.e. the resulting energies are approximately
\begin{equation}
\label{specLO}
    E_{J} = E_{0} + a J(J+1) \ ,
\end{equation}
and each energy has degeneracy $2J+1$. Here, $a$ is the rotational constant (and proportional to the inverse moment of inertia). 

Thus, a very simple physics picture results from a possibly quite complicated microscopic Hamiltonian. The microscopic details are all contained in the ground-state energy $E_0$ and the rotational constant $a$ (and vary from nucleus to nucleus), while the rotational $J(J+1)$ pattern is universal. We note that the computation of the matrix elements is only simple for product states, i.e. for the $n=1$ approximation. In general, {\it ab initio} computations of $E_0$ and $a$ are somewhat challenging~\citep{qiu2017,hagen2022,sun2024}. An important insight gained from such calculations is that $E_{(n)}-E_0 \ll |E_0|$ and that $a \ll |E_0|$: Both the energy gained from the calculation and the spacing of the resulting levels are small compared to the ground-state energy, and both or of the size ${\cal O}(a)$. Thus, $H(\Omega',\Omega)$ is a low-energy Hamiltonian. 

Let us assume that we had a set of microscopic Hamiltonians that differ in their cutoffs and thus exhibit very different short-range physics. If these Hamiltonians are accurate, they will all yield the spectrum~(\ref{specLO}) to a good approximation. Thus, it must be possible to approach the low-energy physics of the very complicated (and non-local) Hamiltonian $H(\Omega',\Omega)$ of equation~(\ref{H_omega}) via an effective (field) theory, i.e. by constructing a Hamiltonian $H_{\rm EFT}$. Postulating locality, it is clear that $H_{\rm EFT}$ cannot depend on $\Omega$ itself because of rotational invariance. This leaves us with derivatives. As the parameter space is the two-sphere, the derivative is~\citep{varshalovich1988}
\begin{equation}
\label{nablaOmega}
\nabla_\Omega \equiv \mathbf{e}_\theta \partial_\theta + \frac{\mathbf{e}_\phi}{\sin\theta} \partial_\phi.
\end{equation}
Here
\begin{align}
\label{etheta}
     \mathbf{e}_\theta(\phi,\theta)\equiv\left(
     \begin{array}{c}
       \cos\phi\cos\theta \\
       \sin\phi\cos\theta\\
       -\sin\theta
     \end{array}\right)
\end{align}
and 
\begin{align}
\label{ephi}
     \mathbf{e}_\phi(\phi,\theta)\equiv\left(
     \begin{array}{c}
       -\sin\phi \\
       \cos\phi\\
       0
     \end{array}\right)
\end{align}
are the usual tangential unit vectors on the sphere at $\Omega$. Together with the radial unit vector
\begin{equation}
\label{er}
    \mathbf{e}_r(\phi,\theta) \equiv \left(
    \begin{array}{c}
    \cos\phi\sin\theta \\
    \sin\phi\sin\theta \\
    \cos\theta
    \end{array} \right), 
\end{equation}
which denotes the direction of the symmetry axis of the state $|\psi_{(n)},\Omega\rangle$, the set $(\mathbf{e}_\theta,\mathbf{e}_\phi,\mathbf{e}_r)$ forms a right-handed coordinate system. The most simple Hamiltonian one can write down is then 
\begin{align}
\label{HEFTLO}
    H_{\rm EFT} &= E_0 + a \nabla_\Omega\cdot \nabla_\Omega \nonumber\\
    &= E_0 - a \left( \frac{1}{\sin\theta}\partial_\theta\sin\theta\partial_\theta +\frac{1}{\sin^2\theta}\partial^2_\phi \right) \ . 
\end{align}
The corresponding eigenfunctions are spherical harmonics and the energies are given by equation~(\ref{specLO}). Of course,  $E_0$ and $a$ are low-energy constants of the effective field theory and need to be adjusted to data. 

This example shows how simple and powerful the construction of an effective theory can be. The steps involved were (i) the identification of the angles parameterizing the coset space SO(3)/SO(2) dictated by the pattern of symmetry breaking as the relevant degrees of freedom, and (ii) the insight that rotational invariance only allows derivatives to appear in the effective Hamiltonian. Usually, the spectrum~\eqref{specLO} is presented as a result of the rigid rotor model or the variable moment of inertia model~\citep{scharff1976}. However, one does not need any model to arrive at the spectrum; symmetry arguments alone are sufficient.

\subsection{Superfluid nuclei}
\label{sub:superfluid}
Let us now consider a semi-magic nucleus, for example, an isotope of tin or lead. We also assume an even number of neutrons. 
In what follows we will use the neutron number expectation value $N_{(n)}\equiv \braket{\psi_{(n)}|N|\psi_{(n)}}$
and the variance $\Delta N^2_{(n)} \equiv \braket{\psi_{(n)}|N^2|\psi_{(n)}}-N_{(n)}^2$ at step $n$ of the calculation. 

The starting point in this case is a Hartree-Fock-Bogoliubov computation. While the resulting product state $|\psi_{(1)}\rangle$ now exhibits good angular momentum, its neutron number, denoted as $N_{(1)}$, is not a good quantum number. Thus, the number variance fulfills $\Delta N^2_{(1)} > 0$. A second step could be a Bogoliubov coupled-cluster computation~\citep{signoracci2015,tichai2023} yielding the state $|\psi_{(2)}\rangle$, for which $\Delta N_{(2)}^2 < \Delta N_{(1)}^2$, but $\Delta N^2_{(2)} > 0$. As for deformed nuclei, the symmetry breaking costs only little in energy. This situation is illustrated in table~\ref{tab:bogol}, based on the data from~\citep{tichai2023}. We see again that the (estimated) ground-state energy $E_0$ is close to $E_{(2)}$, i.e. symmetry projection yields little gain in energy.

\begin{table}
\caption{\label{tab:bogol}Energies $E_{(n)}$ and particle number variation $\Delta N_{(n)}^2$ from Hartree-Fock Bogoliubov ($n=1$) and coupled-cluster theory including up to two-particle--two hole excitations ($n=2$) compared to the estimated energies $E_0$ from symmetry restoration, for the ground-states of nuclei as indicated.} 
\begin{indented}
\lineup
\item[]\begin{tabular}{@{}*{7}{l}}
\br                              
          & \m$E_{(1)}$& \m$E_{(2)}$ & \m$E_0$ & \m$E_{\rm exp}$  &$\Delta N_{(1)}^2$ &$\Delta N_{(2)}^2$\\
\mr
$^{74}$Ni  & $-447.7$ & \0$-608.3$ &\0$-609.0$ &\0$-624.04$ &  \0$5.1$  &  \0$4.9$ \\
$^{124}$Sn & $-759.9$ & $-1034.3$  & $-1034.7$ & $-1049.96$ &  \0$6.0$  &  \0$6.0$  \\
\br
\end{tabular}
\end{indented}
\end{table}

The particle-number breaking product state~\citep{bogoliubov1958,valatin1958} points into a definite direction of gauge space. Since the microscopic Hamiltonian $H_{\rm mic}$ preserves neutron number, the action of (global) gauge transformations
\begin{equation}
\label{gaugetransf}
    g(\alpha)\equiv e^{-i\alpha \hat{N}} 
\end{equation}
onto $|\psi_{(n)}\rangle$ introduces 
states $|\psi_{(n)},\alpha\rangle \equiv g(\alpha)|\psi_{(n)}\rangle$ with identical energy expectation values as $g^{-1}(\alpha) H_{\rm mic}g(\alpha) = H_{\rm mic}$.

The reader now sees where this journey is heading: One can again diagonalize the microscopic Hamiltonian in the subset of degenerate states $|\psi_{(n)},\alpha\rangle$ with $0\le \alpha< 2\pi$. This yields so-called pairing rotational bands, and ground states in nuclei that differ by pairs of neutrons are members of such a band~\citep{bohr1969,brink-broglia2005}. 
The energy scale associated with the pairing rotational band and the gain from the ensuing particle-number restoration are again small when compared to $E_{(2)}$. 

Similarly as in the case of deformed nuclei, one can construct an effective field theory, and the corresponding Hamiltonian is based on the derivative $\partial_\alpha$ that acts on the unit circle. The effective field theory entirely rests on the fact that the approximate states $|\psi_{(n)}\rangle$ break particle number, i.e., a U(1) phase symmetry of the Hamiltonian.  More details are presented in section~\ref{sec:pair}.

\subsection{Discussion}
\label{sub:discussion}
We have seen that the effective field theories of deformed and of superfluid nuclei have a microscopic foundation. They naturally arise whenever approximations of nuclear states break a symmetry of the Hamiltonian. While we have based our arguments on microscopic Hamiltonians, we see that the universality of these phenomena holds for any nuclear model that exhibits the symmetry breakings described above. From the low-energy perspective, any such model falls into  a ``universality class'' that is entirely determined by the pattern of the symmetry breaking. Thus, the effective field theory truly is model independent.

Given the simplicity of the parameter spaces -- the unit sphere in the case of deformed, axially symmetric nuclei and the unit circle in the case of pairing -- the reader might wonder about how complex the corresponding phenomena can possibly be. As we will see below, interesting phenomena will enter because of non-trivial topological effects. In the case of the unit sphere, radially symmetric ``monopole-like'' gauge potentials are consistent with rotational invariance, and the similar effects are possible for the unit circle. This will introduce Berry-phase physics and explain interesting phenomena.     

The construction of Hamiltonians within effective field theory is based on symmetry breaking and only uses derivative (and possibly gauge) couplings. This rings familiar from quantum field theory: In the presence of spontaneous symmetry breaking of a group $G$ to a subgroup $S$, Nambu-Goldstone bosons are the relevant low-energy degrees of freedom. They parameterize the coset $G/S$ and only derivative and gauge couplings are allowed. This connection to spontaneous symmetry breaking will be discussed in the following section.

\section{Emergent symmetry breaking}
\label{sec:ssb}
\subsection{Symmetry projection and spontaneous symmetry breaking}
\label{subsec:symproj}

The connection between rotational bands and symmetry restoration was made soon after the collective models~\citep{bohr1952,bohr1953} arrived. The Nilsson model~\citep{nilsson1955} exposed the shell structure of axially symmetric, deformed nuclei. A diagonalization of the Hamiltonian in the degenerate set of symmetry-breaking states then led to rotational bands, and this approach combined shell-model and collective aspects~\citep{peierls1957,peierls1962,villars1965}. Similarly, the understanding of superconductivity within BCS theory, and its usage in nuclear physics\citep{bohr1958,migdal1959} introduced pairing rotations as a consequence of particle-number restoration~\citep{bohr1969,bes1970,broglia1973}.  

The development of BCS theory was also most fruitful in particle physics. \citet{nambu1960} and \citet{goldstone1961} discovered that massless excitations (now referred to as Nambu-Goldstone bosons) accompany spontaneous symmetry breaking. \citet{nambu1961,nambu1961b} presented a model where pions emerged as the very light bosons of the spontaneously broken chiral symmetry, and \citet{weinberg1968} introduced chiral effective field theory as a model-independent approach that exploits spontaneous symmetry breaking of the strong force.   
\citet{coleman1969} and \citet{callan1969} generalized Weinberg's approach from the spontaneous breaking of SU(2) symmetry to other  continuous groups. Thus, there were parallel developments regarding symmetry restoration in nuclear physics and spontaneous symmetry breaking in particle physics.

\citet{bohr1975nobel} pointed out the connection between nuclear rotation, spontaneous symmetry breaking and Goldstone bosons in his Nobel lecture.  This picture has been emphasized by several authors~\citep{ui1983,fujikawa1986,nazarewicz1993,nazarewicz1994,frauendorf2001,broglia2000,papenbrock2011}. However, significant differences exist: Spontaneous symmetry breaking only happens in infinite systems while nuclei are finite. Nambu-Goldstone modes are excitations with arbitrary small energies while rotational bands and pairing rotational bands have finite spacings. To emphasize the difference \citet{koma1994} and \citet{yannouleas2007} introduced the expressions ``obscured symmetry breaking''  and ``emergent symmetry breaking'', respectively. We adopt the latter and want to discuss commonalities and differences between spontaneous and emergent symmetry breaking. 

Let us consider the breaking 
of SO(3) rotational symmetry down to SO(2) axial symmetry, and take ferromagnets (where the spins point into the direction of the $z$ axis) and deformed nuclei (as discussed in section~\ref{sub:deformed}) as respective examples. For the ferromagnet the ground state $|{\rm gs}\rangle$ spontaneously breaks the symmetry while we take the correlated state $|\psi_{(2)}\rangle$ as the symmetry breaking state for the deformed nucleus. 

There is a fundamental difference between the Hilbert spaces of finite and infinite systems that exhibit emergent and spontaneous symmetry breaking, respectively~\citep{ui1983}. 
To see this, let us return to equation~(\ref{specLO}), valid for a finite system. Here, the rotational constant $a$ is proportional to the inverse moment of inertia and vanishes in the limit of infinite particle number. As the ground state of the infinite system cannot be infinitely degenerate, one must introduce inequivalent Hilbert spaces and exclude rotations of the whole system.

Let us also present an alternative argument, and this time start from the symmetry-breaking  state. In the case of nuclei the symmetry-breaking state $|\psi_{(2)}\rangle$ and its rotated kin have a nonzero overlap, i.e. $\langle \psi_{(2)}|R(\phi,\theta)|\psi_{(2)}\rangle \ne 0$ for almost all angles. In the case of the ferromagnet's ground state $|{\rm gs}\rangle$, however, we have  $\langle {\rm gs}|R(\phi,\theta)|{\rm gs}\rangle = 0$ for all finite rotation angles. The latter is so because the overlap is an infinite product of single-spin overlaps that all have magnitudes smaller than one. Thus, for infinite systems a global rotation yields a state that is orthogonal to the symmetry breaking state. The rotated state belongs to an inequivalent Hilbert space, and there is no symmetry restoration. 

For ferreomagnets,  Nambu-Goldstone modes $|\phi(\mathbf{x},t),\theta(\mathbf{x},t)\rangle$ are generated by acting with the space- and time-dependent rotation operator 
$R(\phi(\mathbf{x},t),\theta(\mathbf{x},t))$ of equation~(\ref{rotationop}) onto the ground state, i.e. 
\begin{equation}
    |\phi(\mathbf{x},t),\theta(\mathbf{x},t)\rangle = R\left(\phi(\mathbf{x},t),\theta(\mathbf{x},t)\right) \, |{\rm gs}\rangle \ .
\end{equation}
The quantum fields $\phi(\mathbf{x},t)$ and $\theta(\mathbf{x},t)$ generate spin waves. They can have arbitrarily long wave length and arbitrarily low energy. Spatially constant, i.e. $\mathbf{x}$-independent, fields are forbidden because a rotation of the infinite system is not allowed (because it leads to an inequivalent Hilbert space). The Nambu-Goldstone states $|\phi(\mathbf{x},t),\theta(\mathbf{x},t)\rangle$ are orthogonal to the ground state because they involve infinite products of individual overlaps that are almost all smaller than unity. 

In the case of nuclei the Nambu-Goldstone modes are symmetry restoring and can be purely time-dependent.  They are generated by acting with $R(\phi(t),\theta(t))$ onto the symmetry-breaking state $|\psi_{(2)}\rangle$ which gives
\begin{equation}
    |\phi(t),\theta(t)\rangle = R\left(\phi(t),\theta(t)\right) \,|\psi_{(2)}\rangle \ .
\end{equation}
These states generally are not orthogonal to the  state $|\psi_{(2)}\rangle$. 

This discussion shows how rotational excitations differ from Nambu-Goldstone modes. However, rotational excitations and Nambu-Goldstone modes both arise from the action of rotation operators whose angles are time-dependent variables and fields, respectively. In both cases, the angles parameterize the coset SO(3)/SO(2) that reflects the pattern of the symmetry breaking. (We note that the coset space SO(3)/SO(2) is isomorph to the surface of the unit sphere.) This common technical aspect allows one to use the coset approach~\citep{coleman1969,callan1969} to develop effective Lagrangians for collective excitations in finite systems~\citep{leutwyler1987,chandrasekharan2008,papenbrock2011}. We briefly discuss this approach next. 

\subsection{Coset approach}
\label{sub:coset}
The collective degrees of freedom involved in symmetry projection and the Nambu-Goldstone bosons in spontaneous symmetry breaking parameterize the coset $G/S$ when the symmetry is broken from a group $G$ to a subgroup $S$. 
This allows one to construct nonlinear realizations of the symmetry group $G$~\citep{coleman1969,callan1969}. In the example considered so far, $G= {\rm SO}(3)$ and $S={\rm SO}(2)$, and the coset $G/S$ is the two-sphere. Each point on the sphere can be parameterized by the usual azimuth and polar angles $(\phi,\theta)$. A rotation maps the point with coordinates $(\phi,\theta)$  to a new point $(\phi',\theta')$, and the new angles are nonlinear functions of the old ones. This then constitutes the nonlinear (or Nambu-Goldstone) realization of the symmetry group $G$. To nuclear physicists, these may be somewhat less familiar than the usual linear (Wigner-Weyl) realizations. Nonlinear realizations apply in cases of spontaneous and emergent symmetry breaking. The nonlinear transformation properties also allow one to introduce quantities that are invariant under symmetry operations and thereby to construct effective Lagrangians. As usual the Noether  theorem allows one  to identify the corresponding conserved quantities.  The original arguments and derivations of this approach are by \citet{coleman1969,callan1969}.  Excellent expositions can be found in references~\citep{weinberg_v2_1996,brauner2010}. In section~\ref{sub:coset-def} we briefly  display  the main arguments for deformed nuclei~\citep{papenbrock2011}.

\section{Axially symmetric even-even nuclei} 
\label{sec:even}
Describing the ground-state rotational bands  of even-even nuclei with axial symmetry is the simplest application of the effective theory. The emergent symmetry breaking from the spherical group SO(3) to the axial  SO(2) identifies the degrees of freedom as those parameterizing the coset SO(3)/SO(2). These are azimuthal and polar angles ($\phi$, $\theta$) of the two-sphere. The formal construction of the theory was presented by~\citet{papenbrock2011}, and followed the steps presented in chapter~19 of the textbook~\citep{weinberg_v2_1996}. First, one derives the invariant terms that enter the effective Lagrangian. Second, one introduces a power counting and systematically constructs effective Lagrangians. For emergent symmetry breaking one then performs a Legendre transformation to obtain the effective Hamiltonian and solves the Schr\"odinger equation. 
This last step is usually facilitated by computing the conserved quantities (total angular momentum in our case) via Noether's theorem and expressing the Hamiltonian in terms of these quantities instead of the canonical momenta.
More recent derivations can be found in~\citep{papenbrock2014, papenbrock2015, coelloperez2015}.

\subsection{Leading-order theory}

Here we follow the more geometric approach by~\citet{papenbrock2020} as it simplifies steps (i) and (ii) above, as well as the construction of invariants via Noether's theorem. This approach combines the ($\phi$, $\theta$) angles into the radial unit vector~\eqref{er}, oriented along the symmetry axis of the nucleus~\footnote{Ground states of axially symmetric nuclei are often invariant under rotations by $\pi$ around an axis perpendicular to the symmetry axis, i.e. they exhibit ${\cal R}$ invariance~\citep{bohr1975} and thus are nematics~\citep{mermin1979}. Then, one only needs a preferred axis but no direction. This identifies opposite points on the sphere and reduces the coset space to half of the sphere.}. The velocity of this vector,
\begin{equation}
    \mathbf{v} \equiv \frac{d}{dt}\mathbf{e}_r(\phi,\theta)
    = \dot{\theta}\mathbf{e}_\theta(\phi,\theta)
    +\dot{\phi}\sin\theta\mathbf{e}_\phi(\phi,\theta) 
    \label{vel}
\end{equation}
is the building block of the effective theory. Here and in what follows the dot denotes the time derivative. The polar and azimuthal unit vectors were introduced in equations~(\ref{etheta}) and (\ref{ephi}).

The leading contribution to the effective Lagrangian is the simplest term built from the above velocity that is invariant under rotations
\begin{equation}
    L_{\rm LO} = \frac{C_0}{2}\mathbf{v}^2
    = \frac{C_0}{2}\left( \dot{\theta}^2 +\dot{\phi}^2\sin^2\theta\right).
    \label{LagLO}
\end{equation}
Here, $C_0$ is a low-energy constant that must be fit to data. The Legendre transformation of this Lagrangian yields the leading-order effective Hamiltonian
\begin{equation}
    H_{\rm LO} = \frac{1}{2C_0}\left(p_\theta^2 +\frac{p_\phi^2}{\sin^2\theta}\right) \ .
\end{equation}
Here we used the usual canonical momenta
\begin{equation}
\begin{aligned}
    p_\theta &= \frac{\partial L_{\rm LO}}{\partial\dot{\theta}} \ , \\
    p_\phi &= \frac{\partial L_{\rm LO}}{\partial\dot{\phi}} \ .
    \end{aligned}
\end{equation}

Application of Noether's theorem yields the total angular momentum $\mathbf{I}$ with components
\begin{equation}
\begin{aligned}
    I_x &= -\sin\phi p_\theta - \cos\phi\cot\theta p_\phi \ ,
    \\
    I_y &= \cos\phi p_\theta -\sin\phi\cot\theta p_\phi \ , \\
    I_z &= p_\phi \ , 
    \end{aligned}
\end{equation}
as the conserved quantity. One can combine these expressions into 
\begin{equation}
\label{Ibody}
    \mathbf{I} = -\frac{p_\phi}{\sin\theta}\mathbf{e}_\theta + p_\theta\mathbf{e}_\phi \ ,
\end{equation}
and see that the angular momentum has no component in direction of the symmetry axis. One can now rewrite 
the Hamiltonian as 
\begin{equation}
\label{HamLO}
    H_{\rm LO}=\frac{\mathbf{I}^2}{2C_0} \ . 
\end{equation}
Its quantization yields the energy spectrum
\begin{equation}
\label{ELO}
    E_{\rm LO}(I) = \frac{I(I+1)}{2C_0} \ .
\end{equation}
Thus, the well-known rigid-rotor spectrum results from the assumption of emergent symmetry breaking from SO(3) to SO(2). 

Several comments are in order. First, we note that the SO(3) symmetry is realized nonlinearly, as rotations transform the angles ($\phi$, $\theta$) nonlinearly. Transformation laws were presented in~\citep{papenbrock2020}. Second,  one quantizes the canonical momentum according to
\begin{equation}
    \mathbf{p} \equiv p_\theta\mathbf{e}_\theta + \frac{p_\phi}{\sin\theta}\mathbf{e}_\phi
    = -i \nabla_\Omega \ , 
\end{equation}
where $\nabla_\Omega$ is the angular derivative~(\ref{nablaOmega}). Writing the angular momentum as $\mathbf{I}=\mathbf{e}_r\times\mathbf{p}$, and noticing that $\mathbf{I}^2=-\nabla_\Omega^2$ yields
\begin{equation}
    \label{Hnabla} H_{\rm LO} = -\frac{\nabla_\Omega^2}{2C_0} \ .
\end{equation}
The eigenfunctions of this Hamiltonian are the usual spherical harmonics $Y_{IM}(\theta,\phi)$ and $M\in\{-I,-I+1,\ldots I\}$ is the eigenvalue of $I_z$. Finally,  ground-state bands in even-even nuclei only contain states with even spins. 
This pattern arises from the ${\cal R}$ invariance~\citep{bohr1975}.

\subsection{Power counting}
\label{subsec:power}
We can now consider more general Lagrangians. The rotational invariance permits only powers of $\mathbf{v}^2$ and this yields Hamiltonians in powers or $\mathbf{I}^2$.  Thus, the most general spectrum is a polynomial in $I(I+1)$ with coefficients that must be adjusted to data for a given  rotational band. As we will now discuss, this is indeed the power counting of the effective theory.

We introduce the low-energy (or small-frequency) scale $\xi$ that is typical for nuclear rotations. Then, the angular velocity~\eqref{vel} is slow, i.e. 
\begin{equation}
\begin{aligned}
\label{scaleXi}
    \dot{\theta} &\sim \xi \\
    \dot{\phi} &\sim \xi \ .
    \end{aligned}
\end{equation}
We also have  $E(2^+_1)\sim \xi$ in the leading-order spectrum~(\ref{ELO}) and this yields the estimate
\begin{equation}
\label{scaleC0}
    C_0\sim\xi^{-1} \ .
\end{equation}

\begin{figure}[htb]
\centering
\includegraphics[width=0.7\textwidth]{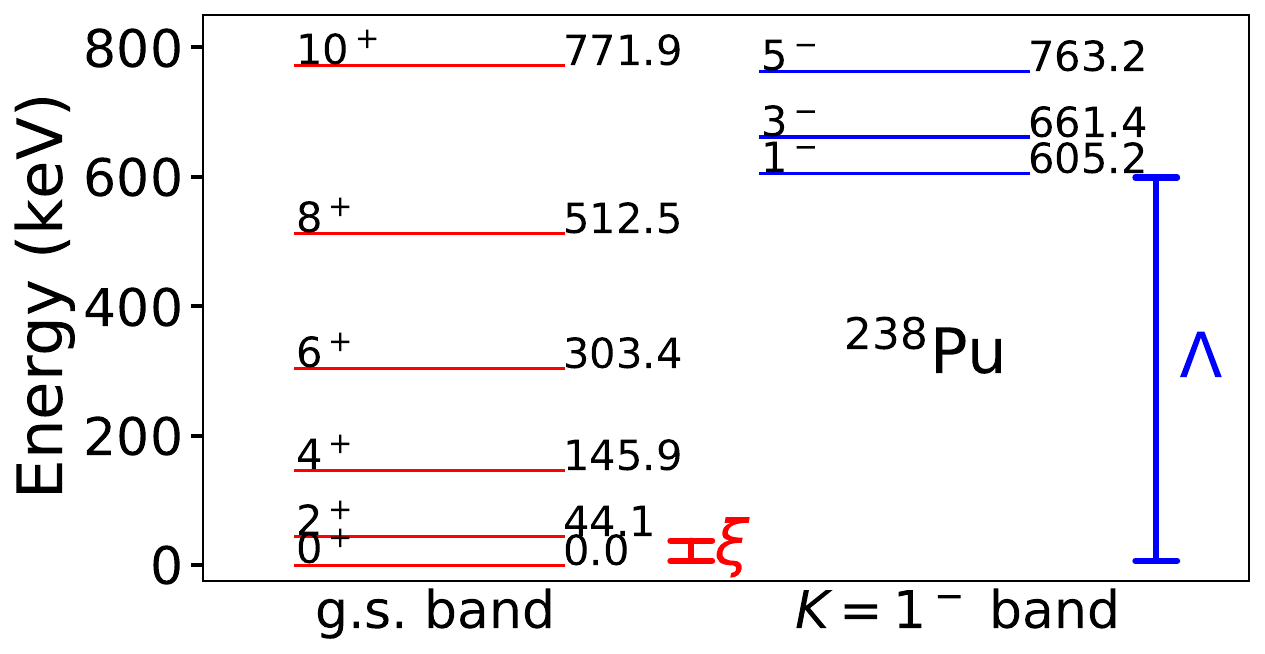}
\caption{The two rotational bands in $^{238}$Pu that are lowest in energies, with spin/parities $I^\pi$ and energies as indicated. Also shown is the low-energy scale $\xi$ and the breakdown energy $\Lambda$. Figure taken from arXiv:2005.11865 with permission from the authors, see also \citep{papenbrock2020}.}
\label{fig:pu238}
\end{figure}

Next we introduce a high-energy  scale $\Lambda$ at which the effective theory breaks down. This scale is due to neglected degrees of freedom that appear at an energy $\Lambda$. An example is shown in figure~\ref{fig:pu238}. The ground-state band closely follows the leading spectrum~\eqref{ELO}, setting the scale $\xi$ as the first level spacing. Clearly, the description of $^{238}$Pu as a single rotational band breaks down at the energy $\Lambda$ where the second rotational band starts. We have $\xi\ll\Lambda$, and this separation of scales allows us to introduce a power counting. 
 
One can, of course, introduce additional degrees of freedom to describe the second rotational band depicted in figure~\ref{fig:pu238}, but this is not what we want to consider here. Instead, the mere existence of other degrees of freedom impacts the low-energy theory. As there is a separation of scales between the excluded degrees of freedoms and the low-energy ones, the effect of the former on the latter can be captured by effective Lagrangians.   

Let us return to the most general Lagrangian
\begin{equation}
\label{genLag}
    L=\sum_{n=1}^\infty \frac{C_{n-1}}{2n} \mathbf{v}^{2n}.
\end{equation}
Here, the factors $1/(2n)$ are introduced out of convenience.
If we set the breakdown energy as $\Lambda$, we must have $C_0\mathbf{v}^2\sim\Lambda$ and thus find  $\mathbf{v}^2\sim \xi\Lambda$. This establishes the breakdown velocity. The key idea is now that  each term in the Lagrangian~(\ref{genLag}) yields an equal contribution at the breakdown scale. Thus $C_{n-1} \mathbf{v}^{2n}\sim \Lambda$ and our estimate for the size of the low-energy coefficients is 
\begin{equation}
C_{n-1} \sim \xi^{-n}\Lambda^{1-n} \ .
\end{equation}

This now  establishes a power counting. At low energies, where  $\mathbf{v}^2\sim\xi^2$,  the contribution of each term in the Lagrangian~(\ref{genLag}) is then 
\begin{equation}
    C_{n-1}\mathbf{v}^{2n} \sim \xi \left(\frac{\xi}{\Lambda}\right)^n \ . 
    \label{eq:rotcorrections}
\end{equation}
These clearly are increasingly smaller corrections~\footnote{One can also use breakdown velocity rather than a breakdown energy to establish a power counting, and that approach was taken in  \citep{papenbrock2011, zhang2013, coelloperez2015, papenbrock2020}.}. Based on this counting scheme, terms containing higher powers of $\mathbf{v}^2$ are higher orders of the effective theory. 

When higher orders are included in the effective Lagrangian, the exact Legendre transformation to the Hamiltonian is not anymore possible because one cannot easily solve for the velocities in terms of the conjugate momenta. Instead, one pursues a perturbative inversion where one expands around the leading-order result~\citep{fukuda1988}. This then yields a Hamiltonian expansion~\citep{zhang2013}
\begin{equation}
\label{genHam}
    H=\sum_{n=1}^\infty g_n \mathbf{I}^{2n} \ ,
\end{equation}
where the couplings $g_n$ are given in terms of the low-energy coefficients $C_{k-1}$. The resulting energy spectrum is then
\begin{equation}
    \label{genspec}
    E(I) = \sum_{n=1}^\infty g_n \left[I(I+1)\right]^{n} \ .
\end{equation}
From the power counting one finds that  $g_n\sim \xi(\xi/\Lambda)^{n-1}$. \citet{coelloperez2015} confirmed that low-energy coefficients for molecules and deformed nuclei scale as estimated by the power counting.

\subsection{Nonlinear realization of SO(3) symmetry}
\label{sub:nonlinear}
We also want to discuss the key concepts behind the nonlinear realization of spontaneously broken symmetries~\citep{weinberg1968, coleman1969, callan1969}. This topic is presented in detail in Weinberg's textbook~(\citeyear{weinberg_v2_1996}) and Brauner's review~(\citeyear{brauner2010}), and for deformed nuclei in references~\citep{papenbrock2011,papenbrock2014}. We let the angles $(\phi,\theta)$ 
set the orientation of the nucleus's symmetry axis. Under a rotation $R(\alpha,\beta,\gamma)$ with Euler angles $(\alpha,\beta,\gamma)$, the $(\phi,\theta)$ angles transform into $(\phi',\theta')$, which are complicated nonlinear functions of the original angles. This 
nonlinear representation of SO(3) 
is in contrast to the usual linear representation 
where spherical tensor  transforms linearly via multiplication by a Wigner-$D$ matrix~\citep{varshalovich1988}.   
 
The nonlinear realization has important consequences. Under a a rotation
\begin{equation}
    \begin{aligned}
        (\phi,\theta) &\to (\phi',\theta') \ , \\
        \mathbf{e}_r(\theta,\phi) & \to \mathbf{e}_r(\theta',\phi') \ ,
    \end{aligned}
\end{equation}
and 
\begin{equation}
\label{eta_angle}
\begin{aligned}
    \mathbf{e}_\theta(\phi,\theta) & \to +\mathbf{e}_\theta(\phi',\theta')\cos\eta + \mathbf{e}_\phi(\phi',\theta')\sin\eta \ , \\
    \mathbf{e}_\phi(\phi,\theta) & \to -\mathbf{e}_\theta(\phi',\theta')\sin\eta + \mathbf{e}_\phi(\phi',\theta')\cos\eta \ , 
    \end{aligned}
\end{equation}
where $\eta$ is an angle that depends on the Euler angles of the rotation (and the original angles $(\phi,\theta)$ of the nuclear symmetry axis).
Thus, the rotated body-fixed coordinate system differs from the basis vectors $(\mathbf{e}_\theta, \mathbf{e}_\phi, \mathbf{e}_r)$ at the rotated point $(\phi',\theta')$ by a rotation around the symmetry axis $\mathbf{e}_r(\phi',\theta')$ with the angle $\eta$. This has two important consequences. First, under a rotation any degrees of freedom defined in the body-fixed coordinate system transform linearly by a SO(2) rotation around the nucleus's symmetry axis. Second, any terms constructed from such degrees of freedom that are invariant under SO(2) rotations around the nucleus's symmetry axis are in fact invariant under full rotations of the system. 

\subsection{Coset approach to deformed nuclei}
\label{sub:coset-def}
The coset approach starts from the rotation operator~(\ref{rotationop}) with time-dependent Euler angles. 
One computes the expression 
\begin{equation}
    R^{-1} \partial_t R = a_x J_x + a_y J_y + a_z J_z 
\end{equation}
via the Baker-Campbell-Hausdorff expansion. This defines functions 
\begin{equation}
\label{afuncts}
\begin{aligned}
    a_x &= - \dot\phi \sin\theta   \ , \\
    a_y &=  \dot\theta  \, \\
    a_z &= \dot\phi \cos\theta  \ .
\end{aligned}
\end{equation}
These quantities are the building blocks for effective Lagrangians because they exhibit definite transformation properties under rotations.   
The components $a_x$ and $a_y$ transform linearly, i.e. by a rotation around the $z'$ axis with a rotation angle $\eta$ of equation~(\ref{eta_angle}) in the body-fixed coordinate system. They are readily identified with the components of the velocity vector, see equation~(\ref{vel}). The quantity $a_z$ is part of the covariant derivative 
\begin{equation}
    D_t = \partial_t -ia_z J_{z}
\end{equation}
and comes into play when other degrees of freedom are coupled to the rotor.

\section{Internal degrees of freedom}
\label{sec:coup}
So far, we have reviewed how to construct an effective theory in the presence of emergent symmetry breaking, and we have only dealt with rotational  degrees of freedom, leading to the physics of an isolated rotational band. 
Nuclei, of course, are finite systems with internal degrees of freedom and their description as rigid rotors must break down eventually. In this section we review how to construct effective field theories for deformed systems with internal degrees of freedom. 

\subsection{Effective theory for quadrupole degrees of freedom}
\label{sub:vib-rot}

The early work~\citep{papenbrock2011} followed \citet{bohr1952} and employed quadrupole degrees of freedom $\Phi_\mu$ with $\mu=-2,-1,\ldots,2$ modeling the shape of the nuclear surface. In the presence of emergent symmetry breaking one works in the co-rotating coordinate system. The component $\Phi_0$ acquires a vacuum expectation value and small oscillations around this expectation value introduce the $\beta$ vibrations. The modes $\Phi_{\pm 1}$  become replaced by the angles $(\phi,\theta)$ that determine the orientation of the symmetry axis. Finally, the modes $\Phi_{\pm 2}$ are the $\gamma$ vibrations. The rotational excitations are assumed to be at lowest energy and well separated from the $\beta$ and $\gamma$ vibrations. This allows one to set up a power counting. 

\citet{papenbrock2011} derived the theory up to next-to-next-to-leading order. At leading order one only deals with (harmonic) vibrations; at next-to-leading order, the ground-state rotational band and the bands on top of the $\beta$ and $\gamma$ vibrational band heads appear and add small corrections. All three bands have identical moments of inertia. At next-to-next-to-leading order, couplings between the different rotational bands appear, adding finer details. At even higher order, the different bands become non-rigid, i.e. they deviate from the $I(I+1)$ pattern~\citep{zhang2013}.
We briefly review these developments in what follows.

The spherical components of the velocity~\eqref{vel}
\begin{equation}
    v_{\pm1} = (v_\theta \pm i v_\phi) \ ,
\end{equation}
are the remnants of the components $\Phi_{\pm 1}$ in the case of emergent symmetry breaking of the quadrupole oscillator. We have ${v}_{\pm 1}\sim \xi$. The remaining components of the quadrupole field can be parameterized as
\begin{equation}
\begin{aligned}
    \Phi_0(t)&\equiv\upsilon_0 + \varphi_0(t) \ , \\
    \Phi_{\pm 2}(t)& \equiv \varphi_2(t) e^{\pm i2\gamma(t)} \ .
    \label{bigPhi}
\end{aligned}
\end{equation}
Here $\phi_2$ and $\gamma$ are real functions and $\Phi_{2}=\Phi_{-2}^{*}$ (because the nuclear surface must be a real function), $\upsilon_0$ is the constant vacuum expectation value, and (being related to the emergent symmetry breaking) scales as an inverse power of the rotational energy scale
\begin{equation}
    \upsilon_0 \sim \xi^{-1/2} \ .
\end{equation}
The other quantities in equation~(\ref{bigPhi}) scale as the vibrational energy scale $\Omega$. We have $\Omega \gg\xi$, and 
\begin{align}
\varphi_0 &\sim \varphi_2 \sim \omega^{-1/2} \ , \nonumber\\
 \dot{\varphi}_0 &\sim \dot{\varphi}_2 \sim \omega^{1/2} \ ,  \\
     \dot{\gamma} &\sim \omega \ .\nonumber  
\end{align}

Under a general SO(3) rotation, the components of the quadrupole field transform as
\begin{equation}
    \Phi_\mu\rightarrow\exp(i\mu\eta)\Phi_\mu \ ,
\end{equation}
where $\eta$ is a complicated angle of the rotation angles and the angles that define the orientation of the body-fixed symmetry axis~\citep{papenbrock2011}.

This simple transformation allows for the construction of  effective Lagrangians. Terms that appear to be invariant under SO(2) are in fact invariant under SO(3).  The effective Lagrangian at the high-energy vibrational scale is 
\begin{align}
    L_{\rm LO} &= \frac{1}2 \left(D_t\Phi_0\right)^2 + \frac{1}{2} D_t\Phi_{+2}D_t\Phi_{-2} - \frac{\omega_0^2}{2} \Phi_0^2 - \frac{\omega_2^2}{2} \Phi_{+2}\Phi_{-2} \nonumber\\
    &\approx \frac{1}{2}\dot{\varphi}_0^2 + \frac{1}{2}\dot{\varphi}_2^2 + 2\varphi_2^2\dot{\gamma}^2
    - \frac{\omega_0^2}{2}\varphi_0^2 - \frac{\omega_2^2}{2}\varphi_2^2 \ .
    \label{eq:vibrotlo}
\end{align}
Here $\omega_0$ and $\omega_2$ are low-energy constants. The approximation in the second line neglects terms of the order $\xi\ll \omega$ coming from the time derivatives of the rotational angles or the expectation value $\upsilon_0$. This yield the Lagrangian of uncoupled harmonic oscillators. The low-energy constants scale as the energies of the excited bandheads, i.e. 
\begin{gather}
    \omega_0 \sim \omega_2 \sim \Omega \ .
\end{gather}
At next-to-leading order, the smaller details of the rotational scale $\xi$ enter via the Lagrangian
\begin{align}
    L_{\rm NLO} &= \frac{C_0}{2} v_{+1}v_{-1} + 4\varphi_2^2\dot{\gamma}\dot{\phi}\cos{\theta} \nonumber \\
    &= \frac{C_0}{2} \left( \dot{\theta}^2 + \dot{\phi}^2\sin^2{\theta} \right) + 4\varphi_2^2\dot{\gamma}\dot{\phi}\cos{\theta} \ .
\end{align}
Here the low-energy constant $C_0$ scales as $C_0\sim \xi^{-1}$, see equation~(\ref{scaleC0}).

A Legendre transformation of the effective Lagrangian at this order yields a Hamiltonian of the form $H=H_{\rm LO}+H_{\rm NLO}$, where
\begin{align}
    H_{\rm LO} &= \frac{p_0^2}{2} + \frac{\omega_0}{2}\varphi_0^2 + \frac{p_2^2}{2} + \frac{p_{\gamma}^2}{8\varphi_2^2} + \frac{\omega_2^2}{2}\varphi_2^2, \nonumber \\
    H_{\rm NLO} &= \frac{\mathbf{p}_{\Omega\gamma}^2}{2C_0} = \frac{\left(\mathbf{I}^2 - p_\gamma^2\right)}{2C_0},
\end{align}
with $p_i\equiv-i\partial_i$ and
\begin{align}
    \mathbf{p}_{\Omega\gamma}\equiv \mathbf{e}_{\theta}p_\theta + \mathbf{e}_\phi \frac{p_\phi-\cos{\theta}p_\gamma}{\sin{\theta}}.
\end{align}
The eigenstates $\ket{n_0n_2IMK}$ of this Hamiltonian, with $n_0$, $n_2$ and $K/2$ the number of quanta for the different oscillation modes, have eigenenergies $E=E_{\rm LO}+E_{\rm NLO}$ with
\begin{equation}
\begin{aligned}
    E_{\rm LO} &= \omega_0 n_0 + \omega_2 \left(2n_2+\frac{K}{2}\right) \ , \\
    E_{\rm NLO} &= \frac{1}{2C_0}\left[ I(I+1) - K^2 \right] \ , 
\label{eq:quadrotenergy}
\end{aligned}
\end{equation}
where the next-to-leading contribution is the rotational energy of the system.

Thus, spectra consists of rigid rotational bands on top of harmonic excitations. Deviations from the harmonic behavior of the band heads can be accounted for by terms containing only vibrational degrees of freedom yielding a correction that can be expanded as a series in powers of $\Omega/\Lambda$ (where $\Lambda$ is the breakdown scale). Since the theory focuses only in the lowest $0^+$ and $2^+$ bands, traditionally known as the $\beta$ and $\gamma$ bands, those contributions are neglected in what follows. The next-to-next-to-leading order Lagrangian
\begin{align}
    L_{\rm N^2LO} &= \frac{C_\beta}{2} \Phi_0 v_{+1}v_{-1} + \frac{C_\gamma}{4}\left(\Phi_{+2}v_{-1}v_{-1} + \Phi_{-2}v_{+1}v_{+1} \right) \nonumber \\
    &= \frac{C_\beta}{2}\varphi_0 \left(\dot{\theta}^2 + \dot{\phi}^2\sin^2{\theta} \right) + \frac{C_{\gamma}}{2}\varphi_2 \left[ \left(\dot{\theta}^2 - \dot{\phi}^2\sin^2{\theta} \right)\cos{2\gamma} + 2\dot{\theta}\dot{\phi}\sin\theta\sin{2\gamma} \right] 
\label{eq:LN2LO}
\end{align}
is off-diagonal and does not impact energies at that order. It will, however, play an important role in describing electromagnetic transitions between different rotational bands, see section~\ref{subsec:interband}. 
Thus, one has to go to one higher order to see dynamical modifications of the rotational moment of inertia.

\citet{zhang2013} showed that the next-to-next-to-next-to-leading order contributions are 
\begin{equation}
\label{eq:Lag-N2LO}
    L_{\rm N^3LO} = D_0 \Phi_0^2 v_{+1}v_{-1} + D_2 \Phi_{+2}\Phi_{-2} v_{+1}v_{-1} + D_{02} \Phi_0 \left(\Phi_{+2}v_{-1}v_{-1} + {\rm h.c.}\right) + \ldots \ .
\end{equation}
Here, the dots denote terms that are not coupled to any rotations (and not of interest to us as they only model vibrational interactions). The Lagrangian~(\ref{eq:Lag-N2LO}) corrects the rotational bands via 
\begin{equation}
\label{erg_high}
    E_{\rm N^3LO} = E_{\rm NLO} \times \left[1+ a_0 n_0 + a_2 \left( 2n_2 + \frac{K}{2} \right) \right] \ .
\end{equation}
Thus, one has a small shift in each band's moment of inertia that is linear in the number of excited phonons $n_0$ and $n_2$ of the modes $\varphi_0$ and $\varphi_2$, respectively.  The parameters $a_0$ and $a_2$ in equation~(\ref{erg_high}) are functions of the low-energy constants of the Lagrangian up to and including equation~(\ref{eq:Lag-N2LO}). 

\begin{figure}[h]
    \centering
    \includegraphics[width=0.7\textwidth]{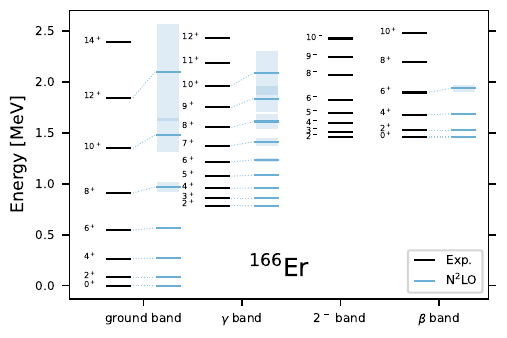}
    \caption{Four lowest rotational bands in $^{166}{\rm Er}$. Next-to-next-to-leading order predictions for states in the ground, $\beta$ and $\gamma$ bands (blue lines) are compared with experimental data (black lines) below 2~MeV. Theory errors at this order (blue boxes) were estimated as $(\varepsilon I^2)^2$ times the predicted energies, using the breakdown scale $\Lambda=3\omega_0$. The description of the $2^-$ band would require us to couple a $K^\pi=2^-$ internal degree of freedom to the rotor.}
    \label{fig:166er}
\end{figure}

Figure~\ref{fig:166er} compares predicted energies for states in the ground, $\beta$ and $\gamma$ bands in $^{166}{\rm Er}$ (blue lines) to experimental data below $2$~MeV (black lines).
The low-energy constants describing the spectra at this order were fitted to the energies of the first excited $2^+$ state, and the energies of the band heads and first excited states in the $\beta$ and $\gamma$ bands. The errors shown as blue boxes were estimated as to be $\varepsilon^2$ times smaller than the next-to-leading energies, with $\varepsilon\equiv E_{\rm NLO}/\Lambda$ defined in terms of the rotational energy and the breakdown scale, the latter lifted due to the explicit inclusion of quadrupole modes to $\Lambda\approx3\Omega$, with $\Omega\equiv\max(\omega_0,\omega_2)$. These results show that the assumptions about the energy scales and the power counting are consistent.

The effective theory for quadrupole degrees of freedom in the presence of emergent symmetry breaking essentially casts the  Bohr Hamiltonian (in the deformed limit or SU(3) limit) into an effective theory. While the leading couplings between vibrational and rotational degrees of freedom are model independent, the resulting effective theory is -- of course -- ultimately based on a model of quadrupole degrees of freedom. It is unable, for example, to account for any low-lying negative parity bands (see figure~\ref{fig:166er}).  On the plus side, however, it makes clear how one could make systematic corrections to the Bohr Hamiltonian.

\subsection{Effective field theory of a deformed droplet}

\citet{papenbrock2014, papenbrock2015} considered the nucleus as a deformed liquid drop with axial symmetry. The construction of an effective field theory for this finite system differs from the one used for infinite systems such as (anti)ferromagnets~\citep{leutwyler1994,roman1999,hofmann1999,baer2004,kampfer2005}.  In infinite systems Nambu-Goldstone bosons are based on the fields $\phi(\mathbf{x},t)$ and $\theta(\mathbf{x},t)$ that describe the local rotations of spins via the rotation operator~(\ref{rotationop}).  The  building blocks of the effective field theory are $a_{k,\mu}(\phi,\theta)$ with $k=x,y,z$ and $\mu=t,x,y,z$ that are derived from 
\begin{equation}
    R^{-1}\partial_\mu R = a_{x,\mu}(\phi,\theta) J_x + a_{y,\mu}(\phi,\theta) J_y +a_{z,\mu}(\phi,\theta) J_z \ .
\end{equation}
It is particularly important that the angles $\phi(\mathbf{x},t)$ and $\theta(\mathbf{x},t)$ and the quantities $a_{k,\mu}(\mathbf{x},t)$ really depend on position. Purely time-dependent angles that lack position dependence would induce an overall rotation of the (anti)ferromagnet. In an infinite system, the rotated state has zero overlap with the state one started from. Thus, the rotated state is really in an inequivalent Hilbert space. For this reason, overall rotations of the system must be excluded and purely time-dependent fields are forbidden. 

In contrast, purely time-dependent fields are allowed in finite systems. Then one must single out the purely time-dependent angles~\citep{leutwyler1987}; we denote them as $\alpha(t)$ and $\beta(t)$. The key rotation operator then becomes the product 
\begin{equation}
\label{Umat2R}
    U(\alpha,\beta;\phi,\theta)\equiv R(\alpha,\beta)R(\phi,\theta) \ ,
\end{equation}
where $\phi(\mathbf{x},t)$ and $\theta(\mathbf{x},t)$ are fields that depend on both time and position. Clearly, this operator induces local distortions in the liquid drop via the fields $\phi(\mathbf{x},t)$ and $\theta(\mathbf{x},t)$ followed by overall rotations of the whole drop via $\alpha(t)$ and $\beta(t)$. The building blocks for the effective field theory, i.e. combinations $a_{k,\mu}(\alpha,\beta;\phi,\theta)$ of degrees of freedom that transform properly under rotations appear on the right-hand-side of
\begin{equation}
    U^{-1}\partial_\mu U = a_{x,\mu}(\alpha,\beta;\phi,\theta) J_x + a_{y,\mu}(\alpha,\beta;\phi,\theta) J_y +a_{z,\mu}(\alpha,\beta;\phi,\theta) J_z \ .
\end{equation}
The derivation of that effective field theory was presented by \citet{papenbrock2014} and followed the coset approach. For simplicity, the position was expressed in polar coordinates, i.e. as $\mathbf{x}=(r,\varphi,\vartheta)$, and the dependence on $r$ was dropped (because radial compression modes were assumed to be high in energy). Finally, the fields $\phi(\varphi,\vartheta,t)$ and $\theta(\varphi,\vartheta,t)$ were decomposed in terms of normal modes and a cutoff was imposed.  The resulting effective field theory consisted of vibrational modes (from the decomposition of the fields $\phi$ and $\theta$) and the rotation angles $\alpha,\beta$. 

That theory naturally explained how a large number of vibrations arise (these quickly become anharmonic as the energy increases), and how these are coupled to the rotations of the whole droplet. This latter point is very important because the coupling of internal degrees of freedom to the rotations is universal. It consists of a Lagrangian term
\begin{equation}
    L_{\rm coup}=K_{z'} \dot\alpha \cos\beta 
\end{equation}
that originates from a covariant derivative and couples the angular-momentum projection $K_{z'}$ of an internal degree of freedom to the rotation angles. As we will see below, this term can be re-written as a gauge potential. 

An important assumption in the construction of the effective field theory~\citep{papenbrock2014,papenbrock2015} was that energies from vibrations are much larger than those from rotations. This is certainly so for sufficiently large droplets. Here, the radius scales as $A^{1/3}$ for a droplet with mass number $A$. Thus,  vibrational energies scale as $A^{-2/3}$ while the rotational energies scale as $A^{-5/3}$. This shows that internal degrees of freedom are always ``fast'' when compared to rotations in sufficiently heavy nuclei. 

As a concrete model for a given nucleus, the effective field theory of the liquid drop  is probably less useful because a considerable number of low-energy coefficients need to be adjusted to the many vibrational states of the system. This corresponds to a modeling of the internal dynamics. Another concern is that modeling the nucleus as a liquid drop is too simple because it neglects superfluidity.

\subsection{Spins as internal degrees of freedom}
\label{sec:internal}
Instead of modeling the physics of a liquid droplet, it is simpler to introduce internal degrees of freedom as spins $\chi_S$ of rank $S$. We allow $S$ to be half integer or integer.  
A key point is the assumption that the internal degrees of freedom are fast compared to the velocity~(\ref{vel}) of the nucleus's symmetry axis. (Otherwise there would be no separation of scales between rotations and internal motions.) This means that one can work in an adiabatic approximation where the fast spin instantly follows the slow motion of the symmetry axis $\mathbf{e}_r(\phi,\theta)$. (This approximation is also known as the ``strong coupling'' regime in the physics of deformed nuclei.) It is then  natural to expand $\chi_S$ in terms of body-fixed helicity basis functions $\chi_{S\lambda}(\theta,\phi)$ 
that fulfill~\citep{varshalovich1988}
\begin{equation}
\begin{aligned}
\label{helicityspin}
        \hat{S}^2\chi_{S\lambda}(\theta,\phi)&=S(S+1)\chi_{S\lambda}(\theta,\phi) \ , \\
        \hat{S}_{z'}\chi_{S\lambda} (\theta,\phi)&=\lambda\chi_{S\lambda}(\theta,\phi) \ .
\end{aligned}
\end{equation}
Here $\hat{S}_{z'}\equiv \mathbf{e}_r\cdot\hat{S} $ and it is clear that the helicity basis functions are quantized with respect to their projection $\lambda$ onto the nucleus's symmetry axis.   

In the body-fixed system, the interaction between the spins and that between the spins and the deformed rotor is only invariant under rotations around the nucleus's symmetry axis. While we do not want to model these interactions, the adiabatic approximation allows us to compute the eigenstates of the Hamiltonian that governs the spins at a given orientation of the nucleus's symmetry axis. The eigenfunctions are superpositions 
\begin{equation}
\label{eigenpsi}
    \psi_{K q}(\theta,\phi) \equiv \sum_S U_{q S}^{(K)}\chi_{S K}(\theta,\phi) \ .
\end{equation}
Here, $K$ is the spin projection of the eigenfunction $\psi_{K  q}(\theta,\phi)$ onto the nucleus's symmetry axis (and we followed nuclear-physics conventions in using this label instead of $\lambda$), and $U_{q S}^{(K)}$ are admixture coefficients with $\sum_S \vert U^{(K)}_{qS}\vert^2=1$ The label $q$ is for other good quantum numbers such as parity and possibly isospin. Clearly, the eigenfunctions~(\ref{eigenpsi}) are superpositions of spherical tensors with different spin $S$ and not eigenstates of the operator $\hat{S}^2$. We only have
\begin{equation}
    \hat{S}_{z'} \psi_{K q}(\theta,\phi)  = K \psi_{K q}(\theta,\phi) \ ,
\end{equation}
and
\begin{equation}
    \psi^\dagger_{L p}(\theta,\phi) \hat{S}_{\pm 1'}\psi_{K q}(\theta,\phi)    \propto \delta_{L}^{K\pm 1}\delta_p^q \ . 
\end{equation}
Here, the proportionality constant can be computed from knowledge of the matrix elements $U_{q S}^{(K)}$.  

Time reversal invariance demands that the eigenfunctions~(\ref{eigenpsi}) come in pairs $\pm K$ for $K>0$. For half integer $K$, this is Kramer's degeneracy. For integer $K$, all states with $K\ne 0$ come in doublets and  a state with $K=0$ is a singlet. Thus, we can denote the eigenenergies of the internal degrees of freedom as $E_{|K|q}$. We note that the eigenfunctions are orthogonal
\begin{equation}
    \psi_{K' q'}^\dagger(\theta,\phi)\psi_{K q}(\theta,\phi) = \delta_{K'}^K\delta_{q'}^q
\end{equation}
and complete
\begin{equation}
    \sum_{K q}\psi_{K q}(\theta,\phi)\psi_{K q}^\dagger(\theta,\phi) = 1 \ , 
\end{equation}
because the matrix $U_{q S}^{(K)}$ in equation~(\ref{eigenpsi}) is unitary for each $K$. 

The energies and quantum numbers $(K, q)$ and energies $E_{|K|q}$ could be taken from data. Alternatively, they could also come from computations that start from deformed  reference states (and do not perform symmetry projections). 

We now introduce time-dependent basis functions (and drop the angular dependence for simplicity), i.e.  
\begin{equation}
\label{psiKq}
    \psi_{K q}(t) \equiv  \psi_{K q}(\theta,\phi;t)
\end{equation}
describes the time dependence of an internal state. 
Then, the Lagrangian for the internal degrees of freedom simply  is
\begin{equation}
\label{Lagpsi}
    L_\psi = \sum_{K q}  \psi_{Kq}^\dagger(t) \left(i \partial_t - E_{|K|,q}\right)\psi_{Kq}(t) \ .
\end{equation}
The arguments we made in section~\ref{sub:nonlinear} imply that under a rotation the coefficient functions  transform linearly via 
\begin{equation}
    \psi_{K q}(t) \to e^{-iK \eta } \psi_{K q}(t) \ , 
\end{equation}
with the angle $\eta$ introduced in equations~(\ref{eta_angle}). 
Thus, terms such as $\psi_{K q}(t)\psi_{-K q}(t)$ or $\psi_{K q}^\dagger(t)\psi_{K q}(t)$ (which are scalars under rotations around the body-fixed symmetry axis) are indeed scalars under full rotations thanks to the non-linear realization of the SO(3) symmetry.
It is then clear that any axially-symmetric Lagrangian or Hamiltonian in the helicity components is admissible to construct a rotationally invariant effective theory. 

Let us discuss two examples. First, for  half-integer $K$ empirical guidance or the Nilsson model~\citep{nilsson1955} would identify which quantum numbers $(K q)$ are closest to the Fermi surface of a deformed odd-mass nucleus. In this case, one could limit the Lagrangian~(\ref{Lagpsi}) to the few pairs of fermion orbitals that are of interest. 
Second, for even-even and odd-odd nuclei $K$ is integer and one would use heuristics to select the helicity components that are lowest in energy and thus most relevant for the construction of effective Lagrangians. For the description of the two lowest-energy rotational bands in $^{238}$Pu, for instance, one would include $\psi_{0^+}$ for the ground-state band with $K^\pi=0^+$, and $\psi_{\pm 1^-}$ for the $K^\pi=1^-$ band (with the signs denoting the time-reversed partners), as can be seen in figure~\ref{fig:pu238}. 

The question arises now how do the internal degrees of freedom couple to the slow degrees of freedom $(\phi,\theta)$ of the nucleus's symmetry axis.  These interactions are interesting because they  are model independent. We discuss them next. 

\subsection{Vector potentials couple internal degrees of freedom to the rotor}
\label{sub:vectorpotentials}
There is a single model-independent (i.e. parameter free) coupling between the internal degrees of freedom and the rotor. In the literature one can find various ways to derive this coupling. The coset approach~\citep{weinberg1968,coleman1969,callan1969,brauner2010} via the nonlinear realization of a spontaneously broken symmetry is probably the most general; it is also a bit technical. The result is that, for fast degrees of freedom, a covariant derivative replaces the usual derivative. For the case of deformed nuclei the derivation was presented in~\citep{papenbrock2011}.

This applies to our case as well, because the fast degrees of freedom $\psi_{Kq}$ are defined in the body-fixed and co-rotating system.  
Thus, their time derivative now involves $\dot\psi_{Kq}$ and also the change of the helicity basis functions $\chi_{SK}(\theta,\phi)$ (because the angles  $\theta$ and $\phi$ are time-dependent as well). This leads to the introduction of the covariant derivative 
\begin{equation}
\begin{aligned}
\label{covariant}
    D_t\psi_{K q}(t) &=
    \dot\psi_{K  q}(t) -iK \psi_{K q}(t) \dot\phi\cos\theta  \\
    &=  \dot\psi_{K q}(t) -i \mathbf{v} \cdot \mathbf{A}_{\rm uni} \psi_{K q}(t)\ .
    \end{aligned}
\end{equation}
In the last line of equation~(\ref{covariant}), we used the velocity~\eqref{vel} to introduce the universal vector potential
\begin{equation}
\label{gaugepot}
    \mathbf{A}_{\rm uni}\equiv \hat{S}_{z'} \cot\theta \mathbf{e}_\phi \ .
\end{equation}
Here is it implied that $\hat{S}_{z'}$ acts on a helicity component via
\begin{equation}
    S_{z'} \psi_{K q}(t) = K \psi_{K q}(t) \ .
\end{equation}
The covariant derivative couples the slow rotor velocity $\mathbf{v}$ in a universal way to the fast helicity spin function $\chi_{S\lambda}(\theta,\phi)$ via a vector potential. These velocity-dependent forces are typically referred to as Coriolis forces in the literature~\citep{kerman1956}. 

The vector potential~(\ref{gaugepot}) can be used to introduce the ``magnetic field''
\begin{equation}
\label{Bfield}
    \mathbf{B}_{\rm uni} \equiv \nabla_\Omega\times \mathbf{A}_{\rm uni} = -\hat{S}_{z'} \mathbf{e}_r \ . 
\end{equation}
This is a radially symmetric ``monopole'' field~\citep{fierz1944,wu1976}. The resulting magnetic flux is quantized because $\hat{S}_{z'}$ yields integer or half integer values $K$ when acting onto the components $\psi_{K q}$. The appearance of a monopole field is intuitively clear: The fast spin always points into the direction of the symmetry axis and -- in the co-rotating coordinate system -- this corresponding magnetic moment creates a magnetic field that is radially symmetric. 

In contrast to the magnetic field~(\ref{Bfield}) the vector potential~(\ref{gaugepot}) is not invariant under rotations. However, after a rotation one can bring the vector potential back into the original form~(\ref{gaugepot}) by performing a gauge transformation $\mathbf{A}_{\rm uni}\to\mathbf{A}_{\rm uni} +\nabla_\Omega \gamma(\phi,\theta)$, see \citep{fierz1944} for the probably earliest discussion of this point.  The gauge freedom exists because the body-fixed coordinate system is arbitrary with respect to rotations by an angle $\gamma(\phi,\theta)$ around the nucleus's symmetry axis. While we fixed the gauge by using the usual polar basis vectors $(\mathbf{e}_\theta, \mathbf{e}_\phi,\mathbf{e}_r)$ of equations~(\ref{etheta}) to (\ref{er}) to define the body-fixed system, any combination $(\mathbf{e}_1, \mathbf{e}_2,\mathbf{e}_r)$ with 
\begin{equation}
\begin{aligned}
       \mathbf{e}_1 &\equiv +\mathbf{e}_\theta \cos\gamma(\phi,\theta) + \mathbf{e}_\phi \sin\gamma(\phi,\theta)\ , \\
       \mathbf{e}_2 &\equiv -\mathbf{e}_\theta \sin\gamma(\phi,\theta) + \mathbf{e}_\phi \cos\gamma(\phi,\theta) \ , 
       \end{aligned}
\end{equation}
could have been used as well \citep{papenbrock2020}. We note that the appearance of gauge potentials is not limited to axial symmetry. The review by \citet{littlejohn1997} shows that gauge potentials naturally enter in many-body systems when a separation between rotations and internal motions is sought, because one cannot unambiguously define internal coordinates.

The universal Lagrangian that couples the  internal degrees of freedom to the rotor consists of the sum of the Lagrangians (\ref{LagLO}) for the rotor and (\ref{Lagpsi}) for the internal degrees of freedom. In the latter,  the time derivative must be replaced by the the covariant derivative~(\ref{covariant}). We thus have
\begin{equation}
\label{LagK}
        L = \frac{C_0}{2}\mathbf{v}^2 +  \sum_{K q}  \psi_{Kq}^* \left(i \partial_t - E_{|K|,q}\right)\psi_{Kq} + \mathbf{v}\cdot\sum_{KL q}  \psi_{Kq}^* \mathbf{A}\psi_{Lq} \ ,
\end{equation}
with $\mathbf{A}=\mathbf{A}_{\rm uni}$ being the universal vector potential from equation~(\ref{gaugepot}). Here, the last term is actually diagonal  (i.e. the vector potential is such that only $K=L$ contributes to the sum) but we left the notation more general, because
the Lagrangian~(\ref{LagK}) is not yet complete. 

One can write down another interaction term between the rotor and the internal degrees of freedom that is linear in the angular velocity $\mathbf{v}$. The complete Lagrangian is obtained by replacing the vector potential in equation~(\ref{LagK}) by  
\begin{equation}
\label{Atot}
\mathbf{A} = \mathbf{A}_{\rm uni}+\mathbf{A}_{\rm non}  \ . 
\end{equation}
Here the non-Abelian (and non-universal) gauge potential
\begin{equation}
\label{Anon}
    \mathbf{A}_{\rm non} = g \mathbf{e}_r\times \hat{S}  
\end{equation}
depends on the dimensionless low-energy constant $g$.   Naturalness arguments imply that $g\sim{\cal O}(1)$. 
We note that the non-Abelian gauge potential~(\ref{Anon}) can mix internal degrees of freedom whose $K$ quantum numbers differ by one unit. In particular, it mixes the  time-reversed partners of a fermionic internal state with $K=\pm 1/2$.   

The total gauge potential~(\ref{Atot})
then leads to the total magnetic monopole field
\begin{equation}
        \mathbf{B} \equiv \nabla_\Omega\times \mathbf{A} -i \mathbf{A} \times \mathbf{A} \\
        = (g^2-1)\hat{S}_{z'} \mathbf{e}_r \ , 
\end{equation}
which is invariant under rotations. 

For the power counting we remind the reader (see section~\ref{subsec:power}) that $|\mathbf{v}|\sim \xi$ and $C_0\sim\xi^{-1}$  are related to the low-energy scale $\xi$. The vector potential is dimensionless (and of order one), while the internal energies $E_{|K|,q}\sim\Omega$. Usually one has $\Omega\gg\xi$. However, differences if the energies of the internal degrees of freedom can be small. In that case, the interaction between internal degrees of freedom and the rotor can also strongly couple internal degrees of freedom whose $K$ quantum numbers differ by one unit.

\subsection{Total angular momentum}
Starting with the Lagrangian~(\ref{LagK}) Noether's theorem yields that the total angular momentum $\mathbf{I}$ with components
\begin{equation}
\begin{aligned}
        I_x&=-\sin\phi p_\theta -\cos\phi\cot\theta p_\phi +K_{\rm tot}\frac{\cos\phi}{\sin\theta} \ , \\
        I_y&=+\cos\phi p_\theta -\sin\phi\cot\theta p_\phi +K_{\rm tot}\frac{\sin\phi}{\sin\theta}\ , \\
        I_z&=p_\phi
    \end{aligned}
\end{equation}
is conserved under rotations. Here, 
\begin{equation}
    K_{\rm tot} \equiv \sum_{Kq} \psi_{Kq}^\dagger\hat{S}_{z'}\psi_{Kq}
\end{equation}
is the angular momentum projection of the internal degrees of freedom onto the symmetry axis. One can rewrite the total angular momentum  as
\begin{equation}
\label{Ibody}
    \mathbf{I} = -\frac{p_\phi}{\sin\theta}\mathbf{e}_\theta + p_\theta\mathbf{e}_\phi +K_{\rm tot} \mathbf{e}_r\ .
\end{equation}
This equation makes clear that the angular momentum in direction of the rotor's symmetry axis is entirely carried by the internal degree of freedom.

The introduction of the total angular momentum helps in the solution of the quantum mechanical problem posed by the Lagrangian~(\ref{LagK}). After performing a Legendre transformation one arrives at the Hamiltonian and re-expressing the canonical momenta in terms of angular momentum facilitates the quantization and solution, see \citep{papenbrock2020}.  

\subsection{Leading-order Hamiltonian}
Performing a Legendre transformation of the Lagrangian~(\ref{LagK}), with the gauge potential~(\ref{Atot}), yields the Hamiltonian. It is useful to replace the canonical momenta by the angular momentum~(\ref{Ibody}), and one obtains
\begin{equation}
\label{HamK}
    H_{\rm LO} = \sum_{Kq} E_{|K|q} \hat{\psi}_{Kq}^\dagger \hat{\psi}_{Lq}+\frac{g^2}{2C_0} \left(\hat{K}_{x'}^2 + \hat{K}_{y'}^2\right) + \frac{\mathbf{I}^2-\hat{K}_{z'}^2}{2C_0} + \frac{g}{C_0} \left(I_{+1}\hat{K}_{-1} + I_{-1}\hat{K}_{+1} \right) \ . 
\end{equation}
Here, 
\begin{equation}
    \hat{\mathbf{K}} \equiv \sum_{KLq} \hat{\psi}_{Kq}^\dagger \hat{\mathbf{S}}\hat{\psi}_{Lq} 
\end{equation}
is the spin operator for the internal degrees of freedom, and all components (Cartesian $K_{x'}$, $K_{y'}$, $K_{z'}$ or spherical $K_{\pm 1}$, $K_{0}$) are with respect to the body-fixed coordinate system. We have quantized the field $\psi_{Kq}$ such that $\hat{\psi}_{Kq}^\dagger$ creates the internal  mode with quantum numbers $(Kq)$, i.e. $\hat{\psi}^\dagger_{Kq}|0\rangle = |Kq\rangle$. 

The Hilbert space is spanned the products $D^I_{M,-K}(\phi,\theta,0)|Kq\rangle$ of Wigner $D$ matrices and intrinsic states. The first three terms in the Hamiltonian~(\ref{HamK}) are diagonal in this basis. It is the last term that mixes basis states whose $K$ quantum numbers differ by one unit. Combinations 
\begin{equation}
 D^I_{MK}\ket{-K} + (-1)^{I+K}D^I_{M-K}\ket{K}
\end{equation}
are invariant under ${\cal R}$ symmetry~\citep{bohr1975}, i.e. under rotations of the nucleus by $\pi$ around  an axis perpendicular of the symmetry axis. 

\section{Odd-mass nuclei} 
\label{sec:odd}

We now review applications \citep{papenbrock2020,alnamlah2021,alnamlah2022} of effective theories to odd-mass nuclei. In leading order, these theories recover the particle-rotor model. 

\subsection{Effective theory for a nucleon coupled to a rotor}

\citet{alnamlah2021} focused on a single pair of time-reversed states for the nucleon and derived a leading-order Lagrangian similar to equation~(\ref{LagK}) with $\mathbf{A}=\mathbf{A}_{\rm uni}$ from equation~(\ref{gaugepot}). All other rotor-nucleon couplings were ordered by the number of powers of the angular velocity $\mathbf{v}$. Thus, the contribution from the non-Abelian gauge potential~(\ref{Anon}) was treated as a next-to-leading-order correction.

The inclusion of rotor-nucleon couplings containing up to three powers of \textbf{v} resulted in the energy spectrum
\begin{equation}
\begin{aligned}
    E_{\rm N^2LO} = E_{|K|} +& A_K \left[I(I+1)-K^2\right] \\
    +& a_{1/2}(-1)^{I+1/2} \left(I+\frac{1}{2}\right) \delta_{K}^{1/2} \\
    +& b_{1/2}(-1)^{I+1/2} I(I+1) \left(I+\frac{1}{2}\right)\delta_{K}^{1/2} \\
    +& a_{3/2}(-1)^{I+3/2} \left(I-\frac{1}{2}\right) \left(I+\frac{1}{2}\right) \left(I+\frac{3}{2}\right) \delta_{K}^{3/2}.
\end{aligned}
\label{eq:odd_spec}
\end{equation}
This expression, derived within an effective theory, agrees with the corresponding one in the textbook~\citep{bohr1975} when the latter is limited to odd-mass nuclei. 
In equation~(\ref{eq:odd_spec}), the signature splitting for $K=3/2$ bands enters with a strength determined by $a_{3/2}$. The term proportional to $b_{1/2}$ corrects the staggering in $K=1/2$ bands introduced at leading order. Adding terms containing four powers of \textbf{v} yields the band-dependent correction $B_K I^2(I+1)^2$ to the rotor spectrum.

\begin{figure}[b]
    \centering
    \includegraphics[width=0.475\textwidth]{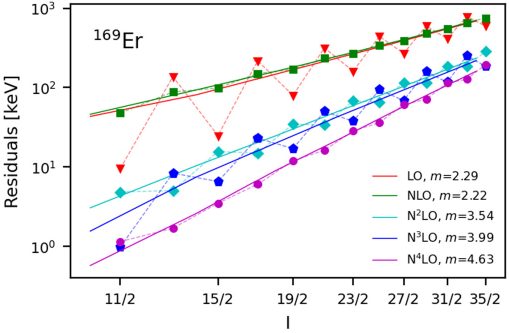}
    \includegraphics[width=0.475\textwidth]{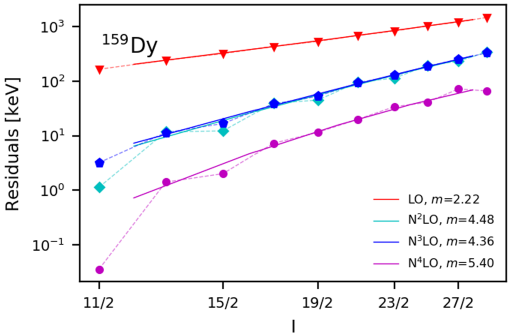}
    \caption{Order-by-order description of the ground-state rotational band in $^{169}{\rm Er}$ (left panel) and $^{159}{\rm Dy}$ (right panel). The residuals, i.e. the absolute difference between theory and data is shown as a function of angular momentum $I$ (data points connected by dashed lines). The full straight lines are proportional to $I^m$ with $m$ as indicated. Figures taken from arXiv:2011.01083 with permission from the authors, see also \citep{alnamlah2021}.}
    \label{fig:oddresiduals}
\end{figure}

Figure~\ref{fig:oddresiduals} shows the systematic improvement of the effective theory describing the $I^\pi=1/2^-$ and $I^\pi=3/2^-$ ground-state bands in $^{169}{\rm Er}$ and $^{159}{\rm Dy}$, respectively.
The residuals, i.e. the difference between theory and data are shown on a double log scale versus angular momentum $I$ (which is approximately the square root of the energy) for various orders of the theory. Full lines show average trends.  A significant reduction in the residuals takes place at even orders ($\rm N^2LO$ and $\rm N^4LO$ in the figure), while the energy staggering is reduced at odd orders (NLO and $\rm N^3LO$ in the figure). Straight full lines are proportional to $I^m$ with $m$ as indicated. We see that the  effective theory fulfills a power counting. However, the power counting for effective Lagrangians is not simply in powers of the angular velocity $\mathbf{v}$, but rather -- at a given order -- one needs to include all terms up to and including powers of $\mathbf{v}^2$.

\begin{figure}[h]
    \centering
    \includegraphics[width=0.9\textwidth]{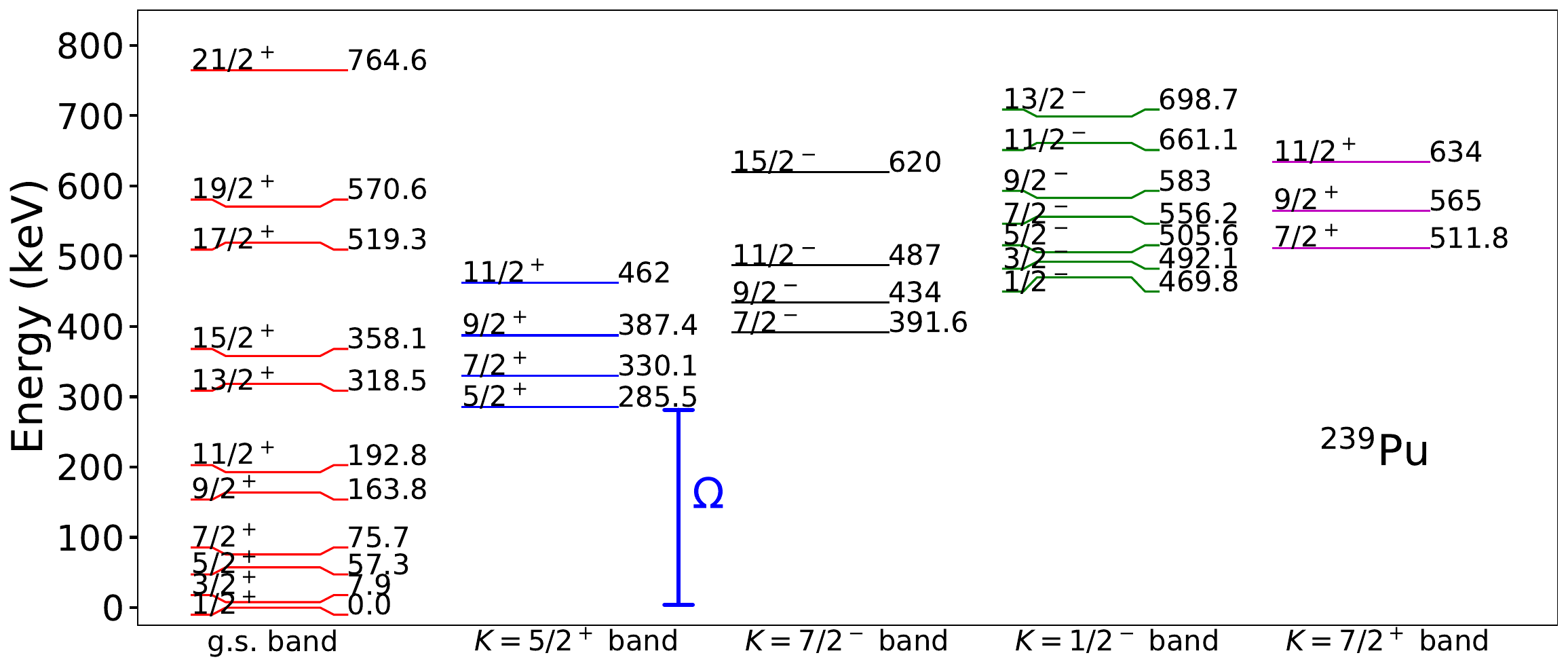}
    \caption{Levels of $^{239}$Pu at low energies, sorted in rotational bands. The fermion-energy scale $\Omega$ separates band heads. The low-energy scale $\xi$ (not shown) is the energy difference between levels in a rotational band. Figure taken from arXiv:2005.11865 with permission from the authors, see also \citep{papenbrock2020}.}
    \label{fig:239Pu}
\end{figure}

\citet{papenbrock2020} derived the leading-order Lagrangian~(\ref{LagK}) which included the total gauge potential~(\ref{Atot}),  and applied it the odd mass nuclei $^{239}$Pu and $^{187}$Os. The interest was in studying how the non-Abelian gauge potential~(\ref{Anon}) couples band heads whose spins differed by one unit in angular momentum. Figure~\ref{fig:239Pu} shows the low-energy spectrum of $^{239}$Pu. Levels can be sorted into rotational bands with band heads as indicated. Also visible is the separation of scale between the fermion energy scale $\Omega$ and the smallest energy scale, $\xi$, that measures energy differences in a rotational band.  

In their approach to $^{239}$Pu, \citet{papenbrock2020} focused on the ground-state band. Then, the fermion degrees of freedom that enter are $\psi_{Kq}$ with $K=\pm 1/2$ and parity $q=+$. Thus, only a single pair of fermion states in time-reversed states $|K=1/2\rangle$ and  $|\overline{K}\rangle\equiv|K=-1/2\rangle$ contribute. 

\begin{figure}[h]
    \centering
    \includegraphics[width=0.5\textwidth]{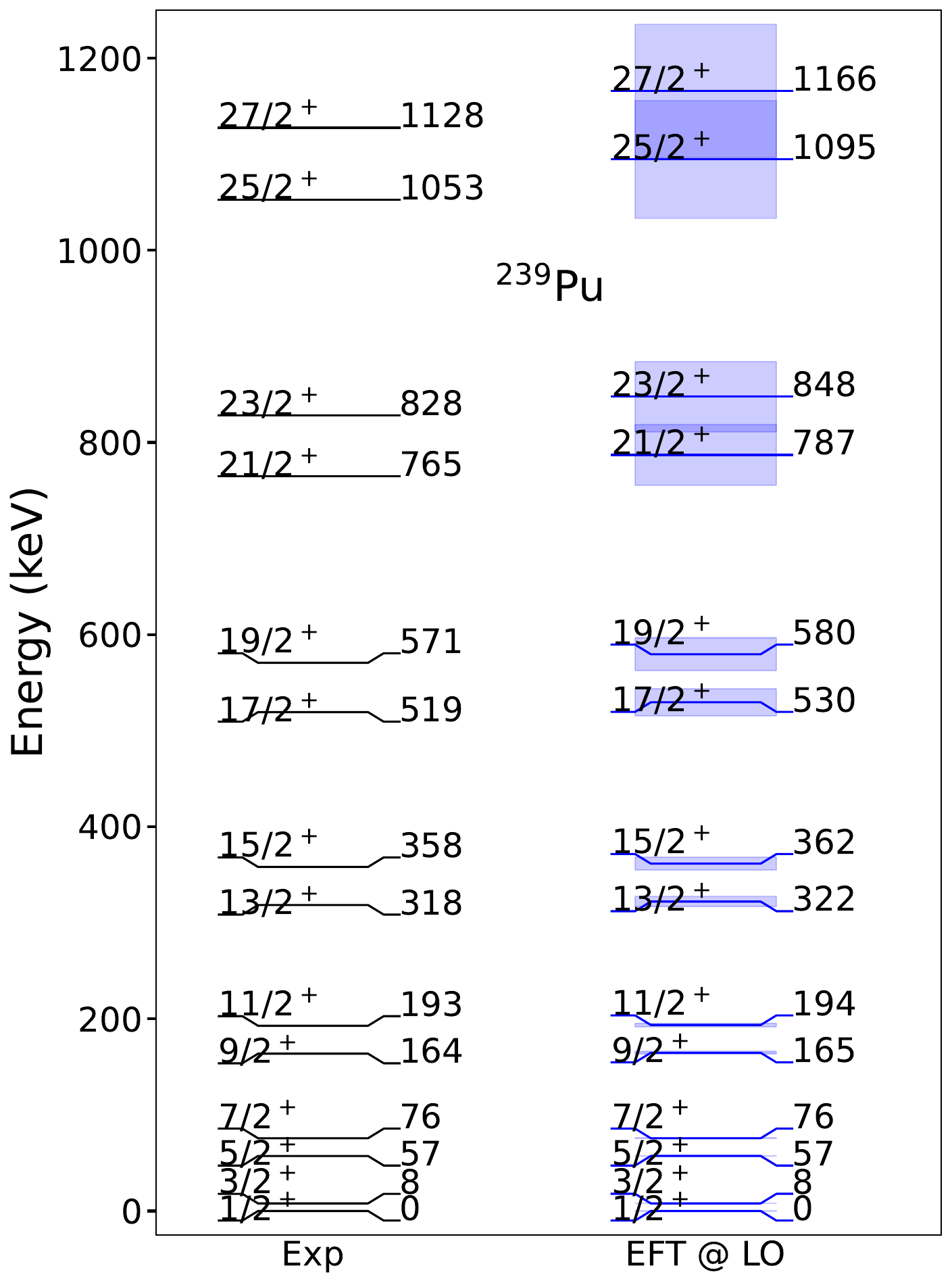}
    \caption{Spectrum of the ground-state band of $^{239}$Pu computed with the effective theory at leading order ``EFT @LO''  and compared to experimental data. The blue areas are uncertainty estimates based on neglected higher orders. Figure adapted from~\citep{papenbrock2020}.}
    \label{fig:239PuEFT}
\end{figure}

The leading-order Hamiltonian can be written as 
\begin{equation}
    H_{\rm LO} = E_{1/2+} + \frac{g^2}{2C_0} \left(\mathbf{K}^2 - K_{z'}^2\right) + \frac{1}{2C_0} \left(\mathbf{I}^2-K_{z'}^2\right) + \frac{g}{C_0} \left(I_{+1}K_{-1} + I_{-1}K_{+1} \right) \ . 
\end{equation}
Here, $I_{\pm 1}$ are spherical components of the total angular momentum~(\ref{Ibody}) in the body-fixed system, and $K_{\mu}$ are the spherical components of the spin operator in the body-fixed system and act on the fermion states. The resulting energy spectrum is
\begin{equation}
\begin{aligned}
    E_{\rm LO}(I,K) = E_{|K|} &+ A_0 \left[I(I+1)-K^2\right]+ a_{1/2}(-1)^{I+1/2}\left(I+\frac{1}{2}\right)\delta_{K}^{1/2},
\end{aligned}
\label{eq:elo_odd}
\end{equation}
where $A_0\equiv(2C_0)^{-1}$ and $a_{1/2}\equiv gC_0^{-1}\braket{1/2|K_{+1}|\overline{1/2}}$. The last term in~\eqref{eq:elo_odd}, known as signature splitting, accounts for the energy staggering in $K=1/2$ bands. Results are shown in figure~\ref{fig:239PuEFT}. The results ``EFT @ LO''  were obtained from adjusting $A_0$ and $a_{1/2}$ to $^{239}$Pu; here $E_{1/2+}$ is fixed such that the spectrum starts at zero energy. Uncertainty estimates (shown as blue bands) reflect estimated contributions from terms beyond the Lagrangian~(\ref{LagK}). 

\begin{figure}[h]
    \centering
    \includegraphics[width=0.9\textwidth]{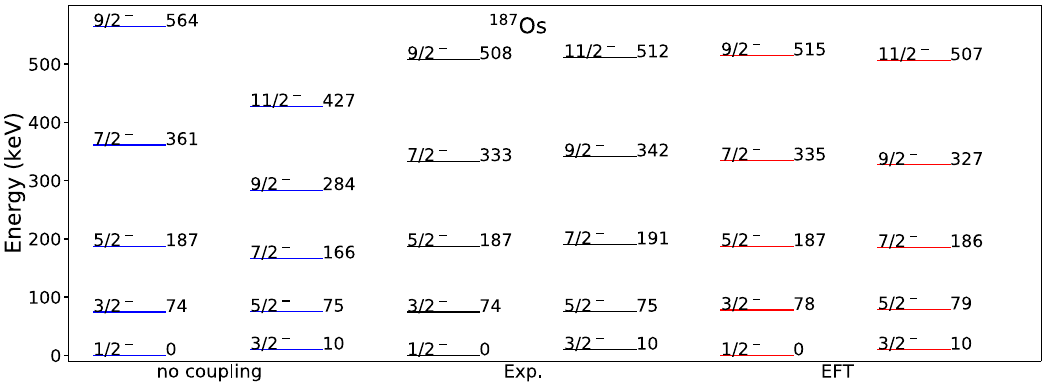}
    \caption{Lowest negative parity bands in $^{187}{\rm Os}$ shown in black at the center. The independent description of each band (blue lines) disagrees with data (black lines). Results allowing band mixing due to the leading Coriolis term (red lines) consistently describe the spectrum. Figure taken from arXiv:2005.11865 with permission from the authors, see also \citep{papenbrock2020}.}
    \label{fig:187Os}
\end{figure}

The nucleus $^{187}$Os exhibits two low-lying rotational bands whose band heads are close in energy and differ by one unit of angular momentum. This is shown in the center of figure~\ref{fig:187Os}. The non-Abelian gauge potential~(\ref{Anon}) will couple these bands. 
The simultaneous description of two low-lying bands is achieved through the diagonalization of the matrix spanned by the nucleon states $\ket{K}$ and $\ket{K+1}$ (and their time reversed partners)  using $K=1/2$. The resulting spectrum can be expressed in terms of the energies~(\ref{eq:elo_odd}) as
\begin{equation}
\begin{aligned}
    E_{\rm LO}(I,K,K+1) = \frac{1}{2} &\left[ E_{\rm LO}(I,K) + E_{\rm LO}(I,K+1) \right] \\
    \pm \frac{1}{2} \Big\{ &\left[ E_{\rm LO}(I,K) - E_{\rm LO}(I,K+1) \right]^2 + \Tilde{g}^2 [I(I+1) - K(K+1)] \Big\}^{1/2} \ .
\end{aligned}
\end{equation}
Here $\Tilde{g}\equiv2gC_0^{-1}\braket{K|K_{-1}|K+1}$, and the sign of the second term is chosen to obtain the energies of the corresponding bandheads when $g=0$. Due to the mixing, the angular momentum projection onto the  rotor's symmetry axis is no longer a good quantum number. This implies a triaxial deformation of the nucleon-rotor system. Figure~\ref{fig:187Os} compares the theoretical results (red levels, labeled ``EFT'') to experimental data.
One might also attempt an independent description of both bands, using the energies~(\ref{eq:elo_odd}). The results, show as blue levels with the label ``no coupling,'' deviate immediately from data above the lowest three and two levels that were adjusted to data in the bands based on the spins $1/2^-$ and $3/2^-$, respectively.

\subsection{Bayesian analysis of the effective theory}

The results reviewed in the previous section were obtained by adjusting low-energy constants to data from the lowest states in the band (or bands) of interest. Such an approach runs the risk of fine-tuning these parameters. \citet{alnamlah2022} used Markov chain Monte Carlo sampling to produce joint posterior distributions of the low energy constants and other parameters encoding the systematic expansion of predicted energies for $K=1/2$ bands. This allowed them to study how the values of low-energy constants change as the number of levels used for their extraction and/or the order of the effective theory is increased.

Uncertainty quantification is now standard for effective field theories and based on the works \citep{schindler2009,cacciari2011,furnstahl2014c,bagnaschi2015,wesolowski2016,wesolowski2019, wesolowski2021}. One uses Bayes' theorem to derive the joint posterior distribution of low-energy constants $\mathbf{a}_k$ and parameters $I_{\rm b}$, $\Bar{c}_{\rm even,odd}$ given the data $\mathbf{y}_{\rm exp}$ and assumptions $P^*$ as
\begin{equation}
\begin{aligned}
    {\rm pr}\left( \mathbf{a}_k, I_{\rm b}, \Bar{c}_{\rm even}, \Bar{c}_{\rm odd} | \mathbf{y}_{\rm exp}, P^* \right) &\propto {\rm pr}\left( \mathbf{y}_{\rm exp} | \mathbf{a}_k, I_{\rm b}, \Bar{c}_{\rm even}, \Bar{c}_{\rm odd}, P^* \right) \\
    &\times {\rm pr}\left( \mathbf{a}_k | I_{\rm b}, \Bar{c}_{\rm even}, \Bar{c}_{\rm odd}, P^* \right) \\
    &\times {\rm pr}\left( I_{\rm b} | \Bar{c}_{\rm even}, \Bar{c}_{\rm odd}, P^* \right) \\
    &\times {\rm pr}\left( \Bar{c}_{\rm even} | P^* \right) {\rm pr}\left( \Bar{c}_{\rm odd} | P^* \right) \ .
\end{aligned}
\label{eq:ppd}
\end{equation}
Here the vector $\mathbf{a}_k$ contains the low-energy constants at order $k$, $I_{\rm b}$ is the breakdown spin (i.e. the high spin at which the effective theory breaks down), and $\Bar{c}_{\rm even,odd}$ are the characteristic sizes of low-energy constants entering at even and odd orders of equation~(\ref{eq:odd_spec}), respectively. The vector $\mathbf{y}_{\rm exp}$ contains the data about energy levels, and $P^*$ is any information one has about the model. The posterior predictive distribution of any low-energy constant or parameter can be obtained from the joint posterior~\eqref{eq:ppd} via marginalization, i.e. by integrating over all other low-energy constants and parameters. 

The posterior predictive distribution of any observable $\mathcal{O}$ can be written as
\begin{equation}
    {\rm pr}\left( \mathcal{O} | \mathbf{y}_{\rm exp}, P^* \right) = \int d\boldsymbol{\theta}_k \delta\left(\mathcal{O}-\mathcal{O}(\boldsymbol{\theta}_k) \right) {\rm pr} \left( \boldsymbol{\theta}_k | \mathbf{y}_{\rm exp}, P^* \right) \ .
\end{equation}
Here $\boldsymbol{\theta}_k$ collectively represents the low-energy constants and parameters describing the observable at order $k$.

The first function in the right-hand side of equation ~\eqref{eq:ppd} is the likelihood of the data given the low-energy-constants and parameters entering their description at order $k$. It is from a sum $\Sigma=\Sigma_{\rm exp}+\Sigma_{\rm theo}$ of an experimental and a theoretical covariance matrix. The latter is written as $\Sigma_k\equiv \delta\mathbf{y}_k\otimes\delta\mathbf{y}_k$ and contains the uncertainties in predicted energies due to the truncation of the effective theory at order $k$. It estimates omitted terms from the orders $k+1$ to a maximum order $k_{\rm max}$ as
\begin{equation}
    (\delta\mathbf{y}_k)_i \equiv A_0 \sum_{l=k+1}^{k_{\rm max}} \Bar{c}_l \frac{P_l(I_i)}{I_{\rm b}^{l-1}} \ .
    \label{eq:kmax}
\end{equation}
Here, $P_l(I)$ is a power of $I(I+1)$ and $(I+1/2)I(I+1)$ for even and odd contributions, respectively. The form of this expansion is based on generalizing equation~(\ref{eq:odd_spec}) to higher orders and -- via the coefficients $\overline{c}_l$ and the breakdown spin $I_{\rm b}$ -- implements the power counting of the effective theory. The experimental covariance matrix is assumed to be diagonal  in terms of the experimental errors $\delta\mathbf{y}_{\rm exp}$
\begin{equation}
    {\rm pr}\left( \mathbf{y}_{\rm exp} | \boldsymbol{\theta}_k, P^* \right) = \sqrt{\frac{1}{(2\pi)^m|\Sigma|}} \exp \left( -\frac{1}{2} \mathbf{r}^{\rm T} \Sigma^{-1} \mathbf{r} \right).
\end{equation}
Here, $m$ is the number of observables entering the analysis and $\mathbf{r}\equiv\mathbf{y}_{\rm exp}-\mathbf{y}_k$ is the residual.

The second factor in equation~\eqref{eq:ppd} is the prior distribution of the low-energy constants given the parameters encoding the systematic expansion of the effective theory. \citet{alnamlah2022} assumed Gaussian priors with zero mean and standard deviation $\sigma_n = A_0\Bar{c}_l W^{n-1}$ for all low-energy constants except $E_K$, for which they allowed a larger one (as its size is not determined by the power counting). This allowed them  to write the prior for the low-energy constants as
\begin{equation}
    {\rm pr}\left( \mathbf{a}_k | W, \Bar{c}_{\rm even}, \Bar{c}_{\rm odd}, P^* \right) = \frac{1}{\sqrt{2\pi}\Bar{E}} \exp\left( -\frac{E_K^2}{2\Bar{E}^2} \right) \prod_{n=1}^k \frac{1}{\sqrt{2\pi}\sigma_n} \exp\left( -\frac{(\mathbf{a}_k)_n^2}{2\sigma_n^2} \right).
\end{equation}
Assuming a flat prior distribution between zero and a maximum $W_{\rm cut}$ for the inverse breakdown spin, and low-energy constants drawn from independent scaled-inverse-$\chi^2$ priors allowed them to extract low-energy constants. This line of arguments shows that the Bayesian approach allows one state and to quantify one's assumptions and (via marginalization) arrive at posterior predictive  distributions.

\begin{figure}[htb]
    \centering
    \includegraphics[width=0.475\textwidth]{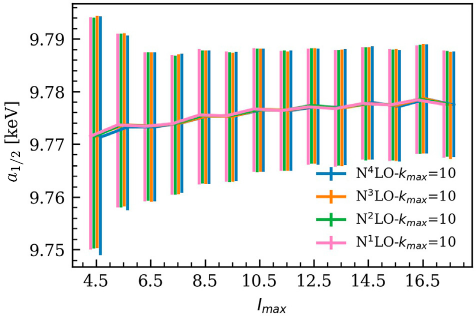}
    \includegraphics[width=0.47\textwidth]{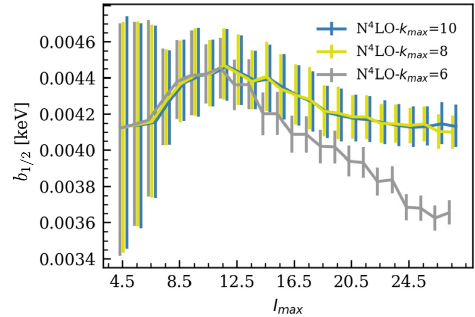}
    \caption{Posterior predictive distributions for low-energy constants describing $K=1/2$ bands. The posterior predictive distributions for $a_{1/2}$ describing the ground-state band in $^{169}{\rm Er}$ (left) were constructed at multiple orders, increasing the number of data points while fixing $k_{\rm max}$. Those for $b_{1/2}$ describing the ground-state band in $^{239}{\rm Pu}$ were constructed at fix order while increasing both the number of data points and $k_{\rm max}$. Figures adapted from~\citep{alnamlah2022}.}
    \label{fig:extraction}
\end{figure}
Results are shown in figure~\ref{fig:extraction}. The left panel shows 68\% degree-of-belief intervals for the low-energy constant $a_{1/2}$ of equation~\eqref{eq:odd_spec} describing the nucleus $^{169}$Er as a function of the maximum spin $I_{\rm max}$ used in the analysis. The different vertical present different orders of the effective theory and levels were included up to $I_{\rm max}$ while fixing $k_{\rm max}=10$ in equation~\eqref{eq:kmax}. 

The right panel of the figure shows similar intervals from posterior distributions for the low-energy constant $b_{1/2}$ describing the ground-state band in $^{239}{\rm Pu}$. Here the order of the effective theory was kept fixed (at N$^4$LO) while increasing the maximum power $k_{\rm max}$ of omitted terms in the theoretical uncertainty. These results indicated that considering multiple omitted contributions in the approximation for the theoretical error (i.e. a sufficiently high $k_{\rm max}$ is required for the stable extraction of low-energy constants.

\section{Electromagnetic transitions}
\label{sec:electro}
The effective theories described in previous sections recover the expressions for the energy spectra predicted by well-known collective models. While such models describe electromagnetic transitions between states in the same rotational band properly, they often struggle to accurately describe the much weaker transitions between states belonging to different bands. 

It is easy to see why in-band transitions are strong and inter-band transitions are weak: The theory of the deformed quadrupole oscillator~\citep{bohr1975} yielded rotational excitations on top of vibrational band heads. In-band transitions naturally are large (because a quadrupole operator has an order-one matrix element between spherical harmonics that differ by two units of angular momentum). In contrast,  inter-band transition vanish in leading order because they connect states that differ in their number of vibrational quanta. Only higher-order terms mix different vibrational states (see  Section~\ref{sub:vib-rot}) and can thereby yield finite transition matrix elements. The approach via effective field theory~\citep{coelloperez2015} revealed that the transition operator also has a systematic expansion. This makes it possible to accurately describe the faint inter-band transitions -- although at the expense of an additional low-energy constant. 

\subsection{In-band electric quadrupole transitions}

\citet{coelloperez2015} computed electric quadrupole  transitions in deformed nuclei with the framework of an effective theory.  They used both minimal coupling and operators involving the electric field to arrive at their results. In an effective theory one needs to write down all operators that involve electromagnetic couplings and apply an ordering scheme, i.e. the power counting, to them. 
Minimal coupling alone does not provide one with an unambiguous approach~\citep{jenkins2013}.

The systematic construction~\citep{coelloperez2015} of the interaction between the rotor and the electromagnetic field started by requiring invariance under local gauge transformations of the rotor wave function
\begin{equation}
    \Psi(\phi,\theta) \rightarrow e^{iq\lambda(\phi,\theta)} \Psi(\phi,\theta) \ .
\end{equation}
Here $q$ is the effective  charge. Gauge invariance is achieved by minimal coupling $-i\nabla_\Omega \to -i\nabla_\Omega-q\mathbf{A}$ where $\mathbf{A}(\Omega)=\nabla_{\Omega}\lambda(\Omega)$ is the vector potential representing the photon. (The angular derivative $\nabla_\Omega$ was defined in equation~(\ref{nablaOmega}). In  the rotor Hamiltonian~\eqref{HamLO} one then employs
\begin{equation}
    \mathbf{I} \rightarrow \mathbf{I} - q \mathbf{e}_r \times \mathbf{A}(\Omega) \ .
\end{equation}
Inserting this into the leading-order Hamiltonian~\eqref{HamLO} of the rotor then generates the leading electromagnetic coupling as
\begin{align}
    H_{\rm LO}^{\mathbf{A}} =& -\frac{q}{2C_0} \left[ (\mathbf{I}\cdot(\mathbf{e}_r\times\mathbf{A}) + (\mathbf{e}_r\times\mathbf{A})\cdot\mathbf{I} \right] \nonumber \\
    =& - i\frac{q}{2} [H_{\rm LO}, \mathbf{A}\cdot\mathbf{e}_r] - i\frac{q}{2} \left(\mathbf{A} \cdot H_{\rm LO}\mathbf{e}_r - \mathbf{e}_r \cdot H_{\rm LO} \mathbf{A} \right) \ .
\label{eq:HALO}
\end{align}
Here, the last line casts this interaction into a form that is attractive for the computation of transition matrix elements. Taking $\mathbf{A}$ as a plane wave with amplitude $|\mathbf{A}|$, polarization $\mathbf{e}_z$, and momentum $\mathbf{k}=k\mathbf{e}_x$ then yields the quadrupole component $\mathbf{A}^{(2)}=|\mathbf{A}|\mathbf{e}_zkr\cos\theta\sin\phi$.  When employed into~\eqref{eq:HALO} one finds
\begin{align}
    H_{\rm LO}^{A(2)} =& -i\frac{q}{2}\left[H_{\rm LO}, \mathbf{A}^{(2)}\cdot\mathbf{e}_r\right] 
\end{align}
with $Y_{IM}\equiv Y_{IM}(\theta,\phi)$ a spherical harmonic. The corresponding transition matrix elements
\begin{equation}
    M_{\rm LO}(E2,i\rightarrow f) = -i\frac{qw}{2} \braket{f|\mathbf{A}^{(2)}\cdot\mathbf{e}_r|i},
\end{equation}
depend on the energy difference 
\begin{equation}
    \omega\equiv E_{\rm LO}(I_f)-E_{\rm LO}(I_i)
\end{equation}
between the states. The matrix element is calculated by integrating products of spherical harmonics over the unit sphere.

Gauging the next-to-leading contribution to the Hamiltonian~\eqref{genHam} yields the interaction term
\begin{equation}
    H_{\rm NLO}^{\mathbf{A}} = 2 g_2 C_0 \left( \mathbf{I}^2 H_{\rm LO}^{\mathbf{A}} + H_{\rm LO}^{\mathbf{A}} \mathbf{I}^2 \right),
\end{equation}
where the low-energy constant $g_2$ must be adjusted to data. The corresponding correction to the transition matrix elements
\begin{equation}
    M_{\rm NLO}(E2,i\rightarrow f) = 2g_2C_0 \left[I_f(I_f+1) + I_i(I_i+1) \right] M_{\rm LO}
\end{equation}
is thus expected to be $\varepsilon^2$ times smaller than the leading contribution.

Besides the minimal couplings, the effective theory must also consider  nonminimal couplings. The simplest of these  is
\begin{equation}
    H_{\rm LO}^{\mathbf{E}} = q d_0 \mathbf{E}\cdot\mathbf{e}_r \ .
\end{equation}
For the electric field corresponding to the plane wave vector potential, $\mathbf{E}=i\omega\mathbf{A}$, this coupling yields a contribution to the transition matrix elements equivalent to that from the leading minimal coupling, and is thus accounted for when fitting the effective charge or quadrupole moment of the rotor. The nonminimal couplings entering at next-to-leading order are
\begin{equation}
    H_{\rm NLO}^{\mathbf{E}} = -\frac{qd_1}{2} \left( \mathbf{I}^2 \mathbf{E}\cdot\mathbf{e}_r + \mathbf{E}\cdot\mathbf{e}_r \mathbf{I}^2 \right) - \frac{qd_2}{2} \left( \mathbf{E} \cdot \mathbf{I}^2\mathbf{e}_r + \mathbf{e}_r\cdot\mathbf{I}^2\mathbf{E} \right).
\end{equation}

\begin{figure}[h]
    \centering
    \includegraphics[width=0.475\textwidth]{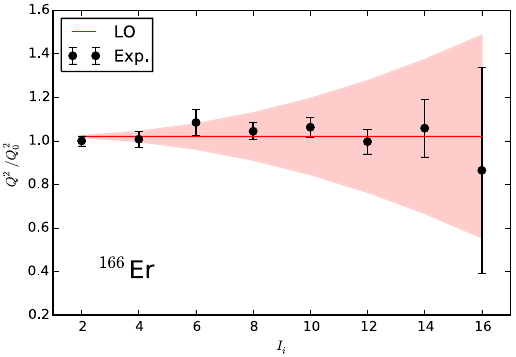}
    \includegraphics[width=0.475\textwidth]{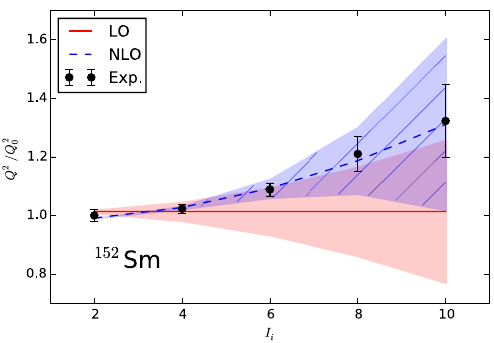}
    \caption{Normalized squared \textit{E}2 transition moments for decays in $^{166}{\rm Er}$ (left panel) and $^{152}{\rm Sm}$ (right panel). The expansion parameter $\varepsilon\sim 1/22$ in $^{166}{\rm Er}$ yields a consistent description of the moments at leading order within uncertainty estimates (red shaded area). For the transitional nucleus $^{152}{\rm Sm}$, next-to-leading corrections (shown as a blue line with blue uncertainty estimates) are required to describe the moments up to $I_i=10$. The energy of this state is taken as the breakdown scale of the theory. Figures taken from arXiv:1502.04405 with permission from the authors, see also \citep{coelloperez2015}.}
    \label{fig:intraq2}
\end{figure}

At next-to-leading order the \textit{E}2 strength for in-band transitions from initial spin $I_i$ to fianl spin $I_f$ is 
\begin{equation}
    B_{\rm NLO}(E2,i\rightarrow f) = a Q_0^2 \left(C_{I_i020}^{I_f0}\right)^2 \left[1 + \frac{b}{a}I_i(I_i-1) \right] \ .
\end{equation}
Here $Q_0$ is the effective quadrupole moment, and we used the short hands $a\equiv 1+4g_2C_0+2d_1$ and $b=4g_2C_0+2d_1+2d_2$. This result, of course, is well known  \citep{bohr1975}. For plots of results it is profitable to remove the Clebsch-Gordan coefficient (because it simply is a geometric factor), and instead look at the  squared \textit{E}2 transition moment
\begin{equation}
    Q^2(E2,i\rightarrow f) = \frac{B(E2,i\rightarrow f)}{\left(C_{I_i020}^{I_f0}\right)^2} \ . 
\label{eq:transmoment}
\end{equation}
In leading order, $Q^2=Q_0^2$ and smaller angular-momentum dependent corrections arise at next-to-leading order.  Figure~\ref{fig:intraq2} shows the normalized squared \textit{E}2 transition moments for decays in $^{166}{\rm Er}$ (left) and $^{152}{\rm Sm}$ (right).
The nucleus $^{166}$Er is a textbook example of a rigid rotor, and the experimental moments (black circles) are close to the leading-order approximation (red line).  Uncertainties estimates reflect the size of omitted contributions. The moments for the transitional nucleus $^{152}{\rm Sm}$ exhibit visible  deviations from the constant behavior. These are accounted for at next-to-leading order (blue line) up to $I_i=10$ at which the leading and next-to-leading uncertainties (red and blue shaded areas) are comparable, signaling the breaking point of the theory.

\subsection{Electric quadrupole transitions between bands}
\label{subsec:interband}

The leading-order spectra resembling  a rigid rotor get modified by higher-order contributions to the Lagrangian that include more than two powers of the velocity \textbf{v}. The terms relevant for out discussion were presented in section~\ref{sub:vib-rot} and are contained in the Lagrangian~(\ref{eq:LN2LO}) at next-to-next-to-leading order. These yield the contribution 
\begin{equation}
    H_{\rm N^2LO} = -\frac{1}{2C_0^2} \left( C_\beta \varphi_0 \mathbf{p}_{\Omega\gamma}^2 + C_\gamma \varphi_2 \mathbf{p}_{\Omega\gamma}^T \Gamma \mathbf{p}_{\Omega\gamma} \right)
\end{equation}
to the Hamiltonian. Here, 
\begin{equation}
    \Gamma \equiv \left( \begin{array}{cc}
    \cos{2\gamma} & \sin{2\gamma} \\
    \sin{2\gamma} & -\cos{2\gamma} 
    \end{array} \right)  \ ,
\end{equation}
corrects the spectra at second-order in perturbation theory. Here, $C_\beta$ and $C_\gamma$ are expected to scale as $\xi^{-1/2}$.

\begin{figure}[htb]
    \centering
    \includegraphics[width=0.7\linewidth]{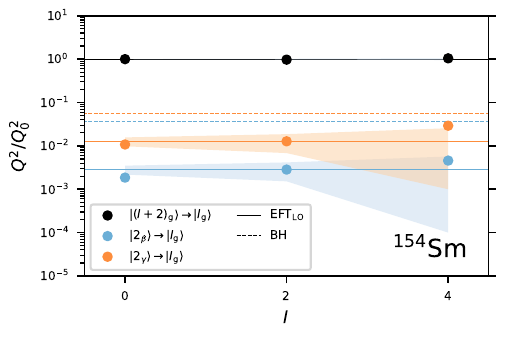}
    \caption{Electric quadrupole transition moments for $^{154}{\rm Sm}$. Transition moments for decays within the ground state band are shown in black. Moments describing decays from the $\beta$ and $\gamma$ bands are shown in blue and orange, respectively. In contrast to the adiabatic Bohr Hamiltonian (BH, dashed lines), the effective theory (solid lines) consistently describes experimental inter-band transition moments (circles) within estimated uncertainties (shaded bands). This improvement comes at the expense of an additional low-energy constant for each band.}
    \label{fig:intermoments}
\end{figure}

Coupling to electromagnetic fields then yields the contribution to the Hamiltonian
\begin{equation}
    H_{\rm N^2LO}^{\mathbf{A}} = i\frac{qC_\beta}{2C_0^2} \varphi_0 \left( \mathbf{A} \cdot \mathbf{p}_{\Omega\gamma} + \mathbf{p}_{\Omega\gamma} \cdot \mathbf{A} \right) + i\frac{qC_\gamma}{2C_0^2} \varphi_2 \left( \mathbf{A}^{\rm T} \Gamma \mathbf{p}_{\Omega\gamma} + \mathbf{p}_{\Omega\gamma}^{\rm T} \Gamma \mathbf{A} \right) \ . 
\end{equation}
The leading \textit{E}2 strengths for transitions from a state with spin $I_i$ in the  excited $\beta$ and $\gamma$ band to a state with spin $I_f$ in the ground-state band are
\begin{equation}
    B(E2,I_i\to I_f) = \frac{(2K+1)C_{\beta,\gamma}^2}{2C_0^2\omega_K} Q_0^2 \left( C_{I_iK2-K}^{I_f0} \right)^2 
\end{equation}
for $K=0$ and 2, respectively. 
The corresponding squared \textit{E}2 transition moment is
\begin{equation}
    Q^2(E2, I_i, K_i\rightarrow I_f) = B(E2, I_i\rightarrow I_f)/ \left(C_{I_iK_i2-K_i}^{I_f0}\right)^2 \ .
\end{equation}
It depends on the effective quadrupole moment $Q_0$ that was adjusted to the in-band transitions, and the low-energy constants $C_\beta$ and $C_\gamma$. In principle, one can adjust these two unknowns to spectra. However, as there are other terms that enter at that order, it is simpler to adjust $C_\beta$ and $C_\gamma$ to a single inter-band transition from the respective band. Then the theory predicts other inter-band transitions. The result is that the effective theory describes inter-band transitions much more accurately than traditional collective models. This is shown in figure~\ref{fig:intermoments} for  \textit{E}2 transition strengths in $^{152}{\rm Sm}$. 

We note that the values $C_\beta=0.091~{\rm keV}^{-1/2}$ and $C_\gamma=0.127~{\rm keV}^{-1/2}$ were adjusted to the transitions from the $2^+$ states in the $\beta$ and $\gamma$ bands, respectively,  to the $2^+$ state in the ground-state band. They are of natural size, i.e. of order  $\xi^{-1/2}=0.110~{\rm keV}^{-1/2}$. While the Bohr model also predicts that the inter-band transitions are much smaller than the in-band transitions, it fails to accurately predict their magnitude. The effective theory in contrast, also expands the transition operator and thereby is able to deliver precision and accuracy. 

\section{Triaxial deformation} 
\label{sec:triax}
Triaxial deformation is an evergreen in nuclear structure physics. When \citet{davydov1958} proposed the triaxial rotor model  most deformed  nuclei were thought to be axially symmetric in their ground states. The more recent computations of binding energies within a triaxially deformed finite droplet model confirmed this result~\citep{moller2006}: There are only a few smaller regions on the nuclear chart that exhibit static triaxial deformation, and the corresponding gain in binding energy is small, ranging from tens to hundreds of keV. This is in contrast to triaxial deformation in excited states which is much more abundant~\citep{frauendorf1997,frauendorf2001}.
The challenge in identifying triaxial deformation in nuclear ground states is as follows. Relying only on spectral signatures, such as a low-lying $K^\pi=2^+$ band, can be misleading because that can also be accommodated in nuclei with axial symmetry~\citep{bohr1975}. Stronger evidence comes from observations of a large number of gamma-ray transitions and use of the Kumar Cline sum rules~\citep{kumar1972,cline1986}. In recent years, increased gamma-ray tracking capabilities made it possible to better study triaxial deformation, and that has led to a renewed interest, see~\citep{doherty2017,ayangeakaa2019} for examples. 

In this Section we review effective theories that deal with triaxial deformation~\citep{chen2017,chen2018,chen2020}. The orientation of a potato is determined by three Euler angles that specify the body-fixed coordinate system, and the effective theory exhibits both richer and simpler aspects than in the axially symmetric case. As we will see, the number of low-energy coefficients increases significantly but the theory becomes simpler because there is no covariant derivative. Within the collective model by \citeauthor{bohr1953}, triaxial deformation is usually associated with a corresponding body-fixed potential that depends on $\beta $ and $\gamma$ degrees of freedom~\citep{fortunato2005}. However, the situation is more complicated and interesting. Coriolis forces (or gauge potentials) induce deviations from axial symmetry. That is emphasized in the following section~\ref{sub:tri-coriolis}. After that clarification we review effective theories of static triaxial deformation in section~\ref{sub:tri-static}.  

\subsection{Breaking of axial symmetry through gauge potentials or Coriolis forces}
\label{sub:tri-coriolis}
Axial symmetry implies that the angular momentum projection $K$ onto the body-fixed symmetry axis is a conserved quantity. The gauge potentials or Coriolis forces we reviewed in section~\ref{sec:odd} clearly mix $K$ quantum numbers. The simplest example is an odd-mass nucleus with a $|K|=1/2$ ground-state rotational band. Here, the eigenstates are superpositions of $K=\pm 1/2$ states and axial symmetry is clearly broken: The angular momentum projection onto the symmetry axis is not any more  conserved but only its magnitude. Similarly, in odd-mass nuclei Coriolis forces can mix low-lying bands whose $K$ quantum numbers differ by one unit~\citep{stephens1975}. Again, this also breaks axial symmetry. Similar statements apply to odd-odd nuclei~\citep{jain1989,jain1998}. Higher-order Coriolis forces could also mix bands that differ by more than one unit in $K$ quantum numbers. However such forces are expected to be small (based on the power counting for axially deformed nuclei). Thus, Coriolis forces are not strong enough to explain deviations from axial symmetries in even-even nuclei.

\subsection{Static triaxial deformation}
\label{sub:tri-static}

The effective theory has been derived for triaxially deformed nuclei in the papers~\citep{chen2017,chen2018,chen2020}. In this case one deals with the complete breaking of SO(3) symmetry and the coset space then becomes SO(3). This space is naturally parameterized by three Euler angles, and the effective theory can be constructed using the coset approach~\citep{chen2017}. We briefly sketch that derivation in what follows.

The time-dependent Euler angles  $(\alpha,\beta,\gamma)$ parameterize the rotation operator
\begin{equation}
    R(\alpha,\beta,\gamma)= e^{-i\alpha J_z} e^{-i\beta J_y} e^{-i\gamma J_z} \ , 
\end{equation}
and one computes the expression 
\begin{equation}
    R^{-1} \partial_t R = a_x J_x + a_y J_y + a_z J_z 
\end{equation}
via the Baker-Campbell-Hausdorff expansion. This defines functions 
\begin{equation}
\begin{aligned}
    a_x &= - \dot\alpha \sin\beta \cos\gamma +\dot\beta \sin\gamma \ , \\
    a_y &= \dot\alpha \sin\beta \sin\gamma +\dot\beta \cos\gamma \, \\
    a_z &= \dot\alpha \cos\beta + \dot\gamma \ .
\end{aligned}
\end{equation}
These expressions are the building blocks for the effective theory. As the Euler angles define the co-rotating, body-fixed coordinate system these building blocks are invariant under rotations. The leading-order Lagrangian becomes
\begin{equation}
    L_{\rm LO} = {1\over 2} \left( {\cal J}_1 a_x^2 + {\cal J}_2 a_y^2 + {\cal J}_3 a_z^2 \right) \ .
\end{equation}
Here, ${\cal J}_i$ with $i=1,2,3$ are low-energy constants and have to be adjusted to data. Note that no mixed terms (such as $a_xa_y$ appear because one works with in a coordinate system spanned by the the principal axes. A Legendre transform  yields the Hamiltonian. As in the axially symmetric case, one introduces  the angular momentum and rewrites the Hamiltonian in terms of it. This yields
\begin{equation}
    H_{\rm LO} = \frac{I_1}{2{\cal J}_1}  + \frac{I_2}{2{\cal J}_2}  + \frac{I_3}{2{\cal J}_3}    \ .
\end{equation}
This is the Hamiltonian of the asymmetric rotor model~\citep{davydov1958,wood2004}. One sees that the asymmetric rotor depends on three low-energy constants (instead of one for the axially symmetric case). 

\citet{chen2017} also discussed some subleading corrections by including $a_k^4$ terms ($k=x,y,z)$; however, this is not a complete next-to-leading-order calculation as mixed terms such as $a_x^2a_y^2$ or $a_x^2a_ya_z$ (and many other combinations) can also enter. The reason such terms might appear is as follows: Higher orders implicitly include effects from neglected degrees of freedom that are active beyond the cutoff scale. This introduces non-rigid rotation, and might perturb the principal axes.  
We see that an effective theory for triaxial rotation involves a larger number of low-energy constants than in the axially symmetric case.

\begin{figure}[h]
\centering
\includegraphics[width=0.8\textwidth]{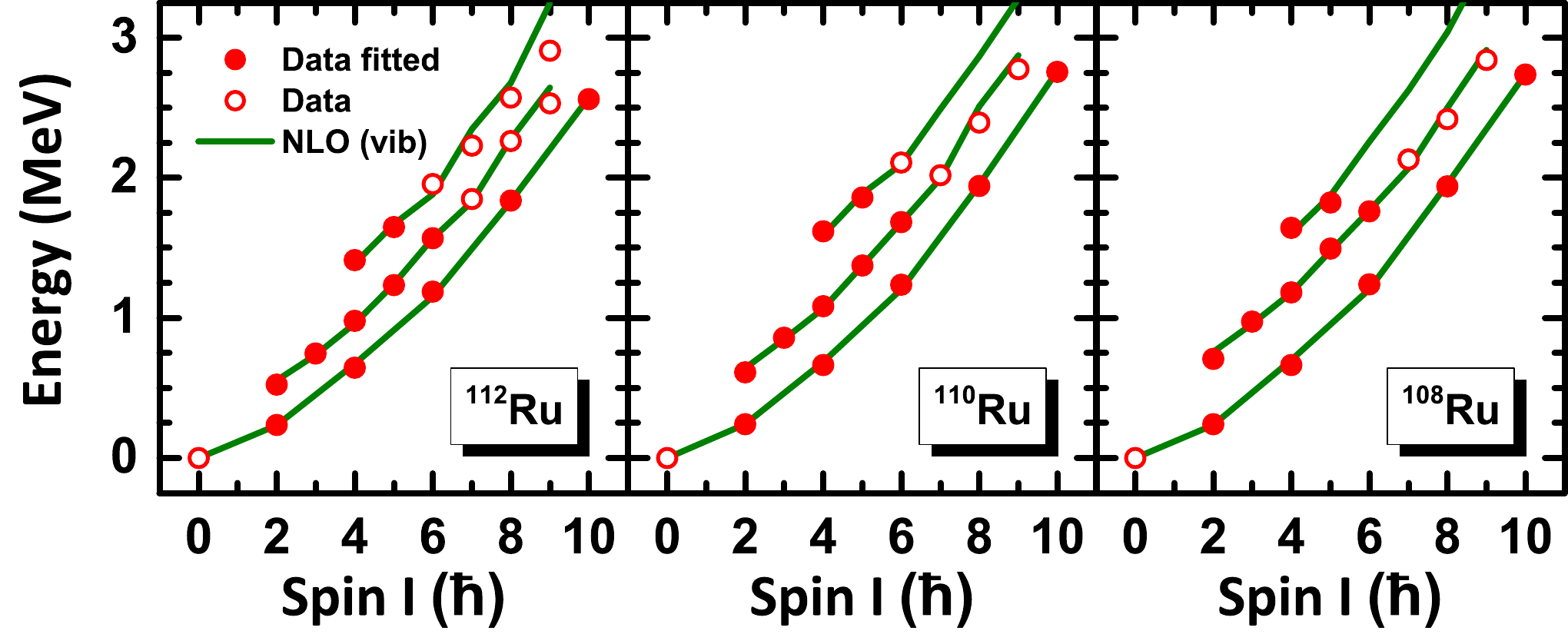}
\caption{Energies of states belonging to three low-lying rotational bands in ruthenium isotopes, shown as a function of spin. The full points show data used in adjusting the low-energy constants of the effective theory. The green lines are predictions and can be compared to the hollow data points.  Figure taken from arXiv:1707.04353 with permission from the authors, see also \citep{chen2018}.}
\label{fig:chen2018}
\end{figure}

\citet{chen2018} added vibrational degrees of freedom to the triaxial rotor.  The effective Hamiltonian is quite general and also contains the familiar  collective model~\citep{bohr1975}. At leading order there are 12 low-energy constants that need to be adjusted to data. Figure~\ref{fig:chen2018} shows how three rotational bands in $^{108,110,112}$Ru are described by the effective theory. A hallmark of static triaxial deformation is that the $2_2^+$ and the $4_1^+$ states are close in energy. 

We finally remark that one could also have followed a more geometric approach. For a tri-axially deformed object, an orthogonal basis is spanned by the eigenvectors of the tensor of the moment of inertia. Using Euler angles $(\alpha,\beta,\gamma)$ the eigenvectors are the basis vectors 
\begin{equation}
\label{gamma_angle}
\begin{aligned}
    \mathbf{e}_1(\alpha,\beta,\gamma) & \equiv +\mathbf{e}_\theta(\alpha,\beta)\cos\gamma + \mathbf{e}_\phi(\alpha,\beta)\sin\gamma \ , \\
    \mathbf{e}_2(\alpha,\beta,\gamma) & \equiv -\mathbf{e}_\theta(\alpha,\beta)\sin\gamma + \mathbf{e}_\phi(\alpha,\beta)\cos\gamma \ , \\
    \mathbf{e}_3(\alpha,\beta,\gamma)&\equiv \mathbf{e}_r(\alpha,\beta) \ .
    \end{aligned}
\end{equation}
We note that for $\gamma=0$, this dreibein becomes the body-fixed coordinate system for axially symmetric nuclei. Indeed, $\gamma$ describes rotations around the axis $\mathbf{e}_r$ (which were not allowed for an axially symmetric nucleus).

\section{Pairing rotations} 
\label{sec:pair}
Nuclei are BCS superconductors except in the rare cases of doubly-magic nuclei where both, protons and neutrons, fully occupy a shell~\citep{bohr1958, migdal1959, brink-broglia2005}. Semi-magic nuclei (where either protons or neutrons fully occupy a shell) consist of one superfluid; other open-shell nuclei exhibit two interacting superfluids. A BCS superconductor breaks particle number and a corresponding U(1) phase symmetry in its mean field. This allows one to construct effective Lagrangians for such systems. The resulting collective excitations are known as pairing rotations~\citep{broglia1968,bohr1969}, and they connect the binding energies of nuclei that differ by pairs of nucleons. This Section describes the simple physics of these systems and summarizes some of the results of reference~\citep{papenbrock2022}.

We build on the example of a semimagic nucleus discussed in section~\ref{sub:superfluid}. The mean field state, i.e. the Hartree-Fock-Bogoliubov state, breaks the global U(1) gauge symmetry of the neutron number. Similar comments apply to {\it ab initio} computations that start from such a reference state (but fail to restore the broken symmetry). So, $G={\rm U}(1)$, the subgroup of the preserved symmetry is (only) the identity, $S=I$, and the coset $G/S\sim {\rm U}(1)$. This is the unit circle in the complex plane and it is parameterized by the gauge angle $\alpha$. 

The elements of the coset are the global gauge transformations~\eqref{gaugetransf}. In analogy to the case of deformed nuclei in section~\ref{sub:coset-def}, the relevant quantity for the construction of effective Lagrangians results from computing 
\begin{equation}
    g^{-1}(\alpha) \partial_t g(\alpha) = -i \dot{\alpha}\hat{N} \ .
\end{equation} 
Thus, effective Lagrangians are functions containing $\dot{\alpha}$. The simplest Lagrangian is 
\begin{equation}
    L={a\over 2} \dot{\alpha}^2 +n_0\dot{\alpha} \ .
\end{equation}
Here, $a$ and $n_0$ are low-energy constants. As $g(\beta)g(\alpha)=g(\alpha+\beta)$, under a gauge transformation by the angle $\beta$, we see that $\alpha\to\alpha+\beta$, and this is indeed a nonlinear realization of the U(1) symmetry. As $\alpha$ is a cyclic variable, the canonical momentum (i.e the number of pairs)  
\begin{equation}
    p_\alpha \equiv = \frac{\partial L}{\partial \dot{\alpha}} 
\end{equation}
is a conserved quantity. The Hamiltonian is 
\begin{equation}
\label{hamLOpair}
    H=\frac{(p_\alpha-n_0)^2}{2a}
\end{equation}
We quantize $p_\alpha =-i\partial_\alpha$.  Thus, the eigenfunctions of the Hamiltonian~(\ref{hamLOpair}) are the wave functions \begin{equation}
\psi_n(\alpha)=\frac{1}{\sqrt{2\pi}}e^{i\alpha n}
\end{equation}
with integer $n$. The corresponding energies are 
\begin{equation}
\label{pair-rot}
\varepsilon_n=\frac{(n-n_0)^2}{2a} \ , 
\end{equation}
and they describe a pairing rotational band. Here, the low-energy constant $a$ is the pairing rotational moment of inertia. The size of $1/(2a)$ ranges from about 1~MeV in $N=82$ isotones to 0.4~MeV in tin isotopes to 0.2~MeV in lead nuclei~\citep{papenbrock2022}.

\begin{figure}[htb]
    \centering
    \includegraphics[width=0.5\textwidth]{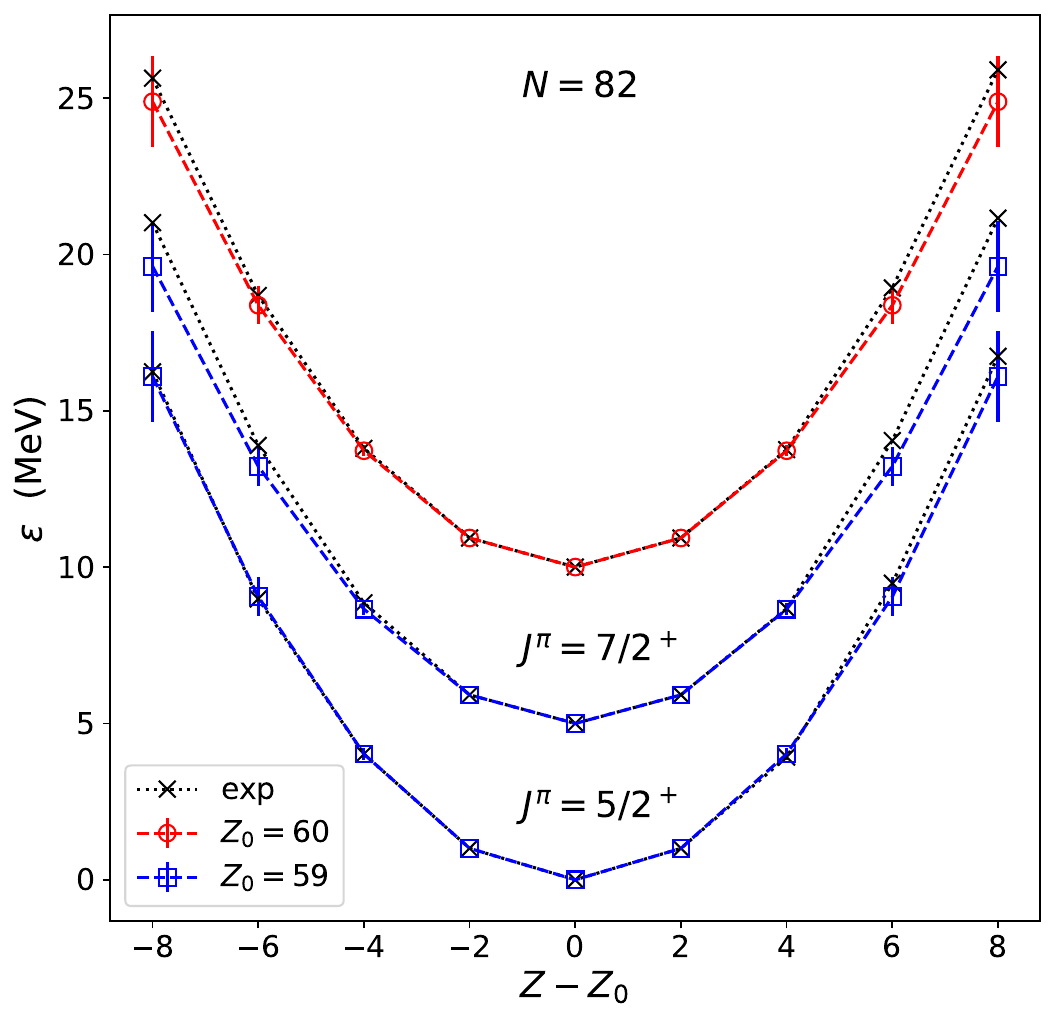}
    \caption{Pairing rotational bands, i.e. the quadratic term when expanding energies in terms of powers of proton pairs away from a reference nucleus, in odd (blue squares) and even
      (red circles) $N=82$ isotones.  The two bands connecting $J^\pi=5/2^+$ and $7/2^+$ states, respectively, in odd nuclei use Pr
      ($Z=59$) as the reference nucleus. The  band connecting ground-states of even nuclei uses Nd ($Z=60$) as the
      reference nucleus. Data are shown as black
      crosses. In each band, the central three points are adjusted to
      data. Bands are shifted by multiples of 5~MeV. Figure taken from arXiv:2202.13146 with permission from the author, see also \citep{papenbrock2022}.}
    \label{fig:N82}
\end{figure}

The theory describes the ground-state energies of semi-magic even-even nuclei (i.e. isotopes of tin and lead or the isotones with neutron number $N=82$). It also applies to odd semi-magic nuclei provided one focuses on states with the same spin. The breakdown scale is set by the maximum number of pairs within such a chain of isotopes or isotones, because superfluidity breaks down in doubly-magic nuclei. This can be translated into an energy scale based on equation~(\ref{pair-rot}). Examples of pairing rotational bands are shown in figure~\ref{fig:N82} for the $N=82$ isotones. Here, $\varepsilon$ denotes the quadratic term when expanding energies of lowest-lying states with spin and parity $J^\pi=0^+$, $5/2^+$ and $7/2^+$ around a nucleus with $N=82$ and $Z=Z_0$. Data is described accurately within theoretical uncertainty estimates.

The effective theory was also extended to the case of two interacting superfluids, as is appropriate for open-shell nuclei~\citep{papenbrock2022}. This introduces two gauge angles, one for protons and one for neutrons, as the dynamical degrees of freedom. The leading-order Lagrangian is a quadratic form in the angular velocities. One then finds that the pairing-rotational bands (parabolas) in semi-magic nuclei are replaced by pairing elliptical paraboloids. The effective theory accurately describes data for doubly-open shell nuclei. We note that pairing
rotational moments of inertia were also studied in Hartree-Fock Bogoliubov computations~\citep{hinohara2015,hinohara2016,hinohara2018}. 

Of course, any more microscopic models of nuclear superfluidity will yield pairing rotational bands. However, the derivation (or computation) of such bands is more complicated in those approaches than in the effective theory, see, e.g.,  \citep{bes1970,broglia1973,broglia2000,potel2011,potel2013,potel2017}. Thus, the potential appeal of the effective theory lies in its clarity and simplicity.  

\section{\bf Nuclear vibrations} 
\label{sec:vib}
The effective theories related to to emergent symmetry breaking are particularly simple and attractive because the pattern of the symmetry breaking identifies the space of the relevant degrees of freedom and also constrains their couplings to other degrees of freedom. Most nuclear models, however, are based on linear realizations of the symmetry, and one might wonder if not all successful models are in some sense leading-order effective theories. In nuclei, there are also collective vibrations at low energies and these are the lowest-lying excitations in nuclei near shell closures. In this section, we review  effective theories for nuclear vibrations.

The purpose for this endeavor is two-fold. \citet{bohr1952} already described the low-lying excitations of nuclei in terms of vibrations and rotations of a liquid droplet, parameterizing its surface in terms of quadrupole deformations. There have been long-standing arguments if, or to what extend, such quadrupole oscillations are realized in nuclei~\citep{bes1969,garrett2010,stuchbery2022}. An approach within an effective theory allows one to present predictions  with quantified uncertainties. Thus, one can unambiguously address the question about quadrupole vibrations~\citep{coelloperez2015b}. 
Second, the material reviewed in this Section is relevant for computations of nuclear matrix elements that govern (neutrino-less) double beta decay. Most candidate nuclei are nearly spherical, and the effective theories ability to quantify uncertainties is a big boon. Those developments will be reviewed in Section~\ref{sec:weak}.

\subsection{Even-even nuclei}
\label{sub:ee-vib}
A large number of even-even nuclei in the vicinity of the shell closures exhibit spectra that at low energies resemble that of the harmonic quadrupole oscillator. The lowest $2^+$ excitation in these systems, characterized as a quadrupole vibration of the nuclear surface by nuclear collective models~\citep{bohr1952, rowe2004, rowe2009, rowe2010}, is frequently followed by positive-parity states with spins $I=0,2,4$ at energies close to that of a double quadrupole excitation. Although in some nuclei states with $I=0,2,3,4,6$ that could be identified as three-phonon excitations have been observed, the appearance of states with octupole and/or single-particle character at the corresponding energy level make it impossible to picture atomic nuclei as pure quadrupole oscillators.

These observations suggest that quadrupole degrees of freedom capture the low-energy physics of nuclear vibrations. The quadrupole creation and annihilation operators, denoted by $d^{\dagger}_\mu$ and $d_\mu$ with $\mu\in\{-2,-1,0,1,2\}$, create and annihilate quadrupole phonons and fulfill the commutation relations
\begin{equation}
    \left[d_\mu,d^\dagger_\nu\right] = \delta_{\mu}^{\nu}.
\end{equation}
While the creation operators behave like the components of a rank-two tensor, the annihilation operators do not and one defines the annihilation tensor as
\begin{equation}
    \Tilde{d}_\mu = (-1)^{\mu}d_{-\mu} \ .
\label{eq:annihilation_tensor}
\end{equation}

The leading-order effective Hamiltonian
is 
\begin{equation}
    \hat{H}_{\rm LO} = \omega \left(d^\dagger \cdot \Tilde{d}\right) = \omega\hat{N},
\label{eq:vib_HLO}
\end{equation}
where the dot product is defined as usual for spherical tensors~\citep{varshalovich1988}. The operator $\hat{N}$ counts the number of phonons in a state, and $\omega$ is a low-energy constant that must be adjusted to data. Eigenstates can be labeled in terms of the number of phonons $N$, the SO(5) angular momentum analog $v$, the radial quantum number $\nu$, and the spin $I$ and spin projection $M$~\citep{rowe2010}.
The spectra of the leading-order  Hamiltonian consist of multiplets with energies
\begin{equation}
    E_{\rm LO}(N) = N\omega \ .
    \label{eq:elo}
\end{equation}

Deviations from this harmonic behavior arise from higher-order contributions to the effective Hamiltonian. The power counting works as follow. At leading order the the matrix element of as single quadrupole operator must scale as 
\begin{equation}
    \braket{d} \sim N^{1/2} \ .
\end{equation}
One assumes that the theory breaks down at the $N_b$-phonon level, i.e., the breakdown energy is $\Lambda_b=N_b\omega$. At the breakdown scale one cannot  distinguish states with $N_b\pm1$ phonons. This then implies that Hamiltonian terms $C_m d^m$ containing $m$ powers of a quadrupole operator have a low-energy constant $C_m$ that scales as   
\begin{equation}
    C_m \sim \left(\omega/\Lambda_b\right)^{m/2} \ .
\end{equation}
Thus, the small expansion parameter of the effective theory is $\varepsilon=\omega/\Lambda_b=N_b^{-1}$. Usually, the breakdown of harmonic vibrations is at the three-phonon level and the expansion parameter is only about $1/3$.   

The power counting suggests that the next-to-leading order Hamiltonian contains operators consisting of three quadrupole fields. However, such operators are off-diagonal and enter only via second-order perturbation theory. This then introduces   operators consisting of four quadruple fields as the next-to-leading order contribution and one has
\begin{equation}
    \hat{H}_{\rm NLO} = g_N \hat{N}^2 + g_v \hat{\Lambda}^2 + g_I \hat{I}^2 \ .
\label{eq:hvib_nlo}
\end{equation}
Here the operators $\hat{N}^2$, $\hat{\Lambda}^2$ and $\hat{I}^2$ are defined as in \citep{rowe2010}, and $g_N$, $g_v$ and $g_I$ are low-energy constants. The next-to-leading contribution to the energy is
\begin{equation}
    E_{\rm NLO}(N,v,I) = g_N N^2 + g_v v(v+3) + g_I I(I+1) \ .
    \label{eq:enlo}
\end{equation}

\subsection{Uncertainty quantification for even-even spectra}
\citet{coelloperez2015b} presented computations with quantified uncertainties  
and used Bayesian methods for that purpose. These tools are particularly suited for effective theories where a power counting informs one about uncertainties coming from a truncation at a given order~\citep{schindler2009,cacciari2011,furnstahl2014c,bagnaschi2015}. Making assumptions about the distribution of low-energy constants one can then marginalize over the parameters of such distribution functions and make quantitative predictions. 

The power counting yields an expansion of any observable in terms of the small expansion parameter $\varepsilon$. For energies of the quadrupole oscillator we have 
\begin{equation}
\label{Epower}
    E(N,v,\nu,I,M) = \omega\sum_i c_i(N,v,\nu,I,M)\varepsilon^i.
\end{equation}
Here, the leading-order energy scale $\omega$ has been factored out and the coefficient functions $c_i$ in equation~(\ref{Epower}) are dimensionless. For nuclear vibrations, for instance, one can easily relate them to the low-energy coefficients that define the leading-order and next-to-leading-order energies~(\ref{eq:elo}) and (\ref{eq:enlo}), respectively. Any practical calculation truncates the sum~(\ref{Epower}) at a finite $i=k$, and one is thus interested in  the contributions from the first $M$ neglected terms (in units of $\omega$)
\begin{equation}
\Delta_k^{(M)} = \sum_{i=k+1}^{k+M} c_i \varepsilon^i \ .
\label{eq:residual}
\end{equation}
Of course, one does not know the size of the neglected $c_i$. This is where Bayesian methods come in. In an effective theory, one can make reasonable (and testable) assumptions about these coefficients. The assumption of naturalness, for instance, implies that all coefficients are of order one. Thus, any probability distribution for these unknown coefficients should have a characteristic scale that can be sampled from a log-normal distribution
\begin{equation}
\label{lognorm}
    {\rm pr}(c) = \frac{1}{\sqrt{2\pi}\sigma c} e^{-\frac{\log^2{c}}{2\sigma^2}} \ .
\end{equation}
Here,  $\sigma$ a hyperparameter that defines intervals $[e^{-n\sigma},e^{n\sigma}]$ containing 68, 95 and 99 percent of the distribution for $n=1,2,3$, respectively. Having set the overall scale $c$, one next has to make assumptions about the prior ${\rm pr}(c_i|c)$. \citet{furnstahl2014c} showed that the specific form of that prior, e.g. being a Gaussian with width $c$ or a uniform distribution between $\pm c$ (``hard-wall'' prior), has only small impacts on degree-of-belief intervals for $\Delta_k^{(M)}$; however one needs to make a choice  to be quantitative. Reference~\citep{coelloperez2015b} presents results from both Gaussian priors
\begin{equation}
\label{Gauss}
    {\rm pr}(c_i|c) = \frac{1}{\sqrt{2\pi}c}e^{-\frac{c_i^2}{2c^2}},
\end{equation}
and hard-wall priors choices. Finally, the last assumption was that the coefficients $c_i$ in equation~(\ref{eq:residual}) are independent from each other. 

According to Bayes' theorem, the distribution for the omitted $M$ higher-order contributions given the first $k$ expansion coefficients then becomes~\citep{furnstahl2014c}
\begin{equation}
    p_M(\Delta|c_0,\ldots,c_k) = \frac{\int_0^\infty
    dc\ {\rm pr}(c) p_M(\Delta|c) \prod_{i=0}^k {\rm pr}(c_i|c)}
    {\int_0^\infty dc\ {\rm pr}(c) \prod_{i=0}^k {\rm pr}(c_i|c)} \ .
\end{equation}
This expression is easily understood:  The
numerator reflects  how the uncertainty
depends on the expansion coefficients given our assumptions about the the prior ${\rm pr}(c)$, and  the denominator is a normalization. 
\citet{coelloperez2015b} derived the simple expression 
\begin{equation}
    p_M(\Delta|c) = \frac{1}{2\pi} \int_{-\infty}^\infty
    dt\ e^{i\Delta t} \prod_{n=k+1}^{k+M} \int_{-\infty}^\infty
    dc_n\ {\rm pr}(c_n|c) e^{-i\varepsilon^n c_n t} \ .
\end{equation}
Thus, one only has to compute the Fourier transform of the prior ${\rm pr}(c_i|c)$ and then perform a single integration over products of Fourier transforms. In the end, one finds with the Gaussian prior~(\ref{Gauss})
\begin{equation}
    p_M(\Delta|c_0,\ldots,c_k) = \frac{1}{\sqrt{2\pi}q}\frac{\int_0^\infty dx\ x^{k+1} e^{-\frac{\log^2x}{2\sigma^2}} e^{-\frac{\gamma^2+\Delta^2/q^2}{2}x^2}}{\int_0^\infty dx\ x^k e^{-\frac{\log^2x}{2\sigma^2}} e^{-\frac{\gamma^2x^2}{2}}} \ .
    \label{eq:deltadistg}
\end{equation}
Here $q^2=\sum_{i=k+1}^{k+M}\varepsilon^{2i}$ and $\gamma^2=\sum_{i=0}^k c_i^2$. 

\begin{figure}[h]
    \centering
    \includegraphics[width=0.7\textwidth]{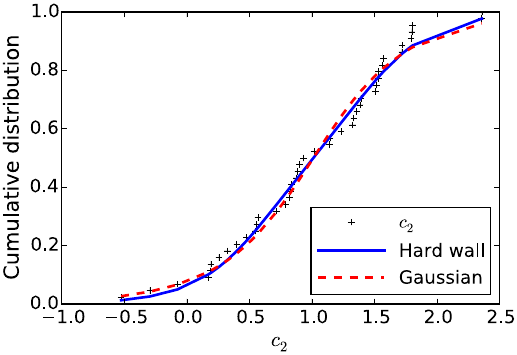}
    \caption{Cumulative distribution for the $c_2$ coefficient. The distribution from next-to-leading calculations for an ensemble of nuclei (black crosses) is in good agreement with the cumulative distributions of the hard wall (blue line) and Gaussian (red dashed line) priors. Figure taken from arXiv:1510.02401 with permission from the authors, see also \citep{coelloperez2015b}.}
    \label{fig:c2dist}
\end{figure}

The beauty of Bayesian uncertainty quantification is that one can test the assumptions being made. Of course, each effective theory, when evaluated at increasingly higher orders, has just one coefficient $c_n$ at order $n$. However, looking at an ensemble of nuclei, \citet{coelloperez2015b} adjusted $c_2$ to data and compared the resulting distribution function with that from a Gaussian and hard-wall prior , i.e. $\int dc \, {\rm pr}(c_2-\overline{c_2}|c) {\rm pr}(c)$ with an appropriately shifted mean $\overline{c_2}$. The comparison is shown  in figure~\ref{fig:c2dist}. They used $\sigma=\log{(3/2)}$ in the prior~(\ref{lognorm})
This shows that assumptions about the distribution functions for priors are consistent with data from an ensemble of nuclei with similar structure and masses.

\begin{figure}[h]
    \centering
    \includegraphics[width=0.7\textwidth]{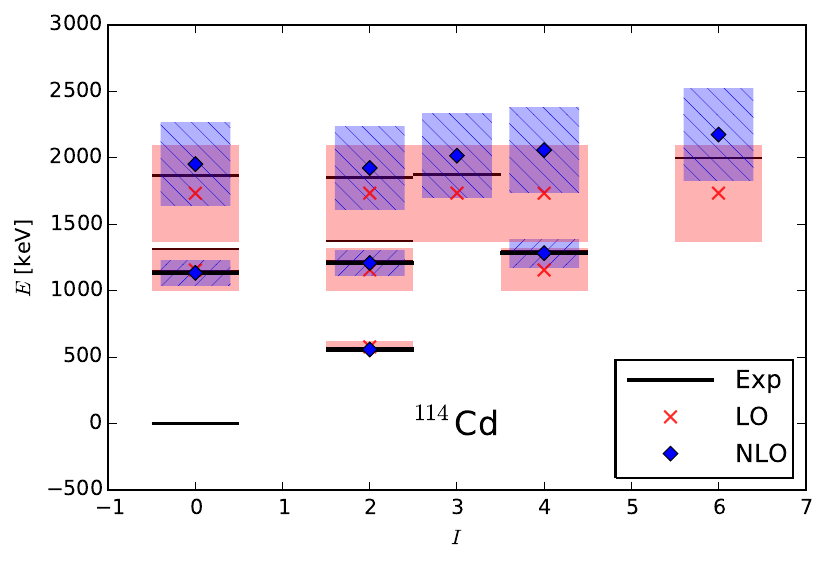}
    \caption{The low-lying states of $^{114}{\rm Cd}$ (black lines) compared with predictions from the effective theory at leading (red crosses) and next-to-leading orders (blue diamonds). The thick lines mark states up to the two-phonon level. The shaded areas mark 68\% degree-of-belief intervals at leading (red) and next-to-leading order. Figure taken from arXiv:1510.02401 with permission from the authors, see also \citep{coelloperez2015b}}
    \label{fig:cd114}
\end{figure}

Figure~\ref{fig:cd114} compares the spectrum of $^{114}$Cd with results from the effective theory. At leading order, the one-phonon state is adjusted to data; at next-to-leading order the two-phonon states are also adjusted.
Shaded areas mark 68\% degree-of-belief intervals (i.e. ``one-sigma'' uncertainties if we dealt with a Gaussian distribution, which we are not). The uncertainties decrease with increasing order of the effective theory. At the three-phonon level, uncertainties for next-to-leading calculations are similar to those from leading order, signaling the breakdown of the effective theory. This is also seen in the proliferation of states at the three-phonon level. 

\begin{figure}[h]
    \centering
    \includegraphics[width=0.7\textwidth]{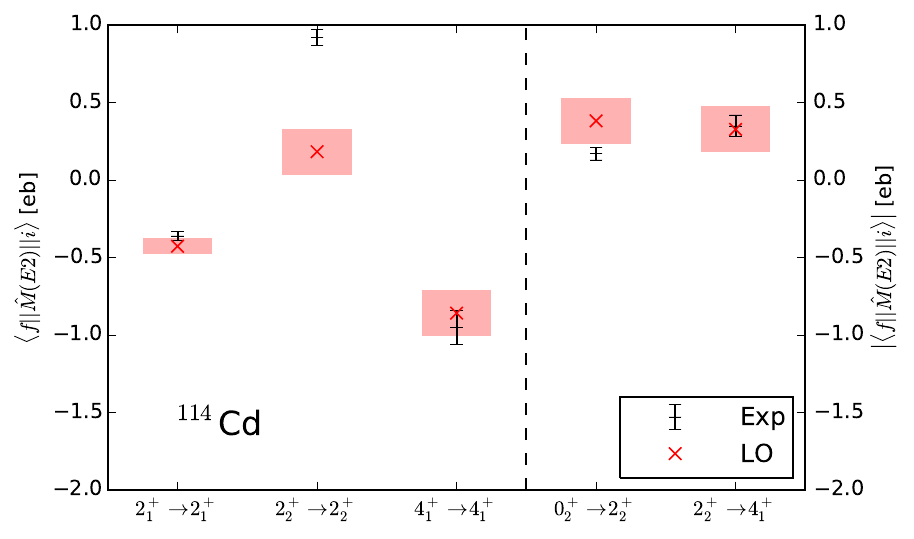}
    \caption{Electric quadrupole moments (left part) and transition matrix elements (right part) between states as indicated. Figure taken from arXiv:1510.02401 with permission from the authors, see also \citep{coelloperez2015b}.}
    \label{fig:cd114be2}
\end{figure}

\citet{garrett2010} particularly questioned the interpretation of cadmium isotopes as (an)harmonic vibrators based on electric quadrupole transition rates and moments. The effective theory~\citep{coelloperez2015b} developed for these observables will be discussed in Section~\ref{sub:vibelectro}. Its results for $^{114}$Cd are shown in figure~\ref{fig:cd114be2}, and the shaded areas again show 68\% degree-of-belief intervals. 
Overall, the effective theory demonstrates that cadmium isotopes  appear as (an)harmonic vibrators when viewed at low resolution. Admittedly, the uncertainties are large because of the low-lying breakdown scale at the three-phonon level. So, the description is accurate yet not very precise. At higher resolution, e.g. within the nuclear shell model, more details emerge and the description becomes more complicated~\citep{garrett2010,stuchbery2016,stuchbery2022}. 

\subsection{Odd-mass nuclei}

Odd-mass nuclei can often be viewed as a nucleon added to an even-even core whose properties are kept intact due to stabilizing effects of pairing.  Such an approach was taken in Section~\ref{sec:odd} for deformed nuclei. It also works for odd-mass neighbors of vibrational nuclei.  The description  of these systems within an effective theory couples a fermion confined to a single $j$ shell to the quadrupole oscillator. This introduces operators creating and annihilating a fermion in the corresponding orbitals. For an orbital with spin and parity $j^\pi$, the fermion operators fulfill the anticommutation relations
\begin{equation}
    \left\{a_\mu,a^\dagger_\nu\right\} = \delta_{\mu}^{\nu}
\end{equation}
with $\mu,\nu\in\{-j,-j+1,\ldots,j\}$. For the construction of spherical tensors we define a fermion annihilation tensor analogous to the quadrupole annihilation tensor in equation~\eqref{eq:annihilation_tensor}.

\citet{coelloperez2016} considered odd nuclei with $I^\pi=1/2^-$ ground states by coupling a $j^\pi=1/2^-$ orbital to the even-even vibrating nucleus. The leading contribution to the effective Hamiltonian consists of the most simple rank-zero operators constructed from either quadrupole or fermion tensors
\begin{equation}
    \hat{H}_{\rm LO} = \omega_0\hat{N} - S\left(a^\dagger\cdot\Tilde{a}\right) = \omega_0\hat{N} - S\hat{n},
\end{equation}
where the operator $\hat{n}$ counts the system's odd fermion. Notice that this fermion represents either a particle on top of the core or a hole in it. Therefore, the unknown constant $S$ scales as the nucleon separation energy. The spectrum of this Hamiltonian is
\begin{equation}
    \hat{H}_{\rm LO}\ket{Nv\nu J;n;IM} = \left(N\omega_0 -S\delta_n^1 \right)\ket{Nv\nu J;n;IM},
\end{equation}
where the state $\ket{Nv\nu J;n;IM}$ is the coupling of the quadrupole harmonic oscillator states and the fermion states $\ket{j}_\mu=a^\dagger_\mu\ket{0}$ to spin $I$ and projection into the $z$-axis $M$.
Since we are only concerned about the spectroscopic information of even-even and odd-mass systems, the constant $S$ is set to zero yielding zero-energy ground states. The constant $\omega_0$ must be fit to data.

The construction of the core-fermion interaction starts from the most simple operators including both quadrupole and fermion fields
\begin{equation}
    \hat{H}_{\rm c-f} = \omega_1\hat{N}\hat{n} + g_{Jj}\hat{J}\cdot\hat{j},
\end{equation}
where the operators $\hat{J}$ and $\hat{j}$ are the core and fermion angular momentum operators, and the low-energy constants $\omega_1$ and $g_{Jj}$ must be fit to data. While the first term shifts the frequency of the odd-mass oscillator, the second one, which may be thought of as a Coriolis interaction, couples the angular momenta of core and fermion splitting states in an odd-mass multiplet with different spins. Data on these effects suggest that the matrix element of an effective operator containing $m$ pairs of fermion fields, $\hat{O}_m$, is approximately a factor $\varepsilon$ smaller than the matrix element of another one, $\hat{O}_{m-1}$, containing $m-1$ pairs, that is,
\begin{equation}
    \braket{\hat{O}_m} \sim \braket{\hat{O}_{m-1}}\varepsilon.
\end{equation}

This power counting suggests the next-to-leading and next-to-next-to-leading contributions to the effective Hamiltonian are
\begin{align}
    \hat{H}_{\rm NLO} &= \hat{H}_{\rm c-f} \nonumber \\
    \hat{H}_{\rm NNLO} &= g_N \hat{N}^2 + g_v \hat{\Lambda}^2 + g_J \hat{J}^2.
\end{align}
Notice that the next-to-next-to-leading contribution is analogous to that in equation~\eqref{eq:hvib_nlo}. The interaction Hamiltonian yields the energy correction
\begin{equation}
    E_{\rm NLO}(I,J,n) = \omega_1 N n + \frac{g_{Jj}}{2}\left[I(I+1)-J(J+1)-\frac{3}{4} \right].
\end{equation}

\begin{figure}[h]
    \centering
    \includegraphics[width=0.475\textwidth]{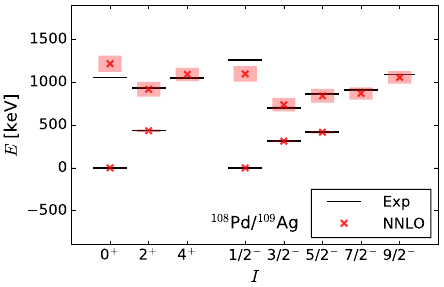}
    \includegraphics[width=0.475\textwidth]{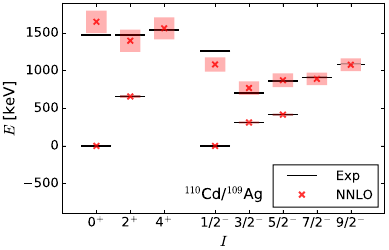}
    \caption{The low-lying states with negative parity of $^{109}{\rm Ag}$ (black lines) compared with predictions from effective theories at next-to-next-to-leading order (red crosses). On the left, $^{109}{\rm Ag}$ is modeled as a proton added to a $^{108}{\rm Pd}$ core. On the right, the same nucleus is modeled as a proton hole in $^{110}{\rm Cd}$. Figures taken from arXiv:1608.02802 with permission from the authors, see also \citep{coelloperez2016}.}
    \label{fig:ag109}
\end{figure}

Figure~\ref{fig:ag109} shows next-to-next-to-leading calculations for $^{109}$Ag and compares it with data on its low-lying negative-parity states.  The resulting theories consistently describe the spectrum of $^{109}$Ag  as a proton added to $^{108}$Pa, or as proton hole in  $^{110}$Cd.

\subsection{Electromagnetic transitions and moments}
\label{sub:vibelectro}

Characterizing a nuclear state as a quadrupole excitation based solely on energetics is often challenging due to the presence of multiple states with suitable spins and parities around the two- and three-phonon levels. Quadrupole (transition) moments have also been used for that purpose. In the past, the predictions based on harmonic vibrations were often deemed ``too large'' to be consistent with data. However, as those predictions lacked any uncertainty estimates what is ``too large'' is hard to quantify. This is where uncertainty quantification again is important. 

The quadrupole fields describe collective effects emerging from the unresolved dynamics of individual nucleons. Therefore, a minimal coupling scheme does not fully captures the interaction between these degrees of freedom and an electromagnetic field. Instead, the electric quadrupole properties are calculated from the most general rank-two operator that can couple to the quadrupole component of an electric field
\begin{equation}
    \hat{Q}_\mu = Q_0 \left( d^\dagger + \Tilde{d} \right)_\mu +
    Q_1 \left( d^\dagger \otimes \Tilde{d} \right)^{(2)}_\mu + \ldots,
    \label{eq:e2operator}
\end{equation}
where the dots stand for omitted terms. While the low-energy constants $Q_m$ cannot be computed within the theory, the power counting suggests a natural size for them relative to $Q_0$. Indeed, assuming that the matrix elements of all contributions to the quadrupole operator scale similarly at the breakdown scale yields $Q_m \sim Q_0\varepsilon^{m/2}$.

The first term in the expansion for the quadrupole operator~\eqref{eq:e2operator} couples states with a phonon difference of one, producing the leading contributions to the reduced matrix elements defining the corresponding transition strengths,
\begin{equation}
    B\left(E2,i\rightarrow f\right) = \frac{\left|\braket{f||\hat{Q}||i}\right|^2}{2I_i+1} \ .
\end{equation}
Corrections to the reduced matrix elements for these transitions arise from operators with an odd number of quadrupole fields, allowing one to write the expansions
\begin{align}
    \braket{f||Q||i} &= \braket{f||\hat{Q}||i}_{\rm LO}\sum_i c_i(i,f) \varepsilon^i,
    \nonumber \\
    B\left(E2,i\rightarrow f\right) &= B\left(E2,i\rightarrow f\right)_{\rm LO} \sum_i \Tilde{c}_i(i,f) \varepsilon^i
\label{eq:e2expansion}
\end{align}
from which uncertainties can be quantified through equation~\eqref{eq:deltadistg}. Expressions for the leading matrix elements governing one-phonon transition strengths from one- and two-phonon states in even-even systems, and odd-mass ones with $I^\pi=1/2^-$ ground states can be found in \citep{coelloperez2015b, coelloperez2016}.

The term proportional to $Q_1$ in the quadrupole operator~\eqref{eq:e2operator} couples states with the same number of phonons, hence contributing to the matrix elements determining transition strengths between states in the same multiplet, and electric quadrupole moments
\begin{equation}
    Q(i) = \sqrt{\frac{16\pi}{5}} \frac{C_{I_iI_i20}^{I_iI_i}}{\sqrt{2I_i+1}} \braket{i||\hat{Q}||i}.
\end{equation}
Since corrections to these matrix elements arise from contributions to the quadrupole operator with an even number of quadrupole fields, the expansions for the corresponding transition strengths and electric quadrupole moments take forms similar to those in equations~\eqref{eq:e2expansion}. 

\begin{figure}[h]
    \centering
    \includegraphics[width=0.7\textwidth]{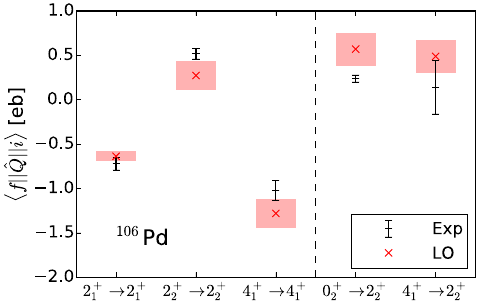}
    \caption{Electric quadrupole properties of low-lying states in $^{106}{\rm Pd}$. Experimental data (black lines) on electric quadrupole moments (left side) were used to fit the value of $Q_1$. Data on phonon-conserving transition strengths (right side) are consistently described by uncertainties quantified from intervals with 68\% degree-of-belief. Figure taken from arXiv:1510.02401 with permission from the authors, see also \citep{coelloperez2015b}.}
    \label{fig:pd106e2rme}
\end{figure}

Figure~\ref{fig:pd106e2rme} shows reduced matrix elements describing low-lying electric quadrupole moments and phonon-conserving transition strengths in $^{106}{\rm Pd}$.
After fitting the value of $Q_1$ to data on the static moments, the strengths of phonon-conserving transitions are predictions of the theory. The bands mark  68\% degree-of-belief intervals. In three out of five cases, theory and data agree within uncertainties. Thus, the theoretical uncertainty bands are consistent with a statistical interpretation.

A similar approach allows for the calculation of magnetic dipole properties within the effective theory. These properties are computed from the most general rank-one operator that can couple to the dipole component of a magnetic field
\begin{equation}
    \hat{\mu}_\nu = \mu_d \hat{J}_\nu + \mu_a \hat{j}_\nu + \mu_{d1} \left(\left(d^\dagger + \Tilde{d}\right)\otimes \hat{J}\right)^{(1)}_\nu + \mu_{a1} \left(\left(d^\dagger + \Tilde{d}\right)\otimes \hat{j}\right)^{(1)}_\nu + \ldots,
    \label{eq:m1operator}
\end{equation}
where the dots denote omitted terms in expression for the operator. The first two terms  preserve the phonon number and  therefore describe transitions between states in the same multiplet. The transition strength and the magnetic dipole moment are
\begin{align}
    B\left(M1,i\rightarrow f\right) &= \frac{\left|\braket{f||\hat{\mu}||i}\right|^2}{2I_i+1}, \nonumber \\
    \mu(i) &= \sqrt{\frac{4\pi}{3}}\frac{C_{I_iI_i10}^{I_iI_i}}{\sqrt{2I_i+1}} \braket{i||\hat{\mu}||i} \ .
\end{align}
Values for the low-energy constants $\mu_d$ and $\mu_a$ can be estimated from data on the magnetic moments of $I^\pi=2^+$ states and the Schmidt moment of a proton in a $j^\pi=1/2^-$ orbital. However, the latter estimate must be taken with a grain of salt as it is based on the assumption that all nucleon pairs in the core are coupled to zero spin. Even small contributions to the odd-mass nuclear state from an unresolved configuration in which a neutron pair with spin $J=2$ couples to the proton in a $j^\pi=3/2^-$ orbital result in a magnetic moment for the ground state of the system that largely deviates from the Schmidt value~\citep{ueno_h_1996}.

The third and fourth terms in the expansion for the magnetic dipole operator~\eqref{eq:m1operator} induce transitions between states with a phonon difference of one. Expressions for the reduced matrix elements governing the discussed magnetic dipole transition strengths and moments were given by~\citet{coelloperez2016}. Their comparison with (the admittedly sparse) data showed that the effective theory provides one with a consistent description.

\section{Matrix elements for neutrinoless double beta decay}
\label{sec:weak}
The nuclear matrix element for neutrinoless double beta decay connects the lifetime of this process -- if observed -- to the neutrino mass scale~\citep{engel2017}. Most candidate nuclei for neutrinoless double beta decay are not deformed and exhibit vibrational characteristics at low energies. This motivates one to apply effective theories developed to describe such nuclei to compute the relevant nuclear matrix elements. In this section we review the papers~\citep{coelloperez2018,brase2022,jokiniemi2023}.

\subsection{Gamow-Teller decays}
The vast majority of unstable nuclei lighter than $^{208}{\rm Pb}$ decay to more stable systems via weak-interaction processes, namely, $\beta$~decay or electron capture. In $\beta^-$~decay
\begin{equation}
    A(Z,N) \overset{\beta^-}{\longrightarrow} A(Z+1,N-1) + e^- + \Bar{\nu}_e \ .
\end{equation}

Measured rates for these decays range from milliseconds to billions of years, making their description a daunting test for any nuclear-structure theory. Furthermore, reliable predictions for the decay rates of experimentally inaccessible neutron-rich systems are paramount for $r$-process calculations to yield nuclear abundances consistent with observations for elements heavier than iron. Considering that most of the latter nuclei are difficult to calculate from first principles, their $\beta$~decay rates are commonly calculated within nuclear models with adjustable parameters, for which uncertainties are more difficult to estimate or quantify.

At low-order, the weak interaction consists of two contributions $\hat{H}_{\rm weak}=g_V \hat{O}_{\rm F}+g_A \hat{O}_{\rm GT}$ with
\begin{align}
    \hat{O}_{\rm F} &= \sum_i \tau_i^\pm, \nonumber \\
    \hat{O}_{\rm GT} &= \sum_i \sigma_i \tau_i^\pm.
\end{align}
Here, $\sigma$ is the spin (vector) operator, $\tau^\pm$ are the isospin rising and lowering operators, and the sum runs over all nucleons. These contributions induce Fermi and Gamow-Teller decays, for which the spins of the lepton pair are coupled to zero and one, respectively. Fermi decays can only couple states with the same spin. Allowed Gamow-Teller decays, on the other hand, are isospin analogous to magnetic dipole transitions and can couple states with spin differences up to one.

The similarities between Gamow-Teller and magnetic dipole transitions suggest that the former can be described within an extension of the effective theory that succeeded describing the latter (which was reviewed in the previous section). This framework employs quadrupole operators together with neutron and proton ones, denoted by $n$ and $p$, that create and annihilate these fermions in single-particle orbitals with spins and parities $j_n^{\pi_n}$ and $j_p^{\pi_p}$, respectively. The neutron and proton operators fulfill fermionic anticommutation relations.
The theory assumes that the fermions have access to one single-particle orbital each, and that the decaying odd-odd nuclei of interest can be modeled as a particle-hole pair on top of an even-even core. For example, $^{80}{\rm Br}$ can be modeled in terms of a neutron, a proton hole, and a $^{80}{\rm Kr}$ core. The spins and parities of the neutron and proton orbitals are inferred from the spectra of adjacent odd-mass nuclei, such that the ground state of odd-odd nucleus,
\begin{equation}
    \ket{n_n;n_p;IM} = \left(n^\dagger \otimes p^\dagger\right)^{(I)}_M\ket{0}
\end{equation}
possesses the proper spin and and parity.

In light of the fact that the theory's degrees of freedom are not fundamental, the Gamow-Teller operator used to describe such decays must be constructed as the most general rank-one operator capable to change isospin by one unit, and one has
\begin{equation}
    \hat{O}_{\rm GT} = C_{\beta} \left( \tilde{n}\otimes\tilde{p} \right)^{(1)}
    + \sum_{\ell} C_{\beta\ell} \left( \left( d^\dagger + \tilde{d} \right) \otimes \left( \tilde{n}\otimes\tilde{p} \right)^{(\ell)} \right)^{(1)} + \ldots \ .
\label{eq:gt_operator}
\end{equation}
Here the fermion annihilation tensors $\tilde{n}$ and $\tilde{p}$, defined analogous to the quadrupole annihilation tensor~\eqref{eq:annihilation_tensor}
simplify the construction of effective operators with specific ranks, and the dots denote omitted contributions. The particle-hole annihilation operation in the Gamow-Teller operator represents different multistep processes. When the fermion operators represent a neutron and a proton hole, their annihilation represents the decay of the neutron on top of the core into a proton, which proceeds to fill the proton hole. However, this operation can also represent the decay of core neutron, followed by the filling of both the proton hole and the newly created neutron hole. The effects of these unresolved processes are captured in the low-energy constants, which must be fit to data.

While the low-energy constants in the Gamow-Teller operator cannot be calculated by the effective theory, the power counting for the quadrupole fields established in section~\ref{sec:vib} allows one to  estimate their relative sizes. According to this counting scheme, each term in the expansion for the Gamow-Teller operator yields a matrix element of size $N^{m/2}$, where $m$ is the number of quadrupole fields. At the phonon level $N_b$ where the effective theory breaks down, all contributions to the Gamow-Teller operator must yield similar-sized matrix elements. Combining these statements yield the scaling
\begin{equation}
    C_m \sim C_\beta \varepsilon^{m/2}.
    \label{eq:lec_estimates}
\end{equation}

The decay rate of a Gamow-Teller decay $1/t$ is related to the reduced nuclear matrix elements of the Gamow-Teller operator through Fermi's golden rule
\begin{equation}
    \frac{1}{t} = \frac{f}{\kappa} \frac{g_A^2 \left|M_{{\rm GT},i\rightarrow f}\right|^2}{2I_i+1} \ .
\end{equation}
Here, $f$ is a phase-space factor containing the lepton kinematics, $\kappa$ is the $\beta$-decay constant, and $g_A$ is the axial-vector coupling constant. Leading expressions for the matrix elements governing decays from $I^\pi=1^+$ odd-odd nuclei to all zero-, one-, and two-phonon even-even states differ by simple factors, but depend each on a different low-energy constant~\citep{coelloperez2018}. While fitting all these low-energy constants to data will devoid the theory of its predictive power, it is possible to appraise this approach to weak processes using data on the decay to the even-even ground state or transition strengths of charge-exchange reactions to fit the constant $C_\beta$. Predictions for decays to excited states are computed from 68\% degree-of-belief intervals of the distributions for the other low-energy constants.

\begin{figure}[h]
    \centering
    \includegraphics[width=0.475\textwidth]{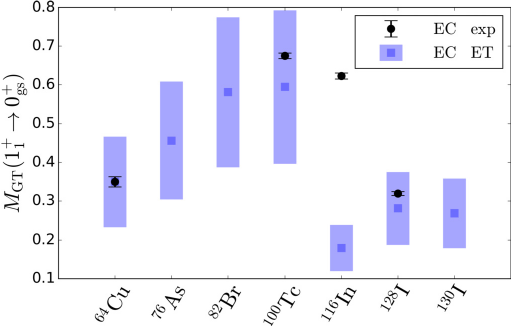}
    \includegraphics[width=0.475\textwidth]{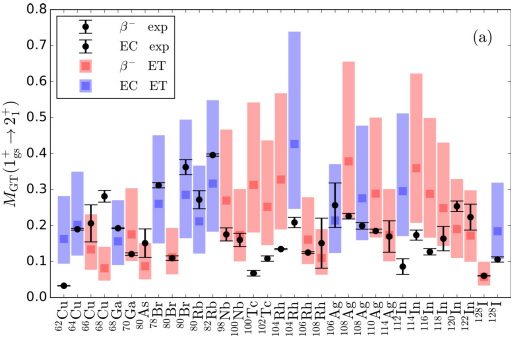}
    \caption{Matrix elements for weak decays from $1^+$ odd-odd ground states to low-lying $0^+$ and $2^+$ even-even states. Data on charge-exchange strengths (left panel) and decays to even-even ground states (right panels) were used to fit $C_\beta$. Figures taken from arXiv:1708.06140 with permission from the authors, see also \citep{coelloperez2018}}
    \label{fig:betadecays}
\end{figure}

Figure~\ref{fig:betadecays} shows predictions for Gamow-Teller decays for initial and final states as indicated. Low-energy constants were extracted from charge exchange reactions (for the left panel) from decays to the corresponding ground states (right panel). Overall, the description is consitent with data, albeit within significant uncertainties.

\subsection{Two-neutrino double beta and electron capture decays}
Second-order weak processes, observed in even-even nuclei for which single weak decays are energetically forbidden, exhibit the longest half-lives measured to date. 
The decay rate $1/t$ of a double-beta decay is related to its Fermi and Gamow-Teller nuclear matrix elements through
\begin{equation}
    \frac{1}{t} = G^{2\nu} g_A^4 \left| M_{{\rm GT},i\rightarrow f}^{2\nu} - \left(\frac{g_V}{g_A}\right)^2 M_{{\rm F},i\rightarrow f}^{2\nu} \right|^2,
\end{equation}
where $G^{2\nu}$ is a phase-space factor containing the lepton kinematics. The Fermi contribution, which can only couple states in the same isospin multiplet with equal spins, does not play a role in the description of decays from initial ground states to low-lying final states, as the excitation energies of isobaric analog states are of the order of tens of MeV. The Gamow-Teller contribution for decays from the $I^\pi=0^+$ ground states of parent nuclei is
\begin{equation}
    M_{{\rm GT},i_{\rm gs}\rightarrow f}^{2\nu} = \sum_n \frac{\braket{f||\hat{O}_{\rm GT}||1^+_n}\braket{1^+_n||\hat{O}_{\rm GT}||i_{\rm gs}}}{\sqrt{s} D_{nf}^s}.
\label{eq:GTcontribution}
\end{equation}
In this expression, the sum runs over all $I^\pi=1^+$ odd-odd states, and the energy denominators $D_{nf}=[E_n-(E_i+E_f)/2]/m_e $ have been defined relative to the electron mass $m_e$ (thus making this contribution dimensionless). The factor $s=1+2\delta_{I_f}^2$ accounts for decays to $I^\pi=2^+$ states.

The computation of these matrix elements poses a challenge to an effective theory that cannot describe contributions from odd-odd states above its breakdown scale. However, the suppression of these contributions by the accompanying energy denominators suggests an approximation for them in which only the contribution from the lowest $1^+$ state is considered.
Assuming that higher $1^+$ odd-odd states can be described as multiphonon excitations of the lowest one, the energies denominators entering higher contributions can be written at leading order as
\begin{equation}
    D_{n+1f} \sim D_{1f} + \frac{n\omega}{m_e}.
\label{eq:approx_denominator}
\end{equation}
Let us discuss decays 
between ground states. 
The numerator of the $n$-th omitted contribution involves two matrix elements each with $n$ quadrupole fields, and is then expected to scale as
\begin{equation}
    \braket{f_{\rm gs}||\hat{O}_{\rm GT}||1^+_{n+1}}\braket{1^+_{n+1}||\hat{O}_{\rm GT}||i_{\rm gs}} \sim \braket{f_{\rm gs}||\hat{O}_{\rm GT}||1^+_1}\braket{1^+_1||\hat{O}_{\rm GT}||i_{\rm gs}} \varepsilon^n.
\label{eq:approx_numerator}
\end{equation}
The Gamow-Teller matrix elements on the right-hand side of this expression are computed from two different models with low-energy constants fitted to data on the corresponding single $\beta$~decays or charge-exchange reaction strengths.
From the approximations~\eqref{eq:approx_denominator} and~\eqref{eq:approx_numerator}, the contribution omitted by the single-state approximation to the Gamow-Teller contribution~\eqref{eq:GTcontribution} relative to its approximate value is expected to scale as
\begin{equation}
    \Delta_{{\rm GT},i_{\rm gs}\rightarrow f_{\rm gs}}^{2\nu(1)} \sim 
    \frac{m_e}{\omega} D_{1f_{\rm gs}} \Phi\left(\varepsilon,1,\frac{D_{1f_{\rm gs}}m_e+\omega}{\omega}\right) \varepsilon,
\end{equation}
where $\Phi(z,s,a)$ is the Lerch transcendent function. The size of the latter, which can be used as an uncertainty estimate, depends on the energy scales of the models describing the nuclei involved in the decay.

\begin{figure}[h]
    \includegraphics[width=0.49\textwidth]{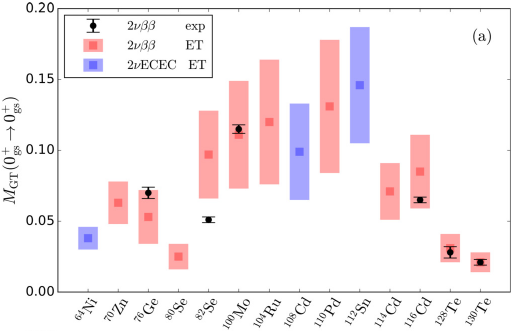}
    \includegraphics[width=0.49\textwidth]{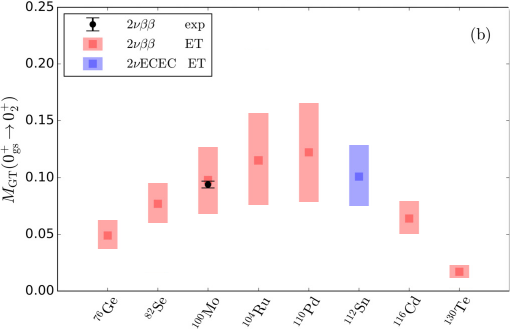}
    \caption{Matrix elements for two-neutrino second-order weak decays from ground states of even-even nuclei. The single-state approximation values for the matrix elements, obtained using data on single beta decays and charge-exchange reaction strengths, are in good agreement with those extracted from the decays' half-lives. Figures adapted from arXiv:1708.06140 with permission from the authors, see also \citep{coelloperez2018}.}
    \label{fig:2nubetadecays}
\end{figure}

\citet{coelloperez2018} computed expressions for the matrix elements governing various two-neutrino weak decays.
Figure~\ref{fig:2nubetadecays} shows matrix elements for two-neutrino second-order weak decays to ground and second $0^+$ states. These results are in good agreement with those extracted from observed half-lives within uncertainties estimated from omitted contributions to the matrix elements, validating the power counting for the quadrupole fields.

\begin{figure}[h]
    \centering
    \includegraphics[height=0.23\textheight]{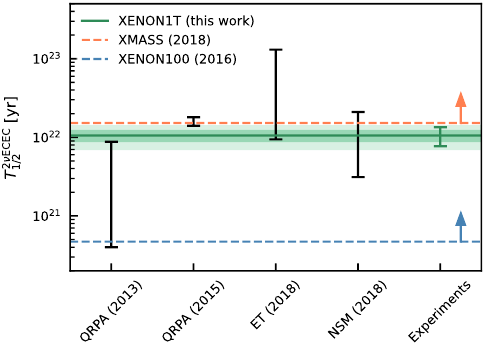}
    \includegraphics[height=0.235\textheight]{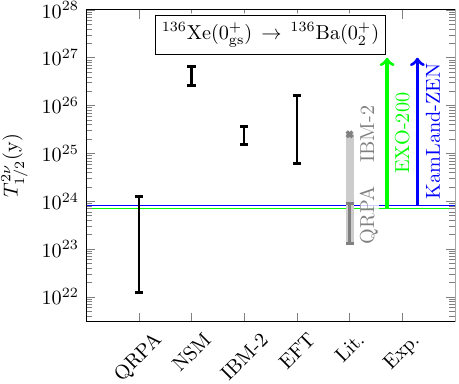}
    \caption{Predicted half-lives for second-order weak decays with emission of neutrinos. On the left, calculations for the double electron capture on $^{124}{\rm Xe}$ are compared with the half-life extracted from the decay's observation with the XENON1T dark-matter detector. Figure taken from arXiv:2205.04158 with permission from the authors, see also~\citep{aprile_e_2022}. The results from the effective theory are labeled as ET. On the right, calculations for the double beta decay of $^{136}{\rm Xe}$ to the first excited $0^+$ state in $^{136}{\rm Ba}$ are compared to current experimental limits. Figure taken from arXiv:2211.03764 with permission from the authors, see also \citep{jokiniemi2023}. Results from the effective field theory are labeled as EFT.}
    \label{fig:2nupredictions}
\end{figure}

Figure~\ref{fig:2nupredictions} compares theoretical results from the effective theory and other nuclear structure models for the half-lives of the two-neutrino double electron capture on $^{124}{\rm Xe}$~\citep{coelloperez_2019}, and the two-neutrino double beta decay of $^{136}{\rm Xe}$ to the second $0^+$ state in $^{136}{\rm Ba}$~\citep{jokiniemi2023}. The first of these predictions is in good agreement with the value subsequently measured by the XENON collaboration~\citep{aprile_e_2019}. The second one is compared to lower limits established by the KamLAND-Zen~\citep{asakura_k_2016} and EXO-200~\citep{albert_jb_2016} experiments. Error bars  represent quantified uncertainties for the effective theory predictions and reflect sensitivity to variation in model parameters for the other models. These results show the ability of the effective theory to estimate these observables, which can be used as a guideline for more involved calculation within more complex frameworks capable of yielding predictions with higher precision.

\subsection{Neutrinoless double beta decays}

Along with the second-order weak decays reviewed in the previous section, large-scale experiments are currently being conducted in search of an alternative decay mode that violates lepton conservation. In this hypothetical mode, the neutrinos emitted during the simultaneous decay of the neutrons are their own antiparticles, i.e. they are Majorana fermions. This is a lepton-number violating process. 
If observed, this decay mode would support theories explaining the matter-antimatter asymmetry in the Universe through leptogenesis~\citep{fukugita_m_1986}. It would also 
give insights into the mass scale of the neutrino. 

Assuming the decay is mediated by the exchange of the light neutrinos we know, and that the impulse and closure approximations are valid, the neutrinoless double beta decay half-life can be related to the Gamow-Teller, Fermi and tensor nuclear matrix elements through
\begin{equation}
    \frac{1}{t_{i\rightarrow f}} = G^{0\nu}_{i\rightarrow f} g_A^4 \left| M_{{\rm GT},i\rightarrow f}^{0\nu} - \left(\frac{g_V}{g_A}\right)^2 M_{{\rm F},i\rightarrow f}^{0\nu} + M_{{\rm T},i\rightarrow f}^{0\nu} \right|^2 m_{\beta\beta}^2 \ ,
\end{equation}
where $G^{0\nu}_{i\rightarrow f}$ is a leptonic phase-space factor and $m_{\beta\beta}$ is the effective neutrino mass $m_{\beta\beta}$.
The nuclear matrix elements are
\begin{equation}
    M_{x,i\rightarrow f}^{0\nu} = \frac{2R}{\pi g_A^2} \int_0^\infty q dq \frac{\braket{f|\hat{O}_x(q)|i}}{D(q)}
\end{equation}
with $x\in\{{\rm GT},{\rm F},{\rm T}\}$. 
Here, the nuclear radius $R$ was introduced to make these matrix element dimensionless. The integrand in the above expression consists of the matrix element of either the Gamow-Teller, Fermi or tensor operator, 
\begin{align}
    \hat{O}_{\rm GT}(q) &= \sum_{ij} j_0(qr_{ij}) h_{\rm GT}(q) \sigma_i\cdot\sigma_j \tau_i^+\tau_j^+ \ , \nonumber \\
    \hat{O}_{\rm F}(q) &= \sum_{ij} j_0(qr_{ij}) h_{\rm F}(q) \tau_i^+\tau_j^+ \ , \nonumber \\
    \hat{O}_{\rm T}(q) &= \sum_{ij} j_2(qr_{ij}) h_{\rm T}(q) \left(3\sigma_i\cdot\hat{r}_{ij}\sigma_j\cdot\hat{r}_{ij} -\sigma_i\cdot\sigma_j\right) \tau_i^+\tau_j^+ \ ,
\end{align}
which act on all neutron pairs of the decaying nucleus, and an energy denominator, $D(q)=q+\Bar{E}-(E_i+E_f)/2$, depending on the closure energy $\Bar{E}$. The spherical Bessel functions $j_n$ and neutrino potentials $h_x$ in the operator depend on the magnitude of the relative momentum $q$ and the relative position $r_{ij}$ of the two neutrons.

\begin{figure}
    \centering
    \includegraphics[width=0.7\textwidth]{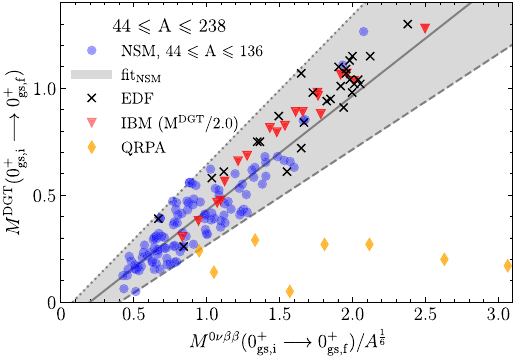}
    \caption{Relation between the neutrinoless double beta decay and double Gamow-Teller matrix elements. Calculations within the nuclear shell model, energy density functional theory and interacting boson model support a linear relation between these matrix elements. Figure taken from arXiv:2108.11805 with permission from the authors, see also \citep{brase2022}.}
    \label{fig:0nurelation}
\end{figure}

The direct calculation of the neutrinoless double beta decay matrix element within an effective theory that cannot resolve individual neutrons would require us to fit unknown constants in the effective operator encoding the effects of neutron pair interactions to data on a not yet observed decay. An alternative is to take advantage of the linear relation suggested by multiple models~\citep{menendez_j_2018, barea_j_2015, rodriguez_tr_2013, menendez_j_2009} between this matrix element and that of the double Gamow-Teller operator, shown in figure~\ref{fig:0nurelation}.
The calculation of this matrix element within the effective theory requires one to write the ground state of the parent even-even nucleus in this framework. A first approximation to this state considers only the configuration in which the neutron and proton-hole pairs on top of the core are coupled to zero spin
\begin{equation}
    \ket{0^+_{\rm gs}} = \frac{1}{2} \left(n^\dagger\otimes n^\dagger \right)^{(0)} \left(p^\dagger\otimes p^\dagger \right)^{(0)} \ket{0}.
\end{equation}
While contributions from configurations in which the neutron and proton pairs couple to non-zero spin are expected to be small, their effects on Gamow-Teller matrix element might be considerable, as is the case for nuclear magnetic moments~\citep{ueno_h_1996}.

The matrix element of the effective double Gamow-Teller operator
\begin{equation}
    \hat{O}_{\rm DGT} = \left(\hat{O}_{\rm GT}\otimes\hat{O}_{\rm GT}\right)^{(0)} = C_{\beta_i}C_{\beta_f} \left( \left( \Tilde{n}\otimes\Tilde{p}\right)^{(1)} \otimes \left( \Tilde{n}\otimes\Tilde{p}\right)^{(1)} \right)^{(0)} + \ldots,
\end{equation}
with $C_{\beta_i}$ and $C_{\beta_f}$ low-energy constants fitted to single beta decays from the lowest $1^+$ state in the intermediate odd-odd nucleus to the initial and final nuclei, respectively, and dots denoting higher-order terms, can be approximated as
\begin{equation}
    M_{{\rm DGT},i_{\rm gs}\rightarrow f_{\rm gs}} \approx C_{\beta_i}C_{\beta_f} \sqrt{\frac{4}{3(2j_n+1)(2j_p+1)}},
\end{equation}
where $j_n$ and $j_p$ are the spins of the orbitals in which the neutron and proton-hole pairs lie. These spins, which must be able to couple to form $1^+$ odd-odd states, are inferred from the low-lying spectra of adjacent odd-mass nuclei. Omitted contributions to this matrix element arise from neglected contributions to the double Gamow-Teller operator including $m$ pairs of quadrupole fields, and $m$-phonon corrections to the lowest $1^+$ odd-odd state used to fit $C_{\beta_i}$ and $C_{\beta_f}$. The latter corrections couple to the ground state of the final nucleus via terms in the Gamow-Teller operator with $m$ quadrupole fields. According to the power counting for the quadrupole fields, and assuming the theory breaks at the three-phonon level, yields an omitted contribution to the double Gamow-Teller matrix element relative to its leading approximation expected to scale as
\begin{equation}
    \Delta_{{\rm DGT},i_{\rm gs}\rightarrow f_{\rm gs}} \sim \sum_{i=1}\epsilon^{i} = \frac{1}{2}.
\end{equation}

\begin{figure}[h]
    \centering
    \includegraphics[height=0.35\textheight]{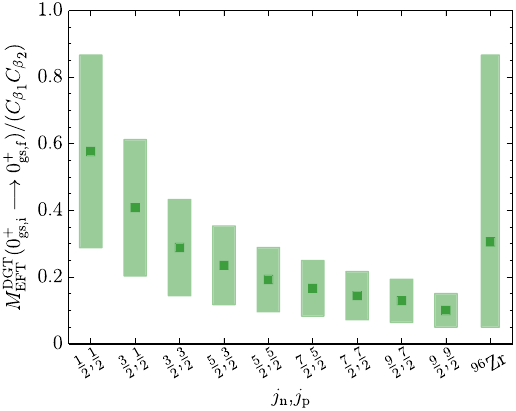}
    \caption{Normalized double Gamow-Teller matrix element. Omitted contributions to the matrix elements are used as uncertainty estimates. The matrix element for $^{96}{\rm Zr}$ is the mean value of the matrix elements for the shown combinations. Its associated uncertainty results from merging the corresponding uncertainties. Figure taken from arXiv:2108.11805 with permission from the authors, see also \citep{brase2022}.}
    \label{fig:dgt_eft}
\end{figure}

This expectation for the omitted contribution can be used as an uncertainty estimate when the neutron and proton spins can be uniquely assigned. For transitions in which the adjacent odd-mass spectra suggest $n_j$ allowed spin combination, the double Gamow-Teller matrix element is assigned a value of
\begin{equation}
    M_{{\rm DGT},i_{\rm gs}\rightarrow f_{\rm gs}} = \frac{C_{\beta_i}C_{\beta_f}}{n_j} \sum_{j_n,j_p}\sqrt{\frac{4}{3(2j_n+1)(2j_p+1)}} \ .
\end{equation}
Its associated uncertainty results from merging the omitted contributions for each spin combination, as shown in figure~\ref{fig:dgt_eft} for the decay of $^{96}{\rm Zr}$, for which all spin combinations $1/2\leq j_n,j_p \leq 9/2$ were considered.

\begin{figure}[h]
    \centering
    \includegraphics[height=0.35\textheight]{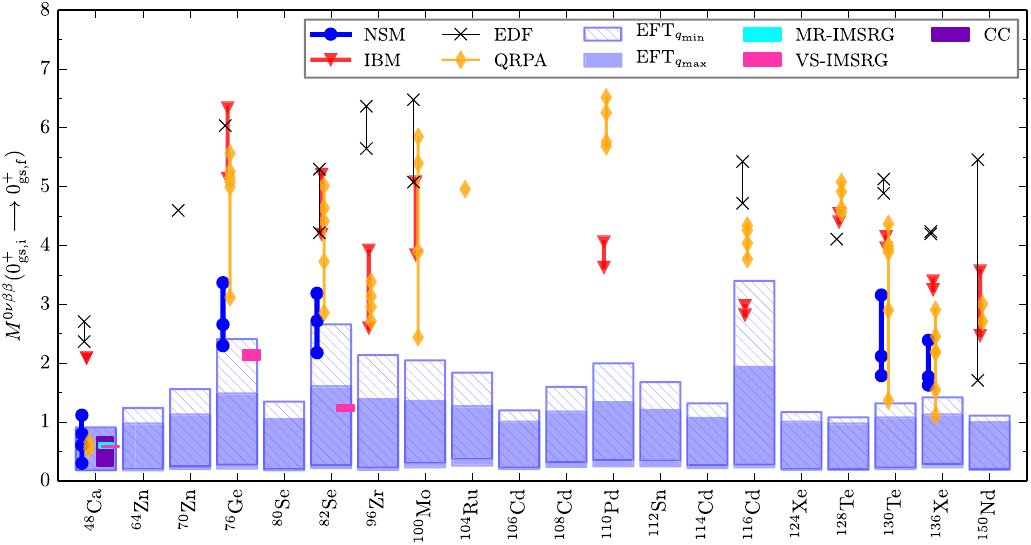}
    \caption{Neutrinoless double beta decay matrix elements. Matrix elements predicted by the effective theory for the decays of $^{48}{\rm Ca}$, $^{76}{\rm Ge}$ and $^{82}{\rm Se}$ are consistent with those obtained from the nuclear shell model (NSM) calculations as well as \textit{ab initio} calculations using coupled-cluster (CC)  and in-medium similarity renormalization group (IMSRG) methods. Predictions for the decays of heavier nuclei underestimate those made within the NSM and other phenomenological models. Figure taken from arXiv:2108.11805 with permission from the authors, see also \citep{brase2022}.}
    \label{fig:0nu}
\end{figure}

Based on a linear fit for the relation between neutrinoless double beta decay and double Gamow-Teller matrix elements, the neutrinoless double beta decay matrix elements are
\begin{equation}
    M^{0\nu}_{i_{\rm gs}\rightarrow f_{\rm gs}} = A^{1/6} \frac{M_{{\rm DGT},i_{\rm gs}\rightarrow f_{\rm gs}}/q^2-n}{m}.
\end{equation}
where $A$ is the nuclear mass number, $q$ is a quenching factor. Values for $m$ and $n$ yielding the grey band in figure~\ref{fig:0nurelation} are given by~\citet{brase2022}. The neutrinoless double beta decay matrix elements computed from double Gamow-Teller ones calculated within the effective theory using two quenching factors (solid and dashed boxes) are compared with results from other nuclear structure models in figure~\ref{fig:0nu}.
It is interesting to notice that predictions by the effective theory for $^{48}{\rm Ca}$, $^{76}{\rm Ge}$ and $^{82}{\rm Se}$ are consistent with those obtained from nuclear shell model (NSM) calculations [values found in reference~\citep{shimizu2018}] as well as \textit{ab initio} calculations using coupled cluster (CC)~\citep{novario2021} and in-medium similarity renormalization group (IMSRG)~\citep{belley2024, belley2021, yao2020} methods. Predictions by the effective theory for the decays of heavier nuclei tend to underestimate values calculated within the NSM and other phenomenological models. 

\section{Comparison with other models}
\label{sec:compare}
Many models have been used to describe and understand nuclear collective modes such as rotations and vibrations, and to connect them with the independent-particle picture of the nuclear shell model, see, e.g. the textbooks~\citep{bohr1975,rowe2010}. The recent focus issue~\citep{dudek2016} presents many developments that built on the ground-breaking work by  \citeauthor{bohr1953}. In this Section, we relate and contrast the effective theories of  this review to some of these models. Finally, we also point out how the effective theories can be useful in {\it ab initio} computations of collective modes. 

One can divide the approaches to nuclear deformation into two sets, namely those that  break rotational symmetry (or use nonlinear realizations of rotational symmetry) and those that do not. The former are conceptually simple and also tend to be computationally much less expensive. The approach by the Copenhagen group \citep{bohr1952,bohr1953,nilsson1955} was the first in this direction. The effective theories of deformed nuclei are closely related to these approaches and highlight their universal properties. The approaches that keep rotational invariance are computationally and conceptually more complicated. The foundational work  by \citet{elliott1958} described deformed many-body states within a valence shell of the harmonic oscillator exploiting its SU(3) symmetry. Several algebraic models are inspired by this work and computations of deformed states in the spherical shell model can also be related to it~\citep{zuker2015}.  

\subsection{Unified collective model and geometric models}
\label{sub:BM}
\citet{bohr1952,bohr1953} modeled the collective behavior of nuclei in terms of the surface oscillations of a liquid. This approach uses the five degrees of freedom associated with quadrupole deformation of the nuclear surface. A decisive step is the transformation to three Euler angles and two shape parameters, $\beta$ and $\gamma$. The former measures the degree of axially symmetric deformation while the latter describes triaxial deformation. This transformation effectively introduces a nonlinear realization of the rotational symmetry, by using the Euler angles, and adds $\beta$ and $\gamma$ as vibrational degrees of freedom.  In deformed even-even nuclei, this allows one to describe the ground-state band (with a $0_1^+$ band head) and neighboring rotational bands with a $0_2^+$ state (for the $\beta$ mode) and a $2_2^+$ state (for the $\gamma$ mode) for the band head, respectively. The collective Bohr Hamiltonian becomes exactly solvable for certain forms of the potential energy,  see \citep{fortunato2005} for a compendium. Such exactly solvable models approximately describe nuclei in various regions of the Segre chart. 

The unified collective model addresses collective and single-particle dynamics. Single-particle aspects are described within the deformed (axially symmetric) \citeauthor{nilsson1955} model. This yields the structure of single-particle excitations in the intrinsic (body-fixed frame). It is understood that suitable superpositions (i.e. rotations) of such states yield a system with good total angular momentum. The model is invaluable in predicting the ground-state spins of odd-mass nuclei.  
These also exhibit rotational bands. Here, the spin projections $K$ of the band heads are based on the Nilsson orbital occupied by the odd nucleon. The spin $j=K$ of the odd nucleon is then coupled to the spin $I$ of the rotor via the ``rotation-particle coupling'' $(I_+j_- +I_-j_+)$~\citep{kerman1956}. Deformed odd-odd nuclei are described by appropriately couple two unpaired nucleons (in the Nilsson model) to the rotor. 

There are several extensions of the unified collective model. These so-called ``geometric models'' all have in common that they describe quantized oscillations  and rotations of a liquid~\citep{faessler1965,eisenberg1970,hess1980}.   
They describes a wealth of data in deformed nuclei for energy levels and electromagnetic transitions, see, e.g., references~\citep{eisenberg1970,bohr1975}.  We still use its expressions such as ``$\beta$ vibrations'' or ``$\gamma$ vibrations'' today. Because of the simple underlying model of an oscillating surface, it is straight forward to introduce other operators (e.g. those for electromagnetic moments and transitions) within the unified model. Of course, we are also aware of its limitations. In cadmium isotopes, for instance, a low-resolution description can indeed be based on $\beta$ vibrations~\citep{coelloperez2015b}, but a high-resolution description reveals that the situation is more complex~\citep{garrett2010,gray2022}. In general, the Nilsson model is successful and most useful in determining the spins of rotational band heads in odd-nuclei; however it is more challenging to view $\beta$ and $\gamma$ excitations in even-even nuclei within this model. Another challenge is posed by deformed even-even actinides where lowest lying rotational band heads with spin/parity $K^\pi=1^-$ appear at lowest energy close to the ground-state rotational band, i.e. below the band heads that could be associated with $\beta$ or $\gamma$ vibrations.  
In contrast, the effective theories can be agnostic regarding the origin of internal degrees of freedom that determine the spins (and $K$ quantum numbers) of band heads; the only assumption about them is that they are fast degrees of freedom.

\subsection{Interacting boson model and algebraic models}
\label{sub:IBM}
Algebraic models can be traced back to \citeauthor{elliott1958}'s SU(3). \citet{elliott1958} pointed out that collective rotational states in the harmonic-oscillator-based shell model have large overlaps with basis states of irreducible representations of SU(3). In other words, the SU(3) symmetry of the three-dimensional harmonic oscillator allows one to introduce basis states that are classified by SU(3) quantum numbers in addition to those of SO(3) reflecting rotational invariance. Collective rotational states appear to be simple in this basis, i.e. they are limited to a single or a few irreducible representations. This seems to suggest that SU(3) could be an approximate symmetry for nuclei when viewed at low resolution. (It is not a symmetry of the nuclear Hamiltonian at higher resolution, e.g. within pion-less effective field theory or chiral effective field theory.) Thus, a low-resolution nuclear Hamiltonian must then be approximately proportional to the SU(3) Casimir operator $\mathbf{Q}\cdot\mathbf{Q}-3\mathbf{L}\cdot\mathbf{L}$ where $\mathbf{Q}$ is the algebraic quadrupole operator and $\mathbf{L}$ the orbital angular momentum. The 
$\mathbf{L}\cdot\mathbf{L}$ term naturally yields the rotational band  with energies proportional to $L(L+1)$. Thus, the SU(3) symmetry is consistent with the leading-order Hamiltonian~(\ref{HamLO}) of the emergent symmetry breaking of SO(3), and we think that this is how the effective theories of this review relate to the SU(3) symmetry of algebraic models. 

The impact of Elliott's work is that it relates the spherical shell model to nuclear rotation and deformed states. However, while the coupling of nucleons to good SU(3) quantum numbers is a solved problem~\citep{hecht1965}, it can be  computationally expensive for many-nucleon systems. It is in the eye of the beholder to what extent such correlated states are simple. The algebraic approach has been extended to more general cases~\citep{rosensteel1977,rosensteel1980} and is also used in {\it ab initio} computations of deformed nuclei~\citep{dytrych2008}. The review~\citep{harvey1968} and the recent textbooks by \citet{rowe2010} and \citet{frank2019} present many aspects of this approach.

The interacting boson model~\citep{iachello1987} describes even-even nuclei in terms of $s$ and $d$-wave bosons. Thus, one deals with a total of six boson creation and six corresponding annihilation operators. While one might associate the former with pairing and the latter with deformation, this is an algebraic model and no reference to any coordinates or momenta needs to be made. It is immediately clear that one has the right degrees of freedom to capture, e.g., the $\beta$ and $\gamma$ vibrations and the rotations of the unified collective model.  When choosing a basis in Hilbert space the six boson operators allow one to label basis states in terms of the U(6) symmetry. There are three different group chains that introduce quantum numbers starting with U(6) and ending with SO(3), i.e. the symmetry of the strong nuclear force that must be used to label energy levels. Thus, one can introduce three different bases in Hilbert space where basis states can be labeled by other quantum numbers besides angular momentum. 

The Hamiltonian of the interacting boson model exhibits six parameters. The parameters can be chosen such that the Hamiltonian is diagonal in one of the three different basis sets. Interestingly, the spectra (and the electromagnetic transitions) in some nuclei are well described using such special combinations of parameters. One then speaks of a ``dynamical symmetry.'' In general, of course, the interacting boson model simply provides one with a basis and eigenstates are superpositions of basis states. We can relate the effective theories of this review to the interacting boson model. Based on \citeauthor{elliott1958}'s SU(3) the group chain $U(6)\supset SU(3)\supset SO(3)$ is consistent with the emergent symmetry breaking of SO(3) and describes deformed nuclei. The group chain  $U(6)\supset U(5)\supset SO(3)$ contains the $U(5)$ symmetry. The leading-order Hamiltonian~(\ref{eq:vib_HLO}) is invariant under U(5) transformations of the $d$-boson operators, i.e. under $d_\mu \to d_\mu'=\sum_\nu U_{\mu\nu} d_\nu$ with a unitary matrix $U$. Thus, harmonic vibrations are consistent with that group chain. Higher-order corrections will break this dynamical symmetry. In this sense, the dynamical symmetries of the interacting boson model might be thought of as resolution-scale dependent symmetries. They are absent at high resolution because they are not symmetries of the strong nuclear force.  The appearance of a higher symmetry than exhibited by the strong nuclear force could be an effect of low resolution: ``From a distance, most things look beautiful.''~\citep{murakami2018}.

The algebraic models seek insights and simplification by classifying nuclear states in terms of higher symmetries than what the strong nuclear force exhibits. While such a classification might be successful at low resolution scales it must break down at sufficiently high resolution.  In any case, they provide one with a basis in Hilbert space. \citet{caprio2011}, for instance employed a basis from an algebraic model to solve the Bohr Hamiltonian numerically. The mathematically inspired approach of the algebraic models, invoking a higher symmetry than exhibited by the strong nuclear force, is in contrast to effective field theories. In the latter approach, nuclear deformation and superfluidity, for instance, are physical phenomena related to emergent symmetry breaking, i.e to a breaking of symmetry in low-energy states.   

\subsection{Shell model}
\label{sub:SM}
The spherical shell model~\citep{mayer1955} provides us with basis of Slater determinants for valence nucleons in a major oscillator shell. By construction the shell model is efficient to understand single-particle aspects of nuclear structure. Collective states can be computed as well; however these  are superpositions of a large number of Slater determinants and thus more complicated. An alternative approach can be based on \citeauthor{elliott1958}'s work. Basis states can be chosen according to irreducible representations of SU(3). A few irreducible representations may suffice to compute rotational bands, and this makes them conceptually simple. However, the SU(3) basis states are superpositions of many Slater determinants and this complicates the computation of Hamiltonian matrix elements for general interactions. The review by~\citet{caurier2005} highlights how single-particle and collective aspects are captured by the spherical shell model, and some SU(3) aspects are emphasized in reference~\citep{zuker2015}. 

The Monte Carlo shell-model~\citep{shimizu2012} employs a huge Hilbert space of symmetry-breaking Slater determinants, restores their symmetry via projection, and uses a Monte Carlo method to determine a relatively small basis in Hilbert space. Here again, symmetry projection is the crucial tool that provides us with an intellectual link with effective theories, see section~\ref{sec:found}.


\subsection{Mean field computations}
\label{sub:MF}
Mean-field approaches describe nuclear properties based on self-consistent computations of a Slater determinants~\citep{bender2003,vretanar2005,niksic2011,robledo2019}. For open-shell nuclei the employed product states break rotational invariance and/or particle-number conservation and symmetry projection becomes relevant~\citep{sheikh2021}. The surveys~\citep{stoitsov2003,delaroche2010,erler2012,kortelainen2012,agbemava2014} explored collective properties from deformation and reference~\citep{hinohara2016} focused on the pairing rotational tensor associated with superfluidity.
A link to effective theories comes from symmetry projection. Projection after variation introduces collective coordinates as described in section~\ref{sec:found}: Euler angles for deformed and gauge angles for superfluid nuclei. Thus, the effective theories of nuclear deformation and superfluidity bring to the fore what is universal in the symmetry projection, namely rotational bands and pairing rotational bands. The success of mean-field approaches show that many aspects of low-energy nuclear structure can be based on symmetry-breaking product states and their symmetry restoration. These aspects link them to effective theories. 

We note finally that effective field theories could also be of some use in mean-field computations. Symmetry projections are problematic for energy functionals~\citep{duguet2010,nazarewicz2014}. Cranking techniques can yield energy spacings in a rotational band (or pairing rotational band). Using effective theories, on can relate such spacings to the gain in ground-state energies from symmetry projection. Similar comments apply to {\it ab initio} computations, where these tools have been used. We present details in the following section. 

\subsection{{\it Ab initio} computations}
\label{sec:abinitio}
The effective theories presented in this review form the lowest rung in a ladder of increasing resolution that ends with {\it ab initio} computations and connects nuclear phenomena that differ by more than two orders of magnitude in energy and resolution~\citep{bontems2021}. 

{\it Ab initio} computations of rotational states in nuclei are somewhat impressive because of the large separation of scales between the ground-state energy and the small level spacing within a rotational band. In $^8$Be, for instance, the binding energy is about 50~MeV while the excitation energy of the $2^+$ state is just 3~MeV, and the {\it ab initio} calculations~\citep{wiringa2013,dytrych2013,caprio2013} reproduced these data. The separation of scale increases with increasing mass, and {\it ab initio} computations have computed accurate binding energies together with rotational bands in increasingly heavier nuclei~\citep{epelbaum2012,frosini2022,hagen2022}. The advance is rapid, with the most recent computations of odd-mass nuclei around mass numbers $A\approx 20$ to 30~\citep{lin2024,sun2024b} and even-even nuclei around mass numbers $A\approx 70$ to 80~\citep{hu2024a,hu2024b}. 
These computations exploit the separation of scales between binding energies and rotational excitations~\citep{sun2024}; they start from symmetry-breaking states and obtain rotational bands from symmetry projection. Such computations  provide us with the microscopic basis of effective theories as we discussed in section~\ref{sec:found}.

The no-core shell model calculations of odd beryllium nuclei \citep{caprio2015,maris2015} include long and short-range physics in a single computation. Then, the interpretation of computed levels, and their placement into rotational bands, is based on the known phenomenological patterns. The  symmetry-adapted no-core-shell model~\citep{dytrych2008} is based on the insight by \citet{elliott1958} that quadrupole deformation in the spherical shell-model can be captured by exploiting the SU(3) symmetry of the harmonic oscillator.  

{\it Ab initio} calculations can yield the low-energy constants that enter the effective Hamiltonians such as equation~({\ref{HEFTLO}). This would then allow one to use the effective theory to compute other collective observables, to potentially extend the reach of {\it ab initio} methods, or to check the quality of {\it ab initio} computations. The latter is possible because effective theories of collective phenomena exhibit a power counting that allows one to estimate uncertainties. Let us for example estimate by how much the energy of a symmetry-breaking state is lowered by angular-momentum projections. For deformed nuclei, all that is needed is are the expectation values $\langle E\rangle$ and   $\langle J^2\rangle$ in the symmetry-broken state. As the low-lying states fulfill equation~(\ref{specLO}), the expectation values are related by
\begin{equation}
    \langle E\rangle = E_0 + a \langle J^2\rangle \ .
\end{equation}
Making assumptions about the rotational constant $a$, e.g. by estimating the energy scale $E_2-E_0=6a$ between the ground state and the $2^+$ state then allows one to estimate the energy gain from angular-momentum projection as 
\begin{equation}
    E_0-\langle E\rangle = -(E_2-E_0) \frac{\langle J^2\rangle}{6} \ .
\end{equation}
Such estimates have been used in reference~\citep{hagen2022}. 

Similar estimates can be obtained for calculations in a Bogoliubov framework where particle-number projection is missing. Using the energy expectation value and the expectation value of the particle number variance, and by estimating the size of the pairing rotational tensor~\citep{hinohara2015,hinohara2016,hinohara2018} then allows one to estimate the energy gain from a particle-number projection. Examples were presented in the works~\citep{papenbrock2022,tichai2023}.

\section{Summary and outlook} 
\label{sec:summary}
We reviewed effective theories for collective excitations in nuclei. These are either based on emergent symmetry breaking or on purely phenomenological degrees of freedom. The main insights and results are as follows. 
\begin{enumerate}
\item Effective theories based on emergent symmetry breaking
\begin{enumerate}
   \item These exploit that most nuclei exhibit deformation and superfluidity, i.e. an emergent breaking of rotational symmetry and of a phase (gauge) symmetry, respectively. The corresponding low-energy excitations are Nambu-Goldstone modes, i.e. rotations and pairing rotations, respectively, are universal. Similarly, the coupling of these excitations to faster degrees of freedom are universal or severely constrained by the patterns of the broken symmetries. Thus, effective theories of these phenomena bring to the fore the essential common properties of a plethora of nuclear models. 
   \item The effective theories for emergent symmetry breaking are based on the well-known coset approach used for spontaneous symmetry breaking but generalize it by including the purely time-dependent mode. This is key in finite systems. 
   \item Gauge potentials naturally enter when fast degrees of freedom are coupled to the Nambu-Goldstone modes. This connects deformed nuclei whose band heads have finite spins to topological phenomena and geometric phases.   
   \item Effective theories provide us with estimates for energy gains from symmetry projection (which are computationally expensive in {\it ab initio} computations) and with model-independent constraints on such computations. They also allow us to potentially extend the reach of {\it ab initio} computations for a description of universal phenomena.
\end{enumerate}
\item Effective theories based on phenomenological degrees of freedom
\begin{enumerate}
   \item In the absence of emergent symmetry breaking there are no universal properties. The effective theories for nuclear vibrations serve as examples. These exploit that one can identify a breakdown energy, which is separated in scale from the low-energy phenomena of interest. This allows one to introduce a power counting and to estimate or quantify uncertainties, and to turn a model (e.g. nuclear vibrations based on quadrupole degrees of freedom) into an effective theory. It opens an avenue to treat other nuclear models this way. 
   \item The effective theory approach demonstrated  that cadmium nuclei (and others) can -- at low resolution -- indeed be viewed as anharmonic vibrators. This interpretations is, of course, resolution dependent, and a more complicated picture might emerge at higher resolution. 
   \item This approach also allows one to quantify uncertainties for nuclear matrix elements of weak decays and neutrinoless double beta decay. The results from effective theories tend to be lower than those of other models and, in $^{48}$Ca and  $^{76}$Ge, are consistent with results from {\it ab initio} methods.
\end{enumerate}
\end{enumerate}

For an outlook we mention a few open problems. First, odd-odd nuclei provide us with a challenging opportunity. On the one hand, the approach of Section~\ref{sec:internal} could be applied to these nuclei. On the other hand, the higher level density of band heads would probably introduce strong inter-band couplings and perhaps also erase the separation of scales between internal degrees of freedom and rotational modes. Second, extending electromagnetic transitions to odd-mass nuclei or those with triaxial deformation might be profitable. Third, halo nuclei with a deformed core  pose an interesting challenge. Here, the low-energy of the halo might be of similar scale as rotational excitations, and one deals with two low-energy scales. Finally, the effective theories for nuclear vibrations might be motivations to turn other nuclear models into effective theories.

\section*{Acknowledgments}
We thank I.~K.~Alnamlah, C.~Brase, J.~Drake, J.~Men{\'e}ndez, D.~R.~Phillips, A.~Schwenk, H.~A.~Weidenm{\"u}ller, and J.~Zhang for their collaboration on the topics of this review, for many insights, and for fruitful discussions.  
We also thank R.~J.~Furnstahl, G.~Hagen, W.~Nazarewicz, and A.~Tichai for stimulating discussions and A.~Tichai for the data shown in table~\ref{tab:bogol}. We thank A.~Baroni for reading of the manuscript. We thank E. Thiriont-Bernolle for bringing the paper~\citep{jenkins2013}  to our attention.
This material is based upon work supported by the U.S.\ Department of Energy, Office of Science, Office of Nuclear Physics, under award number DE-FG02-96ER40963. This work was in part carried out at Oak Ridge National Laboratory, managed by UT-Battelle, LLC for the U.S. Department of Energy under contract DE-AC05-00OR22725.

\addcontentsline{toc}{section}{Bibliography}

\newcommand{\newblock}{} 



\begin{thebibliography}{187}%
\makeatletter
\providecommand \@ifxundefined [1]{%
 \@ifx{#1\undefined}
}%
\providecommand \@ifnum [1]{%
 \ifnum #1\expandafter \@firstoftwo
 \else \expandafter \@secondoftwo
 \fi
}%
\providecommand \@ifx [1]{%
 \ifx #1\expandafter \@firstoftwo
 \else \expandafter \@secondoftwo
 \fi
}%
\providecommand \natexlab [1]{#1}%
\providecommand \enquote  [1]{``#1''}%
\providecommand \bibnamefont  [1]{#1}%
\providecommand \bibfnamefont [1]{#1}%
\providecommand \citenamefont [1]{#1}%
\providecommand \href@noop [0]{\@secondoftwo}%
\providecommand \href [0]{\begingroup \@sanitize@url \@href}%
\providecommand \@href[1]{\@@startlink{#1}\@@href}%
\providecommand \@@href[1]{\endgroup#1\@@endlink}%
\providecommand \@sanitize@url [0]{\catcode `\\12\catcode `\$12\catcode
  `\&12\catcode `\#12\catcode `\^12\catcode `\_12\catcode `\%12\relax}%
\providecommand \@@startlink[1]{}%
\providecommand \@@endlink[0]{}%
\providecommand \url  [0]{\begingroup\@sanitize@url \@url }%
\providecommand \@url [1]{\endgroup\@href {#1}{\urlprefix }}%
\providecommand \urlprefix  [0]{URL }%
\providecommand \Eprint [0]{\href }%
\providecommand \doibase [0]{http://dx.doi.org/}%
\providecommand \selectlanguage [0]{\@gobble}%
\providecommand \bibinfo  [0]{\@secondoftwo}%
\providecommand \bibfield  [0]{\@secondoftwo}%
\providecommand \translation [1]{[#1]}%
\providecommand \BibitemOpen [0]{}%
\providecommand \bibitemStop [0]{}%
\providecommand \bibitemNoStop [0]{.\EOS\space}%
\providecommand \EOS [0]{\spacefactor3000\relax}%
\providecommand \BibitemShut  [1]{\csname bibitem#1\endcsname}%
\let\auto@bib@innerbib\@empty
\bibitem [{\citenamefont {Agbemava}\ \emph {et~al.}(2014)\citenamefont
  {Agbemava}, \citenamefont {Afanasjev}, \citenamefont {Ray},\ and\
  \citenamefont {Ring}}]{agbemava2014}%
  \BibitemOpen
  \bibfield  {author} {\bibinfo {author} {\bibnamefont {Agbemava},
  \bibfnamefont {S~E}}, \bibinfo {author} {\bibnamefont {Afanasjev},
  \bibfnamefont {A~V}}, \bibinfo {author} {\bibnamefont {Ray}, \bibfnamefont
  {D}}, \ and\ \bibinfo {author} {\bibnamefont {Ring}, \bibfnamefont {P}}}
  (\bibinfo {year} {2014}),\ \bibfield  {title} {\enquote {\bibinfo {title}
  {Global performance of covariant energy density functionals: Ground state
  observables of even-even nuclei and the estimate of theoretical
  uncertainties},}\ }\href {\doibase 10.1103/PhysRevC.89.054320} {\bibfield
  {journal} {\bibinfo  {journal} {Phys. Rev. C}\ }\textbf {\bibinfo {volume}
  {89}},\ \bibinfo {pages} {054320}}\BibitemShut {NoStop}%
\bibitem [{\citenamefont {Albert}\ \emph {et~al.}(2016)\citenamefont {Albert}
  \emph {et~al.}}]{albert_jb_2016}%
  \BibitemOpen
  \bibfield  {author} {\bibinfo {author} {\bibnamefont {Albert}, \bibfnamefont
  {J~B}},  \emph {et~al.} (\bibinfo {collaboration} {EXO-200 Collaboration})}
  (\bibinfo {year} {2016}),\ \bibfield  {title} {\enquote {\bibinfo {title}
  {{Search for $2\ensuremath{\nu}\ensuremath{\beta}\ensuremath{\beta}$ decay of
  $^{136}\mathrm{Xe}$ to the ${0}_{1}^{+}$ excited state of $^{136}\mathrm{Ba}$
  with the EXO-200 liquid xenon detector}},}\ }\href {\doibase
  10.1103/PhysRevC.93.035501} {\bibfield  {journal} {\bibinfo  {journal} {Phys.
  Rev. C}\ }\textbf {\bibinfo {volume} {93}},\ \bibinfo {pages}
  {035501}}\BibitemShut {NoStop}%
\bibitem [{\citenamefont {Alnamlah}\ \emph {et~al.}(2021)\citenamefont
  {Alnamlah}, \citenamefont {Coello~P\'erez},\ and\ \citenamefont
  {Phillips}}]{alnamlah2021}%
  \BibitemOpen
  \bibfield  {author} {\bibinfo {author} {\bibnamefont {Alnamlah},
  \bibfnamefont {I~K}}, \bibinfo {author} {\bibnamefont {Coello~P\'erez},
  \bibfnamefont {E~A}}, \ and\ \bibinfo {author} {\bibnamefont {Phillips},
  \bibfnamefont {D~R}}} (\bibinfo {year} {2021}),\ \bibfield  {title} {\enquote
  {\bibinfo {title} {Effective field theory approach to rotational bands in
  odd-mass nuclei},}\ }\href {\doibase 10.1103/PhysRevC.104.064311} {\bibfield
  {journal} {\bibinfo  {journal} {Phys. Rev. C}\ }\textbf {\bibinfo {volume}
  {104}},\ \bibinfo {pages} {064311}}\BibitemShut {NoStop}%
\bibitem [{\citenamefont {Alnamlah}\ \emph {et~al.}(2022)\citenamefont
  {Alnamlah}, \citenamefont {Coello~Pérez},\ and\ \citenamefont
  {Phillips}}]{alnamlah2022}%
  \BibitemOpen
  \bibfield  {author} {\bibinfo {author} {\bibnamefont {Alnamlah},
  \bibfnamefont {I~K}}, \bibinfo {author} {\bibnamefont {Coello~Pérez},
  \bibfnamefont {E~A}}, \ and\ \bibinfo {author} {\bibnamefont {Phillips},
  \bibfnamefont {D~R}}} (\bibinfo {year} {2022}),\ \bibfield  {title} {\enquote
  {\bibinfo {title} {Analyzing rotational bands in odd-mass nuclei using
  effective field theory and bayesian methods},}\ }\href {\doibase
  10.3389/fphy.2022.901954} {\bibfield  {journal} {\bibinfo  {journal}
  {Frontiers in Physics}\ }\textbf {\bibinfo {volume} {10}},\
  10.3389/fphy.2022.901954}\BibitemShut {NoStop}%
\bibitem [{\citenamefont {Altland}\ and\ \citenamefont
  {Sonner}(2021)}]{altland2021}%
  \BibitemOpen
  \bibfield  {author} {\bibinfo {author} {\bibnamefont {Altland}, \bibfnamefont
  {A}}, \ and\ \bibinfo {author} {\bibnamefont {Sonner}, \bibfnamefont {J}}}
  (\bibinfo {year} {2021}),\ \bibfield  {title} {\enquote {\bibinfo {title}
  {{Late time physics of holographic quantum chaos}},}\ }\href {\doibase
  10.21468/SciPostPhys.11.2.034} {\bibfield  {journal} {\bibinfo  {journal}
  {SciPost Phys.}\ }\textbf {\bibinfo {volume} {11}},\ \bibinfo {pages}
  {034}}\BibitemShut {NoStop}%
\bibitem [{\citenamefont {Aprahamian}\ \emph {et~al.}(2006)\citenamefont
  {Aprahamian}, \citenamefont {Wu}, \citenamefont {Lesher}, \citenamefont
  {Warner}, \citenamefont {Gelletly}, \citenamefont {B{\"o}rner}, \citenamefont
  {Hoyler}, \citenamefont {Schreckenbach}, \citenamefont {Casten},
  \citenamefont {Shi}, \citenamefont {Kusnezov}, \citenamefont {Ibrahim},
  \citenamefont {Macchiavelli}, \citenamefont {Brinkman},\ and\ \citenamefont
  {Becker}}]{aprahamian2006}%
  \BibitemOpen
  \bibfield  {author} {\bibinfo {author} {\bibnamefont {Aprahamian},
  \bibfnamefont {A}}, \bibinfo {author} {\bibnamefont {Wu}, \bibfnamefont {X}},
  \bibinfo {author} {\bibnamefont {Lesher}, \bibfnamefont {S~R}}, \bibinfo
  {author} {\bibnamefont {Warner}, \bibfnamefont {D~D}}, \bibinfo {author}
  {\bibnamefont {Gelletly}, \bibfnamefont {W}}, \bibinfo {author} {\bibnamefont
  {B{\"o}rner}, \bibfnamefont {H~G}}, \bibinfo {author} {\bibnamefont {Hoyler},
  \bibfnamefont {F}}, \bibinfo {author} {\bibnamefont {Schreckenbach},
  \bibfnamefont {K}}, \bibinfo {author} {\bibnamefont {Casten}, \bibfnamefont
  {R~F}}, \bibinfo {author} {\bibnamefont {Shi}, \bibfnamefont {Z~R}}, \bibinfo
  {author} {\bibnamefont {Kusnezov}, \bibfnamefont {D}}, \bibinfo {author}
  {\bibnamefont {Ibrahim}, \bibfnamefont {M}}, \bibinfo {author} {\bibnamefont
  {Macchiavelli}, \bibfnamefont {A~O}}, \bibinfo {author} {\bibnamefont
  {Brinkman}, \bibfnamefont {M~A}}, \ and\ \bibinfo {author} {\bibnamefont
  {Becker}, \bibfnamefont {J~A}}} (\bibinfo {year} {2006}),\ \bibfield  {title}
  {\enquote {\bibinfo {title} {Complete spectroscopy of the $^{162}\mathrm{Dy}$
  nucleus},}\ }\href {\doibase 10.1016/j.nuclphysa.2005.09.020} {\bibfield
  {journal} {\bibinfo  {journal} {Nuclear Physics A}\ }\textbf {\bibinfo
  {volume} {764}},\ \bibinfo {pages} {42 }}\BibitemShut {NoStop}%
\bibitem [{\citenamefont {Aprile}\ \emph {et~al.}(2022)\citenamefont {Aprile},
  \citenamefont {Abe}, \citenamefont {Agostini}, \citenamefont
  {Ahmed~Maouloud}, \citenamefont {Alfonsi}, \citenamefont {Althueser},
  \citenamefont {Andrieu}, \citenamefont {Angelino}, \citenamefont {Angevaare},
  \citenamefont {Antochi}, \citenamefont {Ant\'on~Martin}, \citenamefont
  {Arneodo}, \citenamefont {Baudis}, \citenamefont {Baxter}, \citenamefont
  {Bellagamba}, \citenamefont {Biondi}, \citenamefont {Bismark}, \citenamefont
  {Brown}, \citenamefont {Bruenner}, \citenamefont {Bruno}, \citenamefont
  {Budnik}, \citenamefont {Cai}, \citenamefont {Capelli}, \citenamefont
  {Cardoso}, \citenamefont {Cichon}, \citenamefont {Clark}, \citenamefont
  {Colijn}, \citenamefont {Conrad}, \citenamefont {Cuenca-Garc\'{\i}a},
  \citenamefont {Cussonneau}, \citenamefont {D'Andrea}, \citenamefont
  {Decowski}, \citenamefont {Di~Gangi}, \citenamefont {Di~Pede}, \citenamefont
  {Di~Giovanni}, \citenamefont {Di~Stefano}, \citenamefont {Diglio},
  \citenamefont {Eitel}, \citenamefont {Elykov}, \citenamefont {Farrell},
  \citenamefont {Ferella}, \citenamefont {Fischer}, \citenamefont {Fulgione},
  \citenamefont {Gaemers}, \citenamefont {Gaior}, \citenamefont {Gallo~Rosso},
  \citenamefont {Galloway}, \citenamefont {Gao}, \citenamefont {Glade-Beucke},
  \citenamefont {Grandi}, \citenamefont {Grigat}, \citenamefont {Guida},
  \citenamefont {Higuera}, \citenamefont {Hils}, \citenamefont {Hoetzsch},
  \citenamefont {Howlett}, \citenamefont {Iacovacci}, \citenamefont {Itow},
  \citenamefont {Jakob}, \citenamefont {Joerg}, \citenamefont {Joy},
  \citenamefont {Kato}, \citenamefont {Kara}, \citenamefont {Kavrigin},
  \citenamefont {Kazama}, \citenamefont {Kobayashi}, \citenamefont {Koltman},
  \citenamefont {Kopec}, \citenamefont {Landsman}, \citenamefont {Lang},
  \citenamefont {Levinson}, \citenamefont {Li}, \citenamefont {Li},
  \citenamefont {Liang}, \citenamefont {Lindemann}, \citenamefont {Lindner},
  \citenamefont {Liu}, \citenamefont {Loizeau}, \citenamefont {Lombardi},
  \citenamefont {Long}, \citenamefont {Lopes}, \citenamefont {Ma},
  \citenamefont {Macolino}, \citenamefont {Mahlstedt}, \citenamefont {Mancuso},
  \citenamefont {Manenti}, \citenamefont {Manfredini}, \citenamefont
  {Marignetti}, \citenamefont {Marrod\'an~Undagoitia}, \citenamefont {Martens},
  \citenamefont {Masbou}, \citenamefont {Masson}, \citenamefont {Masson},
  \citenamefont {Mastroianni}, \citenamefont {Messina}, \citenamefont {Miuchi},
  \citenamefont {Mizukoshi}, \citenamefont {Molinario}, \citenamefont
  {Moriyama}, \citenamefont {Mor\aa{}}, \citenamefont {Mosbacher},
  \citenamefont {Murra}, \citenamefont {M\"uller}, \citenamefont {Ni},
  \citenamefont {Oberlack}, \citenamefont {Paetsch}, \citenamefont {Palacio},
  \citenamefont {Peres}, \citenamefont {Pienaar}, \citenamefont {Pierre},
  \citenamefont {Pizzella}, \citenamefont {Plante}, \citenamefont {Qi},
  \citenamefont {Qin}, \citenamefont {Ram\'{\i}rez~Garc\'{\i}a}, \citenamefont
  {Reichard}, \citenamefont {Rocchetti}, \citenamefont {Rupp}, \citenamefont
  {Sanchez}, \citenamefont {dos Santos}, \citenamefont {Sarnoff}, \citenamefont
  {Sartorelli}, \citenamefont {Schreiner}, \citenamefont {Schulte},
  \citenamefont {Schulte}, \citenamefont {Schulze~Ei\ss{}ing}, \citenamefont
  {Schumann}, \citenamefont {Scotto~Lavina}, \citenamefont {Selvi},
  \citenamefont {Semeria}, \citenamefont {Shagin}, \citenamefont {Shi},
  \citenamefont {Shockley}, \citenamefont {Silva}, \citenamefont {Simgen},
  \citenamefont {Takeda}, \citenamefont {Tan}, \citenamefont {Terliuk},
  \citenamefont {Thers}, \citenamefont {Toschi}, \citenamefont {Trinchero},
  \citenamefont {Tunnell}, \citenamefont {T\"onnies}, \citenamefont {Valerius},
  \citenamefont {Volta}, \citenamefont {Wei}, \citenamefont {Weinheimer},
  \citenamefont {Weiss}, \citenamefont {Wenz}, \citenamefont {Wittweg},
  \citenamefont {Wolf}, \citenamefont {Xu}, \citenamefont {Yamashita},
  \citenamefont {Yang}, \citenamefont {Ye}, \citenamefont {Yuan}, \citenamefont
  {Zavattini}, \citenamefont {Zerbo}, \citenamefont {Zhong},\ and\
  \citenamefont {Zhu}}]{aprile_e_2022}%
  \BibitemOpen
  \bibfield  {author} {\bibinfo {author} {\bibnamefont {Aprile}, \bibfnamefont
  {E}}, \bibinfo {author} {\bibnamefont {Abe}, \bibfnamefont {K}}, \bibinfo
  {author} {\bibnamefont {Agostini}, \bibfnamefont {F}}, \bibinfo {author}
  {\bibnamefont {Ahmed~Maouloud}, \bibfnamefont {S}}, \bibinfo {author}
  {\bibnamefont {Alfonsi}, \bibfnamefont {M}}, \bibinfo {author} {\bibnamefont
  {Althueser}, \bibfnamefont {L}}, \bibinfo {author} {\bibnamefont {Andrieu},
  \bibfnamefont {B}}, \bibinfo {author} {\bibnamefont {Angelino}, \bibfnamefont
  {E}}, \bibinfo {author} {\bibnamefont {Angevaare}, \bibfnamefont {J~R}},
  \bibinfo {author} {\bibnamefont {Antochi}, \bibfnamefont {V~C}}, \bibinfo
  {author} {\bibnamefont {Ant\'on~Martin}, \bibfnamefont {D}}, \bibinfo
  {author} {\bibnamefont {Arneodo}, \bibfnamefont {F}}, \bibinfo {author}
  {\bibnamefont {Baudis}, \bibfnamefont {L}}, \bibinfo {author} {\bibnamefont
  {Baxter}, \bibfnamefont {A~L}}, \bibinfo {author} {\bibnamefont {Bellagamba},
  \bibfnamefont {L}}, \bibinfo {author} {\bibnamefont {Biondi}, \bibfnamefont
  {R}}, \bibinfo {author} {\bibnamefont {Bismark}, \bibfnamefont {A}}, \bibinfo
  {author} {\bibnamefont {Brown}, \bibfnamefont {A}}, \bibinfo {author}
  {\bibnamefont {Bruenner}, \bibfnamefont {S}}, \bibinfo {author} {\bibnamefont
  {Bruno}, \bibfnamefont {G}}, \bibinfo {author} {\bibnamefont {Budnik},
  \bibfnamefont {R}}, \bibinfo {author} {\bibnamefont {Cai}, \bibfnamefont
  {C}}, \bibinfo {author} {\bibnamefont {Capelli}, \bibfnamefont {C}}, \bibinfo
  {author} {\bibnamefont {Cardoso}, \bibfnamefont {J~M~R}}, \bibinfo {author}
  {\bibnamefont {Cichon}, \bibfnamefont {D}}, \bibinfo {author} {\bibnamefont
  {Clark}, \bibfnamefont {M}}, \bibinfo {author} {\bibnamefont {Colijn},
  \bibfnamefont {A~P}}, \bibinfo {author} {\bibnamefont {Conrad}, \bibfnamefont
  {J}}, \bibinfo {author} {\bibnamefont {Cuenca-Garc\'{\i}a}, \bibfnamefont
  {J~J}}, \bibinfo {author} {\bibnamefont {Cussonneau}, \bibfnamefont {J~P}},
  \bibinfo {author} {\bibnamefont {D'Andrea}, \bibfnamefont {V}}, \bibinfo
  {author} {\bibnamefont {Decowski}, \bibfnamefont {M~P}}, \bibinfo {author}
  {\bibnamefont {Di~Gangi}, \bibfnamefont {P}}, \bibinfo {author} {\bibnamefont
  {Di~Pede}, \bibfnamefont {S}}, \bibinfo {author} {\bibnamefont {Di~Giovanni},
  \bibfnamefont {A}}, \bibinfo {author} {\bibnamefont {Di~Stefano},
  \bibfnamefont {R}}, \bibinfo {author} {\bibnamefont {Diglio}, \bibfnamefont
  {S}}, \bibinfo {author} {\bibnamefont {Eitel}, \bibfnamefont {K}}, \bibinfo
  {author} {\bibnamefont {Elykov}, \bibfnamefont {A}}, \bibinfo {author}
  {\bibnamefont {Farrell}, \bibfnamefont {S}}, \bibinfo {author} {\bibnamefont
  {Ferella}, \bibfnamefont {A~D}}, \bibinfo {author} {\bibnamefont {Fischer},
  \bibfnamefont {H}}, \bibinfo {author} {\bibnamefont {Fulgione}, \bibfnamefont
  {W}}, \bibinfo {author} {\bibnamefont {Gaemers}, \bibfnamefont {P}}, \bibinfo
  {author} {\bibnamefont {Gaior}, \bibfnamefont {R}}, \bibinfo {author}
  {\bibnamefont {Gallo~Rosso}, \bibfnamefont {A}}, \bibinfo {author}
  {\bibnamefont {Galloway}, \bibfnamefont {M}}, \bibinfo {author} {\bibnamefont
  {Gao}, \bibfnamefont {F}}, \bibinfo {author} {\bibnamefont {Glade-Beucke},
  \bibfnamefont {R}}, \bibinfo {author} {\bibnamefont {Grandi}, \bibfnamefont
  {L}}, \bibinfo {author} {\bibnamefont {Grigat}, \bibfnamefont {J}}, \bibinfo
  {author} {\bibnamefont {Guida}, \bibfnamefont {M}}, \bibinfo {author}
  {\bibnamefont {Higuera}, \bibfnamefont {A}}, \bibinfo {author} {\bibnamefont
  {Hils}, \bibfnamefont {C}}, \bibinfo {author} {\bibnamefont {Hoetzsch},
  \bibfnamefont {L}}, \bibinfo {author} {\bibnamefont {Howlett}, \bibfnamefont
  {J}}, \bibinfo {author} {\bibnamefont {Iacovacci}, \bibfnamefont {M}},
  \bibinfo {author} {\bibnamefont {Itow}, \bibfnamefont {Y}}, \bibinfo {author}
  {\bibnamefont {Jakob}, \bibfnamefont {J}}, \bibinfo {author} {\bibnamefont
  {Joerg}, \bibfnamefont {F}}, \bibinfo {author} {\bibnamefont {Joy},
  \bibfnamefont {A}}, \bibinfo {author} {\bibnamefont {Kato}, \bibfnamefont
  {N}}, \bibinfo {author} {\bibnamefont {Kara}, \bibfnamefont {M}}, \bibinfo
  {author} {\bibnamefont {Kavrigin}, \bibfnamefont {P}}, \bibinfo {author}
  {\bibnamefont {Kazama}, \bibfnamefont {S}}, \bibinfo {author} {\bibnamefont
  {Kobayashi}, \bibfnamefont {M}}, \bibinfo {author} {\bibnamefont {Koltman},
  \bibfnamefont {G}}, \bibinfo {author} {\bibnamefont {Kopec}, \bibfnamefont
  {A}}, \bibinfo {author} {\bibnamefont {Landsman}, \bibfnamefont {H}},
  \bibinfo {author} {\bibnamefont {Lang}, \bibfnamefont {R~F}}, \bibinfo
  {author} {\bibnamefont {Levinson}, \bibfnamefont {L}}, \bibinfo {author}
  {\bibnamefont {Li}, \bibfnamefont {I}}, \bibinfo {author} {\bibnamefont {Li},
  \bibfnamefont {S}}, \bibinfo {author} {\bibnamefont {Liang}, \bibfnamefont
  {S}}, \bibinfo {author} {\bibnamefont {Lindemann}, \bibfnamefont {S}},
  \bibinfo {author} {\bibnamefont {Lindner}, \bibfnamefont {M}}, \bibinfo
  {author} {\bibnamefont {Liu}, \bibfnamefont {K}}, \bibinfo {author}
  {\bibnamefont {Loizeau}, \bibfnamefont {J}}, \bibinfo {author} {\bibnamefont
  {Lombardi}, \bibfnamefont {F}}, \bibinfo {author} {\bibnamefont {Long},
  \bibfnamefont {J}}, \bibinfo {author} {\bibnamefont {Lopes}, \bibfnamefont
  {J~A~M}}, \bibinfo {author} {\bibnamefont {Ma}, \bibfnamefont {Y}}, \bibinfo
  {author} {\bibnamefont {Macolino}, \bibfnamefont {C}}, \bibinfo {author}
  {\bibnamefont {Mahlstedt}, \bibfnamefont {J}}, \bibinfo {author}
  {\bibnamefont {Mancuso}, \bibfnamefont {A}}, \bibinfo {author} {\bibnamefont
  {Manenti}, \bibfnamefont {L}}, \bibinfo {author} {\bibnamefont {Manfredini},
  \bibfnamefont {A}}, \bibinfo {author} {\bibnamefont {Marignetti},
  \bibfnamefont {F}}, \bibinfo {author} {\bibnamefont {Marrod\'an~Undagoitia},
  \bibfnamefont {T}}, \bibinfo {author} {\bibnamefont {Martens}, \bibfnamefont
  {K}}, \bibinfo {author} {\bibnamefont {Masbou}, \bibfnamefont {J}}, \bibinfo
  {author} {\bibnamefont {Masson}, \bibfnamefont {D}}, \bibinfo {author}
  {\bibnamefont {Masson}, \bibfnamefont {E}}, \bibinfo {author} {\bibnamefont
  {Mastroianni}, \bibfnamefont {S}}, \bibinfo {author} {\bibnamefont {Messina},
  \bibfnamefont {M}}, \bibinfo {author} {\bibnamefont {Miuchi}, \bibfnamefont
  {K}}, \bibinfo {author} {\bibnamefont {Mizukoshi}, \bibfnamefont {K}},
  \bibinfo {author} {\bibnamefont {Molinario}, \bibfnamefont {A}}, \bibinfo
  {author} {\bibnamefont {Moriyama}, \bibfnamefont {S}}, \bibinfo {author}
  {\bibnamefont {Mor\aa{}}, \bibfnamefont {K}}, \bibinfo {author} {\bibnamefont
  {Mosbacher}, \bibfnamefont {Y}}, \bibinfo {author} {\bibnamefont {Murra},
  \bibfnamefont {M}}, \bibinfo {author} {\bibnamefont {M\"uller}, \bibfnamefont
  {J}}, \bibinfo {author} {\bibnamefont {Ni}, \bibfnamefont {K}}, \bibinfo
  {author} {\bibnamefont {Oberlack}, \bibfnamefont {U}}, \bibinfo {author}
  {\bibnamefont {Paetsch}, \bibfnamefont {B}}, \bibinfo {author} {\bibnamefont
  {Palacio}, \bibfnamefont {J}}, \bibinfo {author} {\bibnamefont {Peres},
  \bibfnamefont {R}}, \bibinfo {author} {\bibnamefont {Pienaar}, \bibfnamefont
  {J}}, \bibinfo {author} {\bibnamefont {Pierre}, \bibfnamefont {M}}, \bibinfo
  {author} {\bibnamefont {Pizzella}, \bibfnamefont {V}}, \bibinfo {author}
  {\bibnamefont {Plante}, \bibfnamefont {G}}, \bibinfo {author} {\bibnamefont
  {Qi}, \bibfnamefont {J}}, \bibinfo {author} {\bibnamefont {Qin},
  \bibfnamefont {J}}, \bibinfo {author} {\bibnamefont
  {Ram\'{\i}rez~Garc\'{\i}a}, \bibfnamefont {D}}, \bibinfo {author}
  {\bibnamefont {Reichard}, \bibfnamefont {S}}, \bibinfo {author} {\bibnamefont
  {Rocchetti}, \bibfnamefont {A}}, \bibinfo {author} {\bibnamefont {Rupp},
  \bibfnamefont {N}}, \bibinfo {author} {\bibnamefont {Sanchez}, \bibfnamefont
  {L}}, \bibinfo {author} {\bibnamefont {dos Santos}, \bibfnamefont {J~M~F}},
  \bibinfo {author} {\bibnamefont {Sarnoff}, \bibfnamefont {I}}, \bibinfo
  {author} {\bibnamefont {Sartorelli}, \bibfnamefont {G}}, \bibinfo {author}
  {\bibnamefont {Schreiner}, \bibfnamefont {J}}, \bibinfo {author}
  {\bibnamefont {Schulte}, \bibfnamefont {D}}, \bibinfo {author} {\bibnamefont
  {Schulte}, \bibfnamefont {P}}, \bibinfo {author} {\bibnamefont
  {Schulze~Ei\ss{}ing}, \bibfnamefont {H}}, \bibinfo {author} {\bibnamefont
  {Schumann}, \bibfnamefont {M}}, \bibinfo {author} {\bibnamefont
  {Scotto~Lavina}, \bibfnamefont {L}}, \bibinfo {author} {\bibnamefont {Selvi},
  \bibfnamefont {M}}, \bibinfo {author} {\bibnamefont {Semeria}, \bibfnamefont
  {F}}, \bibinfo {author} {\bibnamefont {Shagin}, \bibfnamefont {P}}, \bibinfo
  {author} {\bibnamefont {Shi}, \bibfnamefont {S}}, \bibinfo {author}
  {\bibnamefont {Shockley}, \bibfnamefont {E}}, \bibinfo {author} {\bibnamefont
  {Silva}, \bibfnamefont {M}}, \bibinfo {author} {\bibnamefont {Simgen},
  \bibfnamefont {H}}, \bibinfo {author} {\bibnamefont {Takeda}, \bibfnamefont
  {A}}, \bibinfo {author} {\bibnamefont {Tan}, \bibfnamefont {P~L}}, \bibinfo
  {author} {\bibnamefont {Terliuk}, \bibfnamefont {A}}, \bibinfo {author}
  {\bibnamefont {Thers}, \bibfnamefont {D}}, \bibinfo {author} {\bibnamefont
  {Toschi}, \bibfnamefont {F}}, \bibinfo {author} {\bibnamefont {Trinchero},
  \bibfnamefont {G}}, \bibinfo {author} {\bibnamefont {Tunnell}, \bibfnamefont
  {C}}, \bibinfo {author} {\bibnamefont {T\"onnies}, \bibfnamefont {F}},
  \bibinfo {author} {\bibnamefont {Valerius}, \bibfnamefont {K}}, \bibinfo
  {author} {\bibnamefont {Volta}, \bibfnamefont {G}}, \bibinfo {author}
  {\bibnamefont {Wei}, \bibfnamefont {Y}}, \bibinfo {author} {\bibnamefont
  {Weinheimer}, \bibfnamefont {C}}, \bibinfo {author} {\bibnamefont {Weiss},
  \bibfnamefont {M}}, \bibinfo {author} {\bibnamefont {Wenz}, \bibfnamefont
  {D}}, \bibinfo {author} {\bibnamefont {Wittweg}, \bibfnamefont {C}}, \bibinfo
  {author} {\bibnamefont {Wolf}, \bibfnamefont {T}}, \bibinfo {author}
  {\bibnamefont {Xu}, \bibfnamefont {Z}}, \bibinfo {author} {\bibnamefont
  {Yamashita}, \bibfnamefont {M}}, \bibinfo {author} {\bibnamefont {Yang},
  \bibfnamefont {L}}, \bibinfo {author} {\bibnamefont {Ye}, \bibfnamefont {J}},
  \bibinfo {author} {\bibnamefont {Yuan}, \bibfnamefont {L}}, \bibinfo {author}
  {\bibnamefont {Zavattini}, \bibfnamefont {G}}, \bibinfo {author}
  {\bibnamefont {Zerbo}, \bibfnamefont {S}}, \bibinfo {author} {\bibnamefont
  {Zhong}, \bibfnamefont {M}}, \ and\ \bibinfo {author} {\bibnamefont {Zhu},
  \bibfnamefont {T}} (\bibinfo {collaboration} {XENON Collaboration})}
  (\bibinfo {year} {2022}),\ \bibfield  {title} {\enquote {\bibinfo {title}
  {Double-weak decays of $^{124}\mathrm{Xe}$ and $^{136}\mathrm{Xe}$ in the
  xenon1t and xenonnt experiments},}\ }\href {\doibase
  10.1103/PhysRevC.106.024328} {\bibfield  {journal} {\bibinfo  {journal}
  {Phys. Rev. C}\ }\textbf {\bibinfo {volume} {106}},\ \bibinfo {pages}
  {024328}}\BibitemShut {NoStop}%
\bibitem [{\citenamefont {Aprile}\ \emph {et~al.}(2019)\citenamefont {Aprile}
  \emph {et~al.}}]{aprile_e_2019}%
  \BibitemOpen
  \bibfield  {author} {\bibinfo {author} {\bibnamefont {Aprile}, \bibfnamefont
  {E}},  \emph {et~al.} (\bibinfo {collaboration} {XENON})} (\bibinfo {year}
  {2019}),\ \bibfield  {title} {\enquote {\bibinfo {title} {{Observation of
  two-neutrino double electron capture in $^{124}$Xe with XENON1T}},}\ }\href
  {\doibase 10.1038/s41586-019-1124-4} {\bibfield  {journal} {\bibinfo
  {journal} {Nature}\ }\textbf {\bibinfo {volume} {568}}~(\bibinfo {number}
  {7753}),\ \bibinfo {pages} {532}}\BibitemShut {NoStop}%
\bibitem [{\citenamefont {Arima}\ and\ \citenamefont
  {Iachello}(1975)}]{arima1975}%
  \BibitemOpen
  \bibfield  {author} {\bibinfo {author} {\bibnamefont {Arima}, \bibfnamefont
  {A}}, \ and\ \bibinfo {author} {\bibnamefont {Iachello}, \bibfnamefont {F}}}
  (\bibinfo {year} {1975}),\ \bibfield  {title} {\enquote {\bibinfo {title}
  {Collective nuclear states as representations of a su(6) group},}\ }\href
  {\doibase 10.1103/PhysRevLett.35.1069} {\bibfield  {journal} {\bibinfo
  {journal} {Phys. Rev. Lett.}\ }\textbf {\bibinfo {volume} {35}},\ \bibinfo
  {pages} {1069}}\BibitemShut {NoStop}%
\bibitem [{\citenamefont {Asakura}\ \emph {et~al.}(2016)\citenamefont {Asakura}
  \emph {et~al.}}]{asakura_k_2016}%
  \BibitemOpen
  \bibfield  {author} {\bibinfo {author} {\bibnamefont {Asakura}, \bibfnamefont
  {K}},  \emph {et~al.}} (\bibinfo {year} {2016}),\ \bibfield  {title}
  {\enquote {\bibinfo {title} {{Search for double-beta decay of
  $\ensuremath{^{136}}$Xe to excited states of $\ensuremath{^{136}}$Ba with the
  KamLAND-Zen experiment}},}\ }\href {\doibase 10.1016/j.nuclphysa.2015.11.011}
  {\bibfield  {journal} {\bibinfo  {journal} {Nucl. Phys. A}\ }\textbf
  {\bibinfo {volume} {946}},\ \bibinfo {pages} {171}}\BibitemShut {NoStop}%
\bibitem [{\citenamefont {Ayangeakaa}\ \emph {et~al.}(2019)\citenamefont
  {Ayangeakaa}, \citenamefont {Janssens}, \citenamefont {Zhu}, \citenamefont
  {Little}, \citenamefont {Henderson}, \citenamefont {Wu}, \citenamefont
  {Hartley}, \citenamefont {Albers}, \citenamefont {Auranen}, \citenamefont
  {Bucher}, \citenamefont {Carpenter}, \citenamefont {Chowdhury}, \citenamefont
  {Cline}, \citenamefont {Crawford}, \citenamefont {Fallon}, \citenamefont
  {Forney}, \citenamefont {Gade}, \citenamefont {Hayes}, \citenamefont
  {Kondev}, \citenamefont {Krishichayan}, \citenamefont {Lauritsen},
  \citenamefont {Li}, \citenamefont {Macchiavelli}, \citenamefont {Rhodes},
  \citenamefont {Seweryniak}, \citenamefont {Stolze}, \citenamefont {Walters},\
  and\ \citenamefont {Wu}}]{ayangeakaa2019}%
  \BibitemOpen
  \bibfield  {author} {\bibinfo {author} {\bibnamefont {Ayangeakaa},
  \bibfnamefont {A~D}}, \bibinfo {author} {\bibnamefont {Janssens},
  \bibfnamefont {R~V~F}}, \bibinfo {author} {\bibnamefont {Zhu}, \bibfnamefont
  {S}}, \bibinfo {author} {\bibnamefont {Little}, \bibfnamefont {D}}, \bibinfo
  {author} {\bibnamefont {Henderson}, \bibfnamefont {J}}, \bibinfo {author}
  {\bibnamefont {Wu}, \bibfnamefont {C~Y}}, \bibinfo {author} {\bibnamefont
  {Hartley}, \bibfnamefont {D~J}}, \bibinfo {author} {\bibnamefont {Albers},
  \bibfnamefont {M}}, \bibinfo {author} {\bibnamefont {Auranen}, \bibfnamefont
  {K}}, \bibinfo {author} {\bibnamefont {Bucher}, \bibfnamefont {B}}, \bibinfo
  {author} {\bibnamefont {Carpenter}, \bibfnamefont {M~P}}, \bibinfo {author}
  {\bibnamefont {Chowdhury}, \bibfnamefont {P}}, \bibinfo {author}
  {\bibnamefont {Cline}, \bibfnamefont {D}}, \bibinfo {author} {\bibnamefont
  {Crawford}, \bibfnamefont {H~L}}, \bibinfo {author} {\bibnamefont {Fallon},
  \bibfnamefont {P}}, \bibinfo {author} {\bibnamefont {Forney}, \bibfnamefont
  {A~M}}, \bibinfo {author} {\bibnamefont {Gade}, \bibfnamefont {A}}, \bibinfo
  {author} {\bibnamefont {Hayes}, \bibfnamefont {A~B}}, \bibinfo {author}
  {\bibnamefont {Kondev}, \bibfnamefont {F~G}}, \bibinfo {author} {\bibnamefont
  {Krishichayan},}, \bibinfo {author} {\bibnamefont {Lauritsen}, \bibfnamefont
  {T}}, \bibinfo {author} {\bibnamefont {Li}, \bibfnamefont {J}}, \bibinfo
  {author} {\bibnamefont {Macchiavelli}, \bibfnamefont {A~O}}, \bibinfo
  {author} {\bibnamefont {Rhodes}, \bibfnamefont {D}}, \bibinfo {author}
  {\bibnamefont {Seweryniak}, \bibfnamefont {D}}, \bibinfo {author}
  {\bibnamefont {Stolze}, \bibfnamefont {S~M}}, \bibinfo {author} {\bibnamefont
  {Walters}, \bibfnamefont {W~B}}, \ and\ \bibinfo {author} {\bibnamefont {Wu},
  \bibfnamefont {J}}} (\bibinfo {year} {2019}),\ \bibfield  {title} {\enquote
  {\bibinfo {title} {Evidence for rigid triaxial deformation in
  $^{76}\mathrm{Ge}$ from a model-independent analysis},}\ }\href {\doibase
  10.1103/PhysRevLett.123.102501} {\bibfield  {journal} {\bibinfo  {journal}
  {Phys. Rev. Lett.}\ }\textbf {\bibinfo {volume} {123}},\ \bibinfo {pages}
  {102501}}\BibitemShut {NoStop}%
\bibitem [{\citenamefont {Bagnaschi}\ \emph {et~al.}(2015)\citenamefont
  {Bagnaschi}, \citenamefont {Cacciari}, \citenamefont {Guffanti},\ and\
  \citenamefont {Jenniches}}]{bagnaschi2015}%
  \BibitemOpen
  \bibfield  {author} {\bibinfo {author} {\bibnamefont {Bagnaschi},
  \bibfnamefont {E}}, \bibinfo {author} {\bibnamefont {Cacciari}, \bibfnamefont
  {M}}, \bibinfo {author} {\bibnamefont {Guffanti}, \bibfnamefont {A}}, \ and\
  \bibinfo {author} {\bibnamefont {Jenniches}, \bibfnamefont {L}}} (\bibinfo
  {year} {2015}),\ \bibfield  {title} {\enquote {\bibinfo {title} {An extensive
  survey of the estimation of uncertainties from missing higher orders in
  perturbative calculations},}\ }\href {\doibase 10.1007/JHEP02(2015)133}
  {\bibfield  {journal} {\bibinfo  {journal} {J. High Energ. Phys.}\ }\textbf
  {\bibinfo {volume} {2015}}~(\bibinfo {number} {2}),\ \bibinfo {eid} {133},\
  10.1007/JHEP02(2015)133}\BibitemShut {NoStop}%
\bibitem [{\citenamefont {B{\"a}r}\ \emph {et~al.}(2004)\citenamefont
  {B{\"a}r}, \citenamefont {Imboden},\ and\ \citenamefont {Wiese}}]{baer2004}%
  \BibitemOpen
  \bibfield  {author} {\bibinfo {author} {\bibnamefont {B{\"a}r}, \bibfnamefont
  {O}}, \bibinfo {author} {\bibnamefont {Imboden}, \bibfnamefont {M}}, \ and\
  \bibinfo {author} {\bibnamefont {Wiese}, \bibfnamefont {U~J}}} (\bibinfo
  {year} {2004}),\ \bibfield  {title} {\enquote {\bibinfo {title} {{Pions
  versus magnons: from QCD to antiferromagnets and quantum Hall
  ferromagnets}},}\ }\href {\doibase 10.1016/j.nuclphysb.2003.12.041}
  {\bibfield  {journal} {\bibinfo  {journal} {Nucl. Phys. B}\ }\textbf
  {\bibinfo {volume} {686}}~(\bibinfo {number} {3}),\ \bibinfo {pages} {347
  }}\BibitemShut {NoStop}%
\bibitem [{\citenamefont {Barea}\ \emph {et~al.}(2015)\citenamefont {Barea},
  \citenamefont {Kotila},\ and\ \citenamefont {Iachello}}]{barea_j_2015}%
  \BibitemOpen
  \bibfield  {author} {\bibinfo {author} {\bibnamefont {Barea}, \bibfnamefont
  {J}}, \bibinfo {author} {\bibnamefont {Kotila}, \bibfnamefont {J}}, \ and\
  \bibinfo {author} {\bibnamefont {Iachello}, \bibfnamefont {F}}} (\bibinfo
  {year} {2015}),\ \bibfield  {title} {\enquote {\bibinfo {title}
  {$0\ensuremath{\nu}\ensuremath{\beta}\ensuremath{\beta}$ and
  $2\ensuremath{\nu}\ensuremath{\beta}\ensuremath{\beta}$ nuclear matrix
  elements in the interacting boson model with isospin restoration},}\ }\href
  {\doibase 10.1103/PhysRevC.91.034304} {\bibfield  {journal} {\bibinfo
  {journal} {Phys. Rev. C}\ }\textbf {\bibinfo {volume} {91}},\ \bibinfo
  {pages} {034304}}\BibitemShut {NoStop}%
\bibitem [{\citenamefont {{Bedaque}}\ and\ \citenamefont {{van
  Kolck}}(2002)}]{bedaque2002}%
  \BibitemOpen
  \bibfield  {author} {\bibinfo {author} {\bibnamefont {{Bedaque}},
  \bibfnamefont {P~F}}, \ and\ \bibinfo {author} {\bibnamefont {{van Kolck}},
  \bibfnamefont {U}}} (\bibinfo {year} {2002}),\ \bibfield  {title} {\enquote
  {\bibinfo {title} {{Effective field theory for few-nucleon systems}},}\
  }\href {\doibase 10.1146/annurev.nucl.52.050102.090637} {\bibfield  {journal}
  {\bibinfo  {journal} {Annual Review of Nuclear and Particle Science}\
  }\textbf {\bibinfo {volume} {52}},\ \bibinfo {pages} {339}},\ \Eprint
  {http://arxiv.org/abs/nucl-th/0203055} {nucl-th/0203055} \BibitemShut
  {NoStop}%
\bibitem [{\citenamefont {Belley}\ \emph {et~al.}(2021)\citenamefont {Belley},
  \citenamefont {Payne}, \citenamefont {Stroberg}, \citenamefont {Miyagi},\
  and\ \citenamefont {Holt}}]{belley2021}%
  \BibitemOpen
  \bibfield  {author} {\bibinfo {author} {\bibnamefont {Belley}, \bibfnamefont
  {A}}, \bibinfo {author} {\bibnamefont {Payne}, \bibfnamefont {C~G}}, \bibinfo
  {author} {\bibnamefont {Stroberg}, \bibfnamefont {S~R}}, \bibinfo {author}
  {\bibnamefont {Miyagi}, \bibfnamefont {T}}, \ and\ \bibinfo {author}
  {\bibnamefont {Holt}, \bibfnamefont {J~D}}} (\bibinfo {year} {2021}),\
  \bibfield  {title} {\enquote {\bibinfo {title} {Ab initio neutrinoless
  double-beta decay matrix elements for $^{48}\mathrm{Ca}$, $^{76}\mathrm{Ge}$,
  and $^{82}\mathrm{Se}$},}\ }\href {\doibase 10.1103/PhysRevLett.126.042502}
  {\bibfield  {journal} {\bibinfo  {journal} {Phys. Rev. Lett.}\ }\textbf
  {\bibinfo {volume} {126}},\ \bibinfo {pages} {042502}}\BibitemShut {NoStop}%
\bibitem [{\citenamefont {Belley}\ \emph {et~al.}(2024)\citenamefont {Belley},
  \citenamefont {Yao}, \citenamefont {Bally}, \citenamefont {Pitcher},
  \citenamefont {Engel}, \citenamefont {Hergert}, \citenamefont {Holt},
  \citenamefont {Miyagi}, \citenamefont {Rodr\'{\i}guez}, \citenamefont
  {Romero}, \citenamefont {Stroberg},\ and\ \citenamefont
  {Zhang}}]{belley2024}%
  \BibitemOpen
  \bibfield  {author} {\bibinfo {author} {\bibnamefont {Belley}, \bibfnamefont
  {A}}, \bibinfo {author} {\bibnamefont {Yao}, \bibfnamefont {J~M}}, \bibinfo
  {author} {\bibnamefont {Bally}, \bibfnamefont {B}}, \bibinfo {author}
  {\bibnamefont {Pitcher}, \bibfnamefont {J}}, \bibinfo {author} {\bibnamefont
  {Engel}, \bibfnamefont {J}}, \bibinfo {author} {\bibnamefont {Hergert},
  \bibfnamefont {H}}, \bibinfo {author} {\bibnamefont {Holt}, \bibfnamefont
  {J~D}}, \bibinfo {author} {\bibnamefont {Miyagi}, \bibfnamefont {T}},
  \bibinfo {author} {\bibnamefont {Rodr\'{\i}guez}, \bibfnamefont {T~R}},
  \bibinfo {author} {\bibnamefont {Romero}, \bibfnamefont {A~M}}, \bibinfo
  {author} {\bibnamefont {Stroberg}, \bibfnamefont {S~R}}, \ and\ \bibinfo
  {author} {\bibnamefont {Zhang}, \bibfnamefont {X}}} (\bibinfo {year}
  {2024}),\ \bibfield  {title} {\enquote {\bibinfo {title} {Ab initio
  uncertainty quantification of neutrinoless double-beta decay in
  $^{76}\mathrm{Ge}$},}\ }\href {\doibase 10.1103/PhysRevLett.132.182502}
  {\bibfield  {journal} {\bibinfo  {journal} {Phys. Rev. Lett.}\ }\textbf
  {\bibinfo {volume} {132}},\ \bibinfo {pages} {182502}}\BibitemShut {NoStop}%
\bibitem [{\citenamefont {Bender}\ \emph {et~al.}(2003)\citenamefont {Bender},
  \citenamefont {Heenen},\ and\ \citenamefont {Reinhard}}]{bender2003}%
  \BibitemOpen
  \bibfield  {author} {\bibinfo {author} {\bibnamefont {Bender}, \bibfnamefont
  {M}}, \bibinfo {author} {\bibnamefont {Heenen}, \bibfnamefont {P~H}}, \ and\
  \bibinfo {author} {\bibnamefont {Reinhard}, \bibfnamefont {P~G}}} (\bibinfo
  {year} {2003}),\ \bibfield  {title} {\enquote {\bibinfo {title}
  {Self-consistent mean-field models for nuclear structure},}\ }\href {\doibase
  10.1103/RevModPhys.75.121} {\bibfield  {journal} {\bibinfo  {journal} {Rev.
  Mod. Phys.}\ }\textbf {\bibinfo {volume} {75}},\ \bibinfo {pages}
  {121}}\BibitemShut {NoStop}%
\bibitem [{\citenamefont {Berry}(1984)}]{berry1984}%
  \BibitemOpen
  \bibfield  {author} {\bibinfo {author} {\bibnamefont {Berry}, \bibfnamefont
  {M~V}}} (\bibinfo {year} {1984}),\ \bibfield  {title} {\enquote {\bibinfo
  {title} {Quantal phase factors accompanying adiabatic changes},}\ }\href
  {\doibase 10.1098/rspa.1984.0023} {\bibfield  {journal} {\bibinfo  {journal}
  {Proc. Roy. Soc. London. A. Math. Phys. Sciences}\ }\textbf {\bibinfo
  {volume} {392}}~(\bibinfo {number} {1802}),\ \bibinfo {pages}
  {45}}\BibitemShut {NoStop}%
\bibitem [{\citenamefont {B{\`e}s}\ \emph {et~al.}(1970)\citenamefont
  {B{\`e}s}, \citenamefont {Broglia}, \citenamefont {Perazzo},\ and\
  \citenamefont {Kumar}}]{bes1970}%
  \BibitemOpen
  \bibfield  {author} {\bibinfo {author} {\bibnamefont {B{\`e}s}, \bibfnamefont
  {D~R}}, \bibinfo {author} {\bibnamefont {Broglia}, \bibfnamefont {R~A}},
  \bibinfo {author} {\bibnamefont {Perazzo}, \bibfnamefont {R~P~J}}, \ and\
  \bibinfo {author} {\bibnamefont {Kumar}, \bibfnamefont {K}}} (\bibinfo {year}
  {1970}),\ \bibfield  {title} {\enquote {\bibinfo {title} {{Collective
  treatment of the pairing Hamiltonian: (I). Formulation of the model}},}\
  }\href {\doibase 10.1016/0375-9474(70)90677-9} {\bibfield  {journal}
  {\bibinfo  {journal} {Nucl. Phys. A}\ }\textbf {\bibinfo {volume}
  {143}}~(\bibinfo {number} {1}),\ \bibinfo {pages} {1}}\BibitemShut {NoStop}%
\bibitem [{\citenamefont {B{\`e}s}\ and\ \citenamefont
  {Dussel}(1969)}]{bes1969}%
  \BibitemOpen
  \bibfield  {author} {\bibinfo {author} {\bibnamefont {B{\`e}s}, \bibfnamefont
  {D~R}}, \ and\ \bibinfo {author} {\bibnamefont {Dussel}, \bibfnamefont
  {G~G}}} (\bibinfo {year} {1969}),\ \bibfield  {title} {\enquote {\bibinfo
  {title} {Phenomenological treatment of anharmonic effects in cd isotopes},}\
  }\href {\doibase 10.1016/0375-9474(69)90143-2} {\bibfield  {journal}
  {\bibinfo  {journal} {Nucl. Phys. A}\ }\textbf {\bibinfo {volume} {135}},\
  \bibinfo {pages} {1}}\BibitemShut {NoStop}%
\bibitem [{\citenamefont {Bogoljubov}(1958)}]{bogoliubov1958}%
  \BibitemOpen
  \bibfield  {author} {\bibinfo {author} {\bibnamefont {Bogoljubov},
  \bibfnamefont {N~N}}} (\bibinfo {year} {1958}),\ \bibfield  {title} {\enquote
  {\bibinfo {title} {On a new method in the theory of superconductivity},}\
  }\href {\doibase 10.1007/BF02745585} {\bibfield  {journal} {\bibinfo
  {journal} {Il Nuovo Cimento}\ }\textbf {\bibinfo {volume} {7}}~(\bibinfo
  {number} {6}),\ \bibinfo {pages} {794}}\BibitemShut {NoStop}%
\bibitem [{\citenamefont {Bohr}(1952)}]{bohr1952}%
  \BibitemOpen
  \bibfield  {author} {\bibinfo {author} {\bibnamefont {Bohr}, \bibfnamefont
  {A}}} (\bibinfo {year} {1952}),\ \bibfield  {title} {\enquote {\bibinfo
  {title} {The coupling of nuclear surface oscillations to the motion of
  individual nucleons},}\ }\href
  {http://gymarkiv.sdu.dk/MFM/kdvs/mfm%2020-29/mfm-26-14.pdf} {\bibfield
  {journal} {\bibinfo  {journal} {Dan. Mat. Fys. Medd.}\ }\textbf {\bibinfo
  {volume} {26}},\ \bibinfo {pages} {no. 14}}\BibitemShut {NoStop}%
\bibitem [{\citenamefont {Bohr}(1969)}]{bohr1969}%
  \BibitemOpen
  \bibfield  {author} {\bibinfo {author} {\bibnamefont {Bohr}, \bibfnamefont
  {A}}} (\bibinfo {year} {1969}),\ \bibfield  {title} {\enquote {\bibinfo
  {title} {Pair correlations and double transfer reactions},}\ }in\ \href
  {https://www.iaea.org/publications/2149/nuclear-structure-dubna-symposium-1968-dubna-4-11-july-1968}
  {\emph {\bibinfo {booktitle} {Nuclear Structure: Dubna Symposium 1968 (Dubna,
  4-11 July 1968)}}},\ \bibinfo {series and number} {Proceedings Series}\
  (\bibinfo  {publisher} {International Atomic Energy Agency},\ \bibinfo
  {address} {Vienna})\ p.\ \bibinfo {pages} {179}\BibitemShut {NoStop}%
\bibitem [{\citenamefont {Bohr}\ and\ \citenamefont
  {Mottelson}(1953)}]{bohr1953}%
  \BibitemOpen
  \bibfield  {author} {\bibinfo {author} {\bibnamefont {Bohr}, \bibfnamefont
  {A}}, \ and\ \bibinfo {author} {\bibnamefont {Mottelson}, \bibfnamefont
  {B~R}}} (\bibinfo {year} {1953}),\ \bibfield  {title} {\enquote {\bibinfo
  {title} {Collective and individual-particle aspects of nuclear structure},}\
  }\href {http://gymarkiv.sdu.dk/MFM/kdvs/mfm%2020-29/mfm-27-16.pdf} {\bibfield
   {journal} {\bibinfo  {journal} {Dan. Mat. Fys. Medd.}\ }\textbf {\bibinfo
  {volume} {27}},\ \bibinfo {pages} {no. 16}}\BibitemShut {NoStop}%
\bibitem [{\citenamefont {Bohr}\ and\ \citenamefont
  {Mottelson}(1975)}]{bohr1975}%
  \BibitemOpen
  \bibfield  {author} {\bibinfo {author} {\bibnamefont {Bohr}, \bibfnamefont
  {A}}, \ and\ \bibinfo {author} {\bibnamefont {Mottelson}, \bibfnamefont
  {B~R}}} (\bibinfo {year} {1975}),\ \href@noop {} {\emph {\bibinfo {title}
  {Nuclear Structure}}},\ Vol.\ \bibinfo {volume} {II: Nuclear Deformation}\
  (\bibinfo  {publisher} {W. A. Benjamin},\ \bibinfo {address} {Reading,
  Massachusetts, USA})\BibitemShut {NoStop}%
\bibitem [{\citenamefont {Bohr}\ \emph {et~al.}(1958)\citenamefont {Bohr},
  \citenamefont {Mottelson},\ and\ \citenamefont {Pines}}]{bohr1958}%
  \BibitemOpen
  \bibfield  {author} {\bibinfo {author} {\bibnamefont {Bohr}, \bibfnamefont
  {A}}, \bibinfo {author} {\bibnamefont {Mottelson}, \bibfnamefont {B~R}}, \
  and\ \bibinfo {author} {\bibnamefont {Pines}, \bibfnamefont {D}}} (\bibinfo
  {year} {1958}),\ \bibfield  {title} {\enquote {\bibinfo {title} {Possible
  analogy between the excitation spectra of nuclei and those of the
  superconducting metallic state},}\ }\href {\doibase 10.1103/PhysRev.110.936}
  {\bibfield  {journal} {\bibinfo  {journal} {Phys. Rev.}\ }\textbf {\bibinfo
  {volume} {110}},\ \bibinfo {pages} {936}}\BibitemShut {NoStop}%
\bibitem [{\citenamefont {Bohr}(1975)}]{bohr1975nobel}%
  \BibitemOpen
  \bibfield  {author} {\bibinfo {author} {\bibnamefont {Bohr}, \bibfnamefont
  {A~N}}} (\bibinfo {year} {1975}),\ \href
  {http://www.nobelprize.org/nobel_prizes/physics/laureates/1975/bohr-lecture.html}
  {\enquote {\bibinfo {title} {Nobel lecture: Rotational motion in nuclei},}\
  }\BibitemShut {NoStop}%
\bibitem [{\citenamefont {Bontems}\ \emph {et~al.}(2021)\citenamefont
  {Bontems}, \citenamefont {Duguet}, \citenamefont {Hagen},\ and\ \citenamefont
  {Som{\`a}}}]{bontems2021}%
  \BibitemOpen
  \bibfield  {author} {\bibinfo {author} {\bibnamefont {Bontems}, \bibfnamefont
  {V}}, \bibinfo {author} {\bibnamefont {Duguet}, \bibfnamefont {T}}, \bibinfo
  {author} {\bibnamefont {Hagen}, \bibfnamefont {G}}, \ and\ \bibinfo {author}
  {\bibnamefont {Som{\`a}}, \bibfnamefont {V}}} (\bibinfo {year} {2021}),\
  \bibfield  {title} {\enquote {\bibinfo {title} {Topical issue on the tower of
  effective (field) theories and the emergence of nuclear phenomena},}\ }\href
  {\doibase 10.1140/epja/s10050-021-00356-4} {\bibfield  {journal} {\bibinfo
  {journal} {The European Physical Journal A}\ }\textbf {\bibinfo {volume}
  {57}},\ \bibinfo {pages} {42}}\BibitemShut {NoStop}%
\bibitem [{\citenamefont {Brase}\ \emph {et~al.}(2022)\citenamefont {Brase},
  \citenamefont {Men\'endez}, \citenamefont {Coello~P\'erez},\ and\
  \citenamefont {Schwenk}}]{brase2022}%
  \BibitemOpen
  \bibfield  {author} {\bibinfo {author} {\bibnamefont {Brase}, \bibfnamefont
  {C}}, \bibinfo {author} {\bibnamefont {Men\'endez}, \bibfnamefont {J}},
  \bibinfo {author} {\bibnamefont {Coello~P\'erez}, \bibfnamefont {E~A}}, \
  and\ \bibinfo {author} {\bibnamefont {Schwenk}, \bibfnamefont {A}}} (\bibinfo
  {year} {2022}),\ \bibfield  {title} {\enquote {\bibinfo {title} {Neutrinoless
  double-$\ensuremath{\beta}$ decay from an effective field theory for heavy
  nuclei},}\ }\href {\doibase 10.1103/PhysRevC.106.034309} {\bibfield
  {journal} {\bibinfo  {journal} {Phys. Rev. C}\ }\textbf {\bibinfo {volume}
  {106}},\ \bibinfo {pages} {034309}}\BibitemShut {NoStop}%
\bibitem [{\citenamefont {Brauner}(2010)}]{brauner2010}%
  \BibitemOpen
  \bibfield  {author} {\bibinfo {author} {\bibnamefont {Brauner}, \bibfnamefont
  {T}}} (\bibinfo {year} {2010}),\ \bibfield  {title} {\enquote {\bibinfo
  {title} {Spontaneous symmetry breaking and nambu-goldstone bosons in quantum
  many-body systems},}\ }\href {\doibase 10.3390/sym2020609} {\bibfield
  {journal} {\bibinfo  {journal} {Symmetry}\ }\textbf {\bibinfo {volume}
  {2}}~(\bibinfo {number} {2}),\ \bibinfo {pages} {609}}\BibitemShut {NoStop}%
\bibitem [{\citenamefont {Brink}\ and\ \citenamefont
  {Broglia}(2005)}]{brink-broglia2005}%
  \BibitemOpen
  \bibfield  {author} {\bibinfo {author} {\bibnamefont {Brink}, \bibfnamefont
  {D~M}}, \ and\ \bibinfo {author} {\bibnamefont {Broglia}, \bibfnamefont
  {R~A}}} (\bibinfo {year} {2005}),\ \href {\doibase 10.1017/CBO9780511534911}
  {\emph {\bibinfo {title} {Nuclear Superfluidity: Pairing in Finite
  Systems}}}\ (\bibinfo  {publisher} {Cambridge University Press},\ \bibinfo
  {address} {Cambridge, UK})\BibitemShut {NoStop}%
\bibitem [{\citenamefont {Broglia}\ \emph {et~al.}(1968)\citenamefont
  {Broglia}, \citenamefont {Riedel}, \citenamefont {Sørensen},\ and\
  \citenamefont {Udagawa}}]{broglia1968}%
  \BibitemOpen
  \bibfield  {author} {\bibinfo {author} {\bibnamefont {Broglia}, \bibfnamefont
  {R}}, \bibinfo {author} {\bibnamefont {Riedel}, \bibfnamefont {C}}, \bibinfo
  {author} {\bibnamefont {Sørensen}, \bibfnamefont {B}}, \ and\ \bibinfo
  {author} {\bibnamefont {Udagawa}, \bibfnamefont {T}}} (\bibinfo {year}
  {1968}),\ \bibfield  {title} {\enquote {\bibinfo {title} {The pairing
  vibrational model and the analysis of the $^{116,118}\mathrm{Sn}(t, p)$
  reactions},}\ }\href {\doibase 10.1016/0375-9474(68)90004-3} {\bibfield
  {journal} {\bibinfo  {journal} {Nuclear Physics A}\ }\textbf {\bibinfo
  {volume} {115}}~(\bibinfo {number} {2}),\ \bibinfo {pages} {273}}\BibitemShut
  {NoStop}%
\bibitem [{\citenamefont {Broglia}\ \emph {et~al.}(1973)\citenamefont
  {Broglia}, \citenamefont {Hansen},\ and\ \citenamefont
  {Riedel}}]{broglia1973}%
  \BibitemOpen
  \bibfield  {author} {\bibinfo {author} {\bibnamefont {Broglia}, \bibfnamefont
  {R~A}}, \bibinfo {author} {\bibnamefont {Hansen}, \bibfnamefont {O}}, \ and\
  \bibinfo {author} {\bibnamefont {Riedel}, \bibfnamefont {C}}} (\bibinfo
  {year} {1973}),\ \bibfield  {title} {\enquote {\bibinfo {title} {Two-neutron
  transfer reactions and the pairing model},}\ }in\ \href {\doibase
  10.1007/978-1-4615-9041-5_3} {\emph {\bibinfo {booktitle} {Advances in
  Nuclear Physics}}},\ Vol.~\bibinfo {volume} {6},\ \bibinfo {editor} {edited
  by\ \bibinfo {editor} {\bibfnamefont {M}~\bibnamefont {Baranger}}\ and\
  \bibinfo {editor} {\bibfnamefont {E}~\bibnamefont {Vogt}}},\ Chap.~\bibinfo
  {chapter} {3}\ (\bibinfo  {publisher} {Springer},\ \bibinfo {address}
  {Boston, MA})\ p.\ \bibinfo {pages} {287}\BibitemShut {NoStop}%
\bibitem [{\citenamefont {Broglia}\ \emph {et~al.}(2000)\citenamefont
  {Broglia}, \citenamefont {Terasaki},\ and\ \citenamefont
  {Giovanardi}}]{broglia2000}%
  \BibitemOpen
  \bibfield  {author} {\bibinfo {author} {\bibnamefont {Broglia}, \bibfnamefont
  {R~A}}, \bibinfo {author} {\bibnamefont {Terasaki}, \bibfnamefont {J}}, \
  and\ \bibinfo {author} {\bibnamefont {Giovanardi}, \bibfnamefont {N}}}
  (\bibinfo {year} {2000}),\ \bibfield  {title} {\enquote {\bibinfo {title}
  {The anderson-goldstone-nambu mode in finite and in infinite systems},}\
  }\href {\doibase 10.1016/S0370-1573(00)00046-6} {\bibfield  {journal}
  {\bibinfo  {journal} {Physics Reports}\ }\textbf {\bibinfo {volume}
  {335}}~(\bibinfo {number} {1}),\ \bibinfo {pages} {1}}\BibitemShut {NoStop}%
\bibitem [{\citenamefont {Cacciari}\ and\ \citenamefont
  {Houdeau}(2011)}]{cacciari2011}%
  \BibitemOpen
  \bibfield  {author} {\bibinfo {author} {\bibnamefont {Cacciari},
  \bibfnamefont {M}}, \ and\ \bibinfo {author} {\bibnamefont {Houdeau},
  \bibfnamefont {N}}} (\bibinfo {year} {2011}),\ \bibfield  {title} {\enquote
  {\bibinfo {title} {Meaningful characterisation of perturbative theoretical
  uncertainties},}\ }\href {\doibase 10.1007/JHEP09(2011)039} {\bibfield
  {journal} {\bibinfo  {journal} {Journal of High Energy Physics}\ }\textbf
  {\bibinfo {volume} {2011}}~(\bibinfo {number} {9}),\ \bibinfo {eid} {39},\
  10.1007/JHEP09(2011)039}\BibitemShut {NoStop}%
\bibitem [{\citenamefont {Callan}\ \emph {et~al.}(1969)\citenamefont {Callan},
  \citenamefont {Coleman}, \citenamefont {Wess},\ and\ \citenamefont
  {Zumino}}]{callan1969}%
  \BibitemOpen
  \bibfield  {author} {\bibinfo {author} {\bibnamefont {Callan}, \bibfnamefont
  {C~G}}, \bibinfo {author} {\bibnamefont {Coleman}, \bibfnamefont {S}},
  \bibinfo {author} {\bibnamefont {Wess}, \bibfnamefont {J}}, \ and\ \bibinfo
  {author} {\bibnamefont {Zumino}, \bibfnamefont {B}}} (\bibinfo {year}
  {1969}),\ \bibfield  {title} {\enquote {\bibinfo {title} {Structure of
  phenomenological lagrangians. ii},}\ }\href {\doibase
  10.1103/PhysRev.177.2247} {\bibfield  {journal} {\bibinfo  {journal} {Phys.
  Rev.}\ }\textbf {\bibinfo {volume} {177}},\ \bibinfo {pages}
  {2247}}\BibitemShut {NoStop}%
\bibitem [{\citenamefont {Caprio}(2011)}]{caprio2011}%
  \BibitemOpen
  \bibfield  {author} {\bibinfo {author} {\bibnamefont {Caprio}, \bibfnamefont
  {M~A}}} (\bibinfo {year} {2011}),\ \bibfield  {title} {\enquote {\bibinfo
  {title} {Exact diagonalization of the bohr hamiltonian for rotational nuclei:
  Dynamical $\gamma$ softness and triaxiality},}\ }\href {\doibase
  10.1103/PhysRevC.83.064309} {\bibfield  {journal} {\bibinfo  {journal} {Phys.
  Rev. C}\ }\textbf {\bibinfo {volume} {83}},\ \bibinfo {pages}
  {064309}}\BibitemShut {NoStop}%
\bibitem [{\citenamefont {Caprio}\ \emph {et~al.}(2013)\citenamefont {Caprio},
  \citenamefont {Maris},\ and\ \citenamefont {Vary}}]{caprio2013}%
  \BibitemOpen
  \bibfield  {author} {\bibinfo {author} {\bibnamefont {Caprio}, \bibfnamefont
  {M~A}}, \bibinfo {author} {\bibnamefont {Maris}, \bibfnamefont {P}}, \ and\
  \bibinfo {author} {\bibnamefont {Vary}, \bibfnamefont {J~P}}} (\bibinfo
  {year} {2013}),\ \bibfield  {title} {\enquote {\bibinfo {title} {Emergence of
  rotational bands in ab initio no-core configuration interaction calculations
  of light nuclei},}\ }\href {\doibase 10.1016/j.physletb.2012.12.064}
  {\bibfield  {journal} {\bibinfo  {journal} {Phys. Lett. B}\ }\textbf
  {\bibinfo {volume} {719}},\ \bibinfo {pages} {179 }}\BibitemShut {NoStop}%
\bibitem [{\citenamefont {Caprio}\ \emph {et~al.}(2015)\citenamefont {Caprio},
  \citenamefont {Maris}, \citenamefont {Vary},\ and\ \citenamefont
  {Smith}}]{caprio2015}%
  \BibitemOpen
  \bibfield  {author} {\bibinfo {author} {\bibnamefont {Caprio}, \bibfnamefont
  {M~A}}, \bibinfo {author} {\bibnamefont {Maris}, \bibfnamefont {P}}, \bibinfo
  {author} {\bibnamefont {Vary}, \bibfnamefont {J~P}}, \ and\ \bibinfo {author}
  {\bibnamefont {Smith}, \bibfnamefont {R}}} (\bibinfo {year} {2015}),\
  \bibfield  {title} {\enquote {\bibinfo {title} {Collective rotation from ab
  initio theory},}\ }\href {\doibase 10.1142/S0218301315410025} {\bibfield
  {journal} {\bibinfo  {journal} {Int. J. Mod. Phys. E}\ }\textbf {\bibinfo
  {volume} {24}}~(\bibinfo {number} {09}),\ \bibinfo {pages}
  {1541002}}\BibitemShut {NoStop}%
\bibitem [{\citenamefont {Caurier}\ \emph {et~al.}(2005)\citenamefont
  {Caurier}, \citenamefont {Mart{\'i}nez-Pinedo}, \citenamefont {Nowacki},
  \citenamefont {Poves},\ and\ \citenamefont {Zuker}}]{caurier2005}%
  \BibitemOpen
  \bibfield  {author} {\bibinfo {author} {\bibnamefont {Caurier}, \bibfnamefont
  {E}}, \bibinfo {author} {\bibnamefont {Mart{\'i}nez-Pinedo}, \bibfnamefont
  {G}}, \bibinfo {author} {\bibnamefont {Nowacki}, \bibfnamefont {F}}, \bibinfo
  {author} {\bibnamefont {Poves}, \bibfnamefont {A}}, \ and\ \bibinfo {author}
  {\bibnamefont {Zuker}, \bibfnamefont {A~P}}} (\bibinfo {year} {2005}),\
  \bibfield  {title} {\enquote {\bibinfo {title} {The shell model as a unified
  view of nuclear structure},}\ }\href {\doibase 10.1103/RevModPhys.77.427}
  {\bibfield  {journal} {\bibinfo  {journal} {Rev. Mod. Phys.}\ }\textbf
  {\bibinfo {volume} {77}},\ \bibinfo {pages} {427}}\BibitemShut {NoStop}%
\bibitem [{\citenamefont {Chandrasekharan}\ \emph {et~al.}(2008)\citenamefont
  {Chandrasekharan}, \citenamefont {Jiang}, \citenamefont {Pepe},\ and\
  \citenamefont {Wiese}}]{chandrasekharan2008}%
  \BibitemOpen
  \bibfield  {author} {\bibinfo {author} {\bibnamefont {Chandrasekharan},
  \bibfnamefont {S}}, \bibinfo {author} {\bibnamefont {Jiang}, \bibfnamefont
  {F~J}}, \bibinfo {author} {\bibnamefont {Pepe}, \bibfnamefont {M}}, \ and\
  \bibinfo {author} {\bibnamefont {Wiese}, \bibfnamefont {U~J}}} (\bibinfo
  {year} {2008}),\ \bibfield  {title} {\enquote {\bibinfo {title} {Rotor
  spectra, berry phases, and monopole fields: From antiferromagnets to qcd},}\
  }\href {\doibase 10.1103/PhysRevD.78.077901} {\bibfield  {journal} {\bibinfo
  {journal} {Phys. Rev. D}\ }\textbf {\bibinfo {volume} {78}},\ \bibinfo
  {pages} {077901}}\BibitemShut {NoStop}%
\bibitem [{\citenamefont {Chen}\ \emph {et~al.}(2017)\citenamefont {Chen},
  \citenamefont {Kaiser}, \citenamefont {Mei{\ss}ner},\ and\ \citenamefont
  {Meng}}]{chen2017}%
  \BibitemOpen
  \bibfield  {author} {\bibinfo {author} {\bibnamefont {Chen}, \bibfnamefont
  {Q~B}}, \bibinfo {author} {\bibnamefont {Kaiser}, \bibfnamefont {N}},
  \bibinfo {author} {\bibnamefont {Mei{\ss}ner}, \bibfnamefont {U~G}}, \ and\
  \bibinfo {author} {\bibnamefont {Meng}, \bibfnamefont {J}}} (\bibinfo {year}
  {2017}),\ \bibfield  {title} {\enquote {\bibinfo {title} {Effective field
  theory for triaxially deformed nuclei},}\ }\href {\doibase
  10.1140/epja/i2017-12404-5} {\bibfield  {journal} {\bibinfo  {journal} {The
  European Physical Journal A}\ }\textbf {\bibinfo {volume} {53}}~(\bibinfo
  {number} {10}),\ \bibinfo {pages} {204}}\BibitemShut {NoStop}%
\bibitem [{\citenamefont {Chen}\ \emph {et~al.}(2018)\citenamefont {Chen},
  \citenamefont {Kaiser}, \citenamefont {Mei\ss{}ner},\ and\ \citenamefont
  {Meng}}]{chen2018}%
  \BibitemOpen
  \bibfield  {author} {\bibinfo {author} {\bibnamefont {Chen}, \bibfnamefont
  {Q~B}}, \bibinfo {author} {\bibnamefont {Kaiser}, \bibfnamefont {N}},
  \bibinfo {author} {\bibnamefont {Mei\ss{}ner}, \bibfnamefont {U~G}}, \ and\
  \bibinfo {author} {\bibnamefont {Meng}, \bibfnamefont {J}}} (\bibinfo {year}
  {2018}),\ \bibfield  {title} {\enquote {\bibinfo {title} {Effective field
  theory for collective rotations and vibrations of triaxially deformed
  nuclei},}\ }\href {\doibase 10.1103/PhysRevC.97.064320} {\bibfield  {journal}
  {\bibinfo  {journal} {Phys. Rev. C}\ }\textbf {\bibinfo {volume} {97}},\
  \bibinfo {pages} {064320}}\BibitemShut {NoStop}%
\bibitem [{\citenamefont {{Chen}}\ \emph {et~al.}(2020)\citenamefont {{Chen}},
  \citenamefont {{Kaiser}}, \citenamefont {{Mei{\ss}ner}},\ and\ \citenamefont
  {{Meng}}}]{chen2020}%
  \BibitemOpen
  \bibfield  {author} {\bibinfo {author} {\bibnamefont {{Chen}}, \bibfnamefont
  {Q~B}}, \bibinfo {author} {\bibnamefont {{Kaiser}}, \bibfnamefont {N}},
  \bibinfo {author} {\bibnamefont {{Mei{\ss}ner}}, \bibfnamefont {U~G}}, \ and\
  \bibinfo {author} {\bibnamefont {{Meng}}, \bibfnamefont {J}}} (\bibinfo
  {year} {2020}),\ \bibfield  {title} {\enquote {\bibinfo {title} {{Effective
  field theory for triaxially deformed odd-mass nuclei}},}\ }\href@noop {}
  {\bibfield  {journal} {\bibinfo  {journal} {arXiv e-prints}\ ,\ \bibinfo
  {eid} {arXiv:2003.04065}}}\Eprint {http://arxiv.org/abs/2003.04065}
  {2003.04065} \BibitemShut {NoStop}%
\bibitem [{\citenamefont {Cline}(1986)}]{cline1986}%
  \BibitemOpen
  \bibfield  {author} {\bibinfo {author} {\bibnamefont {Cline}, \bibfnamefont
  {D}}} (\bibinfo {year} {1986}),\ \bibfield  {title} {\enquote {\bibinfo
  {title} {Nuclear shapes studied by coulomb excitation},}\ }\href {\doibase
  10.1146/annurev.ns.36.120186.003343} {\bibfield  {journal} {\bibinfo
  {journal} {Annual Review of Nuclear and Particle Science}\ }\textbf {\bibinfo
  {volume} {36}},\ \bibinfo {pages} {683}}\BibitemShut {NoStop}%
\bibitem [{\citenamefont {{Coello P\'erez}}\ \emph {et~al.}(2019)\citenamefont
  {{Coello P\'erez}}, \citenamefont {Men\'endez},\ and\ \citenamefont
  {Schwenk}}]{coelloperez_2019}%
  \BibitemOpen
  \bibfield  {author} {\bibinfo {author} {\bibnamefont {{Coello P\'erez}},
  \bibfnamefont {E}}, \bibinfo {author} {\bibnamefont {Men\'endez},
  \bibfnamefont {J}}, \ and\ \bibinfo {author} {\bibnamefont {Schwenk},
  \bibfnamefont {A}}} (\bibinfo {year} {2019}),\ \bibfield  {title} {\enquote
  {\bibinfo {title} {{Two-neutrino double electron capture on
  $\ensuremath{^{124}}$Xe based on an effective theory and the nuclear shell
  model}},}\ }\href {\doibase 10.1016/j.physletb.2019.134885} {\bibfield
  {journal} {\bibinfo  {journal} {Physics Letters B}\ }\textbf {\bibinfo
  {volume} {797}},\ \bibinfo {pages} {134885}}\BibitemShut {NoStop}%
\bibitem [{\citenamefont {Coello~P\'erez}\ \emph {et~al.}(2018)\citenamefont
  {Coello~P\'erez}, \citenamefont {Men\'endez},\ and\ \citenamefont
  {Schwenk}}]{coelloperez2018}%
  \BibitemOpen
  \bibfield  {author} {\bibinfo {author} {\bibnamefont {Coello~P\'erez},
  \bibfnamefont {E~A}}, \bibinfo {author} {\bibnamefont {Men\'endez},
  \bibfnamefont {J}}, \ and\ \bibinfo {author} {\bibnamefont {Schwenk},
  \bibfnamefont {A}}} (\bibinfo {year} {2018}),\ \bibfield  {title} {\enquote
  {\bibinfo {title} {Gamow-teller and double-$\ensuremath{\beta}$ decays of
  heavy nuclei within an effective theory},}\ }\href {\doibase
  10.1103/PhysRevC.98.045501} {\bibfield  {journal} {\bibinfo  {journal} {Phys.
  Rev. C}\ }\textbf {\bibinfo {volume} {98}},\ \bibinfo {pages}
  {045501}}\BibitemShut {NoStop}%
\bibitem [{\citenamefont {Coello~P\'erez}\ and\ \citenamefont
  {Papenbrock}(2015{\natexlab{a}})}]{coelloperez2015b}%
  \BibitemOpen
  \bibfield  {author} {\bibinfo {author} {\bibnamefont {Coello~P\'erez},
  \bibfnamefont {E~A}}, \ and\ \bibinfo {author} {\bibnamefont {Papenbrock},
  \bibfnamefont {T}}} (\bibinfo {year} {2015}{\natexlab{a}}),\ \bibfield
  {title} {\enquote {\bibinfo {title} {Effective field theory for nuclear
  vibrations with quantified uncertainties},}\ }\href {\doibase
  10.1103/PhysRevC.92.064309} {\bibfield  {journal} {\bibinfo  {journal} {Phys.
  Rev. C}\ }\textbf {\bibinfo {volume} {92}},\ \bibinfo {pages}
  {064309}}\BibitemShut {NoStop}%
\bibitem [{\citenamefont {Coello~P\'erez}\ and\ \citenamefont
  {Papenbrock}(2015{\natexlab{b}})}]{coelloperez2015}%
  \BibitemOpen
  \bibfield  {author} {\bibinfo {author} {\bibnamefont {Coello~P\'erez},
  \bibfnamefont {E~A}}, \ and\ \bibinfo {author} {\bibnamefont {Papenbrock},
  \bibfnamefont {T}}} (\bibinfo {year} {2015}{\natexlab{b}}),\ \bibfield
  {title} {\enquote {\bibinfo {title} {Effective theory for the nonrigid rotor
  in an electromagnetic field: Toward accurate and precise calculations of $e2$
  transitions in deformed nuclei},}\ }\href {\doibase
  10.1103/PhysRevC.92.014323} {\bibfield  {journal} {\bibinfo  {journal} {Phys.
  Rev. C}\ }\textbf {\bibinfo {volume} {92}},\ \bibinfo {pages}
  {014323}}\BibitemShut {NoStop}%
\bibitem [{\citenamefont {Coello~P\'erez}\ and\ \citenamefont
  {Papenbrock}(2016)}]{coelloperez2016}%
  \BibitemOpen
  \bibfield  {author} {\bibinfo {author} {\bibnamefont {Coello~P\'erez},
  \bibfnamefont {E~A}}, \ and\ \bibinfo {author} {\bibnamefont {Papenbrock},
  \bibfnamefont {T}}} (\bibinfo {year} {2016}),\ \bibfield  {title} {\enquote
  {\bibinfo {title} {Effective field theory for vibrations in odd-mass
  nuclei},}\ }\href {\doibase 10.1103/PhysRevC.94.054316} {\bibfield  {journal}
  {\bibinfo  {journal} {Phys. Rev. C}\ }\textbf {\bibinfo {volume} {94}},\
  \bibinfo {pages} {054316}}\BibitemShut {NoStop}%
\bibitem [{\citenamefont {Coleman}\ \emph {et~al.}(1969)\citenamefont
  {Coleman}, \citenamefont {Wess},\ and\ \citenamefont {Zumino}}]{coleman1969}%
  \BibitemOpen
  \bibfield  {author} {\bibinfo {author} {\bibnamefont {Coleman}, \bibfnamefont
  {S}}, \bibinfo {author} {\bibnamefont {Wess}, \bibfnamefont {J}}, \ and\
  \bibinfo {author} {\bibnamefont {Zumino}, \bibfnamefont {B}}} (\bibinfo
  {year} {1969}),\ \bibfield  {title} {\enquote {\bibinfo {title} {Structure of
  phenomenological lagrangians. i},}\ }\href {\doibase
  10.1103/PhysRev.177.2239} {\bibfield  {journal} {\bibinfo  {journal} {Phys.
  Rev.}\ }\textbf {\bibinfo {volume} {177}},\ \bibinfo {pages}
  {2239}}\BibitemShut {NoStop}%
\bibitem [{\citenamefont {Davydov}\ and\ \citenamefont
  {Filippov}(1958)}]{davydov1958}%
  \BibitemOpen
  \bibfield  {author} {\bibinfo {author} {\bibnamefont {Davydov}, \bibfnamefont
  {A~S}}, \ and\ \bibinfo {author} {\bibnamefont {Filippov}, \bibfnamefont
  {G~F}}} (\bibinfo {year} {1958}),\ \bibfield  {title} {\enquote {\bibinfo
  {title} {Rotational states in even atomic nuclei},}\ }\href {\doibase
  10.1016/0029-5582(58)90153-6} {\bibfield  {journal} {\bibinfo  {journal}
  {Nuclear Physics}\ }\textbf {\bibinfo {volume} {8}},\ \bibinfo {pages}
  {237}}\BibitemShut {NoStop}%
\bibitem [{\citenamefont {Delaroche}\ \emph {et~al.}(2010)\citenamefont
  {Delaroche}, \citenamefont {Girod}, \citenamefont {Libert}, \citenamefont
  {Goutte}, \citenamefont {Hilaire}, \citenamefont {P\'eru}, \citenamefont
  {Pillet},\ and\ \citenamefont {Bertsch}}]{delaroche2010}%
  \BibitemOpen
  \bibfield  {author} {\bibinfo {author} {\bibnamefont {Delaroche},
  \bibfnamefont {J~P}}, \bibinfo {author} {\bibnamefont {Girod}, \bibfnamefont
  {M}}, \bibinfo {author} {\bibnamefont {Libert}, \bibfnamefont {J}}, \bibinfo
  {author} {\bibnamefont {Goutte}, \bibfnamefont {H}}, \bibinfo {author}
  {\bibnamefont {Hilaire}, \bibfnamefont {S}}, \bibinfo {author} {\bibnamefont
  {P\'eru}, \bibfnamefont {S}}, \bibinfo {author} {\bibnamefont {Pillet},
  \bibfnamefont {N}}, \ and\ \bibinfo {author} {\bibnamefont {Bertsch},
  \bibfnamefont {G~F}}} (\bibinfo {year} {2010}),\ \bibfield  {title} {\enquote
  {\bibinfo {title} {Structure of even-even nuclei using a mapped collective
  hamiltonian and the d1s gogny interaction},}\ }\href {\doibase
  10.1103/PhysRevC.81.014303} {\bibfield  {journal} {\bibinfo  {journal} {Phys.
  Rev. C}\ }\textbf {\bibinfo {volume} {81}},\ \bibinfo {pages}
  {014303}}\BibitemShut {NoStop}%
\bibitem [{\citenamefont {Doherty}\ \emph {et~al.}(2017)\citenamefont
  {Doherty}, \citenamefont {Allmond}, \citenamefont {Janssens}, \citenamefont
  {Korten}, \citenamefont {Zhu}, \citenamefont {Zielińska}, \citenamefont
  {Radford}, \citenamefont {Ayangeakaa}, \citenamefont {Bucher}, \citenamefont
  {Batchelder}, \citenamefont {Beausang}, \citenamefont {Campbell},
  \citenamefont {Carpenter}, \citenamefont {Cline}, \citenamefont {Crawford},
  \citenamefont {David}, \citenamefont {Delaroche}, \citenamefont {Dickerson},
  \citenamefont {Fallon}, \citenamefont {Galindo-Uribarri}, \citenamefont
  {Kondev}, \citenamefont {Harker}, \citenamefont {Hayes}, \citenamefont
  {Hendricks}, \citenamefont {Humby}, \citenamefont {Girod}, \citenamefont
  {Gross}, \citenamefont {Klintefjord}, \citenamefont {Kolos}, \citenamefont
  {Lane}, \citenamefont {Lauritsen}, \citenamefont {Libert}, \citenamefont
  {Macchiavelli}, \citenamefont {Napiorkowski}, \citenamefont {Padilla-Rodal},
  \citenamefont {Pardo}, \citenamefont {Reviol}, \citenamefont {Sarantites},
  \citenamefont {Savard}, \citenamefont {Seweryniak}, \citenamefont {Srebrny},
  \citenamefont {Varner}, \citenamefont {Vondrasek}, \citenamefont {Wiens},
  \citenamefont {Wilson}, \citenamefont {Wood},\ and\ \citenamefont
  {Wu}}]{doherty2017}%
  \BibitemOpen
  \bibfield  {author} {\bibinfo {author} {\bibnamefont {Doherty}, \bibfnamefont
  {D~T}}, \bibinfo {author} {\bibnamefont {Allmond}, \bibfnamefont {J~M}},
  \bibinfo {author} {\bibnamefont {Janssens}, \bibfnamefont {R~V~F}}, \bibinfo
  {author} {\bibnamefont {Korten}, \bibfnamefont {W}}, \bibinfo {author}
  {\bibnamefont {Zhu}, \bibfnamefont {S}}, \bibinfo {author} {\bibnamefont
  {Zielińska}, \bibfnamefont {M}}, \bibinfo {author} {\bibnamefont {Radford},
  \bibfnamefont {D~C}}, \bibinfo {author} {\bibnamefont {Ayangeakaa},
  \bibfnamefont {A~D}}, \bibinfo {author} {\bibnamefont {Bucher}, \bibfnamefont
  {B}}, \bibinfo {author} {\bibnamefont {Batchelder}, \bibfnamefont {J~C}},
  \bibinfo {author} {\bibnamefont {Beausang}, \bibfnamefont {C~W}}, \bibinfo
  {author} {\bibnamefont {Campbell}, \bibfnamefont {C}}, \bibinfo {author}
  {\bibnamefont {Carpenter}, \bibfnamefont {M~P}}, \bibinfo {author}
  {\bibnamefont {Cline}, \bibfnamefont {D}}, \bibinfo {author} {\bibnamefont
  {Crawford}, \bibfnamefont {H~L}}, \bibinfo {author} {\bibnamefont {David},
  \bibfnamefont {H~M}}, \bibinfo {author} {\bibnamefont {Delaroche},
  \bibfnamefont {J~P}}, \bibinfo {author} {\bibnamefont {Dickerson},
  \bibfnamefont {C}}, \bibinfo {author} {\bibnamefont {Fallon}, \bibfnamefont
  {P}}, \bibinfo {author} {\bibnamefont {Galindo-Uribarri}, \bibfnamefont {A}},
  \bibinfo {author} {\bibnamefont {Kondev}, \bibfnamefont {F~G}}, \bibinfo
  {author} {\bibnamefont {Harker}, \bibfnamefont {J~L}}, \bibinfo {author}
  {\bibnamefont {Hayes}, \bibfnamefont {A~B}}, \bibinfo {author} {\bibnamefont
  {Hendricks}, \bibfnamefont {M}}, \bibinfo {author} {\bibnamefont {Humby},
  \bibfnamefont {P}}, \bibinfo {author} {\bibnamefont {Girod}, \bibfnamefont
  {M}}, \bibinfo {author} {\bibnamefont {Gross}, \bibfnamefont {C~J}}, \bibinfo
  {author} {\bibnamefont {Klintefjord}, \bibfnamefont {M}}, \bibinfo {author}
  {\bibnamefont {Kolos}, \bibfnamefont {K}}, \bibinfo {author} {\bibnamefont
  {Lane}, \bibfnamefont {G~J}}, \bibinfo {author} {\bibnamefont {Lauritsen},
  \bibfnamefont {T}}, \bibinfo {author} {\bibnamefont {Libert}, \bibfnamefont
  {J}}, \bibinfo {author} {\bibnamefont {Macchiavelli}, \bibfnamefont {A~O}},
  \bibinfo {author} {\bibnamefont {Napiorkowski}, \bibfnamefont {P~J}},
  \bibinfo {author} {\bibnamefont {Padilla-Rodal}, \bibfnamefont {E}}, \bibinfo
  {author} {\bibnamefont {Pardo}, \bibfnamefont {R}}, \bibinfo {author}
  {\bibnamefont {Reviol}, \bibfnamefont {W}}, \bibinfo {author} {\bibnamefont
  {Sarantites}, \bibfnamefont {D~G}}, \bibinfo {author} {\bibnamefont {Savard},
  \bibfnamefont {G}}, \bibinfo {author} {\bibnamefont {Seweryniak},
  \bibfnamefont {D}}, \bibinfo {author} {\bibnamefont {Srebrny}, \bibfnamefont
  {J}}, \bibinfo {author} {\bibnamefont {Varner}, \bibfnamefont {R}}, \bibinfo
  {author} {\bibnamefont {Vondrasek}, \bibfnamefont {R}}, \bibinfo {author}
  {\bibnamefont {Wiens}, \bibfnamefont {A}}, \bibinfo {author} {\bibnamefont
  {Wilson}, \bibfnamefont {E}}, \bibinfo {author} {\bibnamefont {Wood},
  \bibfnamefont {J~L}}, \ and\ \bibinfo {author} {\bibnamefont {Wu},
  \bibfnamefont {C~Y}}} (\bibinfo {year} {2017}),\ \bibfield  {title} {\enquote
  {\bibinfo {title} {Triaxiality near the 110ru ground state from coulomb
  excitation},}\ }\href {\doibase 10.1016/j.physletb.2017.01.031} {\bibfield
  {journal} {\bibinfo  {journal} {Physics Letters B}\ }\textbf {\bibinfo
  {volume} {766}},\ \bibinfo {pages} {334}}\BibitemShut {NoStop}%
\bibitem [{\citenamefont {Dudek}(2016)}]{dudek2016}%
  \BibitemOpen
  \bibfield  {author} {\bibinfo {author} {\bibnamefont {Dudek}, \bibfnamefont
  {J}}} (\bibinfo {year} {2016}),\ \bibfield  {title} {\enquote {\bibinfo
  {title} {Focus issue to celebrate the 40-year anniversary of the 1975 nobel
  prize to aage niels bohr, ben roy mottelson and leo james rainwater},}\
  }\href {\doibase 10.1088/0031-8949/91/3/030301} {\bibfield  {journal}
  {\bibinfo  {journal} {Physica Scripta}\ }\textbf {\bibinfo {volume}
  {91}}~(\bibinfo {number} {3}),\ \bibinfo {pages} {030301}}\BibitemShut
  {NoStop}%
\bibitem [{\citenamefont {Duguet}\ and\ \citenamefont
  {Sadoudi}(2010)}]{duguet2010}%
  \BibitemOpen
  \bibfield  {author} {\bibinfo {author} {\bibnamefont {Duguet}, \bibfnamefont
  {T}}, \ and\ \bibinfo {author} {\bibnamefont {Sadoudi}, \bibfnamefont {J}}}
  (\bibinfo {year} {2010}),\ \bibfield  {title} {\enquote {\bibinfo {title}
  {Breaking and restoring symmetries within the nuclear energy density
  functional method},}\ }\href {\doibase 10.1088/0954-3899/37/6/064009}
  {\bibfield  {journal} {\bibinfo  {journal} {Journal of Physics G: Nuclear and
  Particle Physics}\ }\textbf {\bibinfo {volume} {37}}~(\bibinfo {number}
  {6}),\ \bibinfo {pages} {064009}}\BibitemShut {NoStop}%
\bibitem [{\citenamefont {Dukelsky}\ \emph {et~al.}(2004)\citenamefont
  {Dukelsky}, \citenamefont {Pittel},\ and\ \citenamefont
  {Sierra}}]{dukelsky2004}%
  \BibitemOpen
  \bibfield  {author} {\bibinfo {author} {\bibnamefont {Dukelsky},
  \bibfnamefont {J}}, \bibinfo {author} {\bibnamefont {Pittel}, \bibfnamefont
  {S}}, \ and\ \bibinfo {author} {\bibnamefont {Sierra}, \bibfnamefont {G}}}
  (\bibinfo {year} {2004}),\ \bibfield  {title} {\enquote {\bibinfo {title}
  {Colloquium: Exactly solvable richardson-gaudin models for many-body quantum
  systems},}\ }\href {\doibase 10.1103/RevModPhys.76.643} {\bibfield  {journal}
  {\bibinfo  {journal} {Rev. Mod. Phys.}\ }\textbf {\bibinfo {volume} {76}},\
  \bibinfo {pages} {643}}\BibitemShut {NoStop}%
\bibitem [{\citenamefont {Dytrych}\ \emph {et~al.}(2013)\citenamefont
  {Dytrych}, \citenamefont {Launey}, \citenamefont {Draayer}, \citenamefont
  {Maris}, \citenamefont {Vary}, \citenamefont {Saule}, \citenamefont
  {Catalyurek}, \citenamefont {Sosonkina}, \citenamefont {Langr},\ and\
  \citenamefont {Caprio}}]{dytrych2013}%
  \BibitemOpen
  \bibfield  {author} {\bibinfo {author} {\bibnamefont {Dytrych}, \bibfnamefont
  {T}}, \bibinfo {author} {\bibnamefont {Launey}, \bibfnamefont {K~D}},
  \bibinfo {author} {\bibnamefont {Draayer}, \bibfnamefont {J~P}}, \bibinfo
  {author} {\bibnamefont {Maris}, \bibfnamefont {P}}, \bibinfo {author}
  {\bibnamefont {Vary}, \bibfnamefont {J~P}}, \bibinfo {author} {\bibnamefont
  {Saule}, \bibfnamefont {E}}, \bibinfo {author} {\bibnamefont {Catalyurek},
  \bibfnamefont {U}}, \bibinfo {author} {\bibnamefont {Sosonkina},
  \bibfnamefont {M}}, \bibinfo {author} {\bibnamefont {Langr}, \bibfnamefont
  {D}}, \ and\ \bibinfo {author} {\bibnamefont {Caprio}, \bibfnamefont {M~A}}}
  (\bibinfo {year} {2013}),\ \bibfield  {title} {\enquote {\bibinfo {title}
  {Collective modes in light nuclei from first principles},}\ }\href {\doibase
  10.1103/PhysRevLett.111.252501} {\bibfield  {journal} {\bibinfo  {journal}
  {Phys. Rev. Lett.}\ }\textbf {\bibinfo {volume} {111}},\ \bibinfo {pages}
  {252501}}\BibitemShut {NoStop}%
\bibitem [{\citenamefont {Dytrych}\ \emph {et~al.}(2020)\citenamefont
  {Dytrych}, \citenamefont {Launey}, \citenamefont {Draayer}, \citenamefont
  {Rowe}, \citenamefont {Wood}, \citenamefont {Rosensteel}, \citenamefont
  {Bahri}, \citenamefont {Langr},\ and\ \citenamefont {Baker}}]{dytrych2020}%
  \BibitemOpen
  \bibfield  {author} {\bibinfo {author} {\bibnamefont {Dytrych}, \bibfnamefont
  {T}}, \bibinfo {author} {\bibnamefont {Launey}, \bibfnamefont {K~D}},
  \bibinfo {author} {\bibnamefont {Draayer}, \bibfnamefont {J~P}}, \bibinfo
  {author} {\bibnamefont {Rowe}, \bibfnamefont {D~J}}, \bibinfo {author}
  {\bibnamefont {Wood}, \bibfnamefont {J~L}}, \bibinfo {author} {\bibnamefont
  {Rosensteel}, \bibfnamefont {G}}, \bibinfo {author} {\bibnamefont {Bahri},
  \bibfnamefont {C}}, \bibinfo {author} {\bibnamefont {Langr}, \bibfnamefont
  {D}}, \ and\ \bibinfo {author} {\bibnamefont {Baker}, \bibfnamefont {R~B}}}
  (\bibinfo {year} {2020}),\ \bibfield  {title} {\enquote {\bibinfo {title}
  {Physics of nuclei: Key role of an emergent symmetry},}\ }\href {\doibase
  10.1103/PhysRevLett.124.042501} {\bibfield  {journal} {\bibinfo  {journal}
  {Phys. Rev. Lett.}\ }\textbf {\bibinfo {volume} {124}},\ \bibinfo {pages}
  {042501}}\BibitemShut {NoStop}%
\bibitem [{\citenamefont {Dytrych}\ \emph {et~al.}(2008)\citenamefont
  {Dytrych}, \citenamefont {Sviratcheva}, \citenamefont {Draayer},
  \citenamefont {Bahri},\ and\ \citenamefont {Vary}}]{dytrych2008}%
  \BibitemOpen
  \bibfield  {author} {\bibinfo {author} {\bibnamefont {Dytrych}, \bibfnamefont
  {T}}, \bibinfo {author} {\bibnamefont {Sviratcheva}, \bibfnamefont {K~D}},
  \bibinfo {author} {\bibnamefont {Draayer}, \bibfnamefont {J~P}}, \bibinfo
  {author} {\bibnamefont {Bahri}, \bibfnamefont {C}}, \ and\ \bibinfo {author}
  {\bibnamefont {Vary}, \bibfnamefont {J~P}}} (\bibinfo {year} {2008}),\
  \bibfield  {title} {\enquote {\bibinfo {title} {Ab initio symplectic no-core
  shell model},}\ }\href {http://stacks.iop.org/0954-3899/35/i=12/a=123101}
  {\bibfield  {journal} {\bibinfo  {journal} {Journal of Physics G: Nuclear and
  Particle Physics}\ }\textbf {\bibinfo {volume} {35}}~(\bibinfo {number}
  {12}),\ \bibinfo {pages} {123101}}\BibitemShut {NoStop}%
\bibitem [{\citenamefont {Eisenberg}\ and\ \citenamefont
  {Greiner}(1970)}]{eisenberg1970}%
  \BibitemOpen
  \bibfield  {author} {\bibinfo {author} {\bibnamefont {Eisenberg},
  \bibfnamefont {J~M}}, \ and\ \bibinfo {author} {\bibnamefont {Greiner},
  \bibfnamefont {W}}} (\bibinfo {year} {1970}),\ \href@noop {} {\emph {\bibinfo
  {title} {Nuclear Models: Collective and Single-Particle Phenomena}}}\
  (\bibinfo  {publisher} {North-Holland Publishing Company Ltd.},\ \bibinfo
  {address} {London})\BibitemShut {NoStop}%
\bibitem [{\citenamefont {Elliott}(1958)}]{elliott1958}%
  \BibitemOpen
  \bibfield  {author} {\bibinfo {author} {\bibnamefont {Elliott}, \bibfnamefont
  {J~P}}} (\bibinfo {year} {1958}),\ \bibfield  {title} {\enquote {\bibinfo
  {title} {Collective motion in the nuclear shell model. i. classification
  schemes for states of mixed configurations},}\ }\href {\doibase
  10.1098/rspa.1958.0072} {\bibfield  {journal} {\bibinfo  {journal}
  {Proceedings of the Royal Society of London. Series A, Mathematical and
  Physical Sciences}\ }\textbf {\bibinfo {volume} {245}}~(\bibinfo {number}
  {1240}),\ \bibinfo {pages} {128}}\BibitemShut {NoStop}%
\bibitem [{\citenamefont {Engel}\ and\ \citenamefont
  {Men{\'{e}}ndez}(2017)}]{engel2017}%
  \BibitemOpen
  \bibfield  {author} {\bibinfo {author} {\bibnamefont {Engel}, \bibfnamefont
  {J}}, \ and\ \bibinfo {author} {\bibnamefont {Men{\'{e}}ndez}, \bibfnamefont
  {J}}} (\bibinfo {year} {2017}),\ \bibfield  {title} {\enquote {\bibinfo
  {title} {Status and future of nuclear matrix elements for neutrinoless
  double-beta decay: a review},}\ }\href {\doibase 10.1088/1361-6633/aa5bc5}
  {\bibfield  {journal} {\bibinfo  {journal} {Reports on Progress in Physics}\
  }\textbf {\bibinfo {volume} {80}}~(\bibinfo {number} {4}),\ \bibinfo {pages}
  {046301}}\BibitemShut {NoStop}%
\bibitem [{\citenamefont {Epelbaum}\ \emph {et~al.}(2009)\citenamefont
  {Epelbaum}, \citenamefont {Hammer},\ and\ \citenamefont
  {Mei\ss{}ner}}]{epelbaum2009}%
  \BibitemOpen
  \bibfield  {author} {\bibinfo {author} {\bibnamefont {Epelbaum},
  \bibfnamefont {E}}, \bibinfo {author} {\bibnamefont {Hammer}, \bibfnamefont
  {H~W}}, \ and\ \bibinfo {author} {\bibnamefont {Mei\ss{}ner}, \bibfnamefont
  {U~G}}} (\bibinfo {year} {2009}),\ \bibfield  {title} {\enquote {\bibinfo
  {title} {Modern theory of nuclear forces},}\ }\href {\doibase
  10.1103/RevModPhys.81.1773} {\bibfield  {journal} {\bibinfo  {journal} {Rev.
  Mod. Phys.}\ }\textbf {\bibinfo {volume} {81}},\ \bibinfo {pages}
  {1773}}\BibitemShut {NoStop}%
\bibitem [{\citenamefont {Epelbaum}\ \emph {et~al.}(2012)\citenamefont
  {Epelbaum}, \citenamefont {Krebs}, \citenamefont {L\"ahde}, \citenamefont
  {Lee},\ and\ \citenamefont {Mei\ss{}ner}}]{epelbaum2012}%
  \BibitemOpen
  \bibfield  {author} {\bibinfo {author} {\bibnamefont {Epelbaum},
  \bibfnamefont {E}}, \bibinfo {author} {\bibnamefont {Krebs}, \bibfnamefont
  {H}}, \bibinfo {author} {\bibnamefont {L\"ahde}, \bibfnamefont {T~A}},
  \bibinfo {author} {\bibnamefont {Lee}, \bibfnamefont {D}}, \ and\ \bibinfo
  {author} {\bibnamefont {Mei\ss{}ner}, \bibfnamefont {U~G}}} (\bibinfo {year}
  {2012}),\ \bibfield  {title} {\enquote {\bibinfo {title} {Structure and
  rotations of the hoyle state},}\ }\href {\doibase
  10.1103/PhysRevLett.109.252501} {\bibfield  {journal} {\bibinfo  {journal}
  {Phys. Rev. Lett.}\ }\textbf {\bibinfo {volume} {109}},\ \bibinfo {pages}
  {252501}}\BibitemShut {NoStop}%
\bibitem [{\citenamefont {Erler}\ \emph {et~al.}(2012)\citenamefont {Erler},
  \citenamefont {Birge}, \citenamefont {Kortelainen}, \citenamefont
  {Nazarewicz}, \citenamefont {Olsen}, \citenamefont {Perhac},\ and\
  \citenamefont {Stoitsov}}]{erler2012}%
  \BibitemOpen
  \bibfield  {author} {\bibinfo {author} {\bibnamefont {Erler}, \bibfnamefont
  {J}}, \bibinfo {author} {\bibnamefont {Birge}, \bibfnamefont {N}}, \bibinfo
  {author} {\bibnamefont {Kortelainen}, \bibfnamefont {M}}, \bibinfo {author}
  {\bibnamefont {Nazarewicz}, \bibfnamefont {W}}, \bibinfo {author}
  {\bibnamefont {Olsen}, \bibfnamefont {E}}, \bibinfo {author} {\bibnamefont
  {Perhac}, \bibfnamefont {A~M}}, \ and\ \bibinfo {author} {\bibnamefont
  {Stoitsov}, \bibfnamefont {M}}} (\bibinfo {year} {2012}),\ \bibfield  {title}
  {\enquote {\bibinfo {title} {The limits of the nuclear landscape},}\ }\href
  {\doibase 10.1038/nature11188} {\bibfield  {journal} {\bibinfo  {journal}
  {Nature}\ }\textbf {\bibinfo {volume} {486}},\ \bibinfo {pages} {509
  }}\BibitemShut {NoStop}%
\bibitem [{\citenamefont {Estienne}\ \emph {et~al.}(2011)\citenamefont
  {Estienne}, \citenamefont {Haaker},\ and\ \citenamefont
  {Schoutens}}]{estienne2011}%
  \BibitemOpen
  \bibfield  {author} {\bibinfo {author} {\bibnamefont {Estienne},
  \bibfnamefont {B}}, \bibinfo {author} {\bibnamefont {Haaker}, \bibfnamefont
  {S~M}}, \ and\ \bibinfo {author} {\bibnamefont {Schoutens}, \bibfnamefont
  {K}}} (\bibinfo {year} {2011}),\ \bibfield  {title} {\enquote {\bibinfo
  {title} {{Particles in non-Abelian gauge potentials: Landau problem and
  insertion of non-Abelian flux}},}\ }\href {\doibase
  10.1088/1367-2630/13/4/045012} {\bibfield  {journal} {\bibinfo  {journal}
  {New Journal of Physics}\ }\textbf {\bibinfo {volume} {13}}~(\bibinfo
  {number} {4}),\ \bibinfo {pages} {045012}}\BibitemShut {NoStop}%
\bibitem [{\citenamefont {Faessler}\ \emph {et~al.}(1965)\citenamefont
  {Faessler}, \citenamefont {Greiner},\ and\ \citenamefont
  {Sheline}}]{faessler1965}%
  \BibitemOpen
  \bibfield  {author} {\bibinfo {author} {\bibnamefont {Faessler},
  \bibfnamefont {A}}, \bibinfo {author} {\bibnamefont {Greiner}, \bibfnamefont
  {W}}, \ and\ \bibinfo {author} {\bibnamefont {Sheline}, \bibfnamefont {R~K}}}
  (\bibinfo {year} {1965}),\ \bibfield  {title} {\enquote {\bibinfo {title}
  {Rotation vibration interaction in deformed nuclei},}\ }\href {\doibase
  10.1016/0029-5582(65)90224-5} {\bibfield  {journal} {\bibinfo  {journal}
  {Nuclear Physics}\ }\textbf {\bibinfo {volume} {70}}~(\bibinfo {number}
  {1}),\ \bibinfo {pages} {33 }}\BibitemShut {NoStop}%
\bibitem [{\citenamefont {Fierz}(1944)}]{fierz1944}%
  \BibitemOpen
  \bibfield  {author} {\bibinfo {author} {\bibnamefont {Fierz}, \bibfnamefont
  {M}}} (\bibinfo {year} {1944}),\ \bibfield  {title} {\enquote {\bibinfo
  {title} {{Zur Theorie magnetisch geladener Teilchen}},}\ }\href {\doibase
  10.5169/seals-111493} {\bibfield  {journal} {\bibinfo  {journal} {Helvetica
  Physica Acta}\ }\textbf {\bibinfo {volume} {17}},\ \bibinfo {pages}
  {27}}\BibitemShut {NoStop}%
\bibitem [{\citenamefont {Fortunato}(2005)}]{fortunato2005}%
  \BibitemOpen
  \bibfield  {author} {\bibinfo {author} {\bibnamefont {Fortunato},
  \bibfnamefont {L}}} (\bibinfo {year} {2005}),\ \bibfield  {title} {\enquote
  {\bibinfo {title} {Solutions of the bohr hamiltonian, a compendium},}\ }\href
  {\doibase 10.1140/epjad/i2005-07-115-8} {\bibfield  {journal} {\bibinfo
  {journal} {The European Physical Journal A - Hadrons and Nuclei}\ }\textbf
  {\bibinfo {volume} {26}}~(\bibinfo {number} {1}),\ \bibinfo {pages}
  {1}}\BibitemShut {NoStop}%
\bibitem [{\citenamefont {Frank}\ \emph {et~al.}(2019)\citenamefont {Frank},
  \citenamefont {Jolie},\ and\ \citenamefont {Van~Isacker}}]{frank2019}%
  \BibitemOpen
  \bibfield  {author} {\bibinfo {author} {\bibnamefont {Frank}, \bibfnamefont
  {A}}, \bibinfo {author} {\bibnamefont {Jolie}, \bibfnamefont {J}}, \ and\
  \bibinfo {author} {\bibnamefont {Van~Isacker}, \bibfnamefont {P}}} (\bibinfo
  {year} {2019}),\ \href {\doibase 10.1007/978-3-030-21931-4} {\emph {\bibinfo
  {title} {{Symmetries in Atomic Nuclei}}}},\ Vol.\ \bibinfo {volume} {230}\
  (\bibinfo  {publisher} {Springer})\BibitemShut {NoStop}%
\bibitem [{\citenamefont {Frauendorf}(2001)}]{frauendorf2001}%
  \BibitemOpen
  \bibfield  {author} {\bibinfo {author} {\bibnamefont {Frauendorf},
  \bibfnamefont {S}}} (\bibinfo {year} {2001}),\ \bibfield  {title} {\enquote
  {\bibinfo {title} {Spontaneous symmetry breaking in rotating nuclei},}\
  }\href {\doibase 10.1103/RevModPhys.73.463} {\bibfield  {journal} {\bibinfo
  {journal} {Rev. Mod. Phys.}\ }\textbf {\bibinfo {volume} {73}},\ \bibinfo
  {pages} {463}}\BibitemShut {NoStop}%
\bibitem [{\citenamefont {Frauendorf}\ and\ \citenamefont {{Jie
  Meng}}(1997)}]{frauendorf1997}%
  \BibitemOpen
  \bibfield  {author} {\bibinfo {author} {\bibnamefont {Frauendorf},
  \bibfnamefont {S}}, \ and\ \bibinfo {author} {\bibnamefont {{Jie Meng}},}}
  (\bibinfo {year} {1997}),\ \bibfield  {title} {\enquote {\bibinfo {title}
  {Tilted rotation of triaxial nuclei},}\ }\href {\doibase
  10.1016/S0375-9474(97)00004-3} {\bibfield  {journal} {\bibinfo  {journal}
  {Nuclear Physics A}\ }\textbf {\bibinfo {volume} {617}}~(\bibinfo {number}
  {2}),\ \bibinfo {pages} {131}}\BibitemShut {NoStop}%
\bibitem [{\citenamefont {Frosini}\ \emph {et~al.}(2022)\citenamefont
  {Frosini}, \citenamefont {Duguet}, \citenamefont {Ebran}, \citenamefont
  {Bally}, \citenamefont {Mongelli}, \citenamefont {Rodr\'\i{}guez},
  \citenamefont {Roth},\ and\ \citenamefont {Som\`a}}]{frosini2022}%
  \BibitemOpen
  \bibfield  {author} {\bibinfo {author} {\bibnamefont {Frosini}, \bibfnamefont
  {M}}, \bibinfo {author} {\bibnamefont {Duguet}, \bibfnamefont {T}}, \bibinfo
  {author} {\bibnamefont {Ebran}, \bibfnamefont {J~P}}, \bibinfo {author}
  {\bibnamefont {Bally}, \bibfnamefont {B}}, \bibinfo {author} {\bibnamefont
  {Mongelli}, \bibfnamefont {T}}, \bibinfo {author} {\bibnamefont
  {Rodr\'\i{}guez}, \bibfnamefont {T~R}}, \bibinfo {author} {\bibnamefont
  {Roth}, \bibfnamefont {R}}, \ and\ \bibinfo {author} {\bibnamefont {Som\`a},
  \bibfnamefont {V}}} (\bibinfo {year} {2022}),\ \bibfield  {title} {\enquote
  {\bibinfo {title} {{Multi-reference many-body perturbation theory for nuclei.
  II. Ab initio study of neon isotopes via PGCM and IM-NCSM calculations}},}\
  }\href {\doibase 10.1140/epja/s10050-022-00693-y} {\bibfield  {journal}
  {\bibinfo  {journal} {Eur. Phys. J. A}\ }\textbf {\bibinfo {volume} {58}},\
  \bibinfo {pages} {63}}\BibitemShut {NoStop}%
\bibitem [{\citenamefont {Fujikawa}\ and\ \citenamefont
  {Ui}(1986)}]{fujikawa1986}%
  \BibitemOpen
  \bibfield  {author} {\bibinfo {author} {\bibnamefont {Fujikawa},
  \bibfnamefont {K}}, \ and\ \bibinfo {author} {\bibnamefont {Ui},
  \bibfnamefont {H}}} (\bibinfo {year} {1986}),\ \bibfield  {title} {\enquote
  {\bibinfo {title} {{Nuclear Rotation, Nambu-Goldstone Mode and Higgs
  Mechanism}},}\ }\href {\doibase 10.1143/PTP.75.997} {\bibfield  {journal}
  {\bibinfo  {journal} {Progress of Theoretical Physics}\ }\textbf {\bibinfo
  {volume} {75}}~(\bibinfo {number} {5}),\ \bibinfo {pages} {997}}\BibitemShut
  {NoStop}%
\bibitem [{\citenamefont {Fukuda}(1988)}]{fukuda1988}%
  \BibitemOpen
  \bibfield  {author} {\bibinfo {author} {\bibnamefont {Fukuda}, \bibfnamefont
  {R}}} (\bibinfo {year} {1988}),\ \bibfield  {title} {\enquote {\bibinfo
  {title} {Method of calculating nonperturbative effects in quantum
  chromodynamics},}\ }\href {\doibase 10.1103/PhysRevLett.61.1549} {\bibfield
  {journal} {\bibinfo  {journal} {Phys. Rev. Lett.}\ }\textbf {\bibinfo
  {volume} {61}},\ \bibinfo {pages} {1549}}\BibitemShut {NoStop}%
\bibitem [{\citenamefont {Fukugita}\ and\ \citenamefont
  {Yanagida}(1986)}]{fukugita_m_1986}%
  \BibitemOpen
  \bibfield  {author} {\bibinfo {author} {\bibnamefont {Fukugita},
  \bibfnamefont {M}}, \ and\ \bibinfo {author} {\bibnamefont {Yanagida},
  \bibfnamefont {T}}} (\bibinfo {year} {1986}),\ \bibfield  {title} {\enquote
  {\bibinfo {title} {Barygenesis without grand unification},}\ }\href {\doibase
  10.1016/0370-2693(86)91126-3} {\bibfield  {journal} {\bibinfo  {journal}
  {Phys. Lett. B}\ }\textbf {\bibinfo {volume} {174}}~(\bibinfo {number} {1}),\
  \bibinfo {pages} {45}}\BibitemShut {NoStop}%
\bibitem [{\citenamefont {{Furnstahl}}\ \emph {et~al.}(2015)\citenamefont
  {{Furnstahl}}, \citenamefont {{Phillips}},\ and\ \citenamefont
  {{Wesolowski}}}]{furnstahl2014c}%
  \BibitemOpen
  \bibfield  {author} {\bibinfo {author} {\bibnamefont {{Furnstahl}},
  \bibfnamefont {R~J}}, \bibinfo {author} {\bibnamefont {{Phillips}},
  \bibfnamefont {D~R}}, \ and\ \bibinfo {author} {\bibnamefont {{Wesolowski}},
  \bibfnamefont {S}}} (\bibinfo {year} {2015}),\ \bibfield  {title} {\enquote
  {\bibinfo {title} {A recipe for eft uncertainty quantification in nuclear
  physics},}\ }\href {http://stacks.iop.org/0954-3899/42/i=3/a=034028}
  {\bibfield  {journal} {\bibinfo  {journal} {Journal of Physics G: Nuclear and
  Particle Physics}\ }\textbf {\bibinfo {volume} {42}}~(\bibinfo {number}
  {3}),\ \bibinfo {pages} {034028}}\BibitemShut {NoStop}%
\bibitem [{\citenamefont {Garrett}\ and\ \citenamefont
  {Wood}(2010)}]{garrett2010}%
  \BibitemOpen
  \bibfield  {author} {\bibinfo {author} {\bibnamefont {Garrett}, \bibfnamefont
  {P~E}}, \ and\ \bibinfo {author} {\bibnamefont {Wood}, \bibfnamefont {J~L}}}
  (\bibinfo {year} {2010}),\ \bibfield  {title} {\enquote {\bibinfo {title} {On
  the robustness of surface vibrational modes: case studies in the cd
  region},}\ }\href {http://stacks.iop.org/0954-3899/37/i=6/a=064028}
  {\bibfield  {journal} {\bibinfo  {journal} {Journal of Physics G: Nuclear and
  Particle Physics}\ }\textbf {\bibinfo {volume} {37}}~(\bibinfo {number}
  {6}),\ \bibinfo {pages} {064028}}\BibitemShut {NoStop}%
\bibitem [{\citenamefont {{Gasser}}\ and\ \citenamefont
  {{Leutwyler}}(1984)}]{gasser1984}%
  \BibitemOpen
  \bibfield  {author} {\bibinfo {author} {\bibnamefont {{Gasser}},
  \bibfnamefont {J}}, \ and\ \bibinfo {author} {\bibnamefont {{Leutwyler}},
  \bibfnamefont {H}}} (\bibinfo {year} {1984}),\ \bibfield  {title} {\enquote
  {\bibinfo {title} {{Chiral perturbation theory to one loop}},}\ }\href
  {\doibase 10.1016/0003-4916(84)90242-2} {\bibfield  {journal} {\bibinfo
  {journal} {Annals of Physics}\ }\textbf {\bibinfo {volume} {158}},\ \bibinfo
  {pages} {142}}\BibitemShut {NoStop}%
\bibitem [{\citenamefont {Goldstone}(1961)}]{goldstone1961}%
  \BibitemOpen
  \bibfield  {author} {\bibinfo {author} {\bibnamefont {Goldstone},
  \bibfnamefont {J}}} (\bibinfo {year} {1961}),\ \bibfield  {title} {\enquote
  {\bibinfo {title} {Field theories with ``superconductor'' solutions},}\
  }\href {\doibase 10.1007/BF02812722} {\bibfield  {journal} {\bibinfo
  {journal} {Il Nuovo Cimento}\ }\textbf {\bibinfo {volume} {19}}~(\bibinfo
  {number} {1}),\ \bibinfo {pages} {154}}\BibitemShut {NoStop}%
\bibitem [{\citenamefont {Gray}\ \emph {et~al.}(2022)\citenamefont {Gray},
  \citenamefont {Allmond}, \citenamefont {Janssens}, \citenamefont {Korten},
  \citenamefont {Stuchbery}, \citenamefont {Wood}, \citenamefont {Ayangeakaa},
  \citenamefont {Bottoni}, \citenamefont {Bucher}, \citenamefont {Campbell},
  \citenamefont {Carpenter}, \citenamefont {Crawford}, \citenamefont {David},
  \citenamefont {Doherty}, \citenamefont {Fallon}, \citenamefont {Febbraro},
  \citenamefont {Galindo-Uribarri}, \citenamefont {Gross}, \citenamefont
  {Komorowska}, \citenamefont {Kondev}, \citenamefont {Lauritsen},
  \citenamefont {Macchiavelli}, \citenamefont {Napiorkowsi}, \citenamefont
  {Padilla-Rodal}, \citenamefont {Pain}, \citenamefont {Reviol}, \citenamefont
  {Sarantites}, \citenamefont {Savard}, \citenamefont {Seweryniak},
  \citenamefont {Wu}, \citenamefont {Yu},\ and\ \citenamefont
  {Zhu}}]{gray2022}%
  \BibitemOpen
  \bibfield  {author} {\bibinfo {author} {\bibnamefont {Gray}, \bibfnamefont
  {T~J}}, \bibinfo {author} {\bibnamefont {Allmond}, \bibfnamefont {J~M}},
  \bibinfo {author} {\bibnamefont {Janssens}, \bibfnamefont {R~V~F}}, \bibinfo
  {author} {\bibnamefont {Korten}, \bibfnamefont {W}}, \bibinfo {author}
  {\bibnamefont {Stuchbery}, \bibfnamefont {A~E}}, \bibinfo {author}
  {\bibnamefont {Wood}, \bibfnamefont {J~L}}, \bibinfo {author} {\bibnamefont
  {Ayangeakaa}, \bibfnamefont {A~D}}, \bibinfo {author} {\bibnamefont
  {Bottoni}, \bibfnamefont {S}}, \bibinfo {author} {\bibnamefont {Bucher},
  \bibfnamefont {B~M}}, \bibinfo {author} {\bibnamefont {Campbell},
  \bibfnamefont {C~M}}, \bibinfo {author} {\bibnamefont {Carpenter},
  \bibfnamefont {M~P}}, \bibinfo {author} {\bibnamefont {Crawford},
  \bibfnamefont {H~L}}, \bibinfo {author} {\bibnamefont {David}, \bibfnamefont
  {H}}, \bibinfo {author} {\bibnamefont {Doherty}, \bibfnamefont {D~T}},
  \bibinfo {author} {\bibnamefont {Fallon}, \bibfnamefont {P}}, \bibinfo
  {author} {\bibnamefont {Febbraro}, \bibfnamefont {M~T}}, \bibinfo {author}
  {\bibnamefont {Galindo-Uribarri}, \bibfnamefont {A}}, \bibinfo {author}
  {\bibnamefont {Gross}, \bibfnamefont {C~J}}, \bibinfo {author} {\bibnamefont
  {Komorowska}, \bibfnamefont {M}}, \bibinfo {author} {\bibnamefont {Kondev},
  \bibfnamefont {F~G}}, \bibinfo {author} {\bibnamefont {Lauritsen},
  \bibfnamefont {T}}, \bibinfo {author} {\bibnamefont {Macchiavelli},
  \bibfnamefont {A~O}}, \bibinfo {author} {\bibnamefont {Napiorkowsi},
  \bibfnamefont {P}}, \bibinfo {author} {\bibnamefont {Padilla-Rodal},
  \bibfnamefont {E}}, \bibinfo {author} {\bibnamefont {Pain}, \bibfnamefont
  {S~D}}, \bibinfo {author} {\bibnamefont {Reviol}, \bibfnamefont {W}},
  \bibinfo {author} {\bibnamefont {Sarantites}, \bibfnamefont {D~G}}, \bibinfo
  {author} {\bibnamefont {Savard}, \bibfnamefont {G}}, \bibinfo {author}
  {\bibnamefont {Seweryniak}, \bibfnamefont {D}}, \bibinfo {author}
  {\bibnamefont {Wu}, \bibfnamefont {C~Y}}, \bibinfo {author} {\bibnamefont
  {Yu}, \bibfnamefont {C~H}}, \ and\ \bibinfo {author} {\bibnamefont {Zhu},
  \bibfnamefont {S}}} (\bibinfo {year} {2022}),\ \bibfield  {title} {\enquote
  {\bibinfo {title} {E2 rotational invariants of 01+ and 21+ states for 106cd:
  The emergence of collective rotation},}\ }\href {\doibase
  10.1016/j.physletb.2022.137446} {\bibfield  {journal} {\bibinfo  {journal}
  {Physics Letters B}\ }\textbf {\bibinfo {volume} {834}},\ \bibinfo {pages}
  {137446}}\BibitemShut {NoStop}%
\bibitem [{\citenamefont {Hagen}\ \emph {et~al.}(2022)\citenamefont {Hagen},
  \citenamefont {Novario}, \citenamefont {Sun}, \citenamefont {Papenbrock},
  \citenamefont {Jansen}, \citenamefont {Lietz}, \citenamefont {Duguet},\ and\
  \citenamefont {Tichai}}]{hagen2022}%
  \BibitemOpen
  \bibfield  {author} {\bibinfo {author} {\bibnamefont {Hagen}, \bibfnamefont
  {G}}, \bibinfo {author} {\bibnamefont {Novario}, \bibfnamefont {S~J}},
  \bibinfo {author} {\bibnamefont {Sun}, \bibfnamefont {Z~H}}, \bibinfo
  {author} {\bibnamefont {Papenbrock}, \bibfnamefont {T}}, \bibinfo {author}
  {\bibnamefont {Jansen}, \bibfnamefont {G~R}}, \bibinfo {author} {\bibnamefont
  {Lietz}, \bibfnamefont {J~G}}, \bibinfo {author} {\bibnamefont {Duguet},
  \bibfnamefont {T}}, \ and\ \bibinfo {author} {\bibnamefont {Tichai},
  \bibfnamefont {A}}} (\bibinfo {year} {2022}),\ \bibfield  {title} {\enquote
  {\bibinfo {title} {Angular-momentum projection in coupled-cluster theory:
  Structure of $^{34}\mathrm{Mg}$},}\ }\href {\doibase
  10.1103/PhysRevC.105.064311} {\bibfield  {journal} {\bibinfo  {journal}
  {Phys. Rev. C}\ }\textbf {\bibinfo {volume} {105}},\ \bibinfo {pages}
  {064311}}\BibitemShut {NoStop}%
\bibitem [{\citenamefont {Hagen}\ \emph {et~al.}(2014)\citenamefont {Hagen},
  \citenamefont {Papenbrock}, \citenamefont {Hjorth-Jensen},\ and\
  \citenamefont {Dean}}]{hagen2014}%
  \BibitemOpen
  \bibfield  {author} {\bibinfo {author} {\bibnamefont {Hagen}, \bibfnamefont
  {G}}, \bibinfo {author} {\bibnamefont {Papenbrock}, \bibfnamefont {T}},
  \bibinfo {author} {\bibnamefont {Hjorth-Jensen}, \bibfnamefont {M}}, \ and\
  \bibinfo {author} {\bibnamefont {Dean}, \bibfnamefont {D~J}}} (\bibinfo
  {year} {2014}),\ \bibfield  {title} {\enquote {\bibinfo {title}
  {Coupled-cluster computations of atomic nuclei},}\ }\href {\doibase
  10.1088/0034-4885/77/9/096302} {\bibfield  {journal} {\bibinfo  {journal}
  {Rep. Prog. Phys.}\ }\textbf {\bibinfo {volume} {77}}~(\bibinfo {number}
  {9}),\ \bibinfo {pages} {096302}}\BibitemShut {NoStop}%
\bibitem [{\citenamefont {Hammer}\ \emph {et~al.}(2017)\citenamefont {Hammer},
  \citenamefont {Ji},\ and\ \citenamefont {Phillips}}]{hammer2017}%
  \BibitemOpen
  \bibfield  {author} {\bibinfo {author} {\bibnamefont {Hammer}, \bibfnamefont
  {H~W}}, \bibinfo {author} {\bibnamefont {Ji}, \bibfnamefont {C}}, \ and\
  \bibinfo {author} {\bibnamefont {Phillips}, \bibfnamefont {D~R}}} (\bibinfo
  {year} {2017}),\ \bibfield  {title} {\enquote {\bibinfo {title} {Effective
  field theory description of halo nuclei},}\ }\href {\doibase
  10.1088/1361-6471/aa83db} {\bibfield  {journal} {\bibinfo  {journal} {Journal
  of Physics G: Nuclear and Particle Physics}\ }\textbf {\bibinfo {volume}
  {44}}~(\bibinfo {number} {10}),\ \bibinfo {pages} {103002}}\BibitemShut
  {NoStop}%
\bibitem [{\citenamefont {Hammer}\ \emph {et~al.}(2020)\citenamefont {Hammer},
  \citenamefont {K\"onig},\ and\ \citenamefont {van Kolck}}]{hammer2020}%
  \BibitemOpen
  \bibfield  {author} {\bibinfo {author} {\bibnamefont {Hammer}, \bibfnamefont
  {H~W}}, \bibinfo {author} {\bibnamefont {K\"onig}, \bibfnamefont {S}}, \ and\
  \bibinfo {author} {\bibnamefont {van Kolck}, \bibfnamefont {U}}} (\bibinfo
  {year} {2020}),\ \bibfield  {title} {\enquote {\bibinfo {title} {Nuclear
  effective field theory: Status and perspectives},}\ }\href {\doibase
  10.1103/RevModPhys.92.025004} {\bibfield  {journal} {\bibinfo  {journal}
  {Rev. Mod. Phys.}\ }\textbf {\bibinfo {volume} {92}},\ \bibinfo {pages}
  {025004}}\BibitemShut {NoStop}%
\bibitem [{\citenamefont {Harvey}(1968)}]{harvey1968}%
  \BibitemOpen
  \bibfield  {author} {\bibinfo {author} {\bibnamefont {Harvey}, \bibfnamefont
  {M}}} (\bibinfo {year} {1968}),\ \enquote {\bibinfo {title} {The nuclear su3
  model},}\ in\ \href {\doibase 10.1007/978-1-4757-0103-6_2} {\emph {\bibinfo
  {booktitle} {Advances in Nuclear Physics: Volume 1}}},\ \bibinfo {editor}
  {edited by\ \bibinfo {editor} {\bibfnamefont {M}~\bibnamefont {Baranger}}\
  and\ \bibinfo {editor} {\bibfnamefont {E}~\bibnamefont {Vogt}}}\ (\bibinfo
  {publisher} {Springer US},\ \bibinfo {address} {Boston, MA})\ pp.\ \bibinfo
  {pages} {67--182}\BibitemShut {NoStop}%
\bibitem [{\citenamefont {Hecht}(1965)}]{hecht1965}%
  \BibitemOpen
  \bibfield  {author} {\bibinfo {author} {\bibnamefont {Hecht}, \bibfnamefont
  {K}}} (\bibinfo {year} {1965}),\ \bibfield  {title} {\enquote {\bibinfo
  {title} {Su3 recoupling and fractional parentage in the 2s-1d shell},}\
  }\href {\doibase 10.1016/0029-5582(65)90068-4} {\bibfield  {journal}
  {\bibinfo  {journal} {Nuclear Physics}\ }\textbf {\bibinfo {volume}
  {62}}~(\bibinfo {number} {1}),\ \bibinfo {pages} {1}}\BibitemShut {NoStop}%
\bibitem [{\citenamefont {Hess}\ \emph {et~al.}(1980)\citenamefont {Hess},
  \citenamefont {Seiwert}, \citenamefont {Maruhn},\ and\ \citenamefont
  {Greiner}}]{hess1980}%
  \BibitemOpen
  \bibfield  {author} {\bibinfo {author} {\bibnamefont {Hess}, \bibfnamefont
  {P}}, \bibinfo {author} {\bibnamefont {Seiwert}, \bibfnamefont {M}}, \bibinfo
  {author} {\bibnamefont {Maruhn}, \bibfnamefont {J}}, \ and\ \bibinfo {author}
  {\bibnamefont {Greiner}, \bibfnamefont {W}}} (\bibinfo {year} {1980}),\
  \bibfield  {title} {\enquote {\bibinfo {title} {General collective model and
  its application to $^{238}_{92}$u},}\ }\href {\doibase 10.1007/BF01412656}
  {\bibfield  {journal} {\bibinfo  {journal} {Zeitschrift f{\"u}r Physik A
  Atoms and Nuclei}\ }\textbf {\bibinfo {volume} {296}},\ \bibinfo {pages}
  {147}}\BibitemShut {NoStop}%
\bibitem [{\citenamefont {Hinohara}(2015)}]{hinohara2015}%
  \BibitemOpen
  \bibfield  {author} {\bibinfo {author} {\bibnamefont {Hinohara},
  \bibfnamefont {N}}} (\bibinfo {year} {2015}),\ \bibfield  {title} {\enquote
  {\bibinfo {title} {Collective inertia of the nambu-goldstone mode from linear
  response theory},}\ }\href {\doibase 10.1103/PhysRevC.92.034321} {\bibfield
  {journal} {\bibinfo  {journal} {Phys. Rev. C}\ }\textbf {\bibinfo {volume}
  {92}},\ \bibinfo {pages} {034321}}\BibitemShut {NoStop}%
\bibitem [{\citenamefont {Hinohara}(2018)}]{hinohara2018}%
  \BibitemOpen
  \bibfield  {author} {\bibinfo {author} {\bibnamefont {Hinohara},
  \bibfnamefont {N}}} (\bibinfo {year} {2018}),\ \bibfield  {title} {\enquote
  {\bibinfo {title} {Extending pairing energy density functional using pairing
  rotational moments of inertia},}\ }\href {\doibase 10.1088/1361-6471/aa9f8b}
  {\bibfield  {journal} {\bibinfo  {journal} {J. Phys. G: Nucl. Part. Phys.}\
  }\textbf {\bibinfo {volume} {45}}~(\bibinfo {number} {2}),\ \bibinfo {pages}
  {024004}}\BibitemShut {NoStop}%
\bibitem [{\citenamefont {Hinohara}\ and\ \citenamefont
  {Nazarewicz}(2016)}]{hinohara2016}%
  \BibitemOpen
  \bibfield  {author} {\bibinfo {author} {\bibnamefont {Hinohara},
  \bibfnamefont {N}}, \ and\ \bibinfo {author} {\bibnamefont {Nazarewicz},
  \bibfnamefont {W}}} (\bibinfo {year} {2016}),\ \bibfield  {title} {\enquote
  {\bibinfo {title} {Pairing nambu-goldstone modes within nuclear density
  functional theory},}\ }\href {\doibase 10.1103/PhysRevLett.116.152502}
  {\bibfield  {journal} {\bibinfo  {journal} {Phys. Rev. Lett.}\ }\textbf
  {\bibinfo {volume} {116}},\ \bibinfo {pages} {152502}}\BibitemShut {NoStop}%
\bibitem [{\citenamefont {Hofmann}(1999)}]{hofmann1999}%
  \BibitemOpen
  \bibfield  {author} {\bibinfo {author} {\bibnamefont {Hofmann}, \bibfnamefont
  {C~P}}} (\bibinfo {year} {1999}),\ \bibfield  {title} {\enquote {\bibinfo
  {title} {Spin-wave scattering in the effective lagrangian perspective},}\
  }\href {\doibase 10.1103/PhysRevB.60.388} {\bibfield  {journal} {\bibinfo
  {journal} {Phys. Rev. B}\ }\textbf {\bibinfo {volume} {60}},\ \bibinfo
  {pages} {388}}\BibitemShut {NoStop}%
\bibitem [{\citenamefont {Hu}\ \emph {et~al.}(2024{\natexlab{a}})\citenamefont
  {Hu}, \citenamefont {Sun}, \citenamefont {Hagen}, \citenamefont {Jansen},\
  and\ \citenamefont {Papenbrock}}]{hu2024b}%
  \BibitemOpen
  \bibfield  {author} {\bibinfo {author} {\bibnamefont {Hu}, \bibfnamefont
  {B}}, \bibinfo {author} {\bibnamefont {Sun}, \bibfnamefont {Z}}, \bibinfo
  {author} {\bibnamefont {Hagen}, \bibfnamefont {G}}, \bibinfo {author}
  {\bibnamefont {Jansen}, \bibfnamefont {G}}, \ and\ \bibinfo {author}
  {\bibnamefont {Papenbrock}, \bibfnamefont {T}}} (\bibinfo {year}
  {2024}{\natexlab{a}}),\ \bibfield  {title} {\enquote {\bibinfo {title} {Ab
  initio computations from $^{78}\mathrm{Ni}$ towards $^{70}\mathrm{Ca}$ along
  neutron number $n=50$},}\ }\href {\doibase 10.1016/j.physletb.2024.139010}
  {\bibfield  {journal} {\bibinfo  {journal} {Physics Letters B}\ }\textbf
  {\bibinfo {volume} {858}},\ \bibinfo {pages} {139010}}\BibitemShut {NoStop}%
\bibitem [{\citenamefont {Hu}\ \emph {et~al.}(2024{\natexlab{b}})\citenamefont
  {Hu}, \citenamefont {Sun}, \citenamefont {Hagen},\ and\ \citenamefont
  {Papenbrock}}]{hu2024a}%
  \BibitemOpen
  \bibfield  {author} {\bibinfo {author} {\bibnamefont {Hu}, \bibfnamefont
  {B~S}}, \bibinfo {author} {\bibnamefont {Sun}, \bibfnamefont {Z~H}}, \bibinfo
  {author} {\bibnamefont {Hagen}, \bibfnamefont {G}}, \ and\ \bibinfo {author}
  {\bibnamefont {Papenbrock}, \bibfnamefont {T}}} (\bibinfo {year}
  {2024}{\natexlab{b}}),\ \bibfield  {title} {\enquote {\bibinfo {title} {Ab
  initio computations of strongly deformed nuclei near $^{80}\mathrm{Zr}$},}\
  }\href {\doibase 10.1103/PhysRevC.110.L011302} {\bibfield  {journal}
  {\bibinfo  {journal} {Phys. Rev. C}\ }\textbf {\bibinfo {volume} {110}},\
  \bibinfo {pages} {L011302}}\BibitemShut {NoStop}%
\bibitem [{\citenamefont {Iachello}\ and\ \citenamefont
  {Arima}(1987)}]{iachello1987}%
  \BibitemOpen
  \bibfield  {author} {\bibinfo {author} {\bibnamefont {Iachello},
  \bibfnamefont {F}}, \ and\ \bibinfo {author} {\bibnamefont {Arima},
  \bibfnamefont {A}}} (\bibinfo {year} {1987}),\ \href@noop {} {\emph {\bibinfo
  {title} {The Interacting Boson Model}}}\ (\bibinfo  {publisher} {Cambridge
  University Press},\ \bibinfo {address} {Cambridge, UK})\BibitemShut {NoStop}%
\bibitem [{\citenamefont {Jain}\ \emph {et~al.}(1989)\citenamefont {Jain},
  \citenamefont {Kvasil}, \citenamefont {Sheline},\ and\ \citenamefont
  {Hoff}}]{jain1989}%
  \BibitemOpen
  \bibfield  {author} {\bibinfo {author} {\bibnamefont {Jain}, \bibfnamefont
  {A~K}}, \bibinfo {author} {\bibnamefont {Kvasil}, \bibfnamefont {J}},
  \bibinfo {author} {\bibnamefont {Sheline}, \bibfnamefont {R~K}}, \ and\
  \bibinfo {author} {\bibnamefont {Hoff}, \bibfnamefont {R~W}}} (\bibinfo
  {year} {1989}),\ \bibfield  {title} {\enquote {\bibinfo {title} {Coriolis
  coupling in the rotational bands of deformed odd-odd nuclei},}\ }\href
  {\doibase 10.1103/PhysRevC.40.432} {\bibfield  {journal} {\bibinfo  {journal}
  {Phys. Rev. C}\ }\textbf {\bibinfo {volume} {40}},\ \bibinfo {pages}
  {432}}\BibitemShut {NoStop}%
\bibitem [{\citenamefont {Jain}\ \emph {et~al.}(1998)\citenamefont {Jain},
  \citenamefont {Sheline}, \citenamefont {Headly}, \citenamefont {Sood},
  \citenamefont {Burke}, \citenamefont {Hr\ifmmode~\breve{\imath}\else
  \u{\i}\fi{}vn\'acov\'a}, \citenamefont {Kvasil}, \citenamefont {Nosek},\ and\
  \citenamefont {Hoff}}]{jain1998}%
  \BibitemOpen
  \bibfield  {author} {\bibinfo {author} {\bibnamefont {Jain}, \bibfnamefont
  {A~K}}, \bibinfo {author} {\bibnamefont {Sheline}, \bibfnamefont {R~K}},
  \bibinfo {author} {\bibnamefont {Headly}, \bibfnamefont {D~M}}, \bibinfo
  {author} {\bibnamefont {Sood}, \bibfnamefont {P~C}}, \bibinfo {author}
  {\bibnamefont {Burke}, \bibfnamefont {D~G}}, \bibinfo {author} {\bibnamefont
  {Hr\ifmmode~\breve{\imath}\else \u{\i}\fi{}vn\'acov\'a}, \bibfnamefont {I}},
  \bibinfo {author} {\bibnamefont {Kvasil}, \bibfnamefont {J}}, \bibinfo
  {author} {\bibnamefont {Nosek}, \bibfnamefont {D}}, \ and\ \bibinfo {author}
  {\bibnamefont {Hoff}, \bibfnamefont {R~W}}} (\bibinfo {year} {1998}),\
  \bibfield  {title} {\enquote {\bibinfo {title} {Nuclear structure in odd-odd
  nuclei, $144<a<194$},}\ }\href {\doibase 10.1103/RevModPhys.70.843}
  {\bibfield  {journal} {\bibinfo  {journal} {Rev. Mod. Phys.}\ }\textbf
  {\bibinfo {volume} {70}},\ \bibinfo {pages} {843}}\BibitemShut {NoStop}%
\bibitem [{\citenamefont {Jenkins}\ \emph {et~al.}(2013)\citenamefont
  {Jenkins}, \citenamefont {Manohar},\ and\ \citenamefont
  {Trott}}]{jenkins2013}%
  \BibitemOpen
  \bibfield  {author} {\bibinfo {author} {\bibnamefont {Jenkins}, \bibfnamefont
  {E~E}}, \bibinfo {author} {\bibnamefont {Manohar}, \bibfnamefont {A~V}}, \
  and\ \bibinfo {author} {\bibnamefont {Trott}, \bibfnamefont {M}}} (\bibinfo
  {year} {2013}),\ \bibfield  {title} {\enquote {\bibinfo {title} {{On Gauge
  Invariance and Minimal Coupling}},}\ }\href {\doibase
  10.1007/JHEP09(2013)063} {\bibfield  {journal} {\bibinfo  {journal} {JHEP}\
  }\textbf {\bibinfo {volume} {09}},\ \bibinfo {pages} {063}}\BibitemShut
  {NoStop}%
\bibitem [{\citenamefont {Jokiniemi}\ \emph {et~al.}(2023)\citenamefont
  {Jokiniemi}, \citenamefont {Romeo}, \citenamefont {Brase}, \citenamefont
  {Kotila}, \citenamefont {Soriano}, \citenamefont {Schwenk},\ and\
  \citenamefont {Men{\'e}ndez}}]{jokiniemi2023}%
  \BibitemOpen
  \bibfield  {author} {\bibinfo {author} {\bibnamefont {Jokiniemi},
  \bibfnamefont {L}}, \bibinfo {author} {\bibnamefont {Romeo}, \bibfnamefont
  {B}}, \bibinfo {author} {\bibnamefont {Brase}, \bibfnamefont {C}}, \bibinfo
  {author} {\bibnamefont {Kotila}, \bibfnamefont {J}}, \bibinfo {author}
  {\bibnamefont {Soriano}, \bibfnamefont {P}}, \bibinfo {author} {\bibnamefont
  {Schwenk}, \bibfnamefont {A}}, \ and\ \bibinfo {author} {\bibnamefont
  {Men{\'e}ndez}, \bibfnamefont {J}}} (\bibinfo {year} {2023}),\ \bibfield
  {title} {\enquote {\bibinfo {title} {{Two-neutrino $\ensuremath{\beta\beta}$
  decay of $\ensuremath{^{136}}$Xe to the first excited $\ensuremath{0^+}$
  state in $\ensuremath{^{136}}$Ba}},}\ }\href {\doibase
  10.1016/j.physletb.2023.137689} {\bibfield  {journal} {\bibinfo  {journal}
  {Physics Letters B}\ }\textbf {\bibinfo {volume} {838}},\ \bibinfo {pages}
  {137689}}\BibitemShut {NoStop}%
\bibitem [{\citenamefont {K{\"a}mpfer}\ \emph {et~al.}(2005)\citenamefont
  {K{\"a}mpfer}, \citenamefont {Moser},\ and\ \citenamefont
  {Wiese}}]{kampfer2005}%
  \BibitemOpen
  \bibfield  {author} {\bibinfo {author} {\bibnamefont {K{\"a}mpfer},
  \bibfnamefont {F}}, \bibinfo {author} {\bibnamefont {Moser}, \bibfnamefont
  {M}}, \ and\ \bibinfo {author} {\bibnamefont {Wiese}, \bibfnamefont {U~J}}}
  (\bibinfo {year} {2005}),\ \bibfield  {title} {\enquote {\bibinfo {title}
  {Systematic low-energy effective theory for magnons and charge carriers in an
  antiferromagnet},}\ }\href {\doibase 10.1016/j.nuclphysb.2005.09.004}
  {\bibfield  {journal} {\bibinfo  {journal} {Nuclear Physics B}\ }\textbf
  {\bibinfo {volume} {729}}~(\bibinfo {number} {3}),\ \bibinfo {pages} {317
  }}\BibitemShut {NoStop}%
\bibitem [{\citenamefont {Kerman}(1956)}]{kerman1956}%
  \BibitemOpen
  \bibfield  {author} {\bibinfo {author} {\bibnamefont {Kerman}, \bibfnamefont
  {A~K}}} (\bibinfo {year} {1956}),\ \bibfield  {title} {\enquote {\bibinfo
  {title} {Rotational perturbations in nuclei -- application to wolfram 183},}\
  }\href {http://gymarkiv.sdu.dk/MFM/kdvs/mfm%2030-39/mfm-30-15.pdf} {\bibfield
   {journal} {\bibinfo  {journal} {Dan. Mat. Fys. Medd.}\ }\textbf {\bibinfo
  {volume} {30}},\ \bibinfo {pages} {no. 15}}\BibitemShut {NoStop}%
\bibitem [{\citenamefont {Kerman}(1961)}]{kerman1961}%
  \BibitemOpen
  \bibfield  {author} {\bibinfo {author} {\bibnamefont {Kerman}, \bibfnamefont
  {A~K}}} (\bibinfo {year} {1961}),\ \bibfield  {title} {\enquote {\bibinfo
  {title} {Pairing forces and nuclear collective motion},}\ }\href {\doibase
  10.1016/0003-4916(61)90008-2} {\bibfield  {journal} {\bibinfo  {journal}
  {Annals of Physics}\ }\textbf {\bibinfo {volume} {12}}~(\bibinfo {number}
  {2}),\ \bibinfo {pages} {300}}\BibitemShut {NoStop}%
\bibitem [{\citenamefont {van Kolck}(1994)}]{vankolck1994}%
  \BibitemOpen
  \bibfield  {author} {\bibinfo {author} {\bibnamefont {van Kolck},
  \bibfnamefont {U}}} (\bibinfo {year} {1994}),\ \bibfield  {title} {\enquote
  {\bibinfo {title} {{Few-nucleon forces from chiral Lagrangians}},}\ }\href
  {\doibase 10.1103/PhysRevC.49.2932} {\bibfield  {journal} {\bibinfo
  {journal} {Phys. Rev. C}\ }\textbf {\bibinfo {volume} {49}},\ \bibinfo
  {pages} {2932}}\BibitemShut {NoStop}%
\bibitem [{\citenamefont {Koma}\ and\ \citenamefont {Tasaki}(1994)}]{koma1994}%
  \BibitemOpen
  \bibfield  {author} {\bibinfo {author} {\bibnamefont {Koma}, \bibfnamefont
  {T}}, \ and\ \bibinfo {author} {\bibnamefont {Tasaki}, \bibfnamefont {H}}}
  (\bibinfo {year} {1994}),\ \bibfield  {title} {\enquote {\bibinfo {title}
  {Symmetry breaking and finite-size effects in quantum many-body systems},}\
  }\href {\doibase 10.1007/BF02188685} {\bibfield  {journal} {\bibinfo
  {journal} {Journal of Statistical Physics}\ }\textbf {\bibinfo {volume}
  {76}}~(\bibinfo {number} {3-4}),\ \bibinfo {pages} {745}}\BibitemShut
  {NoStop}%
\bibitem [{\citenamefont {Kortelainen}\ \emph {et~al.}(2012)\citenamefont
  {Kortelainen}, \citenamefont {McDonnell}, \citenamefont {Nazarewicz},
  \citenamefont {Reinhard}, \citenamefont {Sarich}, \citenamefont {Schunck},
  \citenamefont {Stoitsov},\ and\ \citenamefont {Wild}}]{kortelainen2012}%
  \BibitemOpen
  \bibfield  {author} {\bibinfo {author} {\bibnamefont {Kortelainen},
  \bibfnamefont {M}}, \bibinfo {author} {\bibnamefont {McDonnell},
  \bibfnamefont {J}}, \bibinfo {author} {\bibnamefont {Nazarewicz},
  \bibfnamefont {W}}, \bibinfo {author} {\bibnamefont {Reinhard}, \bibfnamefont
  {P~G}}, \bibinfo {author} {\bibnamefont {Sarich}, \bibfnamefont {J}},
  \bibinfo {author} {\bibnamefont {Schunck}, \bibfnamefont {N}}, \bibinfo
  {author} {\bibnamefont {Stoitsov}, \bibfnamefont {M~V}}, \ and\ \bibinfo
  {author} {\bibnamefont {Wild}, \bibfnamefont {S~M}}} (\bibinfo {year}
  {2012}),\ \bibfield  {title} {\enquote {\bibinfo {title} {Nuclear energy
  density optimization: Large deformations},}\ }\href {\doibase
  10.1103/PhysRevC.85.024304} {\bibfield  {journal} {\bibinfo  {journal} {Phys.
  Rev. C}\ }\textbf {\bibinfo {volume} {85}},\ \bibinfo {pages}
  {024304}}\BibitemShut {NoStop}%
\bibitem [{\citenamefont {Kumar}(1972)}]{kumar1972}%
  \BibitemOpen
  \bibfield  {author} {\bibinfo {author} {\bibnamefont {Kumar}, \bibfnamefont
  {K}}} (\bibinfo {year} {1972}),\ \bibfield  {title} {\enquote {\bibinfo
  {title} {Intrinsic quadrupole moments and shapes of nuclear ground states and
  excited states},}\ }\href {\doibase 10.1103/PhysRevLett.28.249} {\bibfield
  {journal} {\bibinfo  {journal} {Phys. Rev. Lett.}\ }\textbf {\bibinfo
  {volume} {28}},\ \bibinfo {pages} {249}}\BibitemShut {NoStop}%
\bibitem [{\citenamefont {Kumar}\ and\ \citenamefont
  {Baranger}(1968)}]{kumar1968}%
  \BibitemOpen
  \bibfield  {author} {\bibinfo {author} {\bibnamefont {Kumar}, \bibfnamefont
  {K}}, \ and\ \bibinfo {author} {\bibnamefont {Baranger}, \bibfnamefont {M}}}
  (\bibinfo {year} {1968}),\ \bibfield  {title} {\enquote {\bibinfo {title}
  {Nuclear deformations in the pairing-plus-quadrupole model: (iii). static
  nuclear shapes in the rare-earth region},}\ }\href {\doibase
  10.1016/0375-9474(68)90371-0} {\bibfield  {journal} {\bibinfo  {journal}
  {Nuclear Physics A}\ }\textbf {\bibinfo {volume} {110}}~(\bibinfo {number}
  {3}),\ \bibinfo {pages} {529}}\BibitemShut {NoStop}%
\bibitem [{\citenamefont {Leutwyler}(1987)}]{leutwyler1987}%
  \BibitemOpen
  \bibfield  {author} {\bibinfo {author} {\bibnamefont {Leutwyler},
  \bibfnamefont {H}}} (\bibinfo {year} {1987}),\ \bibfield  {title} {\enquote
  {\bibinfo {title} {Energy levels of light quarks confined to a box},}\ }\href
  {\doibase 10.1016/0370-2693(87)91296-2} {\bibfield  {journal} {\bibinfo
  {journal} {Physics Letters B}\ }\textbf {\bibinfo {volume} {189}},\ \bibinfo
  {pages} {197 }}\BibitemShut {NoStop}%
\bibitem [{\citenamefont {Leutwyler}(1994)}]{leutwyler1994}%
  \BibitemOpen
  \bibfield  {author} {\bibinfo {author} {\bibnamefont {Leutwyler},
  \bibfnamefont {H}}} (\bibinfo {year} {1994}),\ \bibfield  {title} {\enquote
  {\bibinfo {title} {Nonrelativistic effective lagrangians},}\ }\href {\doibase
  10.1103/PhysRevD.49.3033} {\bibfield  {journal} {\bibinfo  {journal} {Phys.
  Rev. D}\ }\textbf {\bibinfo {volume} {49}},\ \bibinfo {pages}
  {3033}}\BibitemShut {NoStop}%
\bibitem [{\citenamefont {Lin}\ \emph {et~al.}(2024)\citenamefont {Lin},
  \citenamefont {Zhou}, \citenamefont {Yao},\ and\ \citenamefont
  {Hergert}}]{lin2024}%
  \BibitemOpen
  \bibfield  {author} {\bibinfo {author} {\bibnamefont {Lin}, \bibfnamefont
  {W}}, \bibinfo {author} {\bibnamefont {Zhou}, \bibfnamefont {E}}, \bibinfo
  {author} {\bibnamefont {Yao}, \bibfnamefont {J}}, \ and\ \bibinfo {author}
  {\bibnamefont {Hergert}, \bibfnamefont {H}}} (\bibinfo {year} {2024}),\
  \bibfield  {title} {\enquote {\bibinfo {title} {Quantum-number projected
  generator coordinate method for $^{21}$ne with a chiral
  two-nucleon-plus-three-nucleon interaction},}\ }\href {\doibase
  10.3390/sym16040409} {\bibfield  {journal} {\bibinfo  {journal} {Symmetry}\
  }\textbf {\bibinfo {volume} {16}}~(\bibinfo {number} {4}),\
  10.3390/sym16040409}\BibitemShut {NoStop}%
\bibitem [{\citenamefont {Lipkin}(1960)}]{lipkin1960}%
  \BibitemOpen
  \bibfield  {author} {\bibinfo {author} {\bibnamefont {Lipkin}, \bibfnamefont
  {H~J}}} (\bibinfo {year} {1960}),\ \bibfield  {title} {\enquote {\bibinfo
  {title} {Collective motion in many-particle systems: Part 1. the violation of
  conservation laws},}\ }\href {\doibase 10.1016/0003-4916(60)90032-4}
  {\bibfield  {journal} {\bibinfo  {journal} {Annals of Physics}\ }\textbf
  {\bibinfo {volume} {9}}~(\bibinfo {number} {2}),\ \bibinfo {pages}
  {272}}\BibitemShut {NoStop}%
\bibitem [{\citenamefont {Littlejohn}\ and\ \citenamefont
  {Reinsch}(1997)}]{littlejohn1997}%
  \BibitemOpen
  \bibfield  {author} {\bibinfo {author} {\bibnamefont {Littlejohn},
  \bibfnamefont {R~G}}, \ and\ \bibinfo {author} {\bibnamefont {Reinsch},
  \bibfnamefont {M}}} (\bibinfo {year} {1997}),\ \bibfield  {title} {\enquote
  {\bibinfo {title} {Gauge fields in the separation of rotations and internal
  motions in the n-body problem},}\ }\href {\doibase 10.1103/RevModPhys.69.213}
  {\bibfield  {journal} {\bibinfo  {journal} {Rev. Mod. Phys.}\ }\textbf
  {\bibinfo {volume} {69}},\ \bibinfo {pages} {213}}\BibitemShut {NoStop}%
\bibitem [{\citenamefont {Machleidt}\ and\ \citenamefont
  {Entem}(2011)}]{machleidt2011}%
  \BibitemOpen
  \bibfield  {author} {\bibinfo {author} {\bibnamefont {Machleidt},
  \bibfnamefont {R}}, \ and\ \bibinfo {author} {\bibnamefont {Entem},
  \bibfnamefont {D}}} (\bibinfo {year} {2011}),\ \bibfield  {title} {\enquote
  {\bibinfo {title} {Chiral effective field theory and nuclear forces},}\
  }\href {\doibase 10.1016/j.physrep.2011.02.001} {\bibfield  {journal}
  {\bibinfo  {journal} {Physics Reports}\ }\textbf {\bibinfo {volume}
  {503}}~(\bibinfo {number} {1}),\ \bibinfo {pages} {1 }}\BibitemShut {NoStop}%
\bibitem [{\citenamefont {Maris}\ \emph {et~al.}(2015)\citenamefont {Maris},
  \citenamefont {Caprio},\ and\ \citenamefont {Vary}}]{maris2015}%
  \BibitemOpen
  \bibfield  {author} {\bibinfo {author} {\bibnamefont {Maris}, \bibfnamefont
  {P}}, \bibinfo {author} {\bibnamefont {Caprio}, \bibfnamefont {M~A}}, \ and\
  \bibinfo {author} {\bibnamefont {Vary}, \bibfnamefont {J~P}}} (\bibinfo
  {year} {2015}),\ \bibfield  {title} {\enquote {\bibinfo {title} {Emergence of
  rotational bands in \textit{ab initio} no-core configuration interaction
  calculations of the be isotopes},}\ }\href {\doibase
  10.1103/PhysRevC.91.014310} {\bibfield  {journal} {\bibinfo  {journal} {Phys.
  Rev. C}\ }\textbf {\bibinfo {volume} {91}},\ \bibinfo {pages}
  {014310}}\BibitemShut {NoStop}%
\bibitem [{\citenamefont {Mayer}\ and\ \citenamefont
  {Jensen}(1955)}]{mayer1955}%
  \BibitemOpen
  \bibfield  {author} {\bibinfo {author} {\bibnamefont {Mayer}, \bibfnamefont
  {M~G}}, \ and\ \bibinfo {author} {\bibnamefont {Jensen}, \bibfnamefont
  {J~H~D}}} (\bibinfo {year} {1955}),\ \href@noop {} {\emph {\bibinfo {title}
  {Elementary Theory of Nuclear Shell Structure}}}\ (\bibinfo  {publisher}
  {John Wiley \& Sons},\ \bibinfo {address} {New York})\BibitemShut {NoStop}%
\bibitem [{\citenamefont {Men{\'e}ndez}(2017)}]{menendez_j_2018}%
  \BibitemOpen
  \bibfield  {author} {\bibinfo {author} {\bibnamefont {Men{\'e}ndez},
  \bibfnamefont {J}}} (\bibinfo {year} {2017}),\ \bibfield  {title} {\enquote
  {\bibinfo {title} {Neutrinoless $\beta\beta$ decay mediated by the exchange
  of light and heavy neutrinos: the role of nuclear structure correlations},}\
  }\href {\doibase 10.1088/1361-6471/aa9bd4} {\bibfield  {journal} {\bibinfo
  {journal} {J. Phys. G Nucl. Part. Phys.}\ }\textbf {\bibinfo {volume}
  {45}}~(\bibinfo {number} {1}),\ \bibinfo {pages} {014003}}\BibitemShut
  {NoStop}%
\bibitem [{\citenamefont {Menéndez}\ \emph {et~al.}(2009)\citenamefont
  {Menéndez}, \citenamefont {Poves}, \citenamefont {Caurier},\ and\
  \citenamefont {Nowacki}}]{menendez_j_2009}%
  \BibitemOpen
  \bibfield  {author} {\bibinfo {author} {\bibnamefont {Menéndez},
  \bibfnamefont {J}}, \bibinfo {author} {\bibnamefont {Poves}, \bibfnamefont
  {A}}, \bibinfo {author} {\bibnamefont {Caurier}, \bibfnamefont {E}}, \ and\
  \bibinfo {author} {\bibnamefont {Nowacki}, \bibfnamefont {F}}} (\bibinfo
  {year} {2009}),\ \bibfield  {title} {\enquote {\bibinfo {title}
  {{Disassembling the nuclear matrix elements of the neutrinoless
  $\ensuremath{\beta\beta}$ decay}},}\ }\href {\doibase
  10.1016/j.nuclphysa.2008.12.005} {\bibfield  {journal} {\bibinfo  {journal}
  {Nucl. Phys. A}\ }\textbf {\bibinfo {volume} {818}}~(\bibinfo {number} {3}),\
  \bibinfo {pages} {139}}\BibitemShut {NoStop}%
\bibitem [{\citenamefont {Mermin}(1979)}]{mermin1979}%
  \BibitemOpen
  \bibfield  {author} {\bibinfo {author} {\bibnamefont {Mermin}, \bibfnamefont
  {N~D}}} (\bibinfo {year} {1979}),\ \bibfield  {title} {\enquote {\bibinfo
  {title} {The topological theory of defects in ordered media},}\ }\href
  {\doibase 10.1103/RevModPhys.51.591} {\bibfield  {journal} {\bibinfo
  {journal} {Rev. Mod. Phys.}\ }\textbf {\bibinfo {volume} {51}},\ \bibinfo
  {pages} {591}}\BibitemShut {NoStop}%
\bibitem [{\citenamefont {Migdal}(1959)}]{migdal1959}%
  \BibitemOpen
  \bibfield  {author} {\bibinfo {author} {\bibnamefont {Migdal}, \bibfnamefont
  {A~B}}} (\bibinfo {year} {1959}),\ \bibfield  {title} {\enquote {\bibinfo
  {title} {Superfluidity and the moments of inertia of nuclei},}\ }\href
  {\doibase 10.1016/0029-5582(59)90264-0} {\bibfield  {journal} {\bibinfo
  {journal} {Nuclear Physics}\ }\textbf {\bibinfo {volume} {13}}~(\bibinfo
  {number} {5}),\ \bibinfo {pages} {655}}\BibitemShut {NoStop}%
\bibitem [{\citenamefont {M\"oller}\ \emph {et~al.}(2006)\citenamefont
  {M\"oller}, \citenamefont {Bengtsson}, \citenamefont {Carlsson},
  \citenamefont {Olivius},\ and\ \citenamefont {Ichikawa}}]{moller2006}%
  \BibitemOpen
  \bibfield  {author} {\bibinfo {author} {\bibnamefont {M\"oller},
  \bibfnamefont {P}}, \bibinfo {author} {\bibnamefont {Bengtsson},
  \bibfnamefont {R}}, \bibinfo {author} {\bibnamefont {Carlsson}, \bibfnamefont
  {B~G}}, \bibinfo {author} {\bibnamefont {Olivius}, \bibfnamefont {P}}, \ and\
  \bibinfo {author} {\bibnamefont {Ichikawa}, \bibfnamefont {T}}} (\bibinfo
  {year} {2006}),\ \bibfield  {title} {\enquote {\bibinfo {title} {Global
  calculations of ground-state axial shape asymmetry of nuclei},}\ }\href
  {\doibase 10.1103/PhysRevLett.97.162502} {\bibfield  {journal} {\bibinfo
  {journal} {Phys. Rev. Lett.}\ }\textbf {\bibinfo {volume} {97}},\ \bibinfo
  {pages} {162502}}\BibitemShut {NoStop}%
\bibitem [{\citenamefont {Murakami}(2018)}]{murakami2018}%
  \BibitemOpen
  \bibfield  {author} {\bibinfo {author} {\bibnamefont {Murakami},
  \bibfnamefont {H}}} (\bibinfo {year} {2018}),\ \href@noop {} {\emph {\bibinfo
  {title} {Killing Commendatore}}}\ (\bibinfo  {publisher} {Alfred A. Knopf},\
  \bibinfo {address} {New York})\BibitemShut {NoStop}%
\bibitem [{\citenamefont {Nambu}(1960)}]{nambu1960}%
  \BibitemOpen
  \bibfield  {author} {\bibinfo {author} {\bibnamefont {Nambu}, \bibfnamefont
  {Y}}} (\bibinfo {year} {1960}),\ \bibfield  {title} {\enquote {\bibinfo
  {title} {Quasi-particles and gauge invariance in the theory of
  superconductivity},}\ }\href {\doibase 10.1103/PhysRev.117.648} {\bibfield
  {journal} {\bibinfo  {journal} {Phys. Rev.}\ }\textbf {\bibinfo {volume}
  {117}},\ \bibinfo {pages} {648}}\BibitemShut {NoStop}%
\bibitem [{\citenamefont {Nambu}\ and\ \citenamefont
  {Jona-Lasinio}(1961{\natexlab{a}})}]{nambu1961}%
  \BibitemOpen
  \bibfield  {author} {\bibinfo {author} {\bibnamefont {Nambu}, \bibfnamefont
  {Y}}, \ and\ \bibinfo {author} {\bibnamefont {Jona-Lasinio}, \bibfnamefont
  {G}}} (\bibinfo {year} {1961}{\natexlab{a}}),\ \bibfield  {title} {\enquote
  {\bibinfo {title} {Dynamical model of elementary particles based on an
  analogy with superconductivity. i},}\ }\href {\doibase
  10.1103/PhysRev.122.345} {\bibfield  {journal} {\bibinfo  {journal} {Phys.
  Rev.}\ }\textbf {\bibinfo {volume} {122}},\ \bibinfo {pages}
  {345}}\BibitemShut {NoStop}%
\bibitem [{\citenamefont {Nambu}\ and\ \citenamefont
  {Jona-Lasinio}(1961{\natexlab{b}})}]{nambu1961b}%
  \BibitemOpen
  \bibfield  {author} {\bibinfo {author} {\bibnamefont {Nambu}, \bibfnamefont
  {Y}}, \ and\ \bibinfo {author} {\bibnamefont {Jona-Lasinio}, \bibfnamefont
  {G}}} (\bibinfo {year} {1961}{\natexlab{b}}),\ \bibfield  {title} {\enquote
  {\bibinfo {title} {Dynamical model of elementary particles based on an
  analogy with superconductivity. ii},}\ }\href {\doibase
  10.1103/PhysRev.124.246} {\bibfield  {journal} {\bibinfo  {journal} {Phys.
  Rev.}\ }\textbf {\bibinfo {volume} {124}},\ \bibinfo {pages}
  {246}}\BibitemShut {NoStop}%
\bibitem [{\citenamefont {Nazarewicz}(1993)}]{nazarewicz1993}%
  \BibitemOpen
  \bibfield  {author} {\bibinfo {author} {\bibnamefont {Nazarewicz},
  \bibfnamefont {W}}} (\bibinfo {year} {1993}),\ \bibfield  {title} {\enquote
  {\bibinfo {title} {Nuclear deformation as a spontaneous symmetry breaking},}\
  }\href {\doibase 10.1142/S0218301393000479} {\bibfield  {journal} {\bibinfo
  {journal} {International Journal of Modern Physics E}\ }\textbf {\bibinfo
  {volume} {02}},\ \bibinfo {pages} {51}}\BibitemShut {NoStop}%
\bibitem [{\citenamefont {Nazarewicz}(1994)}]{nazarewicz1994}%
  \BibitemOpen
  \bibfield  {author} {\bibinfo {author} {\bibnamefont {Nazarewicz},
  \bibfnamefont {W}}} (\bibinfo {year} {1994}),\ \bibfield  {title} {\enquote
  {\bibinfo {title} {Microscopic origin of nuclear deformations},}\ }\href
  {\doibase 10.1016/0375-9474(94)90037-X} {\bibfield  {journal} {\bibinfo
  {journal} {Nuclear Physics A}\ }\textbf {\bibinfo {volume} {574}},\ \bibinfo
  {pages} {27 }}\BibitemShut {NoStop}%
\bibitem [{\citenamefont {Nazarewicz}\ \emph {et~al.}(1984)\citenamefont
  {Nazarewicz}, \citenamefont {Olanders}, \citenamefont {Ragnarsson},
  \citenamefont {Dudek}, \citenamefont {Leander}, \citenamefont {M{\"o}ller},\
  and\ \citenamefont {Ruchowsa}}]{nazarewicz1984}%
  \BibitemOpen
  \bibfield  {author} {\bibinfo {author} {\bibnamefont {Nazarewicz},
  \bibfnamefont {W}}, \bibinfo {author} {\bibnamefont {Olanders}, \bibfnamefont
  {P}}, \bibinfo {author} {\bibnamefont {Ragnarsson}, \bibfnamefont {I}},
  \bibinfo {author} {\bibnamefont {Dudek}, \bibfnamefont {J}}, \bibinfo
  {author} {\bibnamefont {Leander}, \bibfnamefont {G}}, \bibinfo {author}
  {\bibnamefont {M{\"o}ller}, \bibfnamefont {P}}, \ and\ \bibinfo {author}
  {\bibnamefont {Ruchowsa}, \bibfnamefont {E}}} (\bibinfo {year} {1984}),\
  \bibfield  {title} {\enquote {\bibinfo {title} {Analysis of octupole
  instability in medium-mass and heavy nuclei},}\ }\href {\doibase
  10.1016/0375-9474(84)90208-2} {\bibfield  {journal} {\bibinfo  {journal}
  {Nuclear Physics A}\ }\textbf {\bibinfo {volume} {429}}~(\bibinfo {number}
  {2}),\ \bibinfo {pages} {269}}\BibitemShut {NoStop}%
\bibitem [{\citenamefont {{Nazarewicz}}\ \emph {et~al.}(2014)\citenamefont
  {{Nazarewicz}}, \citenamefont {{Reinhard}}, \citenamefont {{Satu{\l}a}},\
  and\ \citenamefont {{Vretenar}}}]{nazarewicz2014}%
  \BibitemOpen
  \bibfield  {author} {\bibinfo {author} {\bibnamefont {{Nazarewicz}},
  \bibfnamefont {W}}, \bibinfo {author} {\bibnamefont {{Reinhard}},
  \bibfnamefont {P~G}}, \bibinfo {author} {\bibnamefont {{Satu{\l}a}},
  \bibfnamefont {W}}, \ and\ \bibinfo {author} {\bibnamefont {{Vretenar}},
  \bibfnamefont {D}}} (\bibinfo {year} {2014}),\ \bibfield  {title} {\enquote
  {\bibinfo {title} {{Symmetry energy in nuclear density functional theory}},}\
  }\href {\doibase 10.1140/epja/i2014-14020-3} {\bibfield  {journal} {\bibinfo
  {journal} {European Physical Journal A}\ }\textbf {\bibinfo {volume} {50}},\
  \bibinfo {eid} {20}}\BibitemShut {NoStop}%
\bibitem [{\citenamefont {Nik{\v s}i{\'c}}\ \emph {et~al.}(2011)\citenamefont
  {Nik{\v s}i{\'c}}, \citenamefont {Vretenar},\ and\ \citenamefont
  {Ring}}]{niksic2011}%
  \BibitemOpen
  \bibfield  {author} {\bibinfo {author} {\bibnamefont {Nik{\v s}i{\'c}},
  \bibfnamefont {T}}, \bibinfo {author} {\bibnamefont {Vretenar}, \bibfnamefont
  {D}}, \ and\ \bibinfo {author} {\bibnamefont {Ring}, \bibfnamefont {P}}}
  (\bibinfo {year} {2011}),\ \bibfield  {title} {\enquote {\bibinfo {title}
  {Relativistic nuclear energy density functionals: Mean-field and beyond},}\
  }\href {\doibase 10.1016/j.ppnp.2011.01.055} {\bibfield  {journal} {\bibinfo
  {journal} {Prog. Part. Nucl. Phys.}\ }\textbf {\bibinfo {volume}
  {66}}~(\bibinfo {number} {3}),\ \bibinfo {pages} {519 }}\BibitemShut
  {NoStop}%
\bibitem [{\citenamefont {Nilsson}(1955)}]{nilsson1955}%
  \BibitemOpen
  \bibfield  {author} {\bibinfo {author} {\bibnamefont {Nilsson}, \bibfnamefont
  {S~G}}} (\bibinfo {year} {1955}),\ \bibfield  {title} {\enquote {\bibinfo
  {title} {Binding states of individual nucleons in strongly deformed
  nuclei},}\ }\href
  {http://gymarkiv.sdu.dk/MFM/kdvs/mfm%2020-29/MFM%2029-16.pdf} {\bibfield
  {journal} {\bibinfo  {journal} {K. Dan. Vidensk. Selsk. Mat. Fys. Medd.}\
  }\textbf {\bibinfo {volume} {29}}~(\bibinfo {number} {no. 16}),\ \bibinfo
  {pages} {no. 16}}\BibitemShut {NoStop}%
\bibitem [{\citenamefont {Novario}\ \emph {et~al.}(2021)\citenamefont
  {Novario}, \citenamefont {Gysbers}, \citenamefont {Engel}, \citenamefont
  {Hagen}, \citenamefont {Jansen}, \citenamefont {Morris}, \citenamefont
  {Navr\'atil}, \citenamefont {Papenbrock},\ and\ \citenamefont
  {Quaglioni}}]{novario2021}%
  \BibitemOpen
  \bibfield  {author} {\bibinfo {author} {\bibnamefont {Novario}, \bibfnamefont
  {S}}, \bibinfo {author} {\bibnamefont {Gysbers}, \bibfnamefont {P}}, \bibinfo
  {author} {\bibnamefont {Engel}, \bibfnamefont {J}}, \bibinfo {author}
  {\bibnamefont {Hagen}, \bibfnamefont {G}}, \bibinfo {author} {\bibnamefont
  {Jansen}, \bibfnamefont {G~R}}, \bibinfo {author} {\bibnamefont {Morris},
  \bibfnamefont {T~D}}, \bibinfo {author} {\bibnamefont {Navr\'atil},
  \bibfnamefont {P}}, \bibinfo {author} {\bibnamefont {Papenbrock},
  \bibfnamefont {T}}, \ and\ \bibinfo {author} {\bibnamefont {Quaglioni},
  \bibfnamefont {S}}} (\bibinfo {year} {2021}),\ \bibfield  {title} {\enquote
  {\bibinfo {title} {Coupled-cluster calculations of neutrinoless
  double-$\ensuremath{\beta}$ decay in $^{48}\mathrm{Ca}$},}\ }\href {\doibase
  10.1103/PhysRevLett.126.182502} {\bibfield  {journal} {\bibinfo  {journal}
  {Phys. Rev. Lett.}\ }\textbf {\bibinfo {volume} {126}},\ \bibinfo {pages}
  {182502}}\BibitemShut {NoStop}%
\bibitem [{\citenamefont {Papenbrock}(2011)}]{papenbrock2011}%
  \BibitemOpen
  \bibfield  {author} {\bibinfo {author} {\bibnamefont {Papenbrock},
  \bibfnamefont {T}}} (\bibinfo {year} {2011}),\ \bibfield  {title} {\enquote
  {\bibinfo {title} {Effective theory for deformed nuclei},}\ }\href {\doibase
  10.1016/j.nuclphysa.2010.12.013} {\bibfield  {journal} {\bibinfo  {journal}
  {Nucl. Phys. A}\ }\textbf {\bibinfo {volume} {852}}~(\bibinfo {number} {1}),\
  \bibinfo {pages} {36 }}\BibitemShut {NoStop}%
\bibitem [{\citenamefont {Papenbrock}(2022)}]{papenbrock2022}%
  \BibitemOpen
  \bibfield  {author} {\bibinfo {author} {\bibnamefont {Papenbrock},
  \bibfnamefont {T}}} (\bibinfo {year} {2022}),\ \bibfield  {title} {\enquote
  {\bibinfo {title} {Effective field theory of pairing rotations},}\ }\href
  {\doibase 10.1103/PhysRevC.105.044322} {\bibfield  {journal} {\bibinfo
  {journal} {Phys. Rev. C}\ }\textbf {\bibinfo {volume} {105}},\ \bibinfo
  {pages} {044322}}\BibitemShut {NoStop}%
\bibitem [{\citenamefont {Papenbrock}\ and\ \citenamefont
  {Weidenm\"uller}(2014)}]{papenbrock2014}%
  \BibitemOpen
  \bibfield  {author} {\bibinfo {author} {\bibnamefont {Papenbrock},
  \bibfnamefont {T}}, \ and\ \bibinfo {author} {\bibnamefont {Weidenm\"uller},
  \bibfnamefont {H~A}}} (\bibinfo {year} {2014}),\ \bibfield  {title} {\enquote
  {\bibinfo {title} {Effective field theory for finite systems with
  spontaneously broken symmetry},}\ }\href {\doibase
  10.1103/PhysRevC.89.014334} {\bibfield  {journal} {\bibinfo  {journal} {Phys.
  Rev. C}\ }\textbf {\bibinfo {volume} {89}},\ \bibinfo {pages}
  {014334}}\BibitemShut {NoStop}%
\bibitem [{\citenamefont {Papenbrock}\ and\ \citenamefont
  {Weidenm\"uller}(2015)}]{papenbrock2015}%
  \BibitemOpen
  \bibfield  {author} {\bibinfo {author} {\bibnamefont {Papenbrock},
  \bibfnamefont {T}}, \ and\ \bibinfo {author} {\bibnamefont {Weidenm\"uller},
  \bibfnamefont {H~A}}} (\bibinfo {year} {2015}),\ \bibfield  {title} {\enquote
  {\bibinfo {title} {Effective field theory of emergent symmetry breaking in
  deformed atomic nuclei},}\ }\href {\doibase 10.1088/0954-3899/42/10/105103}
  {\bibfield  {journal} {\bibinfo  {journal} {Journal of Physics G: Nuclear and
  Particle Physics}\ }\textbf {\bibinfo {volume} {42}}~(\bibinfo {number}
  {10}),\ \bibinfo {pages} {105103}}\BibitemShut {NoStop}%
\bibitem [{\citenamefont {Papenbrock}\ and\ \citenamefont
  {Weidenm\"uller}(2020)}]{papenbrock2020}%
  \BibitemOpen
  \bibfield  {author} {\bibinfo {author} {\bibnamefont {Papenbrock},
  \bibfnamefont {T}}, \ and\ \bibinfo {author} {\bibnamefont {Weidenm\"uller},
  \bibfnamefont {H~A}}} (\bibinfo {year} {2020}),\ \bibfield  {title} {\enquote
  {\bibinfo {title} {Effective field theory for deformed odd-mass nuclei},}\
  }\href {\doibase 10.1103/PhysRevC.102.044324} {\bibfield  {journal} {\bibinfo
   {journal} {Phys. Rev. C}\ }\textbf {\bibinfo {volume} {102}},\ \bibinfo
  {pages} {044324}}\BibitemShut {NoStop}%
\bibitem [{\citenamefont {Peierls}\ and\ \citenamefont
  {Thouless}(1962)}]{peierls1962}%
  \BibitemOpen
  \bibfield  {author} {\bibinfo {author} {\bibnamefont {Peierls}, \bibfnamefont
  {R~E}}, \ and\ \bibinfo {author} {\bibnamefont {Thouless}, \bibfnamefont
  {D~J}}} (\bibinfo {year} {1962}),\ \bibfield  {title} {\enquote {\bibinfo
  {title} {Variational approach to collective motion},}\ }\href {\doibase
  10.1016/0029-5582(62)91025-8} {\bibfield  {journal} {\bibinfo  {journal}
  {Nuclear Physics}\ }\textbf {\bibinfo {volume} {38}},\ \bibinfo {pages}
  {154}}\BibitemShut {NoStop}%
\bibitem [{\citenamefont {Peierls}\ and\ \citenamefont
  {Yoccoz}(1957)}]{peierls1957}%
  \BibitemOpen
  \bibfield  {author} {\bibinfo {author} {\bibnamefont {Peierls}, \bibfnamefont
  {R~E}}, \ and\ \bibinfo {author} {\bibnamefont {Yoccoz}, \bibfnamefont {J}}}
  (\bibinfo {year} {1957}),\ \bibfield  {title} {\enquote {\bibinfo {title}
  {The collective model of nuclear motion},}\ }\href {\doibase
  10.1088/0370-1298/70/5/309} {\bibfield  {journal} {\bibinfo  {journal} {Proc.
  Phys. Soc. A}\ }\textbf {\bibinfo {volume} {70}}~(\bibinfo {number} {5}),\
  \bibinfo {pages} {381}}\BibitemShut {NoStop}%
\bibitem [{\citenamefont {Potel}\ \emph {et~al.}(2011)\citenamefont {Potel},
  \citenamefont {Barranco}, \citenamefont {Marini}, \citenamefont {Idini},
  \citenamefont {Vigezzi},\ and\ \citenamefont {Broglia}}]{potel2011}%
  \BibitemOpen
  \bibfield  {author} {\bibinfo {author} {\bibnamefont {Potel}, \bibfnamefont
  {G}}, \bibinfo {author} {\bibnamefont {Barranco}, \bibfnamefont {F}},
  \bibinfo {author} {\bibnamefont {Marini}, \bibfnamefont {F}}, \bibinfo
  {author} {\bibnamefont {Idini}, \bibfnamefont {A}}, \bibinfo {author}
  {\bibnamefont {Vigezzi}, \bibfnamefont {E}}, \ and\ \bibinfo {author}
  {\bibnamefont {Broglia}, \bibfnamefont {R~A}}} (\bibinfo {year} {2011}),\
  \bibfield  {title} {\enquote {\bibinfo {title} {Calculation of the transition
  from pairing vibrational to pairing rotational regimes between magic nuclei
  $^{100}\mathrm{Sn}$ and $^{132}\mathrm{Sn}$ via two-nucleon transfer
  reactions},}\ }\href {\doibase 10.1103/PhysRevLett.107.092501} {\bibfield
  {journal} {\bibinfo  {journal} {Phys. Rev. Lett.}\ }\textbf {\bibinfo
  {volume} {107}},\ \bibinfo {pages} {092501}}\BibitemShut {NoStop}%
\bibitem [{\citenamefont {Potel}\ \emph {et~al.}(2013)\citenamefont {Potel},
  \citenamefont {Idini}, \citenamefont {Barranco}, \citenamefont {Vigezzi},\
  and\ \citenamefont {Broglia}}]{potel2013}%
  \BibitemOpen
  \bibfield  {author} {\bibinfo {author} {\bibnamefont {Potel}, \bibfnamefont
  {G}}, \bibinfo {author} {\bibnamefont {Idini}, \bibfnamefont {A}}, \bibinfo
  {author} {\bibnamefont {Barranco}, \bibfnamefont {F}}, \bibinfo {author}
  {\bibnamefont {Vigezzi}, \bibfnamefont {E}}, \ and\ \bibinfo {author}
  {\bibnamefont {Broglia}, \bibfnamefont {R~A}}} (\bibinfo {year} {2013}),\
  \bibfield  {title} {\enquote {\bibinfo {title} {Cooper pair transfer in
  nuclei},}\ }\href {\doibase 10.1088/0034-4885/76/10/106301} {\bibfield
  {journal} {\bibinfo  {journal} {Rep. Prog. Phys.}\ }\textbf {\bibinfo
  {volume} {76}}~(\bibinfo {number} {10}),\ \bibinfo {pages}
  {106301}}\BibitemShut {NoStop}%
\bibitem [{\citenamefont {Potel}\ \emph {et~al.}(2017)\citenamefont {Potel},
  \citenamefont {Idini}, \citenamefont {Barranco}, \citenamefont {Vigezzi},\
  and\ \citenamefont {Broglia}}]{potel2017}%
  \BibitemOpen
  \bibfield  {author} {\bibinfo {author} {\bibnamefont {Potel}, \bibfnamefont
  {G}}, \bibinfo {author} {\bibnamefont {Idini}, \bibfnamefont {A}}, \bibinfo
  {author} {\bibnamefont {Barranco}, \bibfnamefont {F}}, \bibinfo {author}
  {\bibnamefont {Vigezzi}, \bibfnamefont {E}}, \ and\ \bibinfo {author}
  {\bibnamefont {Broglia}, \bibfnamefont {R~A}}} (\bibinfo {year} {2017}),\
  \bibfield  {title} {\enquote {\bibinfo {title} {From bare to renormalized
  order parameter in gauge space: Structure and reactions},}\ }\href {\doibase
  10.1103/PhysRevC.96.034606} {\bibfield  {journal} {\bibinfo  {journal} {Phys.
  Rev. C}\ }\textbf {\bibinfo {volume} {96}},\ \bibinfo {pages}
  {034606}}\BibitemShut {NoStop}%
\bibitem [{\citenamefont {Qiu}\ \emph {et~al.}(2017)\citenamefont {Qiu},
  \citenamefont {Henderson}, \citenamefont {Zhao},\ and\ \citenamefont
  {Scuseria}}]{qiu2017}%
  \BibitemOpen
  \bibfield  {author} {\bibinfo {author} {\bibnamefont {Qiu}, \bibfnamefont
  {Y}}, \bibinfo {author} {\bibnamefont {Henderson}, \bibfnamefont {T~M}},
  \bibinfo {author} {\bibnamefont {Zhao}, \bibfnamefont {J}}, \ and\ \bibinfo
  {author} {\bibnamefont {Scuseria}, \bibfnamefont {G~E}}} (\bibinfo {year}
  {2017}),\ \bibfield  {title} {\enquote {\bibinfo {title} {Projected coupled
  cluster theory},}\ }\href {\doibase 10.1063/1.4991020} {\bibfield  {journal}
  {\bibinfo  {journal} {J. Chem. Phys.}\ }\textbf {\bibinfo {volume}
  {147}}~(\bibinfo {number} {6}),\ \bibinfo {pages} {064111}}\BibitemShut
  {NoStop}%
\bibitem [{\citenamefont {Richardson}(1963)}]{richardson1963}%
  \BibitemOpen
  \bibfield  {author} {\bibinfo {author} {\bibnamefont {Richardson},
  \bibfnamefont {R~W}}} (\bibinfo {year} {1963}),\ \bibfield  {title} {\enquote
  {\bibinfo {title} {A restricted class of exact eigenstates of the
  pairing-force hamiltonian},}\ }\href {\doibase 10.1016/0031-9163(63)90259-2}
  {\bibfield  {journal} {\bibinfo  {journal} {Physics Letters}\ }\textbf
  {\bibinfo {volume} {3}}~(\bibinfo {number} {6}),\ \bibinfo {pages}
  {277}}\BibitemShut {NoStop}%
\bibitem [{\citenamefont {Ring}\ and\ \citenamefont
  {Schuck}(1980)}]{ringschuck}%
  \BibitemOpen
  \bibfield  {author} {\bibinfo {author} {\bibnamefont {Ring}, \bibfnamefont
  {P}}, \ and\ \bibinfo {author} {\bibnamefont {Schuck}, \bibfnamefont {P}}}
  (\bibinfo {year} {1980}),\ \href@noop {} {\emph {\bibinfo {title} {The
  Nuclear Many-Body Problem}}}\ (\bibinfo  {publisher} {Springer},\ \bibinfo
  {address} {Heidelberg})\BibitemShut {NoStop}%
\bibitem [{\citenamefont {Robledo}\ \emph {et~al.}(2018)\citenamefont
  {Robledo}, \citenamefont {Rodríguez},\ and\ \citenamefont
  {Rodríguez-Guzmán}}]{robledo2019}%
  \BibitemOpen
  \bibfield  {author} {\bibinfo {author} {\bibnamefont {Robledo}, \bibfnamefont
  {L~M}}, \bibinfo {author} {\bibnamefont {Rodríguez}, \bibfnamefont {T~R}}, \
  and\ \bibinfo {author} {\bibnamefont {Rodríguez-Guzmán}, \bibfnamefont
  {R~R}}} (\bibinfo {year} {2018}),\ \bibfield  {title} {\enquote {\bibinfo
  {title} {Mean field and beyond description of nuclear structure with the
  gogny force: a review},}\ }\href {\doibase 10.1088/1361-6471/aadebd}
  {\bibfield  {journal} {\bibinfo  {journal} {Journal of Physics G: Nuclear and
  Particle Physics}\ }\textbf {\bibinfo {volume} {46}}~(\bibinfo {number}
  {1}),\ \bibinfo {pages} {013001}}\BibitemShut {NoStop}%
\bibitem [{\citenamefont {Rodríguez}\ and\ \citenamefont
  {Martínez-Pinedo}(2013)}]{rodriguez_tr_2013}%
  \BibitemOpen
  \bibfield  {author} {\bibinfo {author} {\bibnamefont {Rodríguez},
  \bibfnamefont {T~R}}, \ and\ \bibinfo {author} {\bibnamefont
  {Martínez-Pinedo}, \bibfnamefont {G}}} (\bibinfo {year} {2013}),\ \bibfield
  {title} {\enquote {\bibinfo {title} {Neutrinoless $\ensuremath{\beta\beta}$
  decay nuclear matrix elements in an isotopic chain},}\ }\href {\doibase
  10.1016/j.physletb.2012.12.063} {\bibfield  {journal} {\bibinfo  {journal}
  {Phys. Lett. B}\ }\textbf {\bibinfo {volume} {719}}~(\bibinfo {number} {1}),\
  \bibinfo {pages} {174}}\BibitemShut {NoStop}%
\bibitem [{\citenamefont {Rom{\'a}n}\ and\ \citenamefont
  {Soto}(1999)}]{roman1999}%
  \BibitemOpen
  \bibfield  {author} {\bibinfo {author} {\bibnamefont {Rom{\'a}n},
  \bibfnamefont {J~M}}, \ and\ \bibinfo {author} {\bibnamefont {Soto},
  \bibfnamefont {J}}} (\bibinfo {year} {1999}),\ \bibfield  {title} {\enquote
  {\bibinfo {title} {Effective field theory approach to ferromagnets and
  antiferromagnets in crystalline solids},}\ }\href {\doibase
  10.1142/S0217979299000655} {\bibfield  {journal} {\bibinfo  {journal}
  {International Journal of Modern Physics B}\ }\textbf {\bibinfo {volume}
  {13}}~(\bibinfo {number} {07}),\ \bibinfo {pages} {755}}\BibitemShut
  {NoStop}%
\bibitem [{\citenamefont {Rosensteel}\ and\ \citenamefont
  {Rowe}(1977)}]{rosensteel1977}%
  \BibitemOpen
  \bibfield  {author} {\bibinfo {author} {\bibnamefont {Rosensteel},
  \bibfnamefont {G}}, \ and\ \bibinfo {author} {\bibnamefont {Rowe},
  \bibfnamefont {D~J}}} (\bibinfo {year} {1977}),\ \bibfield  {title} {\enquote
  {\bibinfo {title} {Nuclear $\mathrm{Sp}(3, r)$ model},}\ }\href {\doibase
  10.1103/PhysRevLett.38.10} {\bibfield  {journal} {\bibinfo  {journal} {Phys.
  Rev. Lett.}\ }\textbf {\bibinfo {volume} {38}},\ \bibinfo {pages}
  {10}}\BibitemShut {NoStop}%
\bibitem [{\citenamefont {Rosensteel}\ and\ \citenamefont
  {Rowe}(1980)}]{rosensteel1980}%
  \BibitemOpen
  \bibfield  {author} {\bibinfo {author} {\bibnamefont {Rosensteel},
  \bibfnamefont {G}}, \ and\ \bibinfo {author} {\bibnamefont {Rowe},
  \bibfnamefont {D~J}}} (\bibinfo {year} {1980}),\ \bibfield  {title} {\enquote
  {\bibinfo {title} {On the algebraic formulation of collective models iii. the
  symplectic shell model of collective motion},}\ }\href {\doibase
  10.1016/0003-4916(80)90180-3} {\bibfield  {journal} {\bibinfo  {journal}
  {Annals of Physics}\ }\textbf {\bibinfo {volume} {126}}~(\bibinfo {number}
  {2}),\ \bibinfo {pages} {343}}\BibitemShut {NoStop}%
\bibitem [{\citenamefont {Rowe}(2004)}]{rowe2004}%
  \BibitemOpen
  \bibfield  {author} {\bibinfo {author} {\bibnamefont {Rowe}, \bibfnamefont
  {D~J}}} (\bibinfo {year} {2004}),\ \bibfield  {title} {\enquote {\bibinfo
  {title} {A computationally tractable version of the collective model},}\
  }\href {\doibase 10.1016/j.nuclphysa.2004.02.018} {\bibfield  {journal}
  {\bibinfo  {journal} {Nuclear Physics A}\ }\textbf {\bibinfo {volume}
  {735}},\ \bibinfo {pages} {372 }}\BibitemShut {NoStop}%
\bibitem [{\citenamefont {Rowe}\ \emph {et~al.}(2009)\citenamefont {Rowe},
  \citenamefont {Welsh},\ and\ \citenamefont {Caprio}}]{rowe2009}%
  \BibitemOpen
  \bibfield  {author} {\bibinfo {author} {\bibnamefont {Rowe}, \bibfnamefont
  {D~J}}, \bibinfo {author} {\bibnamefont {Welsh}, \bibfnamefont {T~A}}, \ and\
  \bibinfo {author} {\bibnamefont {Caprio}, \bibfnamefont {M~A}}} (\bibinfo
  {year} {2009}),\ \bibfield  {title} {\enquote {\bibinfo {title} {Bohr model
  as an algebraic collective model},}\ }\href {\doibase
  10.1103/PhysRevC.79.054304} {\bibfield  {journal} {\bibinfo  {journal} {Phys.
  Rev. C}\ }\textbf {\bibinfo {volume} {79}},\ \bibinfo {pages}
  {054304}}\BibitemShut {NoStop}%
\bibitem [{\citenamefont {Rowe}\ and\ \citenamefont {Wood}(2010)}]{rowe2010}%
  \BibitemOpen
  \bibfield  {author} {\bibinfo {author} {\bibnamefont {Rowe}, \bibfnamefont
  {D~J}}, \ and\ \bibinfo {author} {\bibnamefont {Wood}, \bibfnamefont {J~L}}}
  (\bibinfo {year} {2010}),\ \href@noop {} {\emph {\bibinfo {title}
  {Fundamentals of Nuclear Models}}},\ Vol.\ \bibinfo {volume} {I: Foundational
  Models}\ (\bibinfo  {publisher} {World Scientific},\ \bibinfo {address}
  {Singapore})\BibitemShut {NoStop}%
\bibitem [{\citenamefont {Scharff-Goldhaber}\ \emph {et~al.}(1976)\citenamefont
  {Scharff-Goldhaber}, \citenamefont {Dover},\ and\ \citenamefont
  {Goodman}}]{scharff1976}%
  \BibitemOpen
  \bibfield  {author} {\bibinfo {author} {\bibnamefont {Scharff-Goldhaber},
  \bibfnamefont {G}}, \bibinfo {author} {\bibnamefont {Dover}, \bibfnamefont
  {C~B}}, \ and\ \bibinfo {author} {\bibnamefont {Goodman}, \bibfnamefont
  {A~L}}} (\bibinfo {year} {1976}),\ \bibfield  {title} {\enquote {\bibinfo
  {title} {The variable moment of inertia (vmi) model and theories of nuclear
  collective motion},}\ }\href {\doibase 10.1146/annurev.ns.26.120176.001323}
  {\bibfield  {journal} {\bibinfo  {journal} {Annual Review of Nuclear
  Science}\ }\textbf {\bibinfo {volume} {26}}~(\bibinfo {number} {1}),\
  \bibinfo {pages} {239}}\BibitemShut {NoStop}%
\bibitem [{\citenamefont {Schindler}\ and\ \citenamefont
  {Phillips}(2009)}]{schindler2009}%
  \BibitemOpen
  \bibfield  {author} {\bibinfo {author} {\bibnamefont {Schindler},
  \bibfnamefont {M~R}}, \ and\ \bibinfo {author} {\bibnamefont {Phillips},
  \bibfnamefont {D~R}}} (\bibinfo {year} {2009}),\ \bibfield  {title} {\enquote
  {\bibinfo {title} {Bayesian methods for parameter estimation in effective
  field theories},}\ }\href {\doibase 10.1016/j.aop.2008.09.003} {\bibfield
  {journal} {\bibinfo  {journal} {Ann. Phys.}\ }\textbf {\bibinfo {volume}
  {324}}~(\bibinfo {number} {3}),\ \bibinfo {pages} {682 }}\BibitemShut
  {NoStop}%
\bibitem [{\citenamefont {Sheikh}\ \emph {et~al.}(2021)\citenamefont {Sheikh},
  \citenamefont {Dobaczewski}, \citenamefont {Ring}, \citenamefont {Robledo},\
  and\ \citenamefont {Yannouleas}}]{sheikh2021}%
  \BibitemOpen
  \bibfield  {author} {\bibinfo {author} {\bibnamefont {Sheikh}, \bibfnamefont
  {J~A}}, \bibinfo {author} {\bibnamefont {Dobaczewski}, \bibfnamefont {J~J}},
  \bibinfo {author} {\bibnamefont {Ring}, \bibfnamefont {P}}, \bibinfo {author}
  {\bibnamefont {Robledo}, \bibfnamefont {L~M}}, \ and\ \bibinfo {author}
  {\bibnamefont {Yannouleas}, \bibfnamefont {C}}} (\bibinfo {year} {2021}),\
  \bibfield  {title} {\enquote {\bibinfo {title} {Symmetry restoration in
  mean-field approaches},}\ }\href {\doibase 10.1088/1361-6471/ac288a}
  {\bibfield  {journal} {\bibinfo  {journal} {J. Phys. G.}\ }\textbf {\bibinfo
  {volume} {48}},\ \bibinfo {pages} {123001}}\BibitemShut {NoStop}%
\bibitem [{\citenamefont {Shimizu}\ \emph {et~al.}(2012)\citenamefont
  {Shimizu}, \citenamefont {Abe}, \citenamefont {Tsunoda}, \citenamefont
  {Utsuno}, \citenamefont {Yoshida}, \citenamefont {Mizusaki}, \citenamefont
  {Honma},\ and\ \citenamefont {Otsuka}}]{shimizu2012}%
  \BibitemOpen
  \bibfield  {author} {\bibinfo {author} {\bibnamefont {Shimizu}, \bibfnamefont
  {N}}, \bibinfo {author} {\bibnamefont {Abe}, \bibfnamefont {T}}, \bibinfo
  {author} {\bibnamefont {Tsunoda}, \bibfnamefont {Y}}, \bibinfo {author}
  {\bibnamefont {Utsuno}, \bibfnamefont {Y}}, \bibinfo {author} {\bibnamefont
  {Yoshida}, \bibfnamefont {T}}, \bibinfo {author} {\bibnamefont {Mizusaki},
  \bibfnamefont {T}}, \bibinfo {author} {\bibnamefont {Honma}, \bibfnamefont
  {M}}, \ and\ \bibinfo {author} {\bibnamefont {Otsuka}, \bibfnamefont {T}}}
  (\bibinfo {year} {2012}),\ \bibfield  {title} {\enquote {\bibinfo {title}
  {New-generation monte carlo shell model for the k computer era},}\ }\href
  {\doibase 10.1093/ptep/pts012} {\ \textbf {\bibinfo {volume}
  {2012}}~(\bibinfo {number} {1}),\ \bibinfo {pages} {01A205}}\BibitemShut
  {NoStop}%
\bibitem [{\citenamefont {Shimizu}\ \emph {et~al.}(2018)\citenamefont
  {Shimizu}, \citenamefont {Men\'endez},\ and\ \citenamefont
  {Yako}}]{shimizu2018}%
  \BibitemOpen
  \bibfield  {author} {\bibinfo {author} {\bibnamefont {Shimizu}, \bibfnamefont
  {N}}, \bibinfo {author} {\bibnamefont {Men\'endez}, \bibfnamefont {J}}, \
  and\ \bibinfo {author} {\bibnamefont {Yako}, \bibfnamefont {K}}} (\bibinfo
  {year} {2018}),\ \bibfield  {title} {\enquote {\bibinfo {title} {Double
  gamow-teller transitions and its relation to neutrinoless
  $\ensuremath{\beta}\ensuremath{\beta}$ decay},}\ }\href {\doibase
  10.1103/PhysRevLett.120.142502} {\bibfield  {journal} {\bibinfo  {journal}
  {Phys. Rev. Lett.}\ }\textbf {\bibinfo {volume} {120}},\ \bibinfo {pages}
  {142502}}\BibitemShut {NoStop}%
\bibitem [{\citenamefont {Signoracci}\ \emph {et~al.}(2015)\citenamefont
  {Signoracci}, \citenamefont {Duguet}, \citenamefont {Hagen},\ and\
  \citenamefont {Jansen}}]{signoracci2015}%
  \BibitemOpen
  \bibfield  {author} {\bibinfo {author} {\bibnamefont {Signoracci},
  \bibfnamefont {A}}, \bibinfo {author} {\bibnamefont {Duguet}, \bibfnamefont
  {T}}, \bibinfo {author} {\bibnamefont {Hagen}, \bibfnamefont {G}}, \ and\
  \bibinfo {author} {\bibnamefont {Jansen}, \bibfnamefont {G~R}}} (\bibinfo
  {year} {2015}),\ \bibfield  {title} {\enquote {\bibinfo {title} {\textit{Ab
  initio} bogoliubov coupled cluster theory for open-shell nuclei},}\ }\href
  {\doibase 10.1103/PhysRevC.91.064320} {\bibfield  {journal} {\bibinfo
  {journal} {Phys. Rev. C}\ }\textbf {\bibinfo {volume} {91}},\ \bibinfo
  {pages} {064320}}\BibitemShut {NoStop}%
\bibitem [{\citenamefont {Stephens}(1975)}]{stephens1975}%
  \BibitemOpen
  \bibfield  {author} {\bibinfo {author} {\bibnamefont {Stephens},
  \bibfnamefont {F~S}}} (\bibinfo {year} {1975}),\ \bibfield  {title} {\enquote
  {\bibinfo {title} {Coriolis effects and rotation alignment in nuclei},}\
  }\href {\doibase 10.1103/RevModPhys.47.43} {\bibfield  {journal} {\bibinfo
  {journal} {Rev. Mod. Phys.}\ }\textbf {\bibinfo {volume} {47}},\ \bibinfo
  {pages} {43}}\BibitemShut {NoStop}%
\bibitem [{\citenamefont {Stoitsov}\ \emph {et~al.}(2003)\citenamefont
  {Stoitsov}, \citenamefont {Dobaczewski}, \citenamefont {Nazarewicz},
  \citenamefont {Pittel},\ and\ \citenamefont {Dean}}]{stoitsov2003}%
  \BibitemOpen
  \bibfield  {author} {\bibinfo {author} {\bibnamefont {Stoitsov},
  \bibfnamefont {M~V}}, \bibinfo {author} {\bibnamefont {Dobaczewski},
  \bibfnamefont {J}}, \bibinfo {author} {\bibnamefont {Nazarewicz},
  \bibfnamefont {W}}, \bibinfo {author} {\bibnamefont {Pittel}, \bibfnamefont
  {S}}, \ and\ \bibinfo {author} {\bibnamefont {Dean}, \bibfnamefont {D~J}}}
  (\bibinfo {year} {2003}),\ \bibfield  {title} {\enquote {\bibinfo {title}
  {Systematic study of deformed nuclei at the drip lines and beyond},}\ }\href
  {\doibase 10.1103/PhysRevC.68.054312} {\bibfield  {journal} {\bibinfo
  {journal} {Phys. Rev. C}\ }\textbf {\bibinfo {volume} {68}},\ \bibinfo
  {pages} {054312}}\BibitemShut {NoStop}%
\bibitem [{\citenamefont {Stuchbery}\ \emph {et~al.}(2016)\citenamefont
  {Stuchbery}, \citenamefont {Chamoli},\ and\ \citenamefont
  {Kib\'edi}}]{stuchbery2016}%
  \BibitemOpen
  \bibfield  {author} {\bibinfo {author} {\bibnamefont {Stuchbery},
  \bibfnamefont {A~E}}, \bibinfo {author} {\bibnamefont {Chamoli},
  \bibfnamefont {S~K}}, \ and\ \bibinfo {author} {\bibnamefont {Kib\'edi},
  \bibfnamefont {T}}} (\bibinfo {year} {2016}),\ \bibfield  {title} {\enquote
  {\bibinfo {title} {Particle-rotor versus particle-vibration features in $g$
  factors of $^{111}\mathrm{Cd}$ and $^{113}\mathrm{Cd}$},}\ }\href {\doibase
  10.1103/PhysRevC.93.031302} {\bibfield  {journal} {\bibinfo  {journal} {Phys.
  Rev. C}\ }\textbf {\bibinfo {volume} {93}},\ \bibinfo {pages}
  {031302}}\BibitemShut {NoStop}%
\bibitem [{\citenamefont {Stuchbery}\ and\ \citenamefont
  {Wood}(2022)}]{stuchbery2022}%
  \BibitemOpen
  \bibfield  {author} {\bibinfo {author} {\bibnamefont {Stuchbery},
  \bibfnamefont {A~E}}, \ and\ \bibinfo {author} {\bibnamefont {Wood},
  \bibfnamefont {J~L}}} (\bibinfo {year} {2022}),\ \bibfield  {title} {\enquote
  {\bibinfo {title} {To shell model, or not to shell model, that is the
  question},}\ }\href {\doibase 10.3390/physics4030048} {\bibfield  {journal}
  {\bibinfo  {journal} {Physics}\ }\textbf {\bibinfo {volume} {4}}~(\bibinfo
  {number} {3}),\ \bibinfo {pages} {697}}\BibitemShut {NoStop}%
\bibitem [{\citenamefont {{Sun}}\ \emph
  {et~al.}(2024{\natexlab{a}})\citenamefont {{Sun}}, \citenamefont
  {{Dj{\"a}rv}}, \citenamefont {{Hagen}}, \citenamefont {{Jansen}},\ and\
  \citenamefont {{Papenbrock}}}]{sun2024b}%
  \BibitemOpen
  \bibfield  {author} {\bibinfo {author} {\bibnamefont {{Sun}}, \bibfnamefont
  {Z~H}}, \bibinfo {author} {\bibnamefont {{Dj{\"a}rv}}, \bibfnamefont {T~R}},
  \bibinfo {author} {\bibnamefont {{Hagen}}, \bibfnamefont {G}}, \bibinfo
  {author} {\bibnamefont {{Jansen}}, \bibfnamefont {G~R}}, \ and\ \bibinfo
  {author} {\bibnamefont {{Papenbrock}}, \bibfnamefont {T}}} (\bibinfo {year}
  {2024}{\natexlab{a}}),\ \bibfield  {title} {\enquote {\bibinfo {title}
  {{Structure of odd-mass Ne, Na, and Mg nuclei}},}\ }\href {\doibase
  10.48550/arXiv.2409.02279} {\bibinfo  {journal} {arXiv e-prints}\ ,\ \bibinfo
  {pages} {arXiv:2409.02279}}\BibitemShut {NoStop}%
\bibitem [{\citenamefont {{Sun}}\ \emph
  {et~al.}(2024{\natexlab{b}})\citenamefont {{Sun}}, \citenamefont
  {{Ekstr{\"o}m}}, \citenamefont {{Forss{\'e}n}}, \citenamefont {{Hagen}},
  \citenamefont {{Jansen}},\ and\ \citenamefont {{Papenbrock}}}]{sun2024}%
  \BibitemOpen
\bibfield  {journal} {  }\bibfield  {author} {\bibinfo {author} {\bibnamefont
  {{Sun}}, \bibfnamefont {Z~H}}, \bibinfo {author} {\bibnamefont
  {{Ekstr{\"o}m}}, \bibfnamefont {A}}, \bibinfo {author} {\bibnamefont
  {{Forss{\'e}n}}, \bibfnamefont {C}}, \bibinfo {author} {\bibnamefont
  {{Hagen}}, \bibfnamefont {G}}, \bibinfo {author} {\bibnamefont {{Jansen}},
  \bibfnamefont {G~R}}, \ and\ \bibinfo {author} {\bibnamefont {{Papenbrock}},
  \bibfnamefont {T}}} (\bibinfo {year} {2024}{\natexlab{b}}),\ \bibfield
  {title} {\enquote {\bibinfo {title} {{Multiscale physics of atomic nuclei
  from first principles}},}\ }\href {\doibase 10.48550/arXiv.2404.00058}
  {\bibinfo  {journal} {arXiv e-prints}\ ,\ \bibinfo {pages}
  {arXiv:2404.00058}}\BibitemShut {NoStop}%
\bibitem [{\citenamefont {{Tichai}}\ \emph {et~al.}(2023)\citenamefont
  {{Tichai}}, \citenamefont {{Demol}},\ and\ \citenamefont
  {{Duguet}}}]{tichai2023}%
  \BibitemOpen
\bibfield  {journal} {  }\bibfield  {author} {\bibinfo {author} {\bibnamefont
  {{Tichai}}, \bibfnamefont {A}}, \bibinfo {author} {\bibnamefont {{Demol}},
  \bibfnamefont {P}}, \ and\ \bibinfo {author} {\bibnamefont {{Duguet}},
  \bibfnamefont {T}}} (\bibinfo {year} {2023}),\ \bibfield  {title} {\enquote
  {\bibinfo {title} {{Towards heavy-mass ab initio nuclear structure:
  Open-shell Ca, Ni and Sn isotopes from Bogoliubov coupled-cluster theory}},}\
  }\href {\doibase 10.48550/arXiv.2307.15619} {\bibinfo  {journal} {arXiv
  e-prints}\ ,\ \bibinfo {eid} {arXiv:2307.15619}}\BibitemShut {NoStop}%
\bibitem [{\citenamefont {Ueno}\ \emph {et~al.}(1996)\citenamefont {Ueno},
  \citenamefont {Asahi}, \citenamefont {Izumi}, \citenamefont {Nagata},
  \citenamefont {Ogawa}, \citenamefont {Yoshimi}, \citenamefont {Sato},
  \citenamefont {Adachi}, \citenamefont {Hori}, \citenamefont {Mochinaga},
  \citenamefont {Okuno}, \citenamefont {Aoi}, \citenamefont {Ishihara},
  \citenamefont {Yoshida}, \citenamefont {Liu}, \citenamefont {Kubo},
  \citenamefont {Fukunishi}, \citenamefont {Shimoda}, \citenamefont {Miyatake},
  \citenamefont {Sasaki}, \citenamefont {Shirakura}, \citenamefont {Takahashi},
  \citenamefont {Mitsuoka},\ and\ \citenamefont {Schmidt-Ott}}]{ueno_h_1996}%
  \BibitemOpen
\bibfield  {journal} {  }\bibfield  {author} {\bibinfo {author} {\bibnamefont
  {Ueno}, \bibfnamefont {H}}, \bibinfo {author} {\bibnamefont {Asahi},
  \bibfnamefont {K}}, \bibinfo {author} {\bibnamefont {Izumi}, \bibfnamefont
  {H}}, \bibinfo {author} {\bibnamefont {Nagata}, \bibfnamefont {K}}, \bibinfo
  {author} {\bibnamefont {Ogawa}, \bibfnamefont {H}}, \bibinfo {author}
  {\bibnamefont {Yoshimi}, \bibfnamefont {A}}, \bibinfo {author} {\bibnamefont
  {Sato}, \bibfnamefont {H}}, \bibinfo {author} {\bibnamefont {Adachi},
  \bibfnamefont {M}}, \bibinfo {author} {\bibnamefont {Hori}, \bibfnamefont
  {Y}}, \bibinfo {author} {\bibnamefont {Mochinaga}, \bibfnamefont {K}},
  \bibinfo {author} {\bibnamefont {Okuno}, \bibfnamefont {H}}, \bibinfo
  {author} {\bibnamefont {Aoi}, \bibfnamefont {N}}, \bibinfo {author}
  {\bibnamefont {Ishihara}, \bibfnamefont {M}}, \bibinfo {author} {\bibnamefont
  {Yoshida}, \bibfnamefont {A}}, \bibinfo {author} {\bibnamefont {Liu},
  \bibfnamefont {G}}, \bibinfo {author} {\bibnamefont {Kubo}, \bibfnamefont
  {T}}, \bibinfo {author} {\bibnamefont {Fukunishi}, \bibfnamefont {N}},
  \bibinfo {author} {\bibnamefont {Shimoda}, \bibfnamefont {T}}, \bibinfo
  {author} {\bibnamefont {Miyatake}, \bibfnamefont {H}}, \bibinfo {author}
  {\bibnamefont {Sasaki}, \bibfnamefont {M}}, \bibinfo {author} {\bibnamefont
  {Shirakura}, \bibfnamefont {T}}, \bibinfo {author} {\bibnamefont {Takahashi},
  \bibfnamefont {N}}, \bibinfo {author} {\bibnamefont {Mitsuoka}, \bibfnamefont
  {S}}, \ and\ \bibinfo {author} {\bibnamefont {Schmidt-Ott}, \bibfnamefont
  {W~D}}} (\bibinfo {year} {1996}),\ \bibfield  {title} {\enquote {\bibinfo
  {title} {{Magnetic moments of $^{17}\mathrm{N}$ and $^{17}\mathrm{B}$}},}\
  }\href {\doibase 10.1103/PhysRevC.53.2142} {\bibfield  {journal} {\bibinfo
  {journal} {Phys. Rev. C}\ }\textbf {\bibinfo {volume} {53}},\ \bibinfo
  {pages} {2142}}\BibitemShut {NoStop}%
\bibitem [{\citenamefont {Ui}\ and\ \citenamefont {Takeda}(1983)}]{ui1983}%
  \BibitemOpen
  \bibfield  {author} {\bibinfo {author} {\bibnamefont {Ui}, \bibfnamefont
  {H}}, \ and\ \bibinfo {author} {\bibnamefont {Takeda}, \bibfnamefont {G}}}
  (\bibinfo {year} {1983}),\ \bibfield  {title} {\enquote {\bibinfo {title} {{A
  Class of Simple Hamiltonians with Degenerate Ground State. II A Model of
  Nuclear Rotation: Spontaneous Breakdown of Rotation Symmetry and Goldstone
  Theorem for Finite Dimensional System}},}\ }\href {\doibase
  10.1143/PTP.70.176} {\bibfield  {journal} {\bibinfo  {journal} {Progress of
  Theoretical Physics}\ }\textbf {\bibinfo {volume} {70}}~(\bibinfo {number}
  {1}),\ \bibinfo {pages} {176}}\BibitemShut {NoStop}%
\bibitem [{\citenamefont {Valatin}(1958)}]{valatin1958}%
  \BibitemOpen
  \bibfield  {author} {\bibinfo {author} {\bibnamefont {Valatin}, \bibfnamefont
  {J~G}}} (\bibinfo {year} {1958}),\ \bibfield  {title} {\enquote {\bibinfo
  {title} {Comments on the theory of superconductivity},}\ }\href {\doibase
  10.1007/BF02745589} {\bibfield  {journal} {\bibinfo  {journal} {Il Nuovo
  Cimento (1955-1965)}\ }\textbf {\bibinfo {volume} {7}}~(\bibinfo {number}
  {6}),\ \bibinfo {pages} {843}}\BibitemShut {NoStop}%
\bibitem [{\citenamefont {Varshalovich}\ \emph {et~al.}(1988)\citenamefont
  {Varshalovich}, \citenamefont {Moskalev},\ and\ \citenamefont
  {Khersonskii}}]{varshalovich1988}%
  \BibitemOpen
  \bibfield  {author} {\bibinfo {author} {\bibnamefont {Varshalovich},
  \bibfnamefont {D~A}}, \bibinfo {author} {\bibnamefont {Moskalev},
  \bibfnamefont {A~N}}, \ and\ \bibinfo {author} {\bibnamefont {Khersonskii},
  \bibfnamefont {V~K}}} (\bibinfo {year} {1988}),\ \href@noop {} {\emph
  {\bibinfo {title} {Quantum theory of angular momentum}}}\ (\bibinfo
  {publisher} {World Scientific},\ \bibinfo {address} {Singapore})\BibitemShut
  {NoStop}%
\bibitem [{\citenamefont {Villars}(1965)}]{villars1965}%
  \BibitemOpen
  \bibfield  {author} {\bibinfo {author} {\bibnamefont {Villars}, \bibfnamefont
  {F~M}}} (\bibinfo {year} {1965}),\ \bibfield  {title} {\enquote {\bibinfo
  {title} {Elementary quantum theory of nuclear collective rotation},}\ }\href
  {\doibase 10.1016/0029-5582(65)90087-8} {\bibfield  {journal} {\bibinfo
  {journal} {Nuclear Physics}\ }\textbf {\bibinfo {volume} {74}}~(\bibinfo
  {number} {2}),\ \bibinfo {pages} {353}}\BibitemShut {NoStop}%
\bibitem [{\citenamefont {Vretenar}\ \emph {et~al.}(2005)\citenamefont
  {Vretenar}, \citenamefont {Afanasjev}, \citenamefont {Lalazissis},\ and\
  \citenamefont {Ring}}]{vretanar2005}%
  \BibitemOpen
  \bibfield  {author} {\bibinfo {author} {\bibnamefont {Vretenar},
  \bibfnamefont {D}}, \bibinfo {author} {\bibnamefont {Afanasjev},
  \bibfnamefont {A~V}}, \bibinfo {author} {\bibnamefont {Lalazissis},
  \bibfnamefont {G~A}}, \ and\ \bibinfo {author} {\bibnamefont {Ring},
  \bibfnamefont {P}}} (\bibinfo {year} {2005}),\ \bibfield  {title} {\enquote
  {\bibinfo {title} {Relativistic hartree–bogoliubov theory: static and
  dynamic aspects of exotic nuclear structure},}\ }\href {\doibase
  10.1016/j.physrep.2004.10.001} {\bibfield  {journal} {\bibinfo  {journal}
  {Physics Reports}\ }\textbf {\bibinfo {volume} {409}}~(\bibinfo {number}
  {3}),\ \bibinfo {pages} {101}}\BibitemShut {NoStop}%
\bibitem [{\citenamefont {Weinberg}(1968)}]{weinberg1968}%
  \BibitemOpen
  \bibfield  {author} {\bibinfo {author} {\bibnamefont {Weinberg},
  \bibfnamefont {S}}} (\bibinfo {year} {1968}),\ \bibfield  {title} {\enquote
  {\bibinfo {title} {Nonlinear realizations of chiral symmetry},}\ }\href
  {\doibase 10.1103/PhysRev.166.1568} {\bibfield  {journal} {\bibinfo
  {journal} {Phys. Rev.}\ }\textbf {\bibinfo {volume} {166}},\ \bibinfo {pages}
  {1568}}\BibitemShut {NoStop}%
\bibitem [{\citenamefont {Weinberg}(1990)}]{weinberg1990}%
  \BibitemOpen
  \bibfield  {author} {\bibinfo {author} {\bibnamefont {Weinberg},
  \bibfnamefont {S}}} (\bibinfo {year} {1990}),\ \bibfield  {title} {\enquote
  {\bibinfo {title} {Nuclear forces from chiral lagrangians},}\ }\href
  {\doibase 10.1016/0370-2693(90)90938-3} {\bibfield  {journal} {\bibinfo
  {journal} {Phys. Lett. B}\ }\textbf {\bibinfo {volume} {251}}~(\bibinfo
  {number} {2}),\ \bibinfo {pages} {288 }}\BibitemShut {NoStop}%
\bibitem [{\citenamefont {Weinberg}(1996)}]{weinberg_v2_1996}%
  \BibitemOpen
  \bibfield  {author} {\bibinfo {author} {\bibnamefont {Weinberg},
  \bibfnamefont {S}}} (\bibinfo {year} {1996}),\ \href@noop {} {\emph {\bibinfo
  {title} {The Quantum Theory of Fields}}},\ Vol.~\bibinfo {volume} {II}\
  (\bibinfo  {publisher} {Cambridge University Press},\ \bibinfo {address}
  {Cambridge, UK})\BibitemShut {NoStop}%
\bibitem [{\citenamefont {Wesolowski}\ \emph {et~al.}(2019)\citenamefont
  {Wesolowski}, \citenamefont {Furnstahl}, \citenamefont {Melendez},\ and\
  \citenamefont {Phillips}}]{wesolowski2019}%
  \BibitemOpen
  \bibfield  {author} {\bibinfo {author} {\bibnamefont {Wesolowski},
  \bibfnamefont {S}}, \bibinfo {author} {\bibnamefont {Furnstahl},
  \bibfnamefont {R~J}}, \bibinfo {author} {\bibnamefont {Melendez},
  \bibfnamefont {J~A}}, \ and\ \bibinfo {author} {\bibnamefont {Phillips},
  \bibfnamefont {D~R}}} (\bibinfo {year} {2019}),\ \bibfield  {title} {\enquote
  {\bibinfo {title} {Exploring bayesian parameter estimation for chiral
  effective field theory using nucleon{\textendash}nucleon phase shifts},}\
  }\href {\doibase 10.1088/1361-6471/aaf5fc} {\bibfield  {journal} {\bibinfo
  {journal} {J. Phys. G.}\ }\textbf {\bibinfo {volume} {46}}~(\bibinfo {number}
  {4}),\ \bibinfo {pages} {045102}}\BibitemShut {NoStop}%
\bibitem [{\citenamefont {Wesolowski}\ \emph {et~al.}(2016)\citenamefont
  {Wesolowski}, \citenamefont {Klco}, \citenamefont {Furnstahl}, \citenamefont
  {Phillips},\ and\ \citenamefont {Thapaliya}}]{wesolowski2016}%
  \BibitemOpen
  \bibfield  {author} {\bibinfo {author} {\bibnamefont {Wesolowski},
  \bibfnamefont {S}}, \bibinfo {author} {\bibnamefont {Klco}, \bibfnamefont
  {N}}, \bibinfo {author} {\bibnamefont {Furnstahl}, \bibfnamefont {R~J}},
  \bibinfo {author} {\bibnamefont {Phillips}, \bibfnamefont {D~R}}, \ and\
  \bibinfo {author} {\bibnamefont {Thapaliya}, \bibfnamefont {A}}} (\bibinfo
  {year} {2016}),\ \bibfield  {title} {\enquote {\bibinfo {title} {Bayesian
  parameter estimation for effective field theories},}\ }\href {\doibase
  10.1088/0954-3899/43/7/074001} {\bibfield  {journal} {\bibinfo  {journal}
  {Journal of Physics G: Nuclear and Particle Physics}\ }\textbf {\bibinfo
  {volume} {43}}~(\bibinfo {number} {7}),\ \bibinfo {pages}
  {074001}}\BibitemShut {NoStop}%
\bibitem [{\citenamefont {Wesolowski}\ \emph {et~al.}(2021)\citenamefont
  {Wesolowski} \emph {et~al.}}]{wesolowski2021}%
  \BibitemOpen
  \bibfield  {author} {\bibinfo {author} {\bibnamefont {Wesolowski},
  \bibfnamefont {S}},  \emph {et~al.}} (\bibinfo {year} {2021}),\ \bibfield
  {title} {\enquote {\bibinfo {title} {Rigorous constraints on three-nucleon
  forces in chiral effective field theory from fast and accurate calculations
  of few-body observables},}\ }\href {\doibase 10.1103/PhysRevC.104.064001}
  {\bibfield  {journal} {\bibinfo  {journal} {Phys. Rev. C}\ }\textbf {\bibinfo
  {volume} {104}},\ \bibinfo {pages} {064001}}\BibitemShut {NoStop}%
\bibitem [{\citenamefont {Wilczek}\ and\ \citenamefont
  {Shapere}(1989)}]{wilczek1989}%
  \BibitemOpen
  \bibfield  {author} {\bibinfo {author} {\bibnamefont {Wilczek}, \bibfnamefont
  {F}}, \ and\ \bibinfo {author} {\bibnamefont {Shapere}, \bibfnamefont {A}}}
  (\bibinfo {year} {1989}),\ \href {\doibase 10.1142/0613} {\emph {\bibinfo
  {title} {Geometric Phases in Physics}}}\ (\bibinfo  {publisher} {World
  Scientific},\ \bibinfo {address} {Singapore})\ \Eprint
  {http://arxiv.org/abs/https://www.worldscientific.com/doi/pdf/10.1142/0613}
  {https://www.worldscientific.com/doi/pdf/10.1142/0613} \BibitemShut {NoStop}%
\bibitem [{\citenamefont {Wiringa}\ \emph {et~al.}(2013)\citenamefont
  {Wiringa}, \citenamefont {Pastore}, \citenamefont {Pieper},\ and\
  \citenamefont {Miller}}]{wiringa2013}%
  \BibitemOpen
  \bibfield  {author} {\bibinfo {author} {\bibnamefont {Wiringa}, \bibfnamefont
  {R~B}}, \bibinfo {author} {\bibnamefont {Pastore}, \bibfnamefont {S}},
  \bibinfo {author} {\bibnamefont {Pieper}, \bibfnamefont {S~C}}, \ and\
  \bibinfo {author} {\bibnamefont {Miller}, \bibfnamefont {G~A}}} (\bibinfo
  {year} {2013}),\ \bibfield  {title} {\enquote {\bibinfo {title}
  {Charge-symmetry breaking forces and isospin mixing in ${}^{8}$be},}\ }\href
  {\doibase 10.1103/PhysRevC.88.044333} {\bibfield  {journal} {\bibinfo
  {journal} {Phys. Rev. C}\ }\textbf {\bibinfo {volume} {88}},\ \bibinfo
  {pages} {044333}}\BibitemShut {NoStop}%
\bibitem [{\citenamefont {Wood}\ \emph {et~al.}(2004)\citenamefont {Wood},
  \citenamefont {Oros-Peusquens}, \citenamefont {Zaballa}, \citenamefont
  {Allmond},\ and\ \citenamefont {Kulp}}]{wood2004}%
  \BibitemOpen
  \bibfield  {author} {\bibinfo {author} {\bibnamefont {Wood}, \bibfnamefont
  {J~L}}, \bibinfo {author} {\bibnamefont {Oros-Peusquens}, \bibfnamefont
  {A~M}}, \bibinfo {author} {\bibnamefont {Zaballa}, \bibfnamefont {R}},
  \bibinfo {author} {\bibnamefont {Allmond}, \bibfnamefont {J~M}}, \ and\
  \bibinfo {author} {\bibnamefont {Kulp}, \bibfnamefont {W~D}}} (\bibinfo
  {year} {2004}),\ \bibfield  {title} {\enquote {\bibinfo {title} {Triaxial
  rotor model for nuclei with independent inertia and electric quadrupole
  tensors},}\ }\href {\doibase 10.1103/PhysRevC.70.024308} {\bibfield
  {journal} {\bibinfo  {journal} {Phys. Rev. C}\ }\textbf {\bibinfo {volume}
  {70}},\ \bibinfo {pages} {024308}}\BibitemShut {NoStop}%
\bibitem [{\citenamefont {Wu}\ and\ \citenamefont {Yang}(1976)}]{wu1976}%
  \BibitemOpen
  \bibfield  {author} {\bibinfo {author} {\bibnamefont {Wu}, \bibfnamefont
  {T~T}}, \ and\ \bibinfo {author} {\bibnamefont {Yang}, \bibfnamefont {C~N}}}
  (\bibinfo {year} {1976}),\ \bibfield  {title} {\enquote {\bibinfo {title}
  {Dirac monopole without strings: Monopole harmonics},}\ }\href {\doibase
  10.1016/0550-3213(76)90143-7} {\bibfield  {journal} {\bibinfo  {journal}
  {Nuclear Physics B}\ }\textbf {\bibinfo {volume} {107}}~(\bibinfo {number}
  {3}),\ \bibinfo {pages} {365 }}\BibitemShut {NoStop}%
\bibitem [{\citenamefont {Yannouleas}\ and\ \citenamefont
  {Landman}(2007)}]{yannouleas2007}%
  \BibitemOpen
  \bibfield  {author} {\bibinfo {author} {\bibnamefont {Yannouleas},
  \bibfnamefont {C}}, \ and\ \bibinfo {author} {\bibnamefont {Landman},
  \bibfnamefont {U}}} (\bibinfo {year} {2007}),\ \bibfield  {title} {\enquote
  {\bibinfo {title} {Symmetry breaking and quantum correlations in finite
  systems: studies of quantum dots and ultracold bose gases and related nuclear
  and chemical methods},}\ }\href {\doibase 10.1088/0034-4885/70/12/R02}
  {\bibfield  {journal} {\bibinfo  {journal} {Rep. Prog. Phys.}\ }\textbf
  {\bibinfo {volume} {70}}~(\bibinfo {number} {12}),\ \bibinfo {pages}
  {2067}}\BibitemShut {NoStop}%
\bibitem [{\citenamefont {Yao}\ \emph {et~al.}(2020)\citenamefont {Yao},
  \citenamefont {Bally}, \citenamefont {Engel}, \citenamefont {Wirth},
  \citenamefont {Rodr\'{\i}guez},\ and\ \citenamefont {Hergert}}]{yao2020}%
  \BibitemOpen
  \bibfield  {author} {\bibinfo {author} {\bibnamefont {Yao}, \bibfnamefont
  {J~M}}, \bibinfo {author} {\bibnamefont {Bally}, \bibfnamefont {B}}, \bibinfo
  {author} {\bibnamefont {Engel}, \bibfnamefont {J}}, \bibinfo {author}
  {\bibnamefont {Wirth}, \bibfnamefont {R}}, \bibinfo {author} {\bibnamefont
  {Rodr\'{\i}guez}, \bibfnamefont {T~R}}, \ and\ \bibinfo {author}
  {\bibnamefont {Hergert}, \bibfnamefont {H}}} (\bibinfo {year} {2020}),\
  \bibfield  {title} {\enquote {\bibinfo {title} {Ab initio treatment of
  collective correlations and the neutrinoless double beta decay of
  $^{48}\mathrm{Ca}$},}\ }\href {\doibase 10.1103/PhysRevLett.124.232501}
  {\bibfield  {journal} {\bibinfo  {journal} {Phys. Rev. Lett.}\ }\textbf
  {\bibinfo {volume} {124}},\ \bibinfo {pages} {232501}}\BibitemShut {NoStop}%
\bibitem [{\citenamefont {Zhang}\ and\ \citenamefont
  {Papenbrock}(2013)}]{zhang2013}%
  \BibitemOpen
  \bibfield  {author} {\bibinfo {author} {\bibnamefont {Zhang}, \bibfnamefont
  {J}}, \ and\ \bibinfo {author} {\bibnamefont {Papenbrock}, \bibfnamefont
  {T}}} (\bibinfo {year} {2013}),\ \bibfield  {title} {\enquote {\bibinfo
  {title} {Rotational constants of multi-phonon bands in an effective theory
  for deformed nuclei},}\ }\href {\doibase 10.1103/PhysRevC.87.034323}
  {\bibfield  {journal} {\bibinfo  {journal} {Phys. Rev. C}\ }\textbf {\bibinfo
  {volume} {87}},\ \bibinfo {pages} {034323}}\BibitemShut {NoStop}%
\bibitem [{\citenamefont {Zuker}\ \emph {et~al.}(2015)\citenamefont {Zuker},
  \citenamefont {Poves}, \citenamefont {Nowacki},\ and\ \citenamefont
  {Lenzi}}]{zuker2015}%
  \BibitemOpen
  \bibfield  {author} {\bibinfo {author} {\bibnamefont {Zuker}, \bibfnamefont
  {A~P}}, \bibinfo {author} {\bibnamefont {Poves}, \bibfnamefont {A}}, \bibinfo
  {author} {\bibnamefont {Nowacki}, \bibfnamefont {F}}, \ and\ \bibinfo
  {author} {\bibnamefont {Lenzi}, \bibfnamefont {S~M}}} (\bibinfo {year}
  {2015}),\ \bibfield  {title} {\enquote {\bibinfo {title} {Nilsson-su3
  self-consistency in heavy $n=z$ nuclei},}\ }\href {\doibase
  10.1103/PhysRevC.92.024320} {\bibfield  {journal} {\bibinfo  {journal} {Phys.
  Rev. C}\ }\textbf {\bibinfo {volume} {92}},\ \bibinfo {pages}
  {024320}}\BibitemShut {NoStop}%
\end{thebibliography}

%
\end{document}